%% file: thesis.tex
\documentclass[11pt,final]{thesis}

\usepackage{epsfig,graphicx}

\newcommand{\be}{\begin{equation}}
\newcommand{\ee}{\end{equation}}
\newcommand{\bea}{\begin{eqnarray}}
\newcommand{\eea}{\end{eqnarray}}
\newcommand{\bi}{\begin{itemize}}
\newcommand{\ei}{\end{itemize}}
\newcommand{\ben}{\begin{enumerate}}
\newcommand{\een}{\end{enumerate}}
\newcommand{\la}{\left\langle}
\newcommand{\ra}{\right\rangle}
\newcommand{\lc}{\left[}
\newcommand{\rc}{\right]}

\newcommand{\partt}{\mathrm{part}}
\newcommand{\tr}{\mathrm{tr}}
\newcommand{\val}{\mathrm{val}}
\newcommand{\lp}{\left(}
\newcommand{\rp}{\right)}
\newcommand{\aq}{\alpha_s\left( Q^2 \right)}

\newcommand{\aqz}{\alpha_s\left( M_Z^2 \right)}
\newcommand{\aqq}{\alpha_s \left( Q^2_0 \right)}
\newcommand{\qb}{\bar{q}}

\newcommand{\mo}{\mathcal{O}}
\newcommand{\fd}{F_2(x,Q^2)}
\newcommand{\mrexp}{\mathrm{exp}}
\newcommand{\dat}{\mathrm{dat}}
\newcommand{\art}{\mathrm{art}} 
\newcommand{\con}{\mathrm{con}} 
\newcommand{\rep}{\mathrm{rep}}
\newcommand{\net}{\mathrm{net}}
\newcommand{\sys}{\mathrm{sys}}
\newcommand{\stat}{\mathrm{stat}}
\newcommand{\reff}{\mathrm{ref}}
\newcommand{\fit}{\mathrm{fit}}
\newcommand{\sets}{\mathrm{sets}}
\newcommand{\theo}{\mathrm{theo}}

\newcommand{\stopp}{\mathrm{stop}}
\newcommand{\extra}{\mathrm{extra}}
\newcommand{\QCD}{\mathrm{QCD}}
\newcommand{\LO}{\mathrm{LO}}
\newcommand{\NLO}{\mathrm{NLO}}
\newcommand{\NNLO}{\mathrm{NNLO}}

\newcommand{\parr}{\mathrm{par}}
\newcommand{\tth}{\mathrm{th}}
\newcommand{\tot}{\mathrm{tot}}

\def\gsim{\mathrel{\rlap{\lower4pt\hbox{\hskip1pt$\sim$}}
    \raise1pt\hbox{$>$}}}         
\def\lsim{\mathrel{\rlap{\lower4pt\hbox{\hskip1pt$\sim$}}
    \raise1pt\hbox{$<$}}}         
\def\epm#1#2{\hbox{${\lower1pt\hbox{$\scriptstyle +~#1$}}
\atop {\raise1pt\hbox{$\scriptstyle -~#2$}}$}}

\title{\bf \large THE NEURAL NETWORK APPROACH TO PARTON
DISTRIBUTIONS}
 \shorttitle{The neural network approach to parton distributions}
 \name{Juan Rojo Chac\'on}
\degrees{~}
 \home{~} 
 \monthyear{Julio 2006} 
 \numberofmembers{5} 
\advisor{Jos\'e Ignacio Latorre Sent\'is - Universitat
de Barcelona \\ Dr. Stefano Forte - Universit\`a di Milano}
\department{Estructura i Constituents de la Mat\`eria}
\bienni{Programa de Doctorado: F\'isica Avanzada \\
Bienio: 2002-2004}

\begin{document}

\maketitle

~

\newpage

\include{agradecimientos}

~

\newpage

\begin{publications}

\begin{center}
{\large \bf Research articles - published}
\end{center}

\begin{itemize}

\item
J.~Rojo and J.~I.~Latorre \cite{condensates}, 
``Neural network parametrization of spectral functions from hadronic tau decays
and determination of QCD vacuum condensates,''
JHEP {\bf 0401}, 055 (2004)
[arXiv:hep-ph/0401047].

\item
J.~Brugues, J.~Rojo and J.~G.~Russo \cite{strings},
 ``Non-perturbative states in type II superstring theory from classical
 spinning membranes,''
 Nucl.\ Phys.\ B {\bf 710}, 117 (2005),
 [arXiv:hep-th/0408174].

\item
L.~Del~Debbio, S.~Forte, J.~I.~Latorre, A.~Piccione and J.~Rojo
\cite{f2nnp},
``Unbiased determination of the proton structure 
function $F_2^p$ with faithful uncertainty estimation'', 
JHEP {\bf 0503} (2005) 080,
  [arXiv:hep-ph/0501067].

\item
  S.~Forte, G.~Ridolfi, J.~Rojo and M.~Ubiali \cite{landau},
  ``Borel resummation of soft gluon radiation and higher twists,''
  arXiv:hep-ph/0601048,  Phys. Lett. B {\bf 635} (2006) 313.

\item J.~Rojo \cite{bmeson},
``Neural network parametrization
of the lepton energy spectrum in B meson decays'',
 JHEP {\bf 0605} (2006) 040
  [arXiv:hep-ph/0601229]. 

\item
 J.~Mondejar, A.~Pineda and J.~Rojo \cite{rojoscet},
   ``Heavy meson semileptonic differential decay rate in two dimensions in the
   large $N_c$,''
  arXiv:hep-ph/0605248.

\end{itemize}

\begin{center}
{\large \bf Research articles - in preparation}
\end{center}

\begin{itemize}

\item
   L.~Del Debbio, S.~Forte, J.~I.~Latorre, A.~Piccione
and J.~Rojo  [NNPDF
                  Collaboration],
  ``The neural network approach to parton distributions: The
nonsinglet case'', UB-ECM-PF 06/17.

\item
  C.~Garc\'ia-Gonz\'alez, M.~Maltoni and J.~Rojo,
  ``Neural network parametrization of the
atmospheric neutrino flux'',
in preparation.

\item
 L.~Del Debbio, S.~Forte, J.~I.~Latorre, A.~Piccione
and   J.~Rojo  [NNPDF
                  Collaboration],
  ``The neural network approach to parton distributions: The
singlet case'',
in preparation.

\end{itemize}

\begin{center}
{\large \bf Conference proceedings}
\end{center}

\begin{itemize}

\item
J.~Rojo,
``A probability measure in the space of spectral functions and structure
functions'' \cite{proc1}, proceedings of the QCD International Conference,
Montpellier 2004, Nucl. Phys. B (Proc. Suppl.) {\bf 152} (2006) 57,
arXiv:hep-ph/0407147.

\item
  J.~Rojo, L.~Del Debbio, S.~Forte, J.~I.~Latorre and A.~Piccione  [NNPDF
                  Collaboration] \cite{proc2},
  ``The neural network approach to parton fitting,'' 
proceedings of DIS05 workshop,
  arXiv:hep-ph/0505044.

\item
  J.~Rojo, L.~Del Debbio, S.~Forte, J.~I.~Latorre and A.~Piccione  [NNPDF
                  Collaboration] \cite{proc3},
  ``The neural network approach to parton distributions,'' 
contribution to the proceedings of the Hera-LHC workshop, 
hep-ph/0509059.

\item
  J.~Rojo, L.~Del Debbio, S.~Forte, J.~I.~Latorre and A.~Piccione  [NNPDF
                  Collaboration] \cite{procacat},
  ``The neural network approach to parton fitting,'' 
proceedings of the ACAT05 workshop, 
hep-ph/0509067.

\end{itemize}

\end{publications}

\newpage

~

\newpage

\tableofcontents

\newpage

\include{introduccion}

\newpage

\include{resumen}

\newpage

\chapter{Introduction and motivation}
\begin{figure}[ht]
\begin{center}
\epsfig{figure=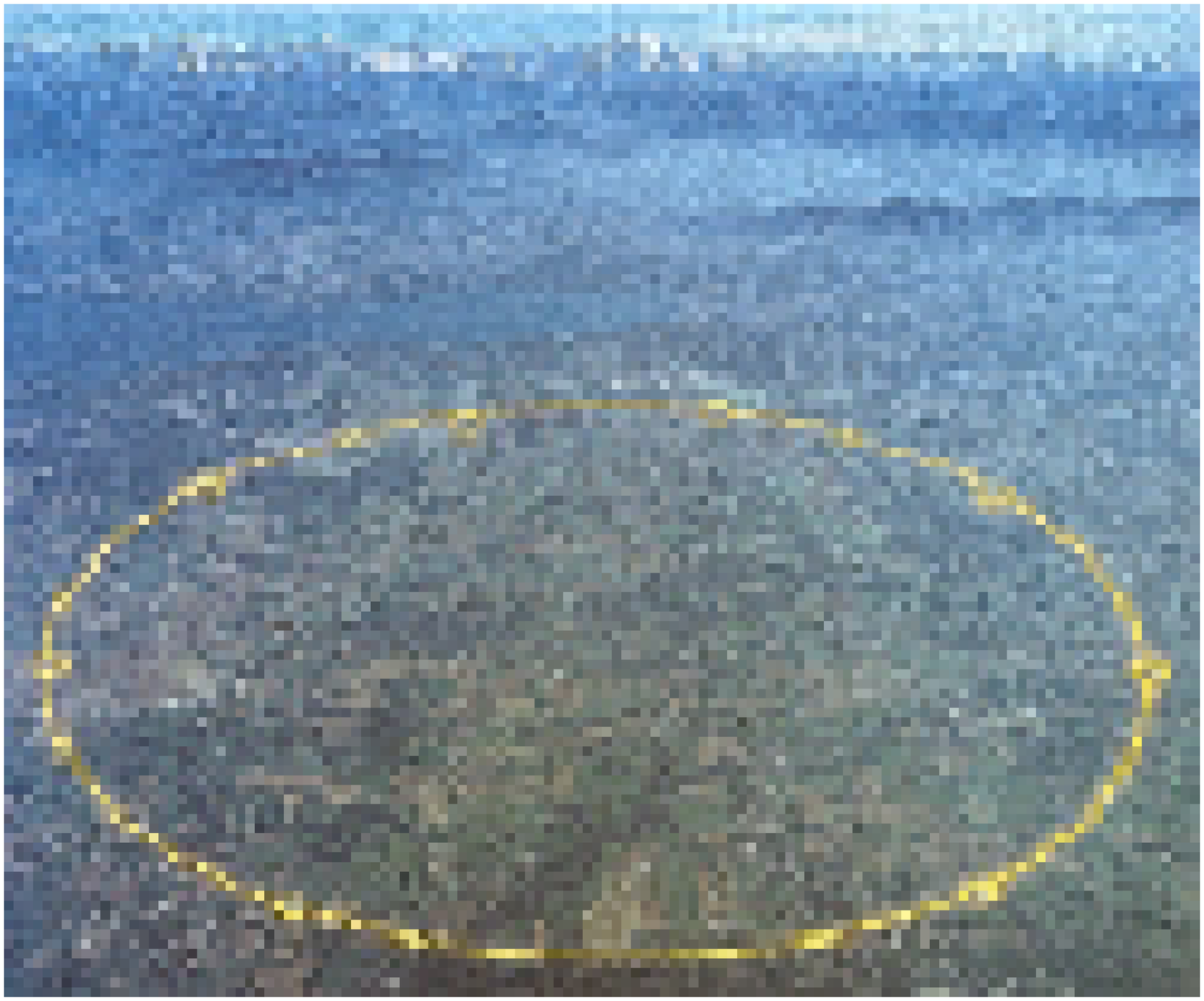,width=0.45\textwidth}
\epsfig{figure=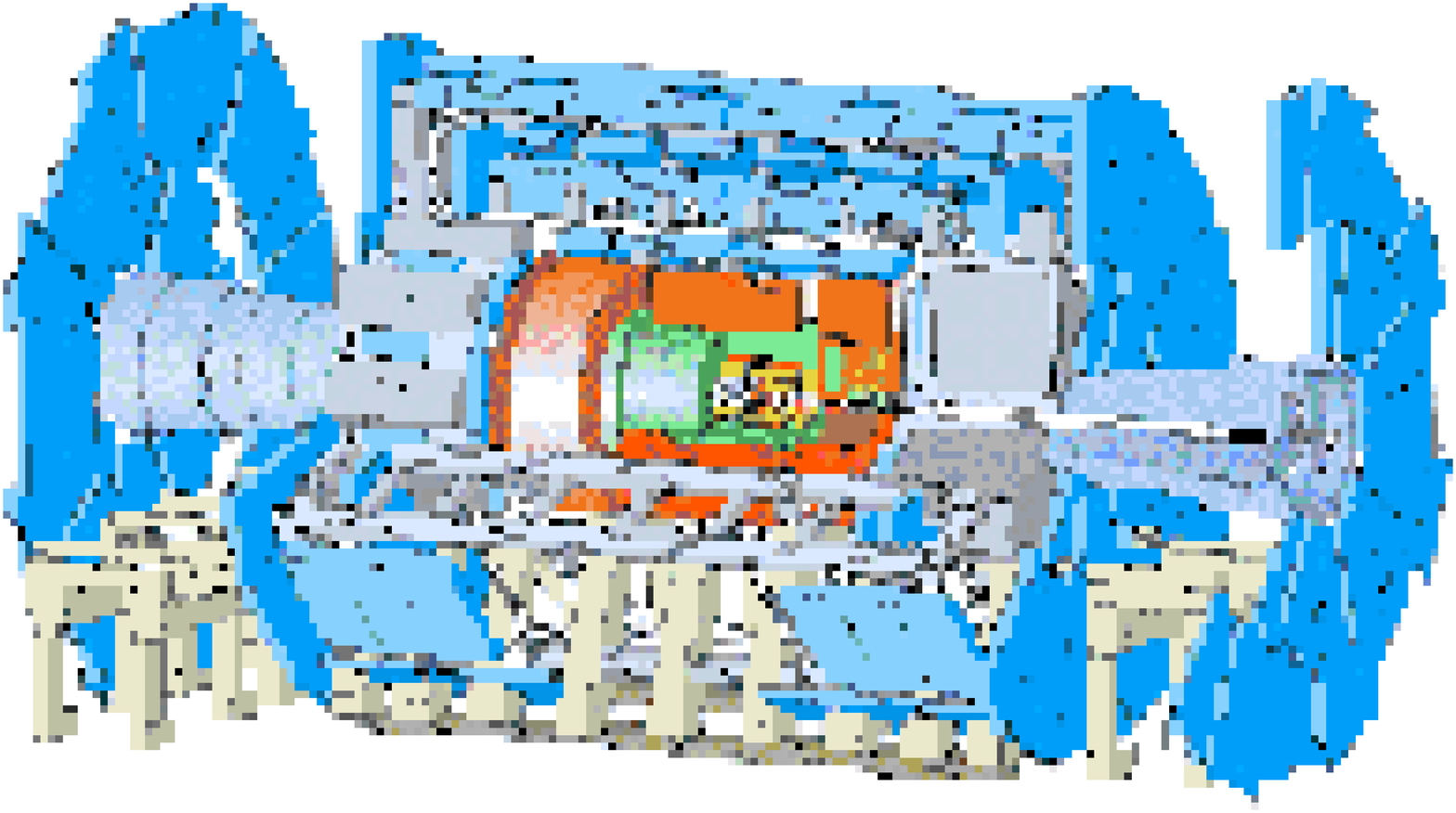,width=0.45\textwidth}
\caption{}{\small The location of the LHC
experiment near G\`eneve (left)
and the one of its detectors, ATLAS (right).}
\label{lhc}
\end{center}
\end{figure}

The Large Hadron Collider (LHC) is
a proton-proton collider at center of mass energy
$\sqrt{s}=14~$TeV which is located at the
Organisation Europ\'eenne pour la Recherche Nucl\'eaire
(CERN), near Gen\`eve in Switzerland (see Fig. \ref{lhc}).
Its construction is almost finished, and it
is scheduled to deliver the first collisions
by the end of 2007.
The energies that will be achieved at the LHC will allow
us to probe the smallest distances ever  and
 will open a new era for particle physics, since it will investigate the true
nature of electroweak symmetry breaking, and hopefully
will deliver  invaluable information of new physics
beyond the Standard Model (SM) of particle
interactions. However, possible new physics signals will
appear together with a far more copious background from
Standard Model  processes, essentially from the
interactions of quarks and gluons within the proton as determined
by Quantum Chromodynamics (QCD), the sector of the SM
that describes the strong interaction.

Therefore, the discovery potential of LHC as well as its
ability to also
perform precision measurements of the new
physics properties
depends crucially on the understanding of the
huge Standard Model background.
Since the LHC is an hadron collider, to be able to perform
precision predictions for different observables one
needs first to understand quantitatively the underlying strong interaction
processes and the associated
uncertainties. 
Among these uncertainties, one of the most important ones
comes from the parton distribution functions, which
measure the momentum distribution of quarks and gluons inside the
protons. Parton distributions cannot be computed 
from first principles, but rather
 they need to be extracted from experimental data from
other hard scattering processes, like for example
deep inelastic scattering.

\begin{figure}[t]
\begin{center}
\epsfig{figure=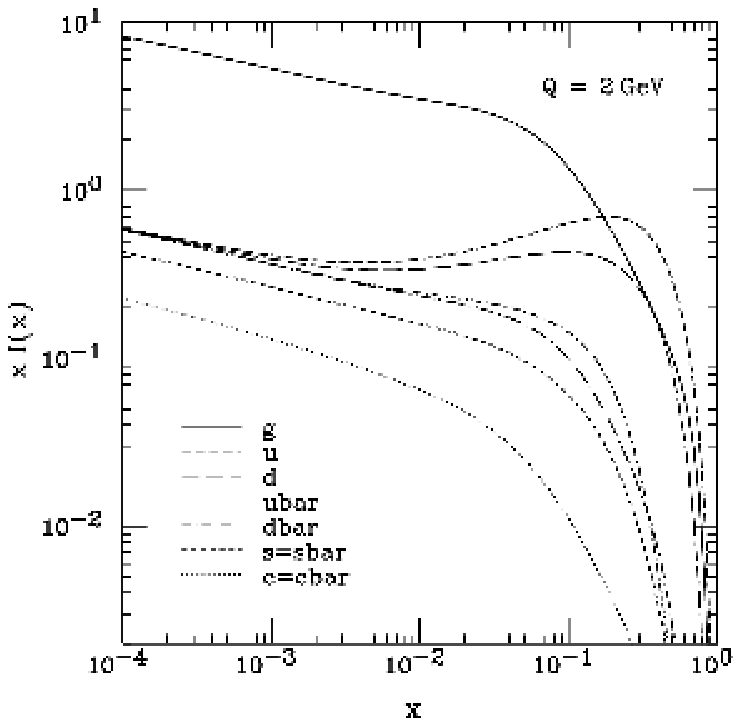,width=0.60\textwidth}
\caption{}{\small Parton distribution functions
$q_i(x,Q_0^2)$ as determined from experimental data
in a recent QCD global analysis \cite{cteq61}.}
\label{pdfsets}
\end{center}
\end{figure}

Parton distribution functions $q_i(x,Q_0^2)$ can
be determined from experimental data
by means of the QCD factorization theorem, which
states that any hard-scattering cross section
can be separated into a process-dependent
coefficient
 and a set of universal, process-independent
parton distribution functions. That is, the factorization theorem relates
observables like deep-inelastic scattering 
structure functions $F(x,Q^2)$
to a convolution of short-distance coefficient
functions, $C_{i}\lp x,\aq\rp$, which can be
computed in perturbation theory, and 
parton distribution functions,
\be
F(x,Q^2)=C_{i}\lp x,\aq\rp \otimes q_i(x,Q^2) \ .
\ee
The dependence with the scale $Q^2$ of the
parton distributions is determined in
QCD perturbation theory from the so-called  parton
evolution equations. Note that in general different
combinations of parton distributions
contribute to different observables. The inclusion of
a wide variety of hard-scattering data is thus crucial to
disentangle the various parton distributions.
Even if the backbone of the determinations of
parton distributions is the high precision deep-inelastic
scattering data, essential experimental input comes
from other measurements like jet production,
heavy boson production or the Drell-Yan process.
In Fig. \ref{pdfsets} we show the set of
parton distribution functions from a recent global
QCD analysis.

\begin{figure}[ht]
\begin{center}
\epsfig{width=0.5\textwidth,figure=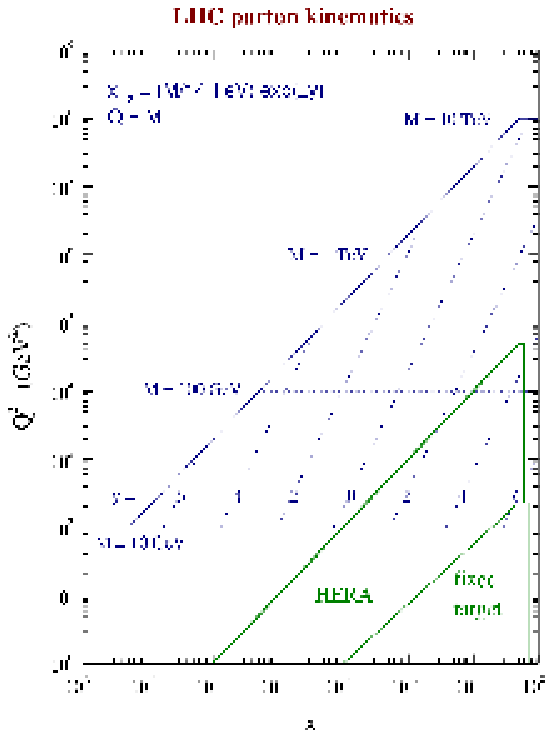}
\caption{}{
\small The kinematical coverage of the LHC compared
to that of the HERA collider and fixed target deep-inelastic
scattering experiments.}
\label{kinheralhc}
\end{center} 
\end{figure}

The requirements of precision physics at hadron colliders,
specially the Large Hadron Collider, determine that
it is now mandatory to
determine accurately not only the 
parton distributions themselves but 
also the uncertainties associated to them. 
Note that a detailed knowledge of parton distributions
and their associated uncertainties is essential
if one wants to perform accurate measurements,
since a typical LHC cross-section $\sigma(x,Q^2)$ reads
\be
\sigma(x,Q^2)= C_{ij}\lp x,\aq\rp \otimes q_i(x,Q^2) \otimes
q_j(x,Q^2) \ ,
\ee
which involves the product of two parton distributions from each of 
 the two partons in the initial state of the
proton-proton collision.
The problem is specially acute since the kinematic
region covered by the LHC overlaps only partially with the
kinematical range of those experiments used to determine the
parton distributions, like HERA, as can be seen in
Fig. \ref{kinheralhc}, and therefore one has
to extrapolate parton distribution into an unknown kinematical
region. Also for this reason it is essential to determine
the uncertainties in parton distributions and propagate them
into the extrapolation region probed by LHC.

The main
problem to be faced here is that one is trying to determine an
uncertainty on a function, i.e., a probability measure on a space of
functions, and to extract it from a finite set of experimental data.
Within the framework of the standard approach to
determine parton distributions from experimental data,
several techniques have been constructed to
assess these  uncertainties.
However, even if all these techniques
have been  useful to estimate the size
of the uncertainties, they suffer from
several drawbacks.
First of all, since in the standard approach parton
distributions are parametrized with relatively simple
functional forms like
\be
\label{functional}
q_i(x,Q_0^2)=A_ix^{b_i}(1-x)^{c_i}\lp 1+d_ix+e_ix^2+\ldots\rp \ ,
\ee
where the parameters $A_i,b_i,\ldots$ are fitted from
experimental data,
the estimation of the uncertainties is restricted
to the space parametrized by Eq.  \ref{functional} and therefore 
depends heavily on the assumptions done for the
non-perturbative
shapes of the parton distributions. Second,
in order to propagate the uncertainties in the
parton distributions to an arbitrary observable,
linearized approximations in the error propagation have
to be used, whose validity is at best doubtful.
Finally, in the presence of experimental data from
different experiments,
it has been argued that incompatibility between
different experiments  force that the
tolerance condition in the $\chi^2$ used to
estimate the errors is not the textbook value
$\Delta\chi^2=1$ but a much larger value
$\Delta \chi^2 =50$ or 100. Even if this choice
can be justified to some extent, in practice 
parton uncertainties
determined with this condition lose  their statistical
meaning, since they depend on the choice of the
free parameter $\Delta\chi^2$.

Therefore, in view of the crucial importance
of parton distribution functions for LHC
precision physics, it is worth investigating
alternative approaches for the determination
of parton distributions and the associated uncertainties
that bypass the problems of the standard approach
discussed above.
In Ref.~\cite{f2ns}  a novel technique was presented to
determine the associated probability measure in the
space of a function, which was applied
to the parametrization of deep
inelastic structure functions.
This technique uses a combination of Monte Carlo samplings
of the experimental data together with artificial neural
networks as universal unbiased interpolants.
The use of neural networks instead of
fixed functional forms as in Eq. \ref{functional}
avoids the bias introduced by the assumption of
 a given functional form,
and the Monte Carlo sampling provides a statistically rigorous
estimation of the function uncertainties, which allows
for a general error propagation to arbitrary observables
without the need of linearized approximations.

In this thesis we extend the
work of Ref. \cite{f2ns}\footnote{See also Ref. \cite{andreathesis}} 
in different ways: first 
we have applied the general strategy to parametrize
experimental data to other processes of interest, in
particular to the case of parton distribution functions.
In all these cases the relation between the parametrized
quantity and the experimental data was completely
different, so it has been shown that the approach
of  Ref. \cite{f2ns} is suited for a variety
of situations.
Second, we have
extended the basic technique with the introduction
of new minimization algorithms, specially genetic
algorithms, as well as new statistical estimators
to assess the stability of the obtained probability
measure in different aspects. Finally, several
improvements of the neural network training
have been introduced to optimize its efficiency.

The outline of the present thesis is as follows.
In Chapter 4 we review the basic elements
of Quantum Chromodynamics, as well as those
high energy
processes that will be used later in the thesis.
We present also a review of the standard approach
to parametrize parton distribution functions with their
associated uncertainties. In Chapter \ref{general} we introduce the
general technique to parametrize experimental data in an
unbiased way with faithful estimation of the
uncertainties, which is the main subject of this thesis.
Then in Chapter \ref{appl} we present four
applications of the general strategy, and special
attention is paid to 
the most important one, the 
parametrization of the nonsinglet parton
distribution. Two appendices summarize 
background material on statistical analysis of experimental data
and on the current status of the standard approach
to global fits of parton distributions.

\chapter{Elements of perturbative QCD and global 
parton distributions analysis }

In this first Chapter we review the basic elements of Quantum
Chromodynamics (QCD), the gauge theory of the strong
interactions.
After a brief description of the
theoretical foundations of QCD, we  describe in some detail
the deep-inelastic scattering process. This
process is the
most important source of information on parton
distribution functions, whose parametrization, as we have
discussed in the Introduction, 
is the main motivation for the set of techniques
developed in the present thesis.
We will consider also two other
high energy processes, since they have
provided testing grounds for our strategy
to parametrize experimental data: 
the semileptonic decays of the B meson and
the hadronic tau decays. 

Parton distribution functions, as has been discussed in the
Introduction, have to extracted from experimental
data, by means of a QCD
analysis of a variety of hard scattering measurements.
In the second
part of this Chapter we present a
summary of the standard approach to global fits
of parton distributions, and we discuss in some detail the
different methods, with their advantages and drawbacks, which
are commonly used to estimate the associated uncertainties of
 parton distribution functions.

\section{Overview  of perturbative QCD}

\subsection{Basics of Quantum Chromodynamics}
Quantum Chromodynamics (QCD) is the gauge theory that describes the
strong interaction (see for example Refs. \cite{collider,
handbook} and references therein). Gauge invariance under the group SU(N)
and renormalizability determine completely the form of the
Lagrangian.
The QCD Lagrangian describes the strong
interaction between quarks and gluons, and it
 reads
\be
\label{lagrangian}
\mathcal{L}_{\QCD}=\sum_{i=1}^{N_f}\bar{q}_i\lp  i\gamma^{\mu}
D_{\mu}-m_i\rp q_i-\frac{1}{4} F_{\mu\nu}^AF_{\mu\nu}^A \ .
\ee
Let us describe the elements of the
above equation. The $N_f$ spinor quark fields 
of different flavor are labeled by $q_i$, each with mass $m_i$. 
The Dirac matrices $\gamma^{\mu}$ appear due to the
fermionic nature of quarks, and are defined by the
anticommutation relation $\{ \gamma^{\mu},\gamma^{\nu}\}=
2g^{\mu\nu}$.
The covariant derivative reads in terms of the
gluon field $A_{\mu}^A$
\be
\lp D_{\mu}\rp_{ab}=\partial_{\mu}\delta_{ab}+ig\lp t^AA^A_{\mu}\rp_{ab} \ ,
\ee
where $\mu$ is a Lorentz index, $\mu=0,1,2,3$,
and $a,b$ are color indices for the
fundamental representation, $a,b=1,\ldots,N$.
The matrices $t^A$, where $A$ is a color
index in the adjoint representation,
 $A=1,\ldots,N^2+1$,  are the SU(N)
generating matrices in the fundamental representation.
The last term in Eq. \ref{lagrangian}
is the field-strength tensor for the gluon field,
\be
\label{fs}
F_{\mu\nu}^A=\partial_{\mu} A_{\nu}^A-\partial_{\nu}A^A_{\mu}
-gf^{ABC}A^B_{\mu}A^C_{\nu} \ ,
\ee
where $f^{ABC}$ are the structure constants of SU(3),
defined by the commutation relation
\be
\lc t^A,t^B\rc=if^{ABC}t^C \ .
\ee
The last term in Eq. \ref{fs} describes the self-interaction
of the gluons, a typical feature of non-abelian gauge
theories like QCD which renders the theory asymptotically free,
as discussed below. Finally, $g$ is the QCD coupling constant, which is,
together with the quark masses, the only free parameter of the
theory.

\begin{figure}[ht]
\begin{center}
\epsfig{width=0.5\textwidth,figure=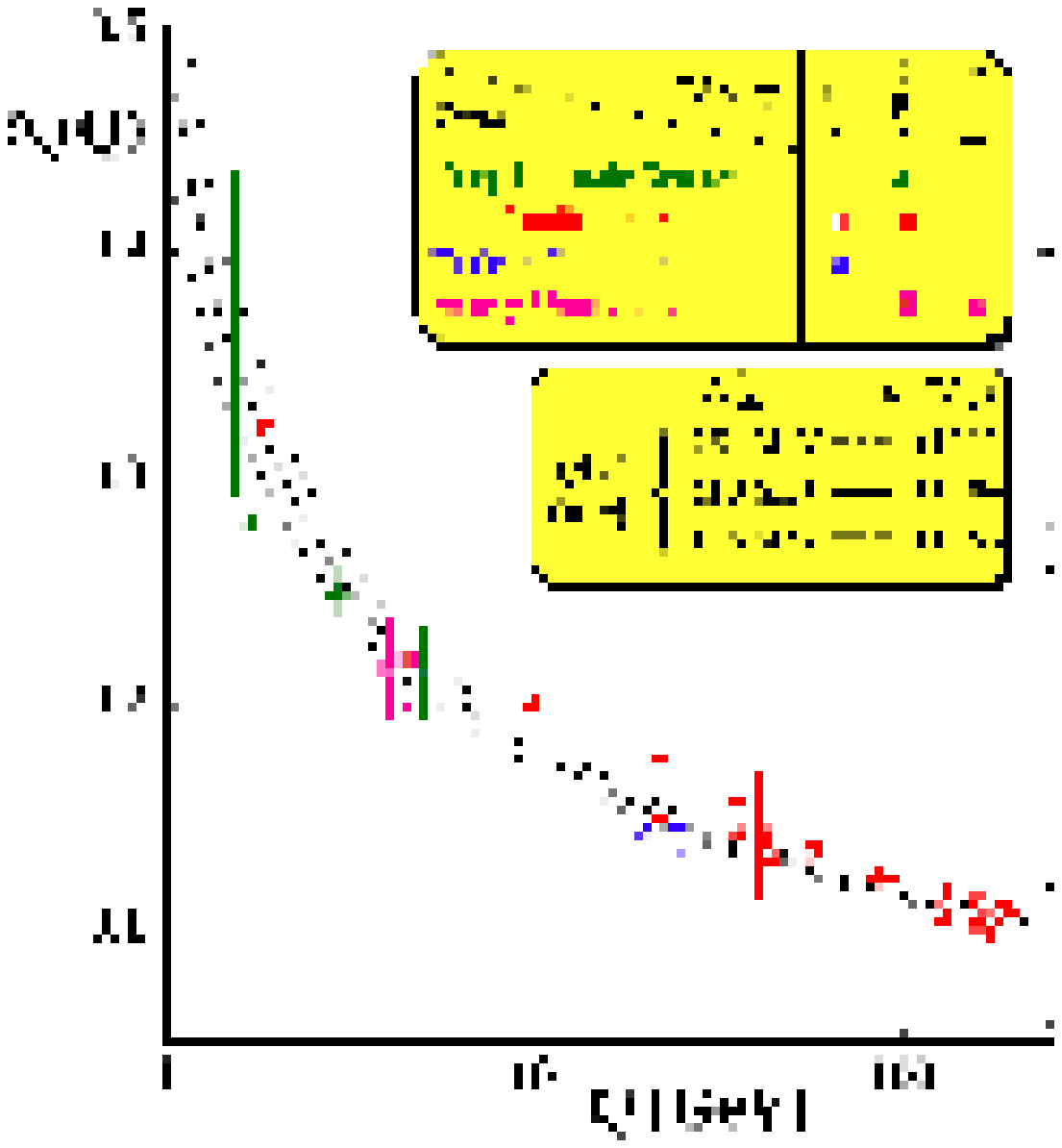}
\caption{}{\small Comparison of different determinations
of the strong coupling $\alpha_s(Q)$ at several energies $Q$
from a variety of processes, as summarized in
Ref. \cite{bethke}.}
\label{alphas}
\end{center} 
\end{figure}

From this Lagrangian, using standard rules of Quantum 
Field Theory \cite{peskin} one can compute several observables, like
cross sections or decay rates, in a perturbative series expansion in powers
of $\alpha_s=g^2/4\pi$.
Radiative quantum corrections induce a dependence of the strong
coupling with respect to the typical energy of the
process $E$, which at lowest order reads
 \be 
\label{alpha}
\alpha_s=\alpha_s(E)=\frac{1}{\beta_0\ln \frac{E^2}{\Lambda^2_{\QCD}}} \ ,
\ee
where $\beta_0$ is the first coefficient of the QCD $\beta$ function that
determines the running of $\alpha_s$ with the energy.
The main feature of QCD can be seen from Eq. \ref{alpha}: the
theory is asymptotically free \cite{gross,politzer}, 
which means that the coupling
constant vanishes when the typical energies of the process become
very large with respect to the typical scale of the theory,
$\Lambda_{\mathrm{QCD}}$. Asymptotic freedom
allows us to apply QCD to many high energy  processes
for which perturbation theory is meaningful, since in this
case $E\gg \Lambda_{\QCD}$ and therefore 
$ \alpha_s(E)\ll 1$.
On the other hand, the same asymptotic freedom 
property implies that the theory
becomes strongly coupled at low
energies, $E\le\Lambda_{\mathrm{QCD}} $, and in this
non-perturbative regime the standard tools
of perturbation theory are useless, and one has to resort
to other methods, like lattice computations \cite{lattice}.
In Fig. \ref{alphas} we show a comparison of different
extractions of the strong coupling $\alpha_s(E)$ with the
theoretical QCD predictions \cite{bethke}.

For single scale observables, that is, for
processes which depend only on a single
hard scale $Q^2$ (where again a hard scale 
is a energy scale which satisfies the condition
$Q^2\gg \Lambda_{\QCD}$), the perturbative expansion
in powers of the strong coupling reads,
for those processes in which we are interested in,
\be
\label{r1}
R(Q^2)=\sum_{n=1}^{\infty}a_n\aq^n \ ,
\ee
where without loss of generality we have assumed that the
perturbative expansion of the
observable $R(Q^2)$ starts at order $\aq$.
The best example of this first case is the
total cross section of $e^+e^-$ going to hadrons,
where in this case the hard scale is identified with the
center of mass energy of the collision, $Q^2=s$.

Perturbative expansions become more 
complicated as the number of hard scales present
in the process begins to increase.
In this case it is not always true that observables can
be expanded in a simple power series in $\aq$. Let us
consider to be definite the deep-inelastic scattering processes,
which is the best known example of a
two-scale process,
and which will be discussed in detail in the next Section.
In this case the two hard scales  will be denoted by
$Q^2$ and $W^2$. 
In deep-inelastic scattering, as we will see in brief,
 the nonsinglet structure function
at leading order is given by
\be
F_2^{NS}(N,Q^2)=\lc \frac{\aq}{\aqq}\rc^{\gamma(N)/\beta_0} \ ,
\ee
where $N$ is the conjugate variable of the ratio $x=Q^2/(Q^2+W^2)$,
and therefore $F_2^{NS}$ 
cannot be expanded in integer powers of $\aq$.
Another example is the singlet  structure function at
small $x$ \cite{das}, which reads
\be
F_2(x,Q^2)\Bigg|_{x\to 0}
\sim \exp\lp\sqrt{\ln \lp 1/x \rp\ln\ln \lp \aq\rp}\rp \ ,
\ee
which again is not expandable in a simple  series in powers
of $\aq$.

In some cases, however,
perturbative computations which depends on two
hard scales have a
perturbative expansion of a similar form of Eq. \ref{r1},
that is, a expansion in integer powers of the strong
coupling, 
like the deep-inelastic scattering
 coefficient functions, which are of the form 
\be
\label{r2}
C(Q^2,W^2)=\sum_{k=1}^{\infty}
\aq^k b_k\lp \frac{Q^2}{W^2}\rp \ .
\ee
Note that unlike Eq. \ref{r1} the coefficients of the expansion
$b_k$ are not pure numbers but functions of the ratio of scales
$Q^2/W^2$.
In these coefficients $b_k$ one can encounter large
logarithmic terms of the form $\lp \ln Q^2/W^2\rp^p$.
These logarithmic terms can become large in some kinematical
regions and need to be resummed to all orders \cite{sterman}
in order
to trust perturbation theory. 
Note that typical perturbative expansions like
Eq. \ref{r1} and \ref{r2} are at best asymptotic
expansions, and often their large order behavior
introduces divergences, leading to
ambiguities in the value of the summed perturbative
expansion related to nonperturbative corrections \cite{renormalons}. 
Note also that since the coupling $\aq$ increases as the
value $Q^2$ decreases, the 
perturbative expansions
Eqs. \ref{r1} and \ref{r2} are
 meaningful only for large enough values of $Q^2\gg \Lambda_{\QCD}$, that
is, for the so-called {\it hard} processes \cite{harddok}.

The foundation for results like Eq. \ref{r1} and \ref{r2}
is the Operator Product Expansion. This expansion
allows to organize any 
quantity in a series in inverse powers of $Q^2$, the
typical energy scale of the process, by means of an expansion
of composite operators in terms of simpler operators
of the appropiate dimension.
The Operator Product
Expansion \cite{svz} can be used to
 parametrize higher-order 
nonperturbative effects that are
mostly relevant at low $Q^2$ in terms of  matrix elements of
local operators, and in  this way to extend the validity range
of theoretical predictions. 
For example, for a single scale observable, the
operator product expansion reads
schematically
\be
\label{ope}
R\lp Q^2\rp=
\sum_{k=0}^{\infty} c_k\lp \aq\rp \frac{\la \mathcal{O}_k\ra}{Q^k} \ ,
\ee
where $\la \mathcal{O}_k\ra$ are nonperturbative expectation
values of operators with the appropriate dimensions, and
where the leading order result corresponds to the unit operator,
$ \mathcal{O}_0= 1$,
\be
 c_0\lp \aq\rp=\sum_{n=1}^{\infty}a_n\aq^n \ .
\ee
It is clear from Eq. \ref{ope} that at large enough values
of $Q^2$ only the leading term in the OPE
is relevant for phenomenological purposes.
The nonperturbative matrix elements $\la \mathcal{O}_k\ra$ 
in   Eq. \ref{ope} can be extracted from experimental data in some
processes and then be used in other processes to increase the
accuracy of the theoretical prediction,
 as in the case of parton distribution functions, which will
be analyzed in detail in Section \ref{dis_theo}.

Now we review  the high energy
processes that will be used in applications
of the general technique to parametrize experimental data which is
the main subject of this thesis: deep-inelastic scattering, semileptonic B
meson decays and hadronic tau decays.

\subsection{Deep-inelastic scattering}
\label{dis_theo}
For a variety of reasons deep-inelastic scattering 
(DIS), the high energy scattering of leptons
and hadrons, is one of the most
important processes in Quantum Chromodynamics. 
The original, and still the most powerful, test of
perturbative QCD is the breaking of Bjorken scaling \cite{bjorken}
in DIS, that is, the logarithmic dependence
of deep-inelastic structure functions
with the momentum transfer $Q^2$.
 Nowadays, deep inelastic structure functions
analyses not only provide some of the most precise tests
of the theory but also determine the momentum
distributions of partons in hadrons, which are an
essential input in predicting cross section in high energy
hadron collisions.

Deep-inelastic scattering is probably the
 best theoretically understood
  process in perturbative QCD. The full 
next-to-next-to-leading order computation
was recently finished \cite{mvvs,mvvns}, and also threshold
resummation at the next-to-next-to-leading logarithmic
accuracy can be implemented \cite{sudakovho}.
On the experimental side, deep inelastic scattering
is the high energy process involving 
strongly interacting particles which has been
  measured experimentally with the 
highest accuracy. For example,
the lepton-proton collider HERA \cite{herarev} 
has measured deep-inelastic scattering
 cross-sections  with 1\% accuracy in a wide
kinematical range.

Moreover, as we have mentioned before, 
deep-inelastic scattering is essential to be able
to use perturbative QCD in other processes involving
hadrons in  the initial state, like
proton-proton collisions at the LHC.
This is so because deep-inelastic scattering
provides the backbone information on the parton distribution functions 
(PDFs) of the nucleon \cite{heralhc}. As will be
discussed in more detail, a detailed
knowledge of parton distribution functions
and its associated uncertainties is  an essential input
 for  precision LHC phenomenology.

\begin{figure}[ht]
\begin{center}
\epsfig{width=0.535\textwidth,figure=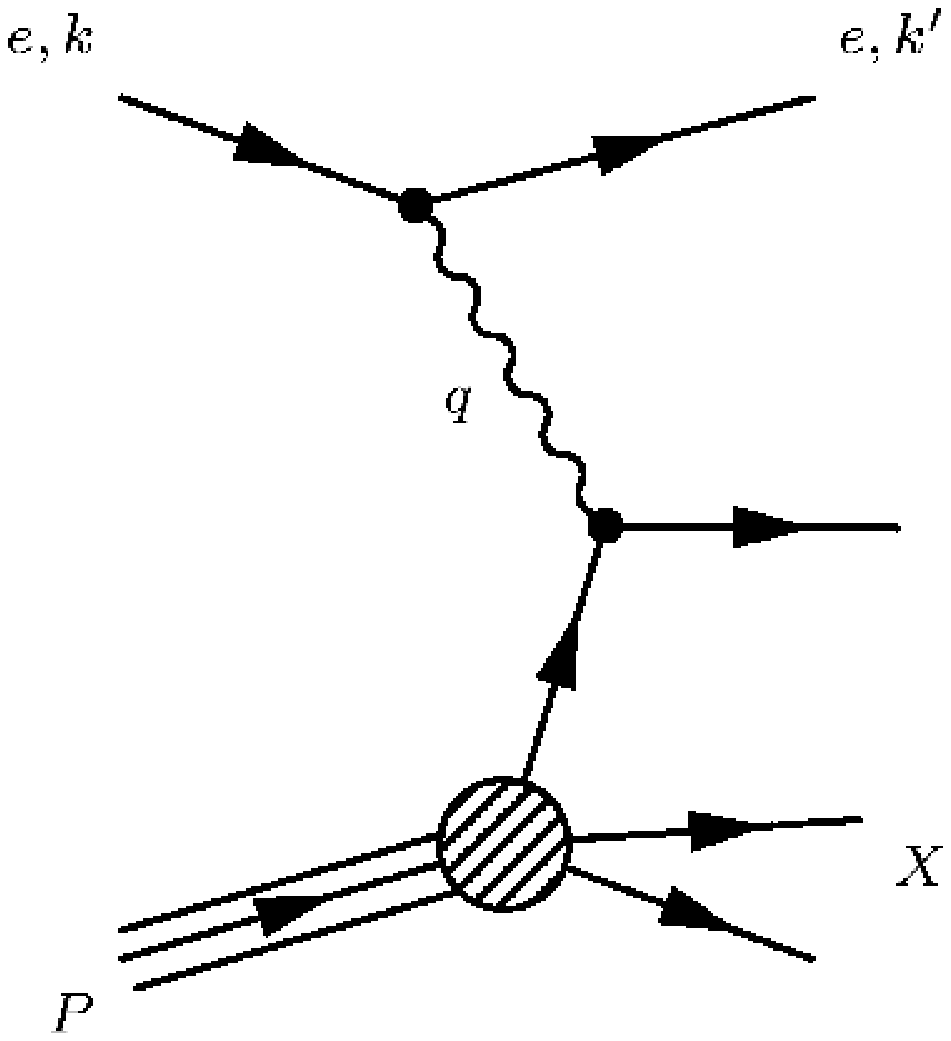}
\caption{}{The deep-inelastic scattering process: the
hard scattering of a lepton off a hadron, typically a proton.
\label{displot}}
\end{center} 
\end{figure}

Now we present the basic formulae that describe deep-inelastic
scattering, and that will be useful in
Chapter \ref{appl}.
Deep-inelastic scattering
is the high-energy collision of a lepton (an electron, a muon or
a neutrino)
against a hadronic target (a proton or a nucleus).
The kinematics of this process (see Fig.  \ref{displot})
 are determined in terms of 
two variables $x$ and $Q^2$, defined as
\be
x=\frac{Q^2}{2p\cdot q} , \qquad Q^2=-q^2 \ , 
\ee
where $q$ is the momentum carried by the virtual 
gauge boson and $p$
is the momentum carried by the incoming proton.
Other important variables are the invariant mass of the
final hadronic state $W^2$ and the
inelasticity $y$, given by
\be
W^2=Q^2 \frac{1-x}{x} \ ,\qquad
 y=\frac{q\cdot p}{
k\cdot p} \ ,
\ee
where $k$
is the momentum carried by the initial state lepton.
The applicability of perturbation theory requires that
both scales $Q^2$ and $W^2$ are large, because if not
either higher twist corrections\footnote{
The twist expansion is the operator
product expansion applied to the deep-inelastic
scattering process.}  are relevant,
or the perturbative expansion breaks down due
to the presence of large logarithms 
of the form $\aq^p \ln^k(1-x)$.
The kinematical cut in $W^2$ can be lowered by including
the effects of threshold resummation \cite{corcellaresum}. Note that
even if $Q^2$ is large, $W^2$ can be small
provided $x$ is large enough. The relation
between perturbative threshold resummation and higher twist 
nonperturbative
corrections is
still an open issue \cite{landau}.

\begin{figure}[ht]
\begin{center}
\epsfig{width=0.535\textwidth,figure=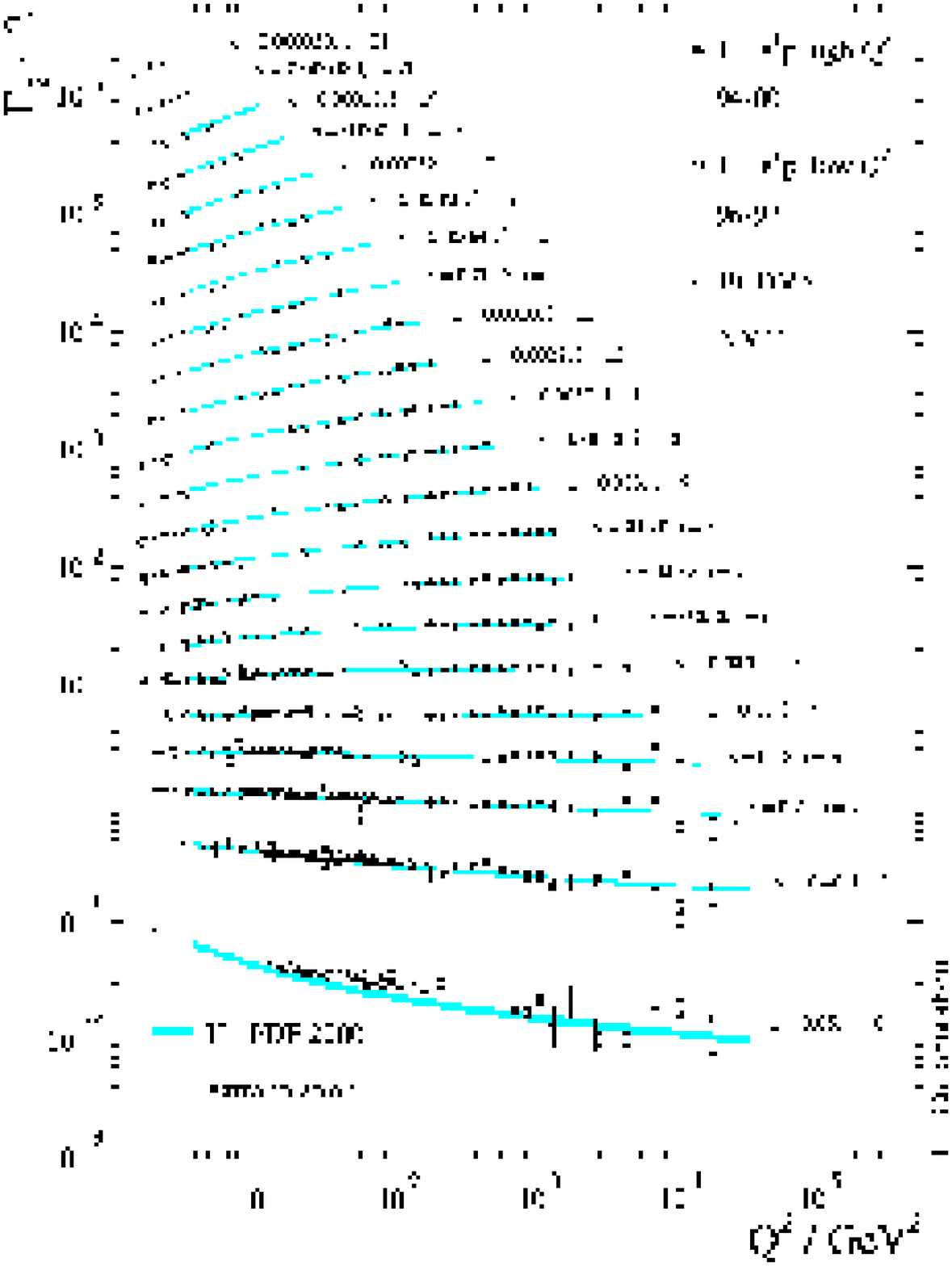}
\caption{}{\small The deep-inelastic
structure function $F_2^p(x,Q^2)$ as measured
by several different experiments (HERA, BCDMS and NMC). 
Note the dependence of $F_2^p(x,Q^2)$ with $Q^2$, as
dictated by perturbative QCD.}
\label{structure_plots}
\end{center}
\end{figure}

The deep-inelastic scattering
 cross section can be decomposed using kinematics and
Lorentz invariance in terms
of structure functions $F_i(x,Q^2)$. These structure
functions parametrize the structure of the proton
as seen by the virtual gauge boson.
If the incoming lepton is a charged lepton
(an electron or a muon) then for $Q^2\ll M_Z^2$
the double differential deep-inelastic scattering
cross section reads
\be
\label{neutral}
\frac{d^2\sigma^{\mathrm{em}}}{dxdQ^2}=\frac{4\pi \alpha^2}{Q^4}
\lc \lp 1+(1-y)^2\rp F_1(x,Q^2)+\frac{1-y}{x}\lp F_2(x,Q^2)-
2xF_1(x,Q^2)\rp\rc \ ,
\ee
where $\alpha$ is the electromagnetic coupling.
If the incoming lepton is a neutrino, then
the cross section reads
\be
\label{disneut}
\frac{d^2\sigma^{\nu p}}{dxdy}=\frac{G_F^2ME}{\pi}\lc 
\lp 1-y-\frac{M}{2E}xy\rp F_2^{\nu}(x,Q^2)+y^2xF_1^{\nu}(x,Q^2)+
y\lp 1-\frac{y}{2}\rp xF_3^{\nu}(x,Q^2)\rc \ .
\ee
where $G_F$ is the Fermi constant, $M$ the mass of the target
hadron and $E$ the neutrino energy in the
hadron rest frame. Note that Eq. \ref{disneut} holds both for
charged current ($W^{\pm}$ exchange) and neutral current
($Z$ exchange) neutrino scattering, even if the decomposition of the
structure functions $F_i^{\nu}(x,Q^2)$ in terms of 
parton distributions is different in the two cases
\cite{neutdis}. 
All the structure functions defined above have
been measured in several experiments, and the
most precisely known is the
charged lepton scattering neutral current
structure function $F_2(x,Q^2)$, thanks to the
high precision measurements at HERA and in fixed target experiments.
In Fig. \ref{structure_plots} we show a summary
of available data on this structure function from
different experiments. Note that the effects of QCD
evolution, that is, the dependence of
 $F_2(x,Q^2)$ with the scale $Q^2$, is clearly
observed in the experimental data, specially in the
small-$x$ region.

Structure functions $F_i(x,Q^2)$ depend on the momentum distribution of
partons (quarks and gluons) inside the proton, which
are determined by low energy nonperturbative dynamics, and therefore
cannot be computed in perturbation theory. However,
each structure function $F_i(x,Q^2)$, using the
 factorization theorem \cite{factorization},
can be written as a convolution of  hard-scattering coefficients
$C_{ij}(x,\alpha_s(Q^2))$, which depend only
on the short-distance (perturbative) physics,
and  parton distribution functions  $q_j(x,Q^2)$ 
\cite{partonsalt}, which parametrize
the (non-perturbative) structure of the proton.
The factorized structure function reads
\be
\label{struccoef}
F_i(x,Q^2)= \int_x^1 \frac{dy}{y}C_{ij}(y,\alpha_s(Q^2))
 q_{j}\lp \frac{x}{y},Q^2\rp
\ .
\ee
The coefficient functions $C_i(x,\alpha_s(Q^2))$ can be
computed in perturbation theory as a power series expansion
in $\alpha_s(Q^2)$. Parton distribution can be formally
defined \cite{collinspdf} by means of suitable operator
matrix elements in the proton,
\be
q_i\lp x,Q^2\rp\equiv\int\frac{dy^-}{4\pi}e^{-ixp^+y^-}
\la p| \bar{\psi}_i\lp 0,y^-\rp W[y,0] \gamma^+\psi(0)|p\ra_R \ ,
\ee
where $W[y]$ is a Wilson line (a path-ordered exponential
of the gluon field) and the parton distributions
are renormalized at the scale $\mu^2=Q^2$.
Since these parton distributions are
non-perturbative,  they must be determined from experimental data.
The techniques used to extract parton distributions from
hard scattering data together with the associated uncertainties will
be discussed in the next Section.

At leading order in the strong coupling $\aq$, the expressions
for the different structure functions in terms of
parton distribution functions are relatively simple.
For example, for charged lepton scattering
off a proton target one has
\be
F_2^{\mathrm{em}}(x,Q^2)=x\lc \frac{4}{9} \lp u+\bar{u}+c
+\overline{c}\rp+
\frac{1}{9} \lp d+\bar{d}
+s
+\overline{s}\rp\rc\lp x, Q^2\rp  \ ,
\ee
\be
2xF_1(x,Q^2)=F_2(x,Q^2) \ ,
\ee
where the last relation is the Callan-Gross relation \cite{callangross}.
The parity violating structure function $xF_3$ in charged lepton
scattering can be neglected in the $Q^2\ll M_Z^2$ limit, since
only contains contributions from Z boson exchange.
For neutrino-proton scattering the appropriate
relations are given by
\be
F_2^{\nu}(x,Q^2)=2x\lc d+s+\bar{u}+\bar{c}\rc\lp x, Q^2\rp \ ,
\ee
\be
xF_3^{\nu}\lp x, Q^2\rp
=2x\lc d+s-\bar{u}-\bar{c}\rc\lp x, Q^2\rp \ .
\ee
In the above expressions, $u(x,Q^2)$ is the parton distribution function
 of the
u-quark in the proton, $d(x,Q^2)$ that of the d-quark, and so on.

At very high energies, $Q^2\ge M_Z^2$, one has to take into
account the contribution of $Z$ boson exchange in neutral
current charged lepton scattering. The generalization
of Eq. \ref{neutral} which incorporates
the complete neutral current exchange is given by
\be
\frac{d^2\sigma}{dxdQ^2}=\frac{4\pi \alpha^2}{xQ^4}
\lc xy^2F_1^{NC}(x,Q^2)+
(1+y)F_2^{NC}(x,Q^2)+y\lp 1-\frac{y}{2} \rp F_3^{NC}(x,Q^2)\rc \ ,
\ee
where in terms of parton distribution functions one has
\be
F_2^{NC}(x)=2xF_1^{NC}(x)=\sum_{i=1}^{N_f} \lc q_i(x)+\bar{q}_i(x)
\rc C_q(Q^2) \ ,
\ee
\be
xF_3^{NC}(x)= \sum_{i=1}^{N_f} \lc q_i(x)-\bar{q}_i(x)\rc D_q(Q^2) \ ,
\ee
\be
C_q(Q^2)=e_q^2-2e_qV_eV_qP_Z+(V_e^2+A_e^2)(V_q^2+A_q^2)P_Z^2 \ ,
\ee
\be
D_q(Q^2)=-2e_q^2A_eA_qP_Z+4V_eA_eV_qA_q P_Z^2, \qquad P_Z=\frac{Q^2}{
Q^2+M_Z^2} \ ,
\ee 
where $e_i$ are the electromagnetic charges
of the quarks and
 $V_i$ and $A_i$ are the vector and axial couplings of the
fermions to the Z boson.
Note the appearance of the parity-violating structure function
$F_3(x,Q^2)$, which is sensitive to the
helicity of the incoming leptons
(that is, it is different for example for an electron and for
a positron).

Even if  parton distribution functions
$q_i(x,Q_0^2)$ are of non-perturbative
origin,
it can be shown that
their  dependence with the scale $Q^2$ 
 is dictated by perturbative QCD, provided
the scale $Q^2$ is large enough. 
The dependence of the parton distributions with the scale $Q^2$,
also known as their {\it evolution} with $Q^2$, is
dictated by the
perturbative
DGLAP \cite{gl,dok,ap} evolution equations.
These equations can be used to evolve with $Q^2$ any
combination of parton distributions, however, 
their form is much simpler if suitable combinations
are defined. For nonsinglet combinations of
parton distributions, defined as differences
between quark distributions, 
\be q_{NS,ij}(x,Q_0^2)\equiv\lp q_i-q_j\rp
(x,Q^2_0) \ ,
\ee
where $i,j$  label either a quark or an antiquark,
the DGLAP evolution equation reads
\be
\label{ap}
\frac{dq_{NS}(x,Q^2)}{d\ln Q^2}=\frac{\alpha_s(Q^2)}{2\pi}
\int_x^1 \frac{dy}{y}P_{NS}\lp y,\alpha_s(Q^2)\rp q_{NS}
\lp \frac{x}{y},Q^2\rp \ ,
\ee
where  $P_{NS}(x,\aq)$ are the non-singlet splitting functions.
These splitting functions can be computed perturbatively as an
expansion in powers of $\aq$. For instance, the
leading order expression for the nonsinglet
splitting function is given by
\be
P_{NS}^{(0)}\lp x\rp =C_F\lp \frac{1+x^2}{(1-x)_+}+\frac{3}{2}
\delta(1-x)\rp\ .
\ee
It is clear from its definition that the gluon is
decoupled from the evolution of nonsinglet
parton distributions.
The remaining independent combination of parton
distribution is called the singlet parton distribution,
defined as the sum of all quark and 
anti-quark flavors,
\be
\Sigma(x,Q^2)\equiv \sum_{i=1}^{N_f}\lp q_i(x,Q^2)+\overline{q}_i(x,Q^2)\rp 
\ .
\ee
In the singlet sector, the DGLAP equation is a 2-dimensional
matrix equation. In this case the singlet distribution
evolves coupled to the gluon parton distribution using
the singlet DGLAP evolution equation,
\be
\label{ap2}
\frac{d}{d\ln Q^2}\lp \begin{array}{c}  \Sigma(x,Q^2) \\g(x,Q^2)
\end{array}\rp=\frac{\alpha_s(Q^2)}{2\pi} 
\int_x^1 \frac{dy}{y}\lp\begin{array}{cc} P_{qq}(y)& 
 P_{qg}(y)\\ P_{gq}(y) & P_{gg}(y)
\end{array} \rp  
\lp \begin{array}{c} \Sigma(x/y,Q^2) \\ g(x/y,Q^2) 
\end{array}\rp \ ,
\ee
in terms of the singlet matrix of splitting functions.

The DGLAP evolution equations, Eqns. \ref{ap} and \ref{ap2} 
can be solved using a wide variety
of techniques. A particularly useful method
is the transformation of  the evolution equations to Mellin space,
also known as  moment space, using
the integral transformation
\be
q_i\lp N,Q^2\rp\equiv\int_0^1 dx x^{N-1}q_i\lp x,Q^2\rp \ .
\ee
In Mellin space, the nonsinglet DGLAP
evolution equation, Eq. \ref{ap} 
is no longer an integro-differential
equation but rather a simple differential
equation,
\be
\label{apns}
\frac{dq_{NS}(N,Q^2)}{d\ln Q^2}=\frac{\alpha_s(Q^2)}{2\pi}
\gamma_{NS}\lp N,\alpha_s(Q^2)\rp q_{NS}
\lp N,Q^2\rp \ ,
\ee
where the anomalous dimension $\gamma_{NS}\lp N,\alpha_s(Q^2)\rp$
is the Mellin transform of the splitting function,
\be
\gamma_{NS}\lp N,\alpha_s(Q^2)\rp=\int_0^1 dx x^{N-1}P_{NS}
\lp x,\alpha_s(Q^2)\rp \ .
\ee
The main advantage of this method is that in Mellin space
the DGLAP equations can be solved analytically. In the
nonsinglet section, for example, Eq. \ref{apns}
has the solution
\be
\label{evfactn}
q_{NS}(N,Q^2)=\Gamma\lp N,\aq,\aqq\rp q_{NS}(N,Q^2_0) \ ,
\ee
in terms of an evolution factor $\Gamma\lp N\rp$ which
is constructed in terms of anomalous dimensions, $\gamma_{NS}(N)$.
Similar results hold for the singlet DGLAP equation
Eq. \ref{ap2} in Mellin space.
Once the evolution equations have been solved in Mellin space,
one needs to invert back to x-space, using the inverse
Mellin transformation,
\be
q\lp x,Q^2\rp=\frac{1}{2\pi i}\int_C dx x^{-N}q\lp N,Q^2\rp \ ,
\ee
where the integration contour $C$ in the complex N plane
is parallel to the imaginary axis and to the right of all the
singularities of the integrand. Except in very special cases,
this inverse transformation can only be performed by numerical integration.

To end with this review of deep-inelastic scattering, let us
describe sum rules of structure functions. 
Sum rules are integrals over
certain combinations of structure functions which have
a particular value in the parton model
\cite{feynmann}. Sum rules
are extremely useful for many purposes, from precision
determinations of the strong coupling, tests of
perturbative QCD and to gain more insight on
non-perturbative dynamics. A first example is
the Gross-Llewellyn Smith sum rule \cite{gls},
which is related to the parity-violating
structure function $F_3^{\nu}(x,Q^2)$ in neutrino-nucleus
scattering. The GLS sum rule is an exact consequence
of QCD in the limit of isospin symmetry, and reads
\be
I_{GLS}=\frac{1}{2}\int_0^1 dx \lp F_3^{\nu p}(x,Q^2)
+ F_3^{\overline{\nu} p}(x,Q^2)\rp
=3\lc 1+\sum_{k=1}^{\infty} d_k\alpha_s^k\rc \ ,
\ee
were the terms of $\mathcal{O}\lp \aq^k\rp$ are 
corrections from perturbative QCD.

The second example of a deep-inelastic scattering
sum rule is the Gottfried sum rule 
\cite{fortegottfried} for charged lepton scattering,
which depends on the difference of structure functions in the
proton and in the neutron, also known as the nonsinglet
structure function $F_2^{NS}(x,Q^2)$,
\be
I_G=\int_0^1 \frac{dx}{x} \lp F_2^{\mu p}(x,Q^2)- F_2^{\mu n}(x,Q^2)\rp
\equiv
\int_0^1 \frac{dx}{x} F_2^{NS}(x,Q^2)
=\frac{1}{3}+\int_0^1 dx \lp \overline{u}-\overline{d}\rp \ .
\ee
The first term in the right side is the naive result
of the parton model. Note that 
oppositely to the case of the GLS sum rule discussed above, the
Gottfried sum rule has no justification in full QCD, and indeed
it is  violated, as was observed from experimental
measurements of the nonsinglet structure function by the
NMC collaboration \cite{nmcgott}.
In this case perturbative corrections turn out
to be negligible. The measurement of this sum rule showed
that isospin symmetry in the sea quarks does
not hold for the proton, that is, $\overline{u}(x)\ne
\overline{d}(x)$.

\subsection{B meson semileptonic decays}
\label{bmeson_theo}
Hadronic states with 
quarks with large masses, the so called heavy quarks,
 are of considerable interest for QCD for a variety
of reasons. Perturbative QCD is not
applicable to strongly interacting bound states composed
of light quarks only, since
all the scales involved in the problem are
of the same size of the typical hadronic
scale, $\Lambda_{\QCD}$. However, soon after the advent of QCD
it was realized that the situation was considerably
different for hadrons with at least one heavy quark,
where heavy means the condition that its
mass $m_H$ satisfies
\be
\label{cond}
m_H\gg \Lambda_{\mathrm{QCD}} \ .
\ee 
This is so because  
the heavy quark mass provides a large mass scale, so that
perturbative computations of a variety of processes related to this
system, like decay rates or spectroscopy, can be computed in 
perturbation theory as an expansion in
$\alpha_s\lp m_H^2\rp\ll 1$. The heavy quark condition Eq. \ref{cond}
is satisfied by the charm, bottom and top quarks, even if
in the latter case it is of no practical interest since the
top quark does not hadronize due to its short lifetime. 

For this reason, hadronic states with
b quarks have become a theoretical
laboratory
for perturbative QCD. It has proved to be
an  useful environment for the
development of several
  effective field theories, like Non Relativistic 
QCD \cite{nrqcd} ,
Heavy Quark Effective Theories \cite{hqet}
and more recently the Soft Collinear 
Effective Theory \cite{scet1,scet2}. All these effective
theories make use, in one form or another, of the condition
Eq. \ref{cond} to simplify the dynamics of the
relevant processes.

\begin{figure}[ht]
\begin{center}
\epsfig{width=0.6\textwidth,figure=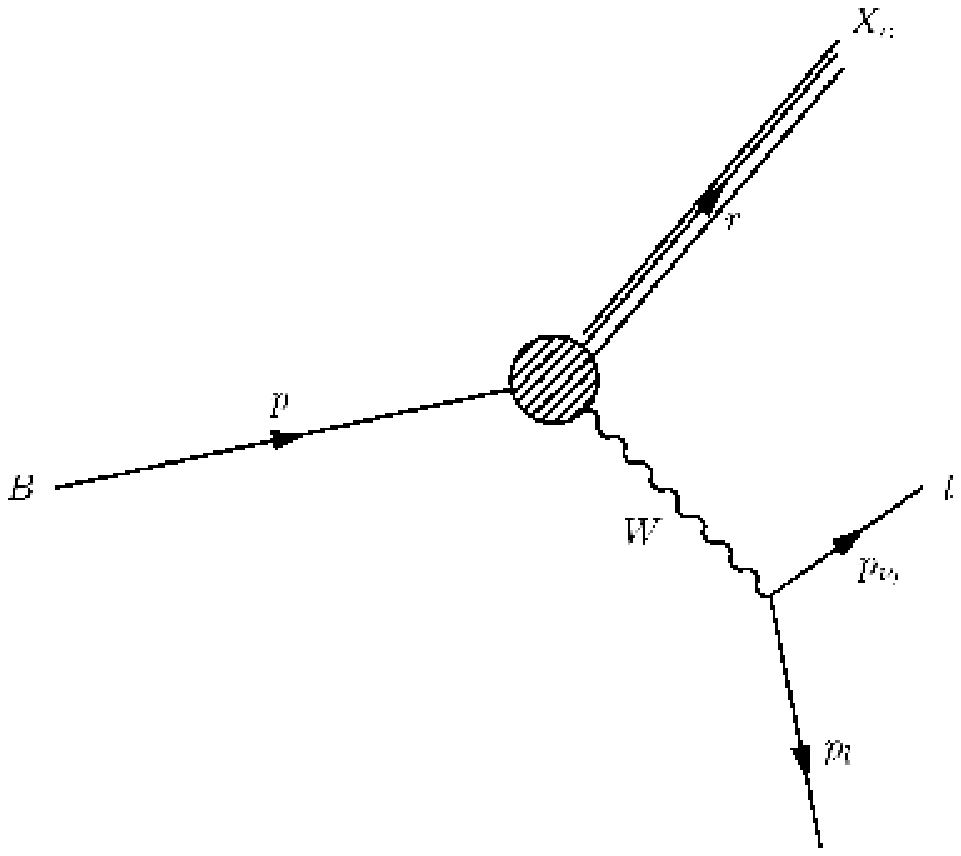}
\caption{}{\small  The semileptonic decay
of a B meson into a charmed hadronic state and 
a lepton-neutrino pair.}
\label{bdecay}
\end{center}
\end{figure}

In this thesis we will focus our attention, for reasons
to be described in the following, to the
 inclusive semileptonic decays of B mesons into charmed
final states.  This process is useful to determine
the CKM matrix element, $V_{cb}$ with high accuracy as
well as the b quark mass $m_b$.
The process, 
$B\to X_c l\nu$, is represented in Fig. \ref{bdecay}.
The most inclusive observable that can be
measured in this process is the total semileptonic decay rate. This decay
rate can be written as perturbative
series
expansion in $\alpha_s(m_b^2)$, where as has been discussed
 before,
the b quark mass $m_b$ plays the role of the hard
scale of the process, as well as an 
expansion in inverse powers of the b quark mass
parametrized by nonperturbative matrix elements of local
operators, in the OPE spirit  \cite{bigitotal}.
This expansion in powers of $1/m_b$ is also known
as the heavy quark expansion \cite{mannel}. The inclusion of the
leading nonperturbative effects through the heavy quark
expansion is crucial to analyze this process, and in general,
heavy meson physics, since the heavy quark masses
are not so large compared to $\Lambda_{\QCD}$.
Therefore, typical nonperturbative corrections of the order
$\mathcal{O}\lp \Lambda_{\QCD}/m_b\rp$ have to be taken into
account for precision theoretical computations.

With this caveats, we can write for the total B meson
semileptonic decay rate the
following expression
\bea
\label{gammamb3}
\Gamma\lp B\to X_c l\nu\rp=\frac{G_F^2m_b^5}{192\pi^3}
|V_{cb}|^2\lp1+A_{\mathrm{ew}}\rp A_{\mathrm{pert}}(\rho) \cdot \nonumber \\
\Bigg[ z_0(\rho)\lp 1-\frac{\lambda_1}{2m_b^2}\rp 
+g(\rho)\frac{\lambda_2}{2m_b^2}+\mathcal{O}\lp \frac{1}{m_b^3}\rp \Bigg] \ ,
\eea
where the phase space factors are given by
\be
\label{zo}
z_0(\rho)= 1-8\rho+8\rho^3-\rho^4-12\rho^2\log \rho, \quad 
\rho=\frac{m_c^2}{m_b^2} \ ,
\ee
\be
g(\rho)=-9+24\rho-72\rho^2+73\rho^3-15\rho^4-26\rho^2\ln\rho \ ,
\ee
and the leading nonperturbative effects are parametrized by 
$\lambda_1$ and $\lambda_2$, which are matrix elements
in the B meson of local operators with the appropriate
dimension,  $A_{\mathrm{ew}}$ stands for the
electroweak radiative corrections and $A_{\mathrm{pert}}(\rho)$,
\be
A_{\mathrm{pert}}(\rho)=\sum_{k=0}^{\infty}c_k\lp \rho
\rp\alpha_s\lp m_b^2\rp^k \ ,
\ee
includes the QCD perturbative
corrections.

Much more detailed information on the
underlying dynamics of the process can be obtained
by analyzing less inclusive observables.
These observables can be constructed from
the triple differential decay rate for this
process,
\be
B(p) \to l(p_l)+\bar{\nu}(p_{\bar{\nu}})+X_c(r) \ ,
\ee
which depends on three different kinematical variables $q^2,r$ and $E_l$,
where $q=p_l+p_{\nu}$ is the total four 
momentum of the leptonic system, $r=p-q$ is the
four-momentum of the charmed hadronic final state, 
with invariant mass $r^2=M_X^2$, and $E_l$ is the lepton
 energy in the rest frame of the
decaying $b$ quark. This triple differential distribution
 can be decomposed, using the kinematics of the process and the symmetries
of the theory,
 in terms of three structure functions, 
\bea
\label{triple}
\frac{d^3\Gamma}{dq^2 dr dE_l}(q^2,r,E_l)=\frac{G_F^2|V_{cb}|^2}{16\pi^4}
\Bigg[ \hat{q}^2W_1(\hat{q}^2,\hat{u})- \nonumber \\
\lp 2v\hat{p}_l-2v\hat{p}_lv\hat{q}+\frac{\hat{q}^2}{2}\rp 
W_2(\hat{q}^2,\hat{u})+\hat{q}^2
\lp 2v\hat{p}_l-v\hat{q}\rp W_3(\hat{q}^2,\hat{u})\Bigg] \ ,
\eea
where $u^2=r^2-m_c^2$, $v=p/m_b$ and
 the quantities with a hat are dimensionless quantities
normalized to $m_b$. All the structure functions $W_i(\hat{q}^2,\hat{u})$ 
have
both a perturbative expansion in powers of $\alpha_s$, and
a nonperturbative expansion in powers of $1/m_b$, which
can be computed in the framework of the heavy quark expansion. 
For example,
the  $\mathcal{O}(\alpha_s)$ 
corrections
for all the differential distributions that can be constructed
from Eq. \ref{triple} have become
available recently \cite{trott,agru}.

Typical observables which are accessible in experiments 
are convolutions of the differential spectra
with suitable weight functions over a large enough
range, with kinematical cuts. A particular case of these
observables 
are the  moments of differential
 decay distributions. In this thesis we will study
the leptonic 
moments, defined as
\be
L_n(E_0,\mu)\equiv \int_{E_0}^{E_{\max}} dE_l \lp E_l -\mu \rp^n 
\int dq^2 dr \frac{d^3\Gamma}{dq^2 dr dE_l}(q^2,r,E_l) \ ,
\ee
where $E_0$ is a lower cut on the lepton energy, required 
experimentally to
select this decay mode from the background from other
B meson decays, and $E_{\max}$ is the maximum energy
allowed from the kinematics of the process that the
lepton can have,
\be
\label{emax}
E_{\max}=\frac{m_B^2-m_D^2}{2m_B} \ ,
\ee
where $m_B$ is the average of the mass of the neutral and charged
B mesons, and similarly for $m_D$.
In particular we are interested in the behavior of the
lepton energy distribution, defined as
 \be  
\frac{d\Gamma}{dE_l}(E_l)\equiv
\int dq^2 dr \frac{d^3\Gamma}{dq^2 dr dE_l}(q^2,r,E_l) \ ,
\label{specdef}
\ee
which is related to the observable leptonic moments via
\be
\label{leptonmom}
L_n(E_0,\mu)=\int_{E_0}^{E_{\max}} dE_l \lp E_l -\mu \rp^n 
\frac{d\Gamma}{dE_l}(E_l) \ .
 \ee
Available published data for this process consist on moments of
the lepton spectrum, Eq. \ref{leptonmom}. In Section
\ref{bdecay_appl} we reconstruct the underlying lepton
spectrum, Eq. \ref{specdef}, from experimental information
on its moments together with additional theoretical
constraints, using the general strategy to
parametrize experimental data described in Chapter
\ref{general}.

\begin{figure}[ht]
\begin{center}
\epsfig{width=0.53\textwidth,figure=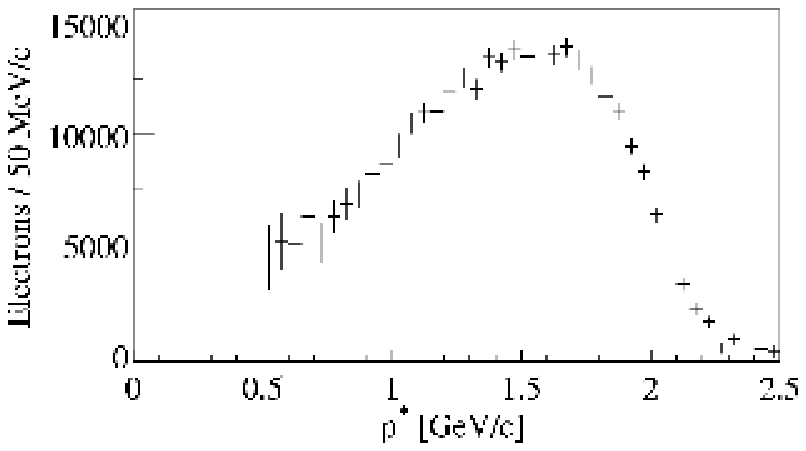}
\epsfig{width=0.41\textwidth,figure=speclo.ps}
\caption{}{\small The folded
lepton energy spectrum as measured by the
Babar collaboration as a function of the
lepton momentum (left) and the
corresponding leading order theoretical
prediction (right).}
\label{specplots}
\end{center}
\end{figure}

The lepton energy spectrum Eq. \ref{specdef} in $B\to X_c l\nu$
decays
can also be expanded in a power series both
in $\alpha_s$ and in $1/m_b$. The leading order spectrum 
in both
expansions 
is given by \cite{ope2}
\be
\label{dgde}
\frac{d\Gamma}{dy}\lp B\to X_cl\nu\rp=\Gamma_0
2y^2\lc (1-f)^2(1+2f)(2-y)+(1-f)^3(1-y)\rc 
\theta(1-y-\rho)\ ,
\ee
where \be
\label{gammalo}
\Gamma_0=\frac{G_F^2m_b^5}{192\pi^3} 
\ee 
is the total semileptonic
decay rate in the parton model for massless final state quarks and
we have defined
\be
f=\frac{\rho}{1-y} \ ,\qquad \rho=\frac{m_c^2}{m_b^2} \ ,\qquad
y=\frac{2E_l}{m_b} \ .
\ee 
The leading perturbative $\mathcal{O}(\alpha_s)$ 
corrections to this spectrum have been known for some
time \cite{kuhn}, and there are estimations of the size of higher
order terms \cite{blm2} 
though the BLM expansion \cite{blm}.
The leading nonperturbative $\mathcal{O}(1/m_b^2)$ 
corrections to the lepton
energy spectrum were computed in Refs. \cite{ope2,manohar}
and the  $\mathcal{O}(1/m_b^3)$ corrections in Ref. \cite{gremm}. 
Similar expansions exist also for the lepton energy moments (see
 Ref. \cite{gambinomom} and references therein), where nonperturbative
corrections have been computed up to $\mathcal{O}\lp 1/m_b^3\rp$.
Note that nonperturbative effects in the
$1/m_b$ expansion ara parametrized by B meson expectation values
that need to be extracted from experimental data. 
In Fig. \ref{specplots} we show the lepton energy
spectrum as measured from the Babar collaboration \cite{Aubert:2004td},
before applying several experimental corrections,
and the corresponding leading order theoretical prediction, 
Eq. \ref{dgde}, for different values of the
b quark mass.

In Section \ref{bdecay_appl} we construct a neural network
parametrization of the leptonic spectrum, Eq. \ref{specdef},
from all the available experimental information on
moments of this spectrum, supplemented by kinematical
constraints. This application
will show that the neural network approach
can be used to reconstruct in a efficient way
a function when the only available information comes
from moments of this function.
 As a byproduct of such parametrization, we will
provide also a determination of the b quark mass in the
$\overline{\mathrm{MS}}$ scheme
$\overline{m}_b\lp \overline{m}_b\rp $.

\subsection{Hadronic tau decays}

\label{tau_theo}

The hadronic decays of the tau lepton \cite{tsai,tauthesis} are for a variety
of reasons a key process to study both the
perturbative and non perturbative regimes of
Quantum Chromodynamics \cite{pich,diberder}. Its importance
for precision studies of hadronic physics
has been extensively studied \cite{physicsaleph},
specially thanks to the high quality data
provided by the LEP accelerator at CERN
\cite{hadrlep1,hadrlep2}.
Not only the hadronic tau decays provide one of the most precise
determinations of the strong coupling
$\alpha_s\lp M_{\tau}\rp$, but since the tau mass is not so large
as compared to $\Lambda_{\QCD}$, the non-perturbative
effects, parametrized by vacuum condensates,
can be extracted from experimental data in a clean way.
In this Section we will briefly review
the theoretical foundations  of the QCD analysis
of hadronic tau decays.

\begin{figure}[ht]
\begin{center}
\epsfig{width=0.535\textwidth,figure=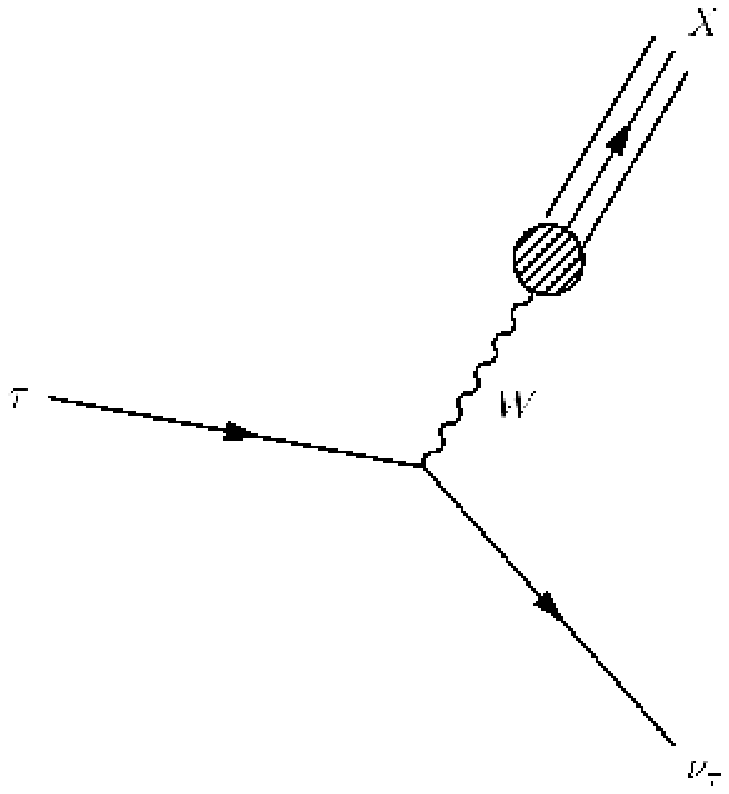}
\caption{}{\small  The hadronic decays
of the tau lepton. The final state of the decay
consists on a hadronic system $X$, composed
mainly by pions, and a undetected neutrino.}
\label{taudecay}
\end{center}
\end{figure}

\begin{figure}[ht]
\begin{center}
\epsfig{width=0.48\textwidth,figure=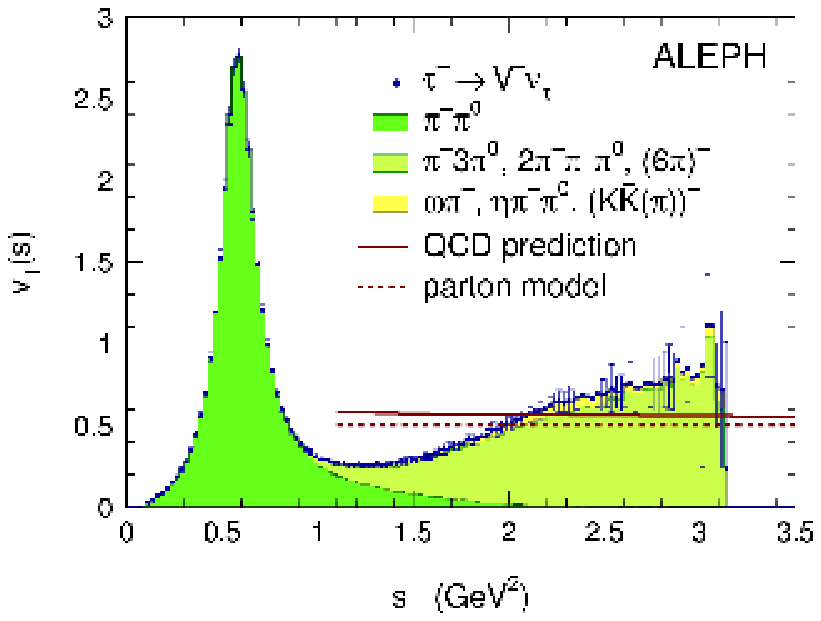}
\epsfig{width=0.48\textwidth,figure=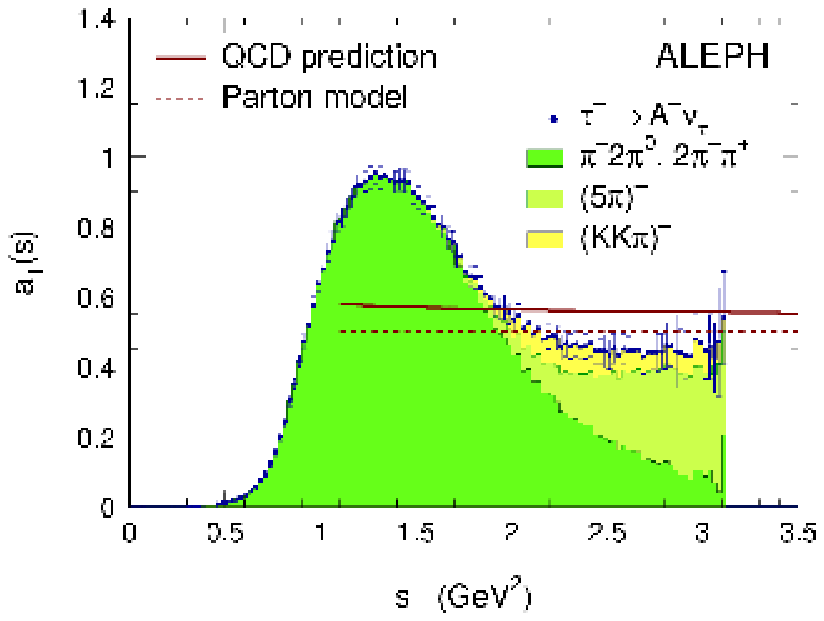}
\caption{}{\small The different contributions to the
total vector (left) and axial-vector (right) spectral
functions from the hadronic decays of the
$\tau$ lepton as measured by the Aleph detector.
\label{tauspecplots}}
\end{center}
\end{figure}

The hadronic decays of the tau lepton (see Fig. \ref{taudecay})
are of the form
\be
\tau \to X \nu_{\tau} \ ,
\ee
where $X$ is an hadronic   system,
composed basically by pions, with vanishing total
strangeness.
The final hadronic state 
can be separated into scalar, vector and axial vector
contributions, since 
parity is maximally violated in
$\tau$ decays.
The hadronic invariant mass-squared $s$ distribution
can be measured for each decay channel.
These invariant mass distributions $dN_{V/A}/ds$ are related to
the so-called spectral functions $\rho_i(s)$ by
\be
\label{specs}
\rho_{V/A}(s)=K_{V/A}(s)\frac{dN_{V/A}(s)}{ds} \ ,
\ee
for vector (V) and axial-vector (A) final states, and
where $K_i(s)$ is a purely kinematic factor.
In Fig. \ref{tauspecplots} we show the
contributions from the different decay modes to
the vector and the axial vector spectral functions \cite{physicsaleph}.

Spectral functions are the observables that give access to the inner 
structure of hadronic tau decays. 
For reasons to be described in the following, we are
in particular interested in the difference between the
vector and the axial vector spectral functions,
\be
\rho_{V-A}(s)\equiv\rho_V(s)-\rho_A(s) \ .
\ee
 As spontaneous chiral symmetry
breaking is a nonperturbative phenomena, the $\rho_{V/A}(s)$ 
spectral functions
are degenerate in perturbative QCD with massless light quarks, so 
any difference
between vector and axial-vector spectral functions is necessarily 
generated by non-perturbative dynamics.
From perturbative QCD one expects that at some value
of $s$ large enough the perturbative result
\be
\label{specpert}
\lim_{s\to \infty}\rho_{V-A}(s)=\rho^{\mathrm{(pert)}}_{V-A}(s)=0 \ ,
\ee
is recovered, but current experimental data do not allow
to draw any conclusion of how large $s$ is in reality.
Therefore, the spectral function $\rho_{V-A}(s)$ is generated entirely 
from nonperturbative QCD dynamics, and provides a laboratory
for the study of these non-perturbative contributions, which have resulted
to be small and therefore
difficult to measure in other processes where the 
perturbative contribution dominates.
As will be discussed in brief, these nonperturbative contributions
are organized by the Operator product expansion and
are parametrized by matrix elements in the vacuum
of local operators, the so-called QCD vacuum condensates.

As it is well known \cite{pich}, the basis of the comparison of
theoretical predictions with experimental data  
in the hadronic decays
of the tau lepton is the fact that unitarity and analyticity connect
the spectral functions of hadronic tau decays to the imaginary part of 
the hadronic vacuum polarization tensor,
\be
\Pi_{ij,U}^{\mu\nu}(q)\equiv \int d^4x~e^{iqx}\la 0|T\lp U^{\mu}_{ij}(x)
 U^{\nu}_{ij}(0)^{\dag}
\rp|0\ra \ ,
\label{corr}
\ee
of vector $ U^{\mu}_{ij}\equiv V^{\mu}_{ij}=\qb_j\gamma^{\mu}q_i$
or axial vector 
$ U^{\mu}_{ij}\equiv A^{\mu}_{ij}=\qb_j\gamma^{\mu}\gamma_5q_i$
color singlet quark currents in corresponding quantum states. 
After Lorentz decomposition is used to separate the correlation function into 
its $J=1$ and $J=0$ components,
\be
\Pi_{ij,U}^{\mu\nu}(q)=\lp-g^{\mu\nu}q^2+q^{\mu}q^{\nu}\rp\Pi^{(1)}_{ij,U}(q^2)
+q^{\mu}q^{\nu}\Pi^{(0)}_{ij,U}(q^2) \ ,
\ee
for non-strange quark currents one 
identifies
\be
\label{anal}
\mathrm{Im} \Pi^{(1)}_{\bar{u}d,V/A}(s)=\frac{1}{2\pi}\rho_{V/A}(s) \ .
\ee
Since as we have shown before the spectral functions 
$\rho_{V/A}(s)$ can be measured experimentally, Eq. \ref{anal}
provides the basis for the comparison between theoretical predictions
in the framework of the operator product expansion for
the hadronic correlator, Eq. \ref{corr}, and 
experimental data in terms
of spectral functions, Eq. \ref{specs}.
This relation then allows us to implement all the technology of QCD vacuum 
correlation functions to hadronic tau decays.

As has been mentioned before,
the basic tool to study in a systematic way the power corrections
introduced by nonperturbative dynamics is the operator product 
expansion. 
Since the approach of Ref. \cite{svz}, the operator product expansion 
has been used to perform calculations with QCD on the ambivalent
energy regions where nonperturbative effects come into play
but still perturbative QCD is relevant. In general, the OPE of a two
point correlation function $\Pi^{(J)}(s)$ takes the form \cite{pich}
\be
\Pi^{(J)}(s)=\sum_{D=0,2,4,\ldots}\frac{1}{(-s)^{D/2}}\sum_{\mathrm{dim}
\mo=D}C^{(J)}(s,\mu)\la\mo(\mu)\ra \ ,
\ee
where the arbitrary parameter $\mu$ separates the long distance 
nonperturbative effects absorbed into the vacuum expectation elements
$\la\mo(\mu)\ra$, from the short distance effects which are included
in the Wilson coefficient $C^{(J)}(s,\mu)$, and $J$ is
the parity of the correlator. The operator of dimension
$D=0$ is the unit operator (the perturbative series)
In the case we are interested in, the vector-axial vector
correlator, the operator product expansion has no
perturbative term and reads
\be
\label{chiralope}
\Pi(Q^2)\Big|_{V-A}\equiv
\sum_{n=1}^{\infty}\frac{1}{Q^{2n+4}}C_{2n+4}(Q^2,\mu^2)
\left\langle \overline{\mathcal{O}}_{2n+4}(\mu^2)\right\rangle \ ,
\ee 
where we observe that the $D=6$ term 
is the first non-vanishing non-perturbative contribution, in the limit of
massless light quarks, to the $\rho_{V-A}(s)$ spectral function and,
moreover, it has been shown to be the dominant one. 
Therefore, 
this spectral function should provide a source for a clean
extraction of the value of the nonperturbative contributions
from experimental data.

Even if the $\rho_{V-A}(s)$ spectral function is
purely non-perturbative, it has to satisfy several sum rules
that can be derived rigorously from QCD.
Sum rules have always been an important tool for
studies of non-perturbative aspects of QCD, and have been applied to a wide
variety of processes, from deep-inelastic scattering, 
as we have seen in Section
\ref{dis_theo}, to heavy quark bound states
\cite{derafael,srnarison}, introduced in Section
\ref{bmeson_theo}. Now we will review one of the 
classical examples of low energy QCD sum rules, 
the chiral sum rules. 
The application of chiral symmetry together with the optical theorem 
leads to low energy sum rules
involving the difference of vector and axial vector spectral functions,
$\rho_{V-A}(s)$.
These sum rules are dispersion relations 
between real and absorptive parts of a two point correlation function
that transforms symmetrically under $SU(2)_{L}\otimes SU(2)_{R}$ in the case
of non strange currents. Corresponding integrals are
the Das-Mathur-Okubo sum rule \cite{dmo},
\be
\label{dmo}
\frac{1}{4\pi}\int_0^{s_0\to\infty}ds\frac{1}{s}
\rho_{V-A}(s)=\frac{f_{\pi}^2 \la r_{\pi}^2 \ra}{3}
-F_A \ ,
\ee
where $f_{\pi}$ is the pion decay constant, 
 $\la r_{\pi}^2 \ra$ the average radius squared
of the pion and $F_A$ the axial-vector pion form factor,
as well as the first and second Weinberg sum rules \cite{weinberg} \ ,
\be
\frac{1}{4\pi^2}\int_0^{s_0\to\infty}ds\rho_{V-A}(s)=f_{\pi}^2 \ ,
\label{wsr1}
\ee
\be
\int_0^{s_0\to\infty}dss\rho_{V-A}(s)=0 \ ,
\label{wsr2}
\ee
where in Eq. \ref{wsr1} the right hand side
 term comes from the integration of the 
spin zero axial contribution, which for massless non-strange quark 
currents consists exclusively of the pion pole. Finally, there is the
chiral sum rule associated with
 the electromagnetic splitting of the pion masses \cite{empion},
\be
\label{empionmass}
\frac{1}{4\pi^2}\int_0^{s_0\to\infty}dss\ln \frac{s}{\lambda^2}\rho_{V-A}(s)=
-\frac{4\pi f_{\pi}^2}{3\alpha}(m_{\pi^{\pm}}^2-m_{\pi^0}^2) \ ,
\ee
where $\lambda^2$ is an arbitrary scale, $\alpha$ the
electromagnetic coupling and $m_{\pi}$ the charged and neutral
pion masses.

In Section \ref{tau_appl} we will use our neural network technique to
construct a parametrization of the $\rho_{V-A}(s)$ 
spectral function from experimental data,
including all the information on error and
correlations, and which incorporates
all the constraints from the chiral sum rules and
from perturbative QCD. Using this parametrization we will
  provide a determination 
of the non-perturbative chiral vacuum condensates,
defined in the operator product expansion of Eq. \ref{chiralope}, whose
extraction from experimental data has been the subject
of active research in the recent years.

\section{Global fits of parton distribution functions}
\label{globalfits}

As has been discussed in Section \ref{dis_theo}, parton distribution
functions
cannot be computed in perturbation theory, but rather they
have to be extracted from experimental data like deep-inelastic scattering. 
In this Section we review  the standard
approach to determine parton distributions 
from experimental data together with their associated
uncertainties. In Appendix \ref{pdfstatus} we present a brief
overview of the current status of the field
of global fits of parton distribution functions.

\subsection{The standard approach}
Let us consider a  set of parton distribution functions
 $q_i(x,Q^2)$,
where $i=1,\ldots,2N_f+1$, since there is
an independent parton distribution for
each quark and antiquark flavor, as well
as one for the gluon. As we have discussed before, the
$Q^2$ dependence of the parton distributions is
determined by the DGLAP \cite{gl,ap,dok} evolution equations,
Eqns. \ref{ap} and \ref{ap2}, so
one needs to parametrize and determine from
data only the $x$ dependence of the
parton distributions at an initial
evolution scale $Q_0^2$. In the standard approach, 
parton distributions are
parametrized with relatively simple functional forms, that in full
generality  can be taken to be
\be
\label{param}
q_i(x,Q_0^2)=A_ix^{b_i}(1-x)^{c_i}P_i\lp x,d_i,e_i,\ldots\rp 
\ ,\qquad i=1,\ldots,2N_f+1 \ .
\ee
The rationale for this functional form is that
parton distributions parametrized this way
follow quark counting rules \cite{brodsky} at large $x$
and Regge behavior \cite{regge} at small $x$, and
 $P_i(x)$ is a smooth polynomial in $x$ that
interpolates between the small-x and the
large-x regions.
Note that neither of the two limiting behavior
(large-x and small-x) of the parton distributions can be
derived in a rigorous way from Quantum Chromodynamics, so they
are more phenomenological expectations rather that firm
theoretical predictions.

 In principle one should parametrize
and extract from experimental data the $2N_f+1$
independent parton distributions. In practice, however, one
has to take some assumptions since available data cannot
constrain all of them. For example, since there is
scarce experimental information of the
valence strange distribution $s-\overline{s}$, it
is typically set to be zero, $s=\overline{s}$. Another 
example of this kind of simplifications was the assumption that the
sea 
$\bar{u}$ and $\bar{d}$ distributions were the same.
As more and better data becomes available, some of this assumptions
are shown not to be true, and one
has to allow more freedom in the
parametrizations of the parton distributions. 
For example, the NMC measurements of the
Gottfried sum rule (see \cite{fortegottfried} and references therein)
showed that for the proton $\bar{u}(x,Q^2) \ne \bar{d}(x,Q^2)$.
Since that measurement, the assumption  
$\bar{u}=\bar{d}$ was not used anymore in global
fits of parton distributions.

To be definite, we show now the explicit parametrizations
from a recent global analysis of parton distributions \cite{mrst_02}
from the MRST collaboration.
The up and down valence distributions are parametrized
as
\be
xu_V(x,Q_0^2)\equiv x\lp u-\overline{u}\rp(x,Q_0^2)=
A_ux^{b_u}(1-x)^{c_u}\lp 1+d_u\sqrt{x}+e_ux\rp \ ,
\ee
\be
xd_V(x,Q_0^2)\equiv x\lp d-\overline{d}\rp(x,Q_0^2)=
A_dx^{b_d}(1-x)^{c_d}\lp 1+d_d\sqrt{x}+e_dx\rp \ ,
\ee
then the sea combination of parton distributions is given by
\be
xS(x,Q_0^2)=
A_sx^{b_s}(1-x)^{c_s}\lp 1+d_s\sqrt{x}+e_s x\rp \ ,
\ee
where the sea parton distribution is defined as
\be
xS(x,Q_0^2)\equiv 2x\lp \overline{u}+\overline{d}+\overline{s}\rp
\lp x,Q_0^2\rp \ ,
\ee
and  the gluon parton distribution is given by
\be
\label{gluonpdf}
xg(x,Q_0^2)=
A_gx^{b_g}(1-x)^{c_g}\lp 1+d_g\sqrt{x}+e_g x\rp-F_gx^{g_g}(1-x)^{h_g} \ .
\ee
 Finally, the structure of the
light quark sea is taken to be
\be
2\overline{u}\ ,\quad 2\overline{d}\  ,\quad 2\overline{s}=0.4S+\Delta\
 ,\quad
0.4S-\Delta\  ,\quad 0.2S \ ,
\ee
\be
x\Delta(x,Q_0^2)=x\lp \overline{d}-\overline{u}\rp\lp x, Q_0^2\rp=
A_{\Delta}x^{b_{\Delta}}(1-x)^{c_{\Delta}}
\lp 1+c_{\Delta}x+d_{\Delta} x^2\rp \ .
\ee
Note that
 the assumption of a vanishing strange valence
distribution
$s=\overline{s}$ is also used.
Not all the parameters in the above equations are left free, in particular
some of them are fixed by quark number conservation sum rules,
\be
\int_0^1 u_V(x,Q_0^2)=2, \qquad \int_0^1 d_V(x,Q_0^2)=1 \ .
\ee
The total number of fitted parameters in this case
is around 20. With the above relatively simple parametrization
one can describe a wealth of hard-scattering processes,
thus showing that QCD factorization 
\cite{factorization} holds
in the majority of high energy processes involving
strongly interacting particles.

Note that Ref. \cite{mrst_02} allows for the gluon parton
distribution to become negative, as can be seen
from Eq. \ref{gluonpdf}. This would appear to be in conflict with
the interpretation of parton distributions 
as the probability distributions of the momentum that quarks and gluon
carry inside the proton. However, it has been emphasized 
\cite{fortepos} that
parton distributions are not physical quantities, in particular
beyond the leading order approximation they depend on the
renormalization scheme. Physical quantities, on the
other hand, like cross sections and structure functions, satisfy 
positivity bounds, that is, even if parton distributions
are allowed to become negative, structure functions are not
since they are observable quantities and therefore are positive
definite.

The next step of the global fitting approach
is to evolve the set of parton distributions that have 
been parametrized, Eq. \ref{param} at the
initial evolution scale $Q_0^2$, to the scale $Q^2$ where there is
experimental data using the solution of the
DGLAP evolution equations, 
Eqns. \ref{ap} and \ref{ap2}. Then for
each specific process one adds the contribution from the perturbative
coefficient functions to construct the corresponding
observable, for example deep-inelastic
structure functions, as in Eq. \ref{struccoef}.
The number of hard-scattering processes that are nowadays
used to constrain the shapes of the different parton
distribution functions is rather large. These include
\begin{itemize}
\item Deep-inelastic scattering structure functions, $F_2(x,Q^2)$,
$F_3(x,Q^2)$ and  $F_L(x,Q^2)$, both in charged lepton and
in neutrino DIS,
\item The Drell-Yan process, $p\bar{p}\to X l\bar{l}$ in
hadronic collisions,
\item Jet production, both in $e^{\pm}p$ collisions and in 
$p\bar{p}$ collisions,
\item Gauge boson production $p\bar{p} \to W(Z) X$,
\end{itemize}
and other processes like prompt photon production and
heavy quark production. In any case, it has
to be emphasized that deep-inelastic scattering
 is and will be in the following years
the most important source of information on parton
distribution functions, specially the
precision structure function measurements
of the HERA collider \cite{herarev}. 
The LHC will not only use
extensively parton distributions, but it will also be useful
to constrain its shape, since it proves
a kinematic region not accessible at available
colliders, for example with
processes like the differential rapidity distribution of gauge
boson production \cite{heralhc}.

The final step of a global fit is to minimize a suitable statistical estimator,
for example the diagonal statistical error
\be
\label{errminim}
\chi^2\lp \{ a_i\}\rp=\sum_{i=1}^{N_{\dat}}\frac{\lp
F_i^{(\exp)}-F_i^{(\mathrm{QCD})} \lp \{ a_l\}\rp\rp^2}{\sigma_{i,\stat}^2} \ .
\ee
where $F_i^{(\exp)}$ is the experimental measurement and 
$F_i^{(\mathrm{QCD})} \lp \{ a_i\}\rp$ the theoretical prediction
as a function of the parameters $\{ a_i\}$ that describe the
set of parton distributions. These parameters are determined
by the condition that the statistical estimator
is minimized, that is, one wants to determine the
set of parameters $\{ a_i^{(0)}\}$ that satisfy the
condition
\be
\chi^2\lp a_l^{(0)}\rp\equiv \mathrm{min}_{\{ a_l\}}\lc 
\chi^2\lp \{ a_l\}\rp\rc \ .
\ee

Until some years ago, in global fits 
of parton distribution functions the error function to be 
minimized was the statistical error function, Eq. \ref{errminim}, 
where $\sigma_{i,\stat}$ is the total uncorrelated 
statistical uncertainty.
However the precision of modern experimental data made compulsory to
consider the effects of the correlated  systematic uncertainties.
This was so both because the statistical accuracy of the data
became higher and the experimental groups became to
provide the contributions from the different sources of 
systematic errors with the experimental data.
The inclusion of correlated systematics
 can be done with two equivalent definitions of the
error function $\chi^2$.
The first one uses the explicit form of the experimental covariance
matrix,
\be
\label{chi2345}
\chi^2=\frac{1}{N_{\dat}}\sum_{i,j=1}^{N_{\dat}}\lp
F_i^{(\exp)}-F_i^{(\QCD)}\rp(\mathrm{cov}^{-1})_{ij}\lp F_j^{(\exp)}
-F_j^{(\QCD)}\rp \ ,
\ee
where the covariance matrix of the experimental data is
defined as
\be
 \mathrm{cov}_{ij}=\rho_{ij}\sigma_{\tot,i}\sigma_{\tot,j} \ ,
\ee
where $\rho_{ij}$ is the correlation matrix
of experimental data and $\sigma_{\tot,j}=\sqrt{\sigma^2_{\stat,j}+
\sigma^2_{\sys,j}}$ is the total experimental uncertainty.
Since the number of data points is typically large,
the inversion of the covariance matrix might lead to
numerical instabilities. For this reason, a equivalent
form of Eq. \ref{chi2345} was proposed that does not involve
explicitely the covariance matrix.
This second equivalent definition is given by
\be
\label{chi23}
\chi^2=\frac{1}{N_{\dat}}\sum_{i=1}^{N_{\dat}}\frac{1}{\sigma_{\stat,i}^2}\lp 
F_i^{(\exp)}-F_i^{(\QCD)}-\sum_{k=1}^K r_k\beta_{ki} \rp^2+
\sum_{k=1}^Kr_k^2 \ ,
\ee
where $\beta_{ki}$ is the contribution 
from the $K$ sources of correlated systematic uncertainties.
In this approach  
one has to minimize this $\chi^2$ both with respect to the parameters $r_k$,
which determine the effect of correlated systematics, 
as well as with respect to the parameters $\{ a_i\}$ defining the 
QCD
model $F^{(\QCD)}$. The problem with this definition is that the
number of correlated systematics can be very large, which leads to a formidable
minimization task. A way to
overcome this difficulty is to perform
 the minimization with respect to the
parameters $r_k$ analytically. In this case
 one obtains a simplified expression,
\be
\label{chi24}
\chi^2(\{ a_l\})=\frac{1}{N_{\dat}}
\sum_{i=1}^{N_{\dat}}\frac{\lp F_i^{(\exp)}-F_i^{(\QCD)}\rp^2}{
\sigma_{i,\stat}^2}-\sum_{k,k'}^K B_k(A^{-1})_{kk'}B_{k'} \ ,
\ee
where we have defined
\be
r_k(\{ a_l\})=\sum_{k'=1}^K(A^{-1})_{kk'}B_{k'}  \ ,
\ee
\be
B_k(\{ a_l\})=\sum_{i=1}^{N_{\dat}}\frac{\beta_{ki}
\lp F_i^{(\exp)}-F_i^{(\QCD)} \rp}{\sigma_{\stat,i}^2}, \quad A_{kk'}=
\delta_{kk'}+\sum_{i=1}^{N_{\dat}}\frac{\beta_{ki}
\beta_{k'i}}{\sigma_{\stat,i}^2} \ .
\ee
 Assuming that the measurements  $F_i^{(\exp)},\sigma_{\stat,i}$
and $\beta_{ik}$ 
are in accord with normal statistics, the $\chi^2$ function defined in
Eq. \ref{chi24} should have the standard probabilistic interpretation.
One can check explicitely that the two
definitions are completely equivalent. The advantages and drawbacks
of the two approaches in fits with covariance matrix error
have been discussed in other contexts, like
global fits of neutrino oscillations \cite{fogli}.
The main advantage of Eq. \ref{chi23} is that it does
not require the inversion of a covariance matrix.

\begin{figure}[ht]
\begin{center}
\epsfig{width=0.7\textwidth,figure=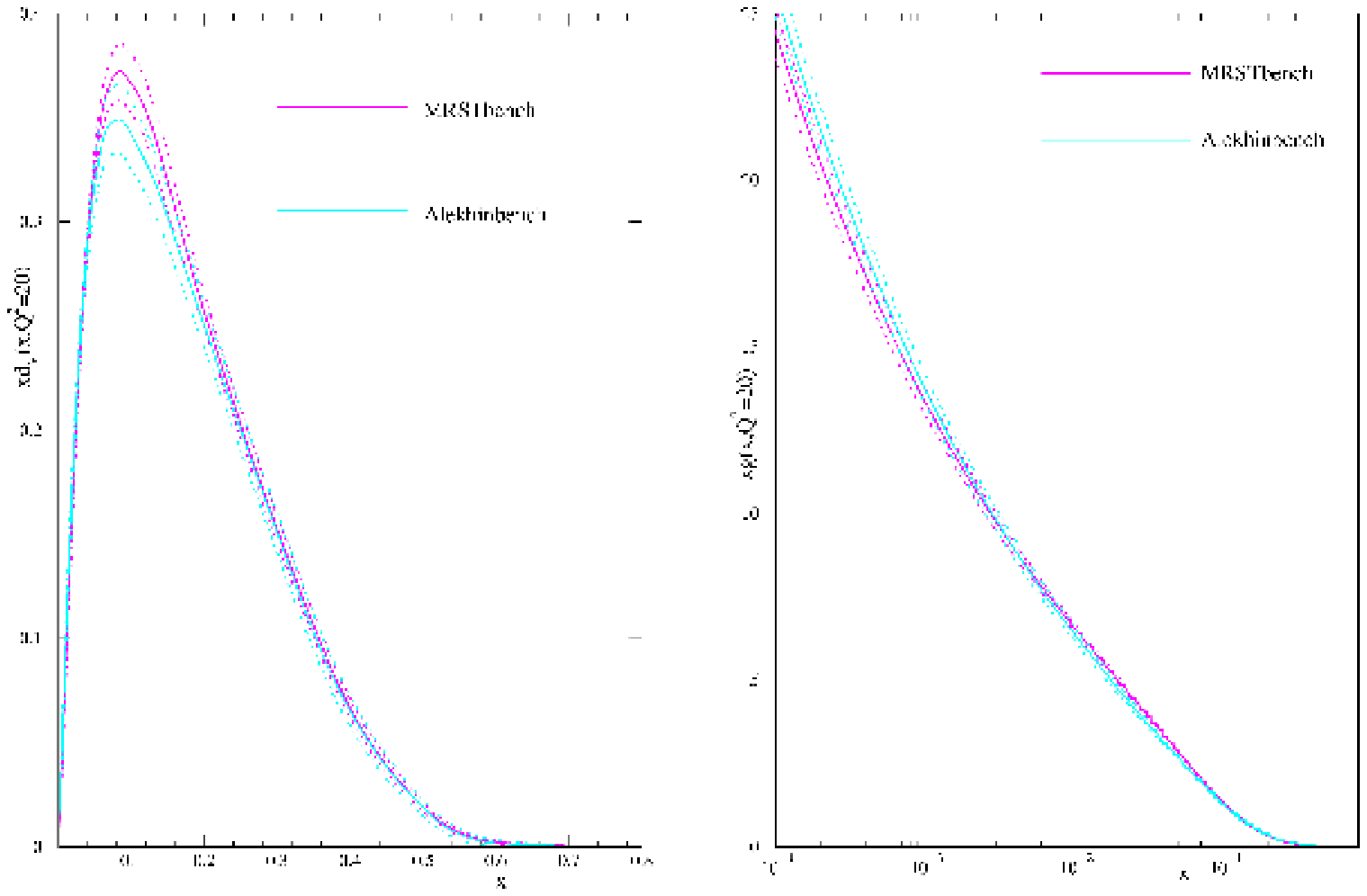}
\caption{}{\small The benchmark parton distribution functions
with associated uncertainties
as described in Ref. \cite{heralhc}: the valence down distribution
(left) and the gluon distribution (right).
\label{pdfbench}}
\end{center}
\end{figure}

In Fig. \ref{pdfbench} we show the result of a recent benchmark \cite{heralhc}
QCD analysis of parton distribution functions. These results
show the characteristic features of the different parton distributions:
while the valence distributions have to vanish at small-x, due
to quark number conservation, the gluon distribution
grows very fast at small-x. The size of this growth depends
on whether or not the parametrization of the gluon
parton distribution at the initial evolution scale $Q_0^2$ is
singular at $x=0$. The error bands for the parton distributions
are computed using some of the techniques discussed in the next
Section.

During some time it was though that the determination of the
{\it best-fit} parton distributions were enough
for practical phenomenological purposes. However, 
as better quality data became available, together
with the progress in higher-order computations, it was realized that
it was necessary to 
 estimate  in addition the uncertainties associated to the
parton distributions. These uncertainties are of two 
classes: experimental uncertainties (since parton distributions
 are extracted from
experimental data with errors) and theoretical
uncertainties (for example the model dependence in the
assumed shapes of the parton distributions, or the possible
effects of non-standard  evolution at low $x$ \cite{fortelowx}).
In the next Section we will review some of the standard methods
used to estimate the uncertainties of the parton distributions
coming from experimental data errors
 \cite{pdferrors}.
Theoretical uncertainties (see for example \cite{mrst_th}) 
are rather more complicated to estimate, and we will not review their
treatment here. This is so because theoretical 
errors are not gaussian, and the best one can do to estimate
their effects is to provide some suitable prescriptions, but no
rigorous statistical analysis is available for these type of
uncertainties.

\subsection{Uncertainties in parton distributions}

The first method to estimate the effects of the experimental 
uncertainties in the parton distributions is called the
 Offset  method \cite{botje}, which has
mostly been used by the QCD analysis of the ZEUS and
H1 collaborations. In this 
 method the systematic uncertainty parameters $r_k$ in Eq. \ref{chi23}
are fixed to zero in the central fit so that the fitted parameters
are as close as possible to the central values of published data.
Then  they
are taken into account for the error analysis.
This is done in the following way:  in addition to the  
usual Hessian matrix,
\be
\label{hessian}
H_{ij}=\frac{\partial^2\chi^2\lp \{a_m\}, 
\{r_n\}\rp}{\partial a_i\partial a_j} \ ,
\ee
defined with respect to the set of parameters $\{a_l\}$ defining the QCD 
model, a second Hessian matrix is defined as
\be
V_{jk}=\frac{\partial^2\chi^2\lp \{a_m\}, 
\{r_n\}\rp}{\partial a_j\partial r_k} \ ,
\ee
which contains information on the variations of the
error function
$\chi^2$ with respect to the systematic uncertainties
parametrized by the $r_k$ parameters.
Then the  covariance matrix for the systematic uncertainties  is
given by
\be
 C^{\sys}=H^{-1}VV^TH^{-1} \ ,
\ee
and the total covariance matrix is constructed
as the sum from the two contributions,
the statistical and the systematic,
\be
C^{\tot}=C^{\stat}+C^{\sys} \ ,
\ee
where $C^{\stat}=H^{-1}$ is the standard statistical
covariance matrix. In this method
the uncertainty in any 
quantity that depends on the
parton distribution functions, $\mathcal{F}\lc \{ q_i\}\rc$ 
is computed from
\be
\lp \Delta \mathcal{F}\rp^2 =\sum_{i,j=1}^{N_{\parr}}\frac{\partial
\mathcal{F} }{\partial a_i}C_{ij}
\frac{\partial \mathcal{F}}{\partial a_j} \ ,
\ee
by substituting $C$ by the
appropriate covariance matrices, $C^{\stat},C^{\sys}$ and $C^{\tot}$ 
to obtain the total
statistical, correlated systematic and total experimental error band
respectively. $N_{\parr}$ is the total number of parameters
used in the parametrization of the set of parton
distributions at the starting evolution scale,
Eq. \ref{param}.
This method  is not statistically rigorous, but it has the virtue that
it does not assume  gaussianly distributed systematic uncertainties. 
It gives a conservative error estimate as compared with other 
methods, like for example the Hessian method with 
$\Delta \chi^2 =1$ rule, to be discussed in brief.
 This method suffers two serious drawbacks:
first of all it assumes that linear error propagation
gives a decent estimate of the total error
propagation, an assumption that has been shown to be
not correct in many cases. Second, as can be seen from the
above formulae, the estimation of the uncertainties depends heavily
on the functional form model than one has assumed for
the set of parton distributions.

Another popular method is the so-called Hessian method.
In the Hessian 
method \cite{hessian} 
one assumes that the deviation in $\chi^2$ for the global fit
from the minimum value is  quadratic in the  
deviation of the parameters
specifying the input parton distributions $\{ a_i \}$ from 
their values at the
minimum $\{ a_i^0 \}$. First one determines the best fit set
of parton distributions
 from the minimization of Eq. \ref{chi23}, that is,
including the contribution from correlated systematic
uncertainties, as opposed to the Offset method. Then to
estimate the associated uncertainty one 
 can write
\be
\Delta \chi^2\equiv\chi^2-\chi_0^2=\sum_{i=1}^n\sum_{j=1}^n
H_{ij}\lp a_i-a_i^0\rp\lp a_i-a_j^0\rp \ ,
\ee
where $H$ is the Hessian matrix, defined in
Eq. \ref{hessian}. 
 Standard linear propagation implies that the error on an observable
$\mathcal{F}\lc \{ q_i\}\rc$ is given by
\be
\label{hess1}
\lp \Delta \mathcal{F} \rp^2=\Delta\chi^2\sum_{i,j=1}^{N_{\parr}}
\frac{\partial \mathcal{F}}{\partial a_i}C_{ij}^{\stat}(\{ a_m\})
\frac{\partial \mathcal{F}}{\partial a_j} \ ,
\ee
where the covariance matrix of the parameters $C^{\stat}$
is again the inverse of the Hessian matrix, Eq. \ref{hessian},
and $\Delta \chi^2$ is the allowed variation in $\chi^2$.
Textbook statistics imply that one should have
$\Delta \chi^2=1$, however it has been argued that a higher value,
$\Delta \chi^2=100$ is required in order to estimate
the uncertainties in a faithful way, due to the
fact that data from different experiments is sometimes
incompatible \cite{incon}.

For practical purposes,
it is more numerically stable to diagonalize the covariance matrix
and work in the basis of eigenvectors, defined by
\be
\sum_{j=1}^{N_{\parr}} H_{ij}\lp \{a_l\}\rp v_{jk}=\lambda_k v_{ik}
\qquad i,k=1,\ldots,N_{\parr} \ .
\ee
One has to take into account that since variations in some 
directions in the parameter space lead to deterioration of the quality of the
fit far more quickly than others, the eigenvalues
$\lambda_k$ span several orders of magnitude.
In the Hessian method, one ends up with a set of $2N_{\parr}$ sets
of parton distributions
$S_i^{\pm}$. One can therefore propagate the
 uncertainty associated to the
parton distributions to any given observable 
$\mathcal{F}\lc \{ q_i \}\rc$ with the
following master formula, equivalent to Eq. \ref{hess1},
\be
\label{hess33}
\lp \Delta \mathcal{F} \rp^2 =\frac{1}{2}\lp \sum_{i=1}^{N_{\parr}} \lc 
\mathcal{F}(\{ S_i^+ \})-\mathcal{F}(\{S_i^-\})\rc^2\rp \ .
\ee
The drawbacks of this method are similar to those
of the Offset method: first of all one assumes
 that the linearized approximation in error propagation
is valid, and second errors estimated with this
method depend heavily on
 the functional form choosen for the
parametrization of parton distributions. Finally the
introduction of non-standard tolerance criteria
$\Delta \chi^2\ > 1$ does not allow to
give a statistically rigorous meaning to the
resulting uncertainties.

\begin{figure}[ht]
\begin{center}
\epsfig{width=0.45\textwidth,figure=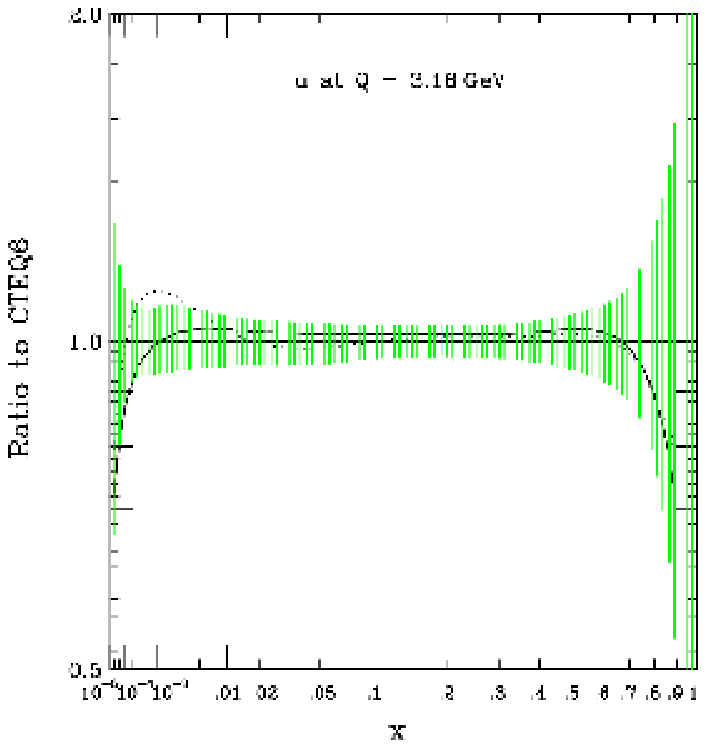}
\epsfig{width=0.51\textwidth,figure=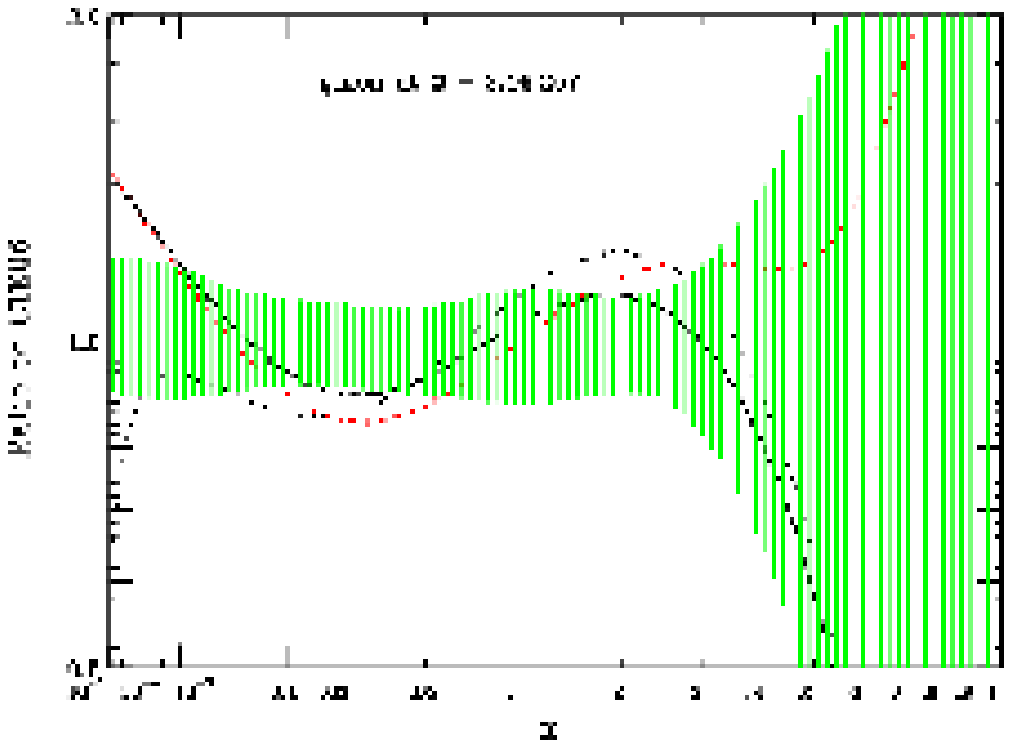}
\caption{}{\small Relative errors of parton distributions
in the recent CTEQ6 analysis \cite{cteq6}: up quark
parton distribution (left) and gluon parton
distribution (right).
\label{pdfunc_cteq}}
\end{center}
\end{figure}
 
The final technique that will be discussed is the
Lagrange multiplier method \cite{lagrange}, which overcomes some of
the drawbacks of the above methods, specially the
linearized approximations.
To investigate the allowed variation of a specific  physical observable
is more rigorous using this method that the previously
discussed ones. In the 
Lagrange multiplier
approach, one performs a global fit while constraining the values
of a physical quantity $\mathcal{F}\lc \{ q_i\}\rc$ 
in the neighborhood of their values
$\mathcal{F}^{(0)}$ obtained in the unconstrained global fit. 
The starting point is the best-fit set of parton distributions
$S_0$ characterized by the parameters $\{ a_i^{(0)}\}$.
The uncertainty associated to the observable $\mathcal{F}$
is estimated in two steps. First we use the Lagrange multiplier method
to determine how the minimum of $\chi^2 \lp\{ a_i \}\rp$
increases, as $\mathcal{F}$ deviates from the best estimate
$\mathcal{F}^{(0)}$, and then one determines the appropriate tolerance
of $\chi^2 \lp\{ a_i \}\rp$, $\Delta \chi^2 $. The first step is then
minimizing the constrained error function
\be
\Psi\lp \lambda,\{ a_i \} \rp=\chi^2 \lp\{ a_i \}\rp+
\lambda \mathcal{F}\lp \{ a_i \}  \rp \ ,
\ee
for various values of $\lambda$, following the chain
\be
\lambda_{\alpha}\to \mathrm{min}\lc \Psi\lp \lambda_{\alpha},\{ a_i \} \rp \rc
\to \{ a_i\lp \lambda_{\alpha}\rp \} \to \mathcal{F}_{\alpha}, \chi^2_{\alpha}.
\ee
This procedure generates a parametric relationship between $\chi^2$ and
the observable, $\mathcal{F}$, 
in terms of the parameter $\lambda$, so that given an 
allowed value of $\Delta \chi^2$, it is straightforward
to derive an allowed range for the observable $\Delta \mathcal{F}$,
without any linearized approximation.
In practice this procedure generates a sample of sets
of parton distributions 
$\{ S_{n}\}$ equal to the size of the parton distribution set
 parameter space $N_{\parr}$, since the minimization
is performed in each direction of the
parameter space, which are then
used to assess the range of variation of $\mathcal{F}$ 
allowed by the data, using Eq. \ref{hess1}.

\begin{figure}[ht]
\begin{center}
\epsfig{width=0.6\textwidth,figure=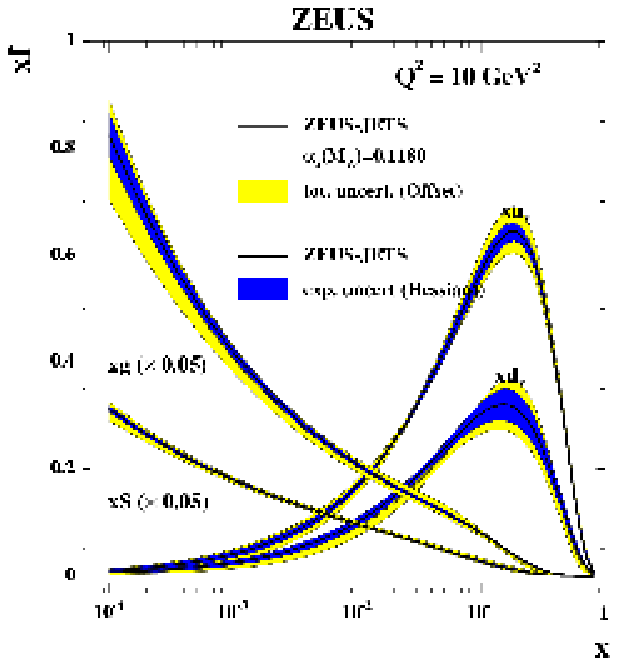}
\caption{}{\small Set of parton distributions with
uncertainties computed with different methods from
a recent QCD analysis of the ZEUS collaboration \cite{pdfzeus}.
\label{pdfunc_zeus}}
\end{center}
\end{figure}

It is clear that the Lagrange multiplier 
 method provides a sample of parton distribution  sets tailored
to assess the uncertainty associated
to the physical problem at hand.
This procedure does not involve some of the approximations involved in the
Hessian  and Offset methods, but
it still has the problem of the introduction of
non-standard tolerance criteria. 
However, this method suffers from a large practical
disadvantage, which is that the
 global fits of parton distributions must be repeated 
every time one needs to determine the
uncertainties of a different observable coming from
parton distributions.

In Figs. \ref{pdfunc_cteq} and \ref{pdfunc_zeus}
we show the errors associated to different parton
distributions from two different QCD global
analysis. Note that the gluon parton distribution
in particular has a rather large uncertainty, specially
at large and small $x$, where there are no direct
constrains from experimental data on its shape.
The valence parton distributions, on the other hand, are
known with higher accuracy since they are directly
constrained from the precise deep-inelastic
scattering data.

It must be emphasized that whatever method is used,
nowadays there is a common format to represent
the uncertainties in parton distribution. This is
accomplished by providing, on top of the {\it best-fit}
parton distribution set, additional sets of
{\it error} parton distributions, that span the
parameter space of the parton distributions allowed
by the corresponding experimental uncertainties, estimated with
some of the methods described above. With the sample of parton
distribution sets, one computes the uncertainty in any physical
quantity $\mathcal{F}\lc q_i\rc$ by means of Eq. \ref{hess33}.
All the modern sets from different global QCD analysis
of parton distribution function, including the error sets,
are available through the LHAPDF library \cite{lhapdf}. 

The standard approach introduced in this section for the
determination of unpolarized parton distributions
and the associated uncertainties is also used
for similar global QCD analysis which
involve different types of parton distributions, like
polarized parton distributions \cite{stefpol,aacpol,
blumpol}, which measure the fraction of the
total spin of the proton carried by the
different partons, or nuclear parton distributions 
\cite{nuclearpdf1,nuclearpdf2,nuclearpdf3}
which measure how parton distributions are 
modified in nucleus within heavy nuclei with respect to
those of the free nucleon. Also for parametrizations
of the photon \cite{photonpdf} and pion
\cite{pionpdf} parton distributions from experimental
data the standard approach is used. However, in all the
above cases experimental data is more scarce and with
larger uncertainties than in the case of
unpolarized structure functions.

In Section \ref{mcerr} we will introduce an alternative
approach to estimate faithfully the uncertainties associated
to a function parametrized from experimental data.
This approach (the Monte Carlo approach) can be
seen to be equivalent to the more
common technique to determine confidence levels
based on the $\Delta\chi^2=1$ condition, assuming
that in the latter case linear error propagation
is a good enough approximation. In Appendix \ref{mcerrequiv}
we show explicitely this equivalence within a simple model.

\newpage

~

\newpage

\chapter{The neural network approach: General strategy}
\label{general}

\begin{figure}[ht]
\begin{center}
\epsfig{width=0.7\textwidth,figure=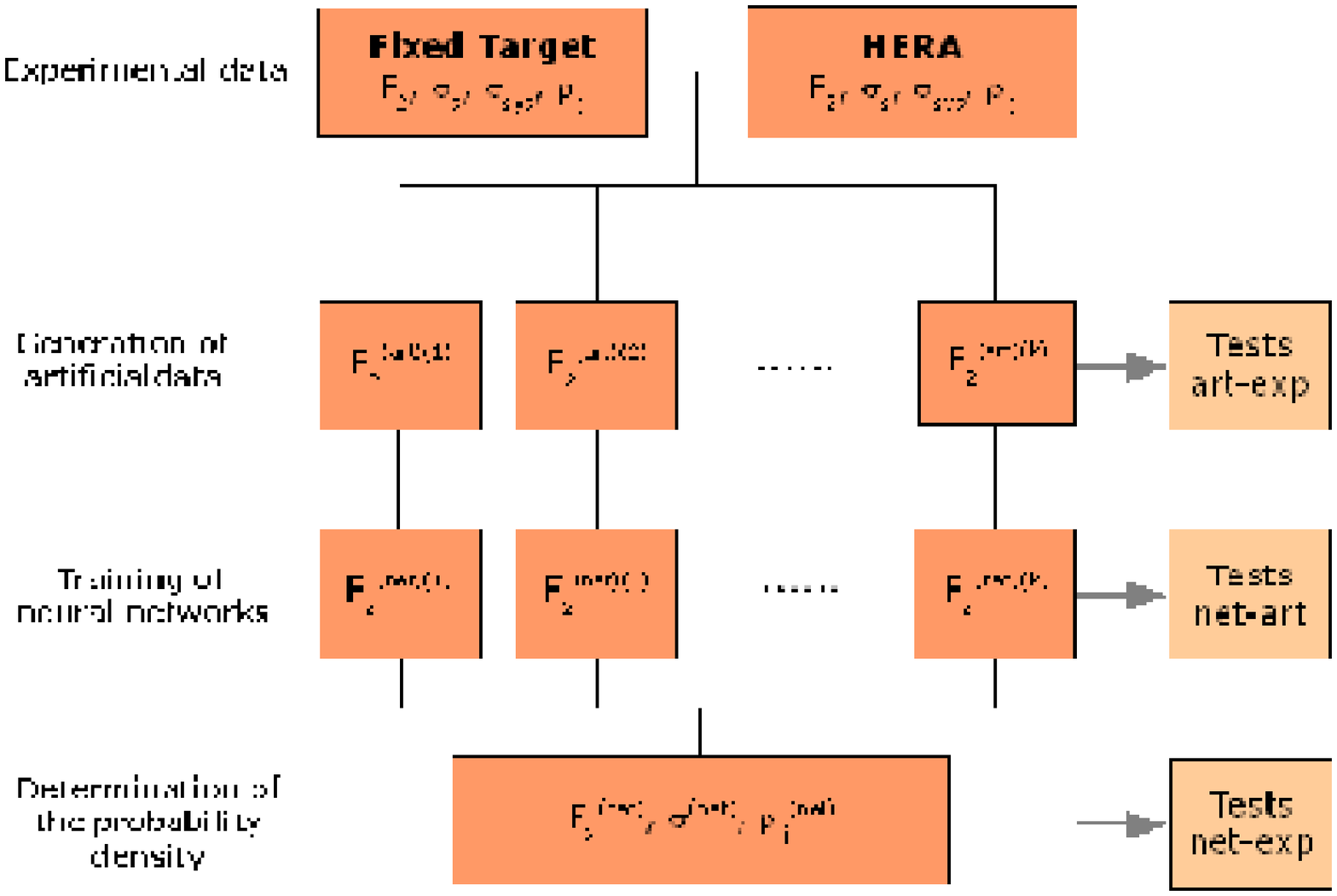}
\caption{}{\small Summary of the strategy presented in this
thesis to parametrize experimental data in an unbiased way with faithful
estimation of the uncertainties for the
particular case of the proton structure function $F_2(x,Q^2)$. 
As can be seen, the three main steps of our
approach are the Monte Carlo generation of replicas of
experimental data, the neural network training and
finally the statistical validation of the results.
\label{gen-strat}}
\end{center}
\end{figure}

This Chapter constitutes the core of the
present thesis: we describe the strategy that will be used
 to parametrize
experimental data in an unbiased way with faithful estimation
of the associated 
uncertainties, using a combination of Monte Carlo techniques
and neural networks as unbiased interpolants. 
This strategy has three main parts. In the first part, one
constructs a Monte Carlo sampling of the experimental data
for a given observable, which determines the probability 
measure of this observable over a finite set of data points.
Then we use neural networks as basic interpolating tools,
to construct a continuous probability density
for the observable under consideration. This
parametrization strategy has the advantage that it does not
require any assumption, rather than continuity, on the functional form
of the underlying law fulfilled by the given
observable. It also provides a faithful
estimation of the uncertainties, which can be then
propagated to other quantities without the
need of any linearized approximation.
Finally the third step  consists on the statistical validation
of the constructed probability measure in the space of the
parametrized function by means of suitable
statistical estimators. In summary, in this Chapter we will describe
how given a measured observable $F$, the associated probability measure
in the space of functions, $\mathcal{P}\lc F\rc$, can be constructed,
so that expectation values of arbitrary functionals of $F$, $\mathcal{F}[F]$
can be computed as with standard probability distributions, that is 
\be
\la \mathcal{F}[F]\ra\equiv\int 
\mathcal{D}F \mathcal{P}\lc F\rc \mathcal{F}[F] \ .
\ee
In Fig. \ref{gen-strat} we show 
a summary of our parametrization strategy for the
particular case of the proton structure function.

\section{Monte Carlo sampling of experimental data}
\label{mcerr}

In this first Section we describe how we construct a Monte Carlo
sample of replicas of the experimental data. 
We discuss also the statistical estimators which are used to validate the
results of the artificial data generation.
It has to be emphasized that the Monte Carlo sampling of the
data allows to estimate in a faithful way the errors
coming from experimental uncertainties associated to the
function to be parametrized. Note that this technique to
assess the uncertainties in a function is completely
independent of the method used to parametrize this
function, and in particular it is independent of the fact
that we use neural networks. This means that for 
example this technique could be used in standard fits
with polynomial functional forms to determine the
associated uncertainties. 

\subsection{Artificial data generation}
Let us 
consider a given observable $F$, which in full generality
can be assumed to depend on 
several kinematical variables $(x_1,x_2,x_3,\ldots)$. 
If $N_{\dat}$ is the number of experimental measurements
of the observable $F$, we define the i-th data
point as
\be
F_i^{(\exp)} 
\equiv F\lp x_{1i},x_{2i},\ldots\rp,\qquad  i=1,\ldots,N_{\dat} \ .
\ee
The first step in our
approach is to construct the probability measure
of the experimental data for these $N_{\dat}$ data points.
This is obtained by means of a Monte Carlo sampling of
 the experimental data in terms of replicas of the
original measurements.

There exist two equivalent techniques to generate the 
replica sample,
depending on how the information
on the experimental uncertainties is
available. In the first situation, information
on all the different independent sources of
uncertainty (statistical errors, correlated and uncorrelated
systematics, and normalization errors) is available.
This is the case for experiments like deep-inelastic
scattering for which
the detectors and the different sources of correlated
uncertainties are well understood.
In this case the k-th replica of the experimental data
$F^{(\art)(k)}$ is generated as
\be
F_i^{(\art)(k)}=\lp 1 +r_N^{(k)}\sigma_N\rp\lc F_i^{(\exp)}
+ r_{t,i}^{(k)}\sigma_{t,i}
+\sum_{j=1}^{N_{\sys}} r^{(k)}_{\sys,j}\sigma_{\sys,ji}\rc,
\quad i=1,\ldots,N_{\dat}, 
\quad k=1,\ldots,N_{\rep} \ .
\label{gen1}
\ee
Let us describe the different elements of the
above equation. 
 $N_{\rep}$ is the number of Monte Carlo replicas that needs
to be generated, while $F_i^{(\exp)}$ are the $N_{\dat}$
experimental values of the observable considered.
 The total uncorrelated uncertainty is defined as
\be
\label{totuncorr}
\sigma_{t,i}^2= \sigma_{\stat,i}^2+\sum_{j=1}^{N_{u}} 
\widetilde{\sigma}_{\sys,ji}^2 \ ,
\ee
as the quadratic sum of the statistical uncertainty $\sigma_{\stat,i}$ 
and the
$N_u$ uncorrelated systematic uncertainties $\widetilde{\sigma}_{\sys,ji}$. 
The contribution to the i-th data point
from the j-th source of correlated systematic
uncertainty is denoted by  $\sigma_{\sys,ji}$,
and $\sigma_N$ is the
total normalization uncertainty. Finally $r_N^{(k)},r_{t,i}^{(k)}$
and $r_{\sys,j}^{(k)}$ are zero mean univariate gaussian random
numbers with the same correlation as the corresponding uncertainties.
For example, within a given experiment, 
for all the data points of the k-th replica we use
the same random number for normalization uncertainties, $r_N^{(k)}$,
since this uncertainty is correlated 
among all these measurements.
The condition that these random numbers are gaussianly distributed
implies that
\be
\lim_{N_{\rep}\to \infty} \la r_i^{(k)}\ra_{\rep}=\mathcal{O}\lp\frac{1}{
N_{\rep}} \rp \ , \qquad \lim_{N_{\rep}\to \infty} 
\la r_i^{(k)2}\ra_{\rep}=1+\mathcal{O}\lp\frac{1}{N_{\rep}} \rp \ .
\ee

The covariance matrix is constructed from this
experimental information as
\be
\label{covmat}
{\rm cov}_{ij}=
\lp\sum_{k=1}^{N_{\sys}}\sigma_{\sys,ki}
\sigma_{\sys,kj}+F^{(\exp)}_iF^{(\exp)}_j
\sigma_N^2\rp+\delta_{ij}\sigma^2_{t,i} \ ,
\ee
and the correlation matrix
is then given by
\be
\rho_{ij}=\frac{ {\rm cov}_{ij}}{\sigma_{\tot,i}\sigma_{\tot,j}} \ ,
\label{cormat}
\ee
where the total error $\sigma_{\tot,i}$ for the i-th point is given by
\be
\sigma_{\tot,i}=\sqrt{\sigma_{t,i}^2+\sigma_{\sys,i}^2+ \lp F_i^{(\exp)}
\rp^2
\sigma_N^2} \ ,
\label{toterr}
\ee
and finally 
the total correlated uncertainty $\sigma_{\sys,i}$ is the sum of all 
correlated systematics,
\be
\sigma_{\sys,i}^2=\sum_{j=1}^{N_{\sys}} \sigma_{\sys,ji}^2 \ .
\label{totsyst}
\ee

The presence of correlated systematic uncertainties
which are not symmetric for some experiments
deserves further comment.
\label{asymm}
As it is well known~\cite{barlow1,barlow2,
dagostini,dagostini1}, asymmetric errors
cannot be combined in a simple multi-gaussian framework, and in
particular they cannot be added to gaussian errors in
quadrature. For
the treatment of  asymmetric uncertainties, we will follow the approach of
Refs.~\cite{dagostini,dagostini1},  which, on top of several 
theoretical advantages, 
is closest to the ZEUS error
analysis and thus adequate for a faithful reproduction of the ZEUS
data on deep-inelastic structure functions, that is were these
asymmetric uncertainties appear. 
In this approach, a data point with central value $x_0$ and
left and right  asymmetric
uncertainties $\sigma_R$ and $\sigma_L$ (not necessarily positive) 
is described by a symmetric gaussian distribution, centered
at 
\be
\la x \ra=x_0+\frac{\sigma_R-\sigma_L}{2}\label{xshift} \ ,
\ee
and with width
\be
\sigma_x^2=\Delta^2=\left(\frac{\sigma_R+\sigma_L}{2} \right)^2 .
\label{Delta}
\ee
The ensuing distribution can then be treated in the standard
gaussian way.

Once the sampling of the experimental data in terms of a set
of replicas of artificial data has been generated, it defines the
probability measure of the observable $F$ in those
data points where there exist experimental measurements.
From this probability density one can compute 
any estimator as with standard probability 
distributions. That is, if $\mathcal{A}\lc F\rc$
is any functional of the observable $F$
(the simplest example is the observable itself for
the i-th data point, $\mathcal{A}\lc F\rc=F_i$ )
then its average as computed from the probability measure
is given by
\be
\label{avrep}
\la A^{(\art)}\ra_{\rep}\equiv\frac{1}{N_{\dat}}
\sum_{k=1}^{N_{\rep}}A^{(\art)(k)} \ .
\ee

Experimental data on the observable $F$ might be
available without information on the
separation of the
different sources of systematic uncertainties, which
are then in general collected together in the correlation matrix 
$\rho_{ij}^{(\exp)}$ of the experimental data.
In this second case, the equivalent of Eq. \ref{gen1}
is given by
\be
\label{gen2}
F_i^{(\art)(k)}= F_i^{(\exp)}+ r_i^{(k)}\sigma_{\tot,i} \ ,
\ee
where $\{r_i^{(k)}\}$ are gaussianly distributed random numbers with the
same correlation as the experimental data points, that is,
they verify
\be
\frac{\la r_i^{(k)}r_j^{(k)}\ra_{\rep}-
\la r_i^{(k)}\ra_{\rep}\la r_j^{(k)}\ra_{\rep}}{
\sqrt{\la r_i^{(k)2}\ra_{\rep}-
\la r_i^{(k)}\ra^2_{\rep}}
\sqrt{\la r_j^{(k)2}\ra_{\rep}-
\la r_j^{(k)}\ra^2_{\rep}}}=\la r_i^{(k)}r_j^{(k)} 
\ra_{\rep}+\mathcal{O}\lp \frac{1}{N_{\rep}}\rp
=\rho_{ij}^{(\exp)} \ ,
\ee
where averages over replicas have been defined in
Eq. \ref{avrep}.
In this situation the covariance matrix can be 
computed using
\be
{\rm cov}_{ij}=\rho_{ij}\sigma_{\tot,i}\sigma_{\tot,j} \ .
\ee

The number of replicas of the experimental data $N_{\dat}$ that
needs to be generated with any of the two
equivalent approaches described above is determined by the condition that
the Monte Carlo sample reproduces the central values,
errors and correlations of the experimental data.
The comparison between experimental data and the
Monte Carlo sample can be made quantitative by the use
of statistical estimators that will be described in the
following Section.

\subsection{Statistical estimators: generated replicas vs.
experimental data}
In this Section we describe the statistical estimators
that are used to validate the results of the
replica generation. Let us recall that the probability measure
constructed in the previous section aims to reproduce
not only central values but also errors and correlations. Therefore
we must also validate how well errors and correlations
are reproduce by the replica sample. These estimators
are the following:

\begin{itemize}
\item Average  for each experimental point
\be
\la
 F_i^{(\art)}\ra_{\rep}=\frac{1}{N_{\rep}}\sum_{k=1}^{N_{\rep}}
F_i^{(\art)(k)}\ .
\ee 
\item Associated variance
\be
\label{var}
\sigma_i^{(\art)}=\sqrt{\la\lp F_i^{(\art)}\rp^2\ra_{\rep}-
\la F_i^{(\art)}\ra^2_{\rep}} \ .
\ee
\item Associated covariance and correlation
\be
\label{ro}
\rho_{ij}^{(\art)}=\frac{\la F_i^{(\art)}F_j^{(\art)}\ra_{\rep}-
\la F_i^{(\art)}\ra_{\rep}\la F_j^{(\art)}\ra_{\rep}}{\sigma_i^{(\art)}
\sigma_j^{(\art)}} \ .
\ee
\be
\label{cov}
\mathrm{cov}_{ij}^{(\art)}=\rho_{ij}^{(\art)}\sigma_i^{(\art)}
\sigma_j^{(\art)} \ .
\ee
The three above quantities provide the estimators of the
experimental central values, errors and correlations
which one extracts from the sample
of experimental data.
\item Mean variance and percentage error on central values 
over the
  $N_{\dat}$ data points.
\be
\la V\lc\la F^{(\art)}\ra_{\rep}\rc\ra_{\dat}=
\frac{1}{N_{\dat}}\sum_{i=1}^{N_{\dat}}\lp \la F_i^{(\art)}\ra_{\rep}-
F_i^{(\mrexp)}\rp^2 \ ,
\ee
\be
\la PE\lc\la F^{(\art)}\ra_{\rep}\rc\ra_{\dat}=
\frac{1}{N_{\dat}}\sum_{i=1}^{N_{\dat}}\lc\frac{ \la F_i^{(\art)}\ra_{\rep}-
F_i^{(\mrexp)}}{F_i^{(\mrexp)}}\rc \ .
\ee
We define analogously $\la V\lc\la
\sigma^{(\art)}\ra_{\rep}\rc\ra_{\dat} $, $\la V\lc\la
\rho^{(\art)}\ra_{\rep}\rc\ra_{\dat}$,  $\la V\lc\la
\mathrm{cov}^{(\art)}\ra_{\rep}\rc\ra_{\dat}$, \\ and $\la PE\lc\la
\sigma^{(\art)}\ra_{\rep}\rc\ra_{\dat}$, 
 $\la PE\lc\la
\rho^{(\art)}\ra_{\rep}\rc\ra_{\dat}$ and
 $\la PE\lc\la
\mathrm{cov}^{(\art)}\ra_{\rep}\rc\ra_{\dat}$, for errors,
correlations and covariances respectively.

These estimators indicate how close the averages over generated data
are to the experimental values.
Note that in averages over correlations and covariances one has
to use the fact that correlation and covariances matrices
are symmetric matrices, and thus one has to be careful to
avoid double counting. For example, the percentage error
on the correlation will be defined as
\be
 \la PE\lc\la
\rho^{(\art)}\ra_{\rep}\rc\ra_{\dat}=\frac{2}{N_{\dat}\lp 
N_{\dat}+1\rp}\sum_{i=1}^{N_{\dat}}
\sum_{j=i}^{N_{\dat}}\lc\frac{ \la \rho_{ij}^{(\art)}\ra_{\rep}-
\rho_{ij}^{(\mrexp)}}{\rho_{ij}^{(\mrexp)}}\rc \ ,
\ee
and similarly for averages over elements of the
covariance matrix.

\item Scatter correlation:
\be
r\lc F^{(\art)}\rc=\frac{\la F^{(\mrexp)}\la F^{(\art)}
\ra_{\rep}\ra_{\dat}-\la F^{(\mrexp)}\ra_{\dat}\la\la F^{(\art)}
\ra_{\rep}\ra_{\dat}}{\sigma_s^{(\mrexp)}\sigma_s^{(\art)}} \ ,
\ee
where the scatter variances are defined as
\be
\sigma_s^{(\mrexp)}=\sqrt{\la \lp F^{(\exp)}\rp^2\ra_{\dat}-
\lp \la  F^{(\exp)}\ra_{\dat}\rp^2} \ ,
\ee
\be
\sigma_s^{(\art)}=\sqrt{\la \lp \la F^{(\art)}\ra_{\rep}\rp^2\ra_{\dat}-
\lp \la  \la F^{(\art)}\ra_{\rep} \ra_{\dat}\rp^2} \ .
\ee 
We define analogously $r\lc\sigma^{(\art)}\rc$, $r\lc\rho^{(\art)}\rc$
and  $r\lc\mathrm{cov}^{(\art)}\rc$. Note that the scatter correlation
and
scatter variance are not related to the variance and correlation
Eqns. \ref{var}-\ref{cov}.
The scatter correlation indicates the size of the spread of data
around a straight line. Specifically $r\lc \sigma^{(\art)}\rc=1$
implies that $\la \sigma_i^{(\art)}\ra$ is proportional
to $\sigma_i^{(\mrexp)}$.
\item Average variance:
\be
\la \sigma^{(\art)}\ra_{\dat}=\frac{1}{N_{\dat}}
\sum_{i=1}^{N_{\dat}}\sigma_i^{(\art)} \ .
\label{avvar}
\ee
We  define analogously $\la\rho^{(\art)}\ra_{\dat}$ and
$\la\mathrm{cov}^{(\art)}\ra_{\dat}$,  as well as the
corresponding experimental quantities.
This quantities are interesting
because even if the scatter correlations
$r$ are close to 1 there could still
be a systematic bias in the estimators Eqns. 
\ref{var}-\ref{cov}. This is so since even if all scatter
correlations are very close to 1, it could be that
some of the  Eqns. 
\ref{var}-\ref{cov} where sizably smaller than its
experimental counterparts, even if being
proportional to them.

\end{itemize}

The typical scaling of the various quantities with the
number of generated replicas $N_{\rep}$
follows the standard behavior of
gaussian Monte Carlo samples \cite{cowan}. For instance, variances on central
values scale as $1/N_{\rep}$, while variances
on errors scale as $1/\sqrt{N_{\rep}}$. Also, because
\be
V[\rho^{(\art)}_{ij}]=\frac{1}{N_{\rep}}\lp1-
\lp\rho^{(\exp)}_{ij}\rp^2\rp^2 \ ,
\ee 
as can be checked using Eq. \ref{varcor} in Appendix
\ref{probrev},
 the
estimated correlation fluctuates more for small values
of $\rho^{(\exp)}$, and thus the average correlation tends
to be larger than the corresponding experimental value.

\section{Neural networks}

\subsection{Neural networks as unbiased interpolants}

\begin{figure}[ht]
\begin{center}
\epsfig{figure=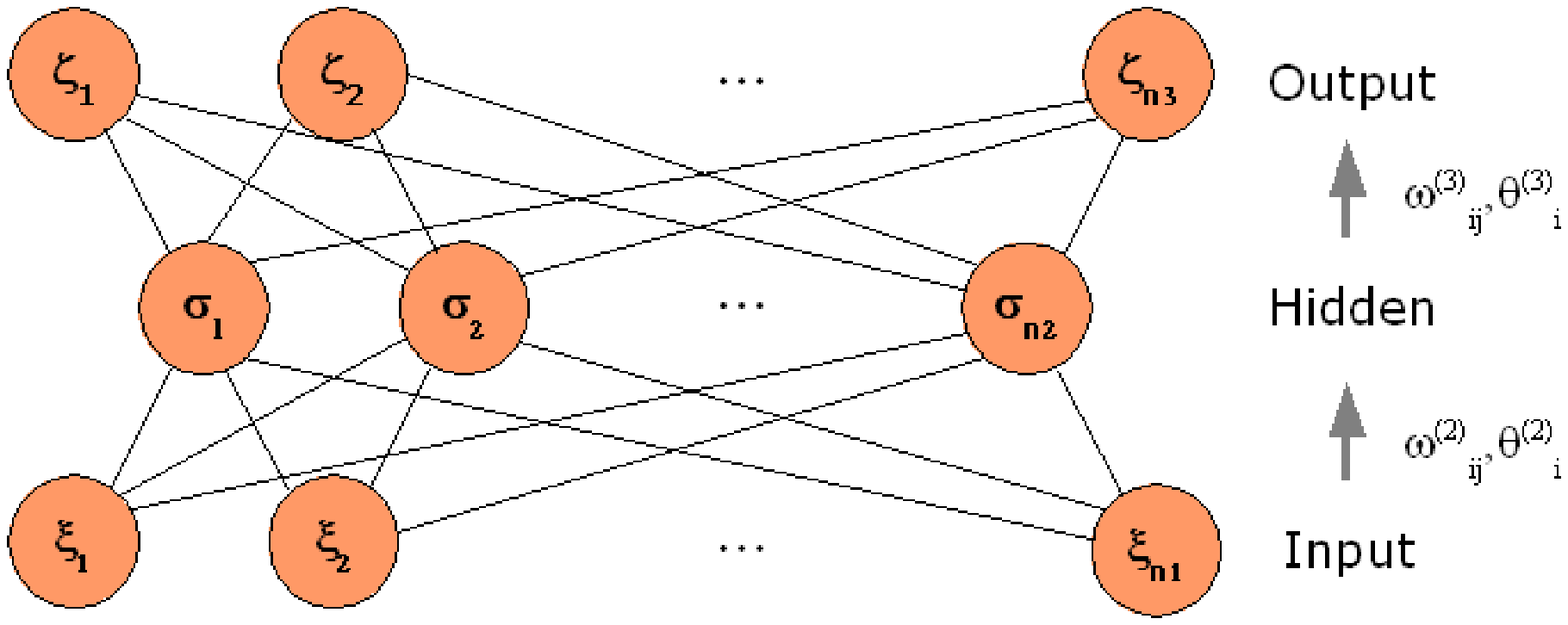,width=0.75\textwidth}
\caption{}{\small  Schematic diagram of a multi-layer feed forward
neural network. \label{nnplot}}
\end{center}
\end{figure}

Artificial neural networks \cite{nnbook,netrev,nnthesis}
 provide unbiased robust
universal approximants to incomplete or noisy data. An artificial neural 
network consists of a set of interconnected units ({\it neurons}).
The {\it activation} state $\xi_i^{(l)}$ of a neuron is determined as
a function of the activation states of the neurons connected to it.
Each pair of neurons $(i,j)$ is connected by a synapsis,
characterized by a {\it weight} $\omega_{ij}$.
In this thesis we will consider only multilayer feed-forward 
Perceptrons. These neural networks are organized in ordered
layers whose neurons only receive input from a previous layer.
In this case  the activation state of a neuron 
in the (l+1)-th layer is given by
\be
\xi^{(l+1)}_i=g\lp h^{(l+1)}_i\rp, 
\qquad i=1,\ldots,n_{l+1}, \qquad l=1,\ldots,
L-1 \ ,
\ee
\be
\label{hfun}
h^{(l+1)}_i= \sum_{j=1}^{n(l)}\omega_{ij}^{(l)}\xi^{(l)}_j+
\theta_{i}^{(l+1)} \ ,
\ee
where $\theta_{i}^{(l)}$ is the {\it activation threshold}
of the given neuron, $n_l$ is the number of neurons
in the l-th layer,
 $L$ is the number of layers that the neural network has, and
 $g(x)$ is the {\it  activation function} of the
neuron, which we will take 
to be a sigmoid, 
\be
g(x)=\frac{1}{1+e^{-x}} \ ,
\ee
except in the last layer, where we use a linear activation
function $g(x)$. This enhances the sensitivity of the neural network,
avoiding the saturation of the neurons in the last layer.
The fact that the activation function $g(x)$ is non-linear allows
the neural network to reproduce nontrivial functions.
An schematic diagram of a feed-forward artificial neural
network can be seen in Fig. \ref{nnplot}.

Therefore multilayer feed-forward neural networks can be viewed
as functions $F:\mathcal{R}^{n_1}\to \mathcal{R}^{n_L}$
parametrized by weights, thresholds and activation functions,
\be
\xi^{L}_j=F\lc \xi^{(1)}_i,\omega_{ij}^{(l)},\theta_i^{(l)},g\rc ,
\qquad 
j=1\ldots,n_L\ .
\ee
It can be proven that any continuous function, no matter
how complex, can be represented by
a multilayer feed-forward neural network.
In particular, it can be shown \cite{nnbook,nnthesis} that two
hidden layers suffice for representing an arbitrary
complicated function \cite{netref1}.

Artificial neural networks are a common tool in experimental
particle physics \cite{netrev,netrev2,netrev3} 
in applications like
pattern recognition, where in this context a {\it pattern}
typically means a experimental measurement as a function of
a set of several variables which characterize this event.
We are interested in artificial neural networks in order to
parametrize physical quantities in an unbiased way, that is,
without the need of any assumptions on its functional
form behavior.
These physical quantities to be parametrized can be either
direct experimental measurements (like in the case
of deep inelastic structure functions), or they can be
related to experimental data through a functional
dependence (like in the case of parton distribution
functions). Neural networks not only provide universal
unbiased interpolants for given physical quantity
in the region were there is experimental information,
they also have the property that their behavior in extrapolation
regions is not determined by their behavior in the data region,
as happens in fits with simple functional forms, so
they are also useful to assess the uncertainty in 
extrapolation regions (that is, in regions without information
from experimental measurements).

Therefore, the performance of neural networks
is worse for extrapolation than for
interpolation. That is, new input patterns
that are not a mixture of some of the
input training patterns cannot be
classified in an efficient way in terms
of the learnt patterns. In particular, for a feed forward
neural network with one hidden layer, the generalization error
(that is, the error in the classification of patterns which
are very different from the patterns used
in the training)  is of the order $\mathcal{O}\lp
N_{\parr}/N_{\dat}\rp$ \cite{netref2}, where $N_{\parr}$
is the number of neural network parameters and 
$N_{\dat}$ the number of training patterns.

Another interesting property of
artificial neural networks is that they
 are efficient in combining in an optimal way
experimental information from different measurements of the same
quantity. That is, when the separation of data points in the
space of possible input patterns is smaller than a certain correlation
length, the neural network manages to combine the corresponding
experimental information, thus leading to a determination of
the output pattern (in the present case the function to
parametrize) which is more accurate than the individual input
patterns (the measurements from the different experiments).

\subsection{Training strategies}
\label{trainstrat}

Training a neural network implies the minimization of
 a suitable statistical estimator. In this Section
we discuss the different estimators that are minimized
in the neural network training. In the next
Section we will analyze the minimization
algorithms that are used to implement this training.

For this thesis three different estimators have been
used:
the central values error function,
\be
\label{er1}
E_1^{(k)}=\frac{1}{N_{\dat}}
\sum_{i=1}^{N_{\dat}}\lp F^{(\net)(k)}_i-F^{(\art)(k)}\rp^2 \ ,
\ee
the diagonal statistical error function, 
\be
\label{er2}
E_2^{(k)}=\frac{1}{N_{\dat}}
\sum_{i=1}^{N_{\dat}}\frac{\lp F^{(\net)(k)}_i-F^{(\art)(k)}\rp^2}{
\sigma_{t,i}^2} \ ,
\ee
where $\sigma_{t,i}$ is the total uncorrelated uncertainty,
Eq. \ref{totuncorr},
and finally the error taking into account correlated systematics,
\be
\label{er3}
E_3^{(k)}=\frac{1}{N_{\dat}}\sum_{i,j=1}^{N_{\dat}} 
\lp F^{(\net)(k)}_i-F^{(\art)(k)}_i\rp 
\lp\lp \overline{\mathrm{cov}}^{(k)}\rp^{-1}\rp_{ij}\lp F^{(\net)(k)}_j-
F^{(\art)(k)}_j\rp \ ,
\ee
where in all the above expressions 
$F^{(\net)(k)}_i$ is the prediction of the
k-th neural network for the i-th data point.
Note that in the above equation we have defined the covariance matrix
${\overline\mathrm{cov}^{(k)}}_{ij}$ 
 with the normalization uncertainty
included as an overall rescaling of the error due to the normalization
offset of that replica,
namely, 
\be
\label{covmatnn}
{\overline\mathrm{cov}^{(k)}}_{ij}\equiv
\lp\sum_{p=1}^{N_{\sys}}{\overline\sigma^{(k)}}_{\sys,pi}{
\overline\sigma^{(k)}}_{\sys,pj}\rp+
\delta_{ij}{\overline\sigma^{(k)}}_{t,i} \ ,
\ee
with 
\be
\overline \sigma_{a,i}^{(k)}= (1+r^{(k)}_N\sigma_N)\sigma_{a,i} \ ,
\ee
where $r_N^{(k)}$ is the same as in Eq.~\ref{gen1}. 
Eq. \ref{covmatnn} is to be compared with the
total covariance matrix, Eq. \ref{covmat}. This is necessary in
order to avoid a biased treatment of the normalization
errors~\cite{dagostini}. 
In Appendix \ref{toymodel} we describe within an
explicitely solvable  model the effects of an
incorrect treatment of normalization uncertainties.

Several relations hold between these estimators.
Properties of minimization using the
 error function defined with the
covariance matrix, Eq. \ref{er3} can be found in
Refs. \cite{dagostini,alekhincme}.
For example, it can be shown that the following relation holds
\be
E_2^{(k)} \ge E_3^{(k)} \ ,
\ee
since in the latter case correlated systematic
uncertainties are included. The above constraint is useful
to cross-check that the experimental covariance matrix  Eq.
\ref{covmatnn} has been
correctly computed and inverted, without numerical instabilities.

The usefulness of neural networks
is due to the availability
of a training algorithm.
This algorithm allows one to select the values of weights
and thresholds such that the neural network
reproduces a given set of input-output data, also known
as {\it patterns}. This procedure is called
{\it learning } since unlike a standard fitting procedure,
there is no need to know in advance
the underlying rule which describes the data.
In our case the patterns that must be learned are those that
minimize any of the estimators described above.
 Therefore we need to define a suitable 
minimization strategy to train the neural networks. During the course
of this thesis we have used  different minimization algorithms,
which are described in
the following Section.

\subsection{Minimization algorithms}
\label{minimstratt}

Three different minimization algorithms have been
used during the course of this thesis: back-propagation,
genetic algorithms and conjugate gradient, which we now
discuss in turn.

\begin{itemize}
\item Back-propagation.

Back-propagation for neural network training was throughly described
in Ref. \cite{f2ns} and we summarize its main features here.
Back-propagation is a minimization strategy specially suited for the
training of neural networks with diagonal
error functions. Let us assume that the error
function used in the minimization is diagonal and a neural
network with a single
output layer, then one has
\be
\chi^2=\sum_{k=1}^{N_{\dat}}\lp F_k^{(\net)}-F_k^{(\exp)} \rp^2 \ ,
\ee
where $F_k^{(\net)}$ is the output of the neural network,
$F_k^{(\net)}=\xi^{(L)}_1$ when the input of the
neural network is $\xi^{(1)}_l=x_{lk}$.
If the error function is non-diagonal, back-propagation still
can be used but the corresponding expressions become  more complicated.

The error function $\chi^2$ is a function of the weights and the
thresholds of the neural network, and can be minimized by looking for
the direction of steepest descent in the space of weights and thresholds,
and modifying the parameters in that direction,
\bea
\label{replace}
\delta \omega_{ij}^{(l)}=-\eta \frac{\partial \chi^2}{\partial
 \omega_{ij}^{(l)} } \ ,
\\
\delta \theta_{i}^{(l)}=-\eta \frac{\partial \chi^2}{\partial
 \theta_{i}^{(l)} } \ ,
\nonumber
\eea
where $\eta$ is the so-called learning rate. The non-trivial 
result on which 
back-propagation training is based is that it can be shown that the
steepest descent direction is given by the following recursive
expression:
\be
\delta \omega_{ij}^{(l)}=\sum_{k=1}^{N_{\dat}}
\Delta_i^{(l)k}\xi_j^{(l-1)k}; \qquad i=1,\ldots,n_l, \qquad
j=1,\ldots,n_{l-1} \ ,
\ee
\be
\delta \theta_{i}^{(l)}=-\sum_{k=1}^{N_{\dat}}\Delta_i^{(l)k}, 
\qquad i=1,\ldots,n_l \ ,
\ee
where the information on the goodness of the fit is fed to the
last layer through
\be
\Delta_1^{(L)k}=g'\lp h_i^{(L)k}\rp\lc F_k^{(\net)}-F_k^{(\exp)}\rc \ ,
\ee
where the $h$ function is defined in Eq. \ref{hfun},
and then back-propagated to the rest of the network by
\be
\Delta_i^{(l-1)k}=g'\lp h_i^{(L)k} \rp
\sum_{j=1}^{n_l}\Delta_j^{(l)k}\omega_{ij}^{(l)} \ .
\ee
The procedure is iterated until a suitable converged criterion
is satisfied, as will be described in brief.
To avoid getting stuck in local minima, it is useful to add
a momentum term in the variations of the parameters. This means
that Eq. \ref{replace} should be replaced by
\bea
\delta \omega_{ij}^{(l)}=-\eta \frac{\partial \chi^2}{\partial
 \omega_{ij}^{(l)} }+\alpha \delta \omega_{ij}^{(l)}(\mathrm{last}) \ ,
\\
\delta \theta_{i}^{(l)}=-\eta \frac{\partial \chi^2}{\partial
 \theta_{i}^{(l)}}+\alpha \delta \theta_{i}^{(l)}(\mathrm{last}) \ ,
\nonumber \eea
where  (last) indicates the use of the last value of the update
for the weights and the thresholds before the current one and
$\alpha$ determines the size of the momentum term.
The drawbacks of back-propagation minimization is that it is not
specially suited for non-linear error functions, or for
error functions which depend on the neural network output
through convolutions.

One major  improvement of this minimization algorithm 
during the course of the present thesis was to implement
weighted training, in order in increase the efficiency of the
algorithm when the experimental data consist on many
different experiments. Since weighted training does not
depend on the minimization algorithm, we discuss it later
in this Section.

\item Genetic Algorithms.

Genetic algorithms\footnote{See for example Refs. 
\cite{gathesis,condensates,quevedo}
and references therein. In particular we follow closely the
description of genetic algorithms of Ref. \cite{quevedo}.}
are the generic name
of function optimization algorithms that do
 not suffer of the drawbacks that 
deterministic minimization strategies\footnote{
The canonical example of a deterministic
minimization algorithm is the MINUIT routine \cite{minuit}
from the CERN software libraries} have when applied to
problems with a large parameter space.
Genetic algorithms for
neural network training were introduced in this
context in \cite{condensates}, and
it is a  minimization strategy  that has been used in different
high energy physics applications \cite{gathesis}.
This method
is specially suitable for finding the global
minima of highly nonlinear problems, as is the one
we are facing in this thesis. Genetic algorithms
have several advantages with respect to deterministic
minimization methods:
\begin{enumerate}
\item They simultaneously work on populations of solutions,
rather than tracing the progress of one point through the
parameter space. This gives them the advantage of checking many regions
of the parameter space at the same time, lowering the
possibility that a global minimum gets missed.
\item No outside knowledge such as local gradient of the
minimized function is required.
\item They have a built-in mix of stochastic elements
applied under deterministic rules, which
improves their behavior in problems with many local
extrema, without the serious performance loss
that a purely random search would bring. 
\end{enumerate}

All the power of genetic algorithms lies in the repeated
application of three basic operations: mutation, crossover and
selection, which we describe in the following.
The first step is to encode the information of the parameter
space of the function we want to minimize into an
ordered chain, called {\it chromosome}. If $N_{\parr}$ is the size
of the parameter space, then a point in this parameter
space  will
be represented by a chromosome,
\be
\label{chain}
{\bf a}=\lp a_1,a_2,a_3,\ldots,a_{N_{\parr}}\rp \ .
\ee
In our case each {\it  bit} $a_i$ of a chromosome corresponds
to either a weight $\omega^{(l)}_{ij}$ or a threshold $\theta^{(l)}_i$
of a neural network.
Once we have the parameters of the neural network written
as a chromosome,
 we replicate that chain until we have a number $N_{\tot}$
of chromosomes.  Each chromosome has an associated  {\it fitness}
$f$, which is a measure of how close it is to the best possible
chromosome (the solution of the minimization problem
under consideration).
In our case, the fitness of a chromosome is given by
the inverse of the function to minimize
\be
f\lp {\bf a} \rp=\frac{1}{\chi^2\lp {\bf a}\rp} \ ,
\ee
so chromosomes with larger fitness correspond to
those with smaller value of the function to minimize 
$\chi^2$.

 Then we apply the three basic operations:
\begin{itemize}
\item Mutation: 

Select  randomly a {\it bit} (an element of the chromosome)
and {\it mutate} it. The size of the mutation is
called {\it mutation rate} $\eta$, and if the
k-th bit has been selected, the mutation is implemented as
\be
a_k \to a_k + \eta \lp r-\frac{1}{2} \rp \ ,
\ee
where $r$ is a uniform random number between 0 and 1. 
Over the span of several generations, even a stagnated 
chromosome position can become reactivated by mutation.
The optimal size of the mutation rate must be determined
for each particular problem, or it can be adjusted dynamically
as a function of the number of iterations.

\item Crossover:

This operation helps in obtaining a child generation which is
genetically different from the parents. Crossover means 
selecting at random pairs of individuals,  for each pair
determine randomly a crossover bit $s$, and from this crossover point
interchange the bits between the two individuals.

\item Selection:

Once mutations and crossover have been performed into the
population of individuals characterized by chromosomes, the selection
operation ensures that individuals with best fitness propagate into
the next generation of genetic algorithms.
Several selection operators can be used. The simplest method
is to select simply the $N_{\mathrm{chain}}$ chromosomes, out of the
total population of $N_{\mathrm{tot}}$ individuals, with best
fitness. Later we will describe a more efficient selection method
based on probabilistic selection.

\end{itemize}
The procedure is repeated iteratively until a suitable
convergence criterion is satisfied. Each iteration
of the procedure is called a {\it generation}.
 A general feature of 
genetic algorithms is that the fitness approaches the
optimal value within a relatively small number of
generations.

The theoretical concept behind the success of genetic algorithms
is the concept of patterns or {\it schemata} within the
chromosomes \cite{gabook}. 
Rather than operating only with $N_{\tot}$ individuals
in each generation, a genetic algorithm works with a much higher
number of schemata that partly match the actual chromosomes.
If for simplicity we consider chromosomes whose elements can take only binary
values, the concept of schemata means that a chromosome like
10110 matches $2^5$ schemata, such as {\tt **11*} or {\tt 1*1*0}.
The generalization of this concept to continuous parameters
is straightforward.
Since fit chromosomes are handled down to the next generation
more often than unfit ones, the number of copies of a certain schema
S associated with fit chromosomes will increase from one generation
to the next,
\be
n_S(t+1)=n_S(t)\frac{\bar{f}(S)}{\bar{f}_{\tot}} \ ,
\ee
where $\bar{f}(S)$ is the average fitness of all individuals whose 
chromosome match schema S and $\bar{f}_{\tot}$ is the average fitness
of all individuals. This implies that if we assume a certain
schema approximately giving all matching chromosomes a constant
fitness advantage $c$ over the average
\be
\bar{f}(S)\equiv \lp 1+c\rp \bar{f}_{\tot} \ ,
\ee
one obtains an exponential growth of the number of this schema
from one generation to the next,
\be
n_S(t)=n_S(0)(1+c)^t \ .
\ee

The above derivation is a rough approximation to the behavior
in realistic cases, and it needs to be corrected for
effects like mutation. To this purpose we define two measures
on schemata:
\begin{enumerate}
\item The defining length $\delta$ is the distance between
the furthest two fixed positions. In the above example,
the defining  length in {\tt 1*1*0} is $\delta=4$.
\item The order $o$ of a shema is the number of fixed positions
it contains. In the above example, $o=3$.
\end{enumerate}
With these measures, if $L$ is the length of a chromosome
(that is, the number of parameters $N_{\parr}$ of the function we 
are minimizing) the following bound can be derived
\be
\label{schemata}
n_S(t+1)\ge n_S(t)\frac{\bar{f}(S)}{\bar{f}_{\tot}}
\lc 1-\frac{\delta(S)}{L-1}-o\lp S\rp p_m\rc \ .
\ee
The first correction includes the effect of crossover and
the second implements the effects of mutations, since in a schema
of order $o$, there is a probability $\lp 1-p_m\rp^o \sim
\lp 1-op_m\rp$ that the schema survives mutation, where
$p_m=1/N_{\parr}$ is the probability to select a bit at random
out out the total $N_{\parr}$ bits of a chromosome chain.
A consequence of Eq. \ref{schemata} is that short, low-order schemata of
high fitness are the building blocks towards a solution of the
problem. During a run of the genetic algorithms,
the selection operator ensures that building blocks associated with fitter
individuals propagate through the population. It can be shown
\cite{gabook}
that in a population of size $N_{\tot}$, approximately 
$\mathcal{O}(N^3_{\tot})$
schemata are processed in each generation.

The basic genetic algorithms that we have introduced above can be
extended in many ways to address specific problems.
Several improvements over this basic version of the
algorithm have been implemented in the course of this
thesis.
The first one is the introduction of
multiple mutations $N_{\mathrm{mut}}\ge 2$. This is helpful 
to
avoid  local minima, thereby increasing the speed of 
training. It is crucial that  rates for these additional mutations are large,
in order to allow for jumps from a local minimum to
a deeper one. That is, after the first mutation is
performed, additional
mutations are performed
\be
a_l=a_l+ \eta_p \lp r-\frac{1}{2} \rp \ , \qquad p=2, \ldots,N_{\mathrm{mut}}
\ ,
\ee
with mutation rate $\eta_p$
and probability $P_p$. Each time a new mutation is performed,
a different bit $a_l$ of the chromosome chain is selected at random.

Second, we have introduced a
probabilistic selection for the mutated
chromosome chains, instead of the basic deterministic selection.
This is helpful in allowing for mutations which only become
beneficial after a combination of several
individual  mutations, and allows for a more efficient
exploration of the whole space of minima. 
Once one has the mutated population of $N_{\tot}$ individuals,
instead of selecting the $N_{\mathrm{chain}}$ individuals with smallest
$\chi^2$, one selects first the chromosome with smallest $\chi^2$,
with value $\chi^2\lp {\bf a}_1\rp $, and then selects the
remaining $N_{\mathrm{chain}}$ individuals according to the
probability
\be
P_i\lp \chi^2\lp {\bf a}_i\rp\rp=
\exp
\lp \frac{
-\lp \chi^2\lp {\bf a}_i\rp
-\chi^2 \lp {\bf a}_1\rp 
\rp}{T}\rp \ , \qquad i=2,\ldots,N_{\tot} \ ,
\ee
where $T$ is the {\it temperature} of the system, in analogy
with the Metropolis algorithm in Monte Carlo
simulations \cite{metropolis}. 
At high temperatures even the chromosomes
with bad fitness have some probability to propagate to the
next generation, while at low temperature one recovers the
deterministic selection rule.
With this selection criteria, individuals with accidentally bad
fitness but with relevant schemata can propagate to the
next generations. Of course, after a number of generations
only the individuals with good fitness will propagate through the
generations.

\item Conjugate Gradient

Conjugate gradient minimization exploits in an efficient way the
information on both the function to minimize $\chi^2\lp {\bf a}\rp$
and its gradient,  $\nabla \chi^2\lp {\bf a} \rp$, where as in 
Eq. \ref{chain} ${\bf a}$ stands for the set of weights and thresholds
that characterize the neural networks.
Starting with an arbitrary initial vector, ${\bf g}_0$ and letting
${\bf h}_0={\bf g}_0$, the conjugate gradient method
constructs two sequences of vectors from the recurrence
\be
\label{cgeq}
{\bf g}_{i+1}={\bf g}_i-\lambda_i {\bf A}\cdot {\bf h}_i, \qquad
{\bf h}_{i+1}={\bf g}_{i+1}+\gamma_i{\bf h}_i, \qquad i=0,1,2,\ldots
\ee
where we have assumed that the function to be minimized
can be approximated by a quadratic form
\be
\chi^2\lp {\bf a}\rp\sim C-{\bf B}\cdot {\bf a}+
\frac{1}{2} {\bf a}\cdot {\bf A}\cdot {\bf a}+\mathcal{O}\lp a^3\rp \ ,
\ee
and the scalars are defined as
\be
\label{hess2}
\lambda_i=\frac{{\bf g}_i \cdot {\bf g}_i}{{\bf h}_i\cdot {\bf A}
\cdot {\bf h}_i} \ ,
\ee
\be
\gamma_i=\frac{{\bf g}_{i+1}\cdot {\bf g}_{i+1}}{{\bf g}_i\cdot {\bf g}_i} ,
\ee
and the dimension of each of the above vectors is
$N_{\parr}$, the size of the parameter space.
With these definitions the following orthogonality and conjugacy
conditions hold,
\be
{\bf g}_i\cdot {\bf g}_j=0, \qquad {\bf h}_i\cdot {\bf A}\cdot
{\bf h}_j=0, \qquad {\bf g}_i\cdot {\bf h}_j=0, \qquad j < i \ .
\ee
If one knew the Hessian matrix ${\bf A}$ in Eq. \ref{hess2}, 
the conjugate gradient
method would take us to the minimum of $\chi^2\lp {\bf a}\rp$,
but in general this Hessian matrix
is not available. However, it can be proven the following property:
suppose we happen to have ${\bf g}_i=-\nabla \chi^2\lp
{\bf a}_i\rp$ at some point ${\bf a_i}$ of the parameter space. Now we proceed
from the point ${\bf a}_i$ along the direction ${\bf h}_i$ to the
local minimum of $\chi^2$ located at the point ${\bf a}_{i+1}$
and then set ${\bf g}_{i+1}=-\nabla \chi^2\lp
{\bf a}_{i+1}\rp$. Then the vector ${\bf g}_{i+1}$ constructed
this way is the same that the one that would have
been constructed from Eq. \ref{cgeq} if the Hessian matrix
had been known.

Once we have constructed a set of conjugate directions, the
minimization of the function $\chi^2\lp {\bf a}\rp$
is straightforward: starting from an initial point in the
parameter space ${\bf a}_1$, one has to find the
quantity $\alpha_k$ that minimizes
\be
\chi^2\lp {\bf a}_k+\alpha_k {\bf g}_k\rp
\ee
and then one sets
\be
{\bf a}_{k+1}={\bf a}_k+\alpha_k {\bf g}_k
\ee
and this procedure is repeated from $k=1$ to $k=N_{\parr}$.
If the function to be minimized was a quadratic form,
conjugate gradient minimization would find
the minimum in a single iterations. Obviously, for real
functions the length of the minimization is
typically larger. Conjugate gradient minimization is
also a suitable minimization algorithm for
nonlinear error functions, and the main reason
for its efficiency is the optimal use of the information
contained in the gradient of the function to be
minimized $\nabla \chi^2$.

\end{itemize}

\begin{figure}[ht]
\begin{center}
\epsfig{width=0.62\textwidth,figure=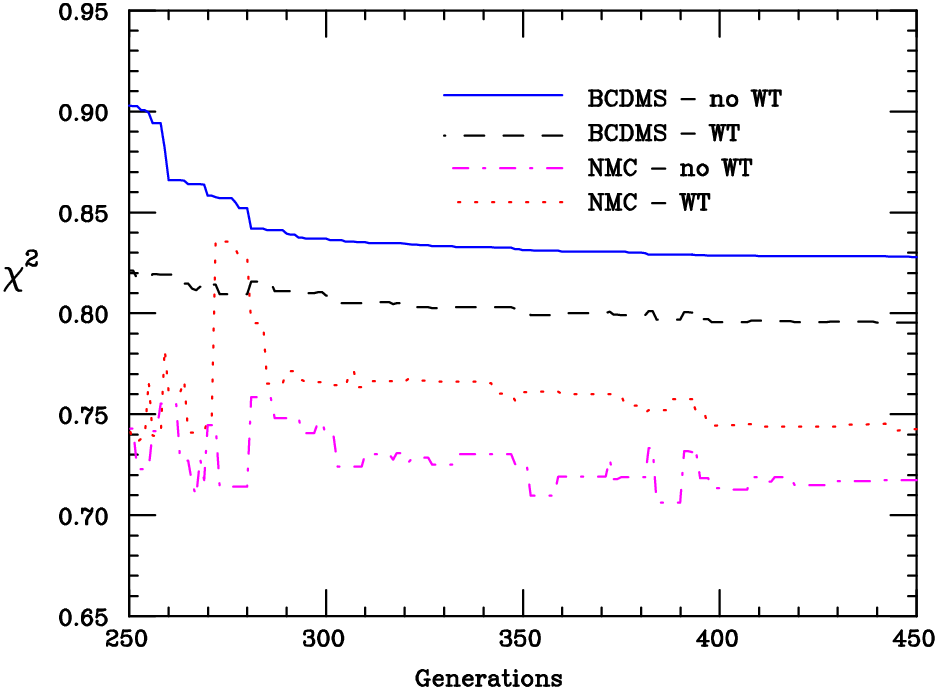}
\end{center}
\caption{}{\small A comparison of a training to  experimental data
with and without weighted training (WT). As can be seen in the
figure, at the end of the neural network training, the $\chi^2$
of the two experiments incorporated in the fit (BCDMS and NMC,
see Section \ref{nnqns_appl}) are more similar in the
case of  weighted training than in the case of
unweighted training.}
\label{weightedtraining}
\end{figure}

Once we have described the different minimization algorithms
that will be used for neural network training, we discuss
the weighted training procedure.
The essential idea about weighted training is that during the
neural network training some experimental points are given
more weight in the error function than others. This is useful
for example to learn in a more efficient way
those data points with a smaller error.
In weighted training, one separates the $N_{\dat}$ data points
into $N_{\sets}$ sets, each with $N_l,l=1,\ldots,N_{\sets}$
data points. A typical partition is that each set
corresponds to each different experiment incorporated
in the fit, $N_{\mathrm{sets}}=N_{\exp}$, but  
arbitrary partitions can be considered, like different
kinematical regions. The cross-correlations between points corresponding
to different sets are neglected.
The weighted error
function that is minimized is then
\be
\chi^2_{\mathrm{minim}}=\frac{1}{N_{\dat}}
\sum_{l=1}^{N_{\mathrm{sets}}}N_lz_l\chi^2_l \ ,
\ee
where $z_l$ is the relative
weight assigned to each of the sets and $\chi^2_l$
is the error function, either Eq. \ref{er2} or \ref{er3},
for the data points that belong to the l-th set.
 There are several ways to select the relative
weights $z_l$ of each experiment. A possible parametrization of the values
of $z_l$ is
\be
z_l=a_l\lp \frac{\chi^2_l}{\chi^2_{\max}}\rp^{b_l} \ ,
\ee
where $\chi^2_{\max}=\mathrm{max}_l \{ \chi^2_l\}$ and
where $a_l,b_l$ are to be determined in a case-by-case basis.
If $b_l=0$ the relative weights of each set is kept fixed during the
training, and if $b_l\ne 0$ the relative weights can be dynamically
adjusted during the training so that sets with larger $\chi^2$ 
have associated a higher relative weight $z_l$.
Note that the final $\chi^2$, Eq. \ref{chi2tot},
 is the standard unweighted one, and that the
minimization of 
$\chi^2_{\mathrm{minim}}$  during the neural
network training training is only a useful strategy
to obtain a more even distribution of $\chi^2_l$ among the
different data sets. In Fig. \ref{weightedtraining}
we show with an example of  Section \ref{nnqns_appl},
the parametrization of the nonsinglet parton distribution
$q_{NS}(x,Q^2_0)$,
how weighted training helps in obtaining more similar
values for the
$\chi^2$ of the different experiments incorporated in the fit.

An important issue to optimize the efficiency of the neural
network training is the choice of the
optimal architecture or {\it topology} of the neural network.
The choice of the
architecture of the neural network (the number of layers $L$ and
the number of neurons $n_l$ in each layer) cannot be derived from
general rules and  must be tailored to each specific 
problem. An essential condition is that
 the neural network has to be redundant, that is,
it has to have a larger number of parameters than the
minimum number required to satisfactorily fit the
patterns we want to learn, in this case 
experimental data.
However, the architecture cannot be arbitrarily large
because then the training length becomes very large.
A suitable criterion to choose the optimal
architecture is to select the architecture which is next to the
first stable architecture, that is, the first architecture that can
fit the data and gives the same fit than an architecture with one
less neuron. This way one is confident that the neural network
is redundant for the problem under consideration. 

A more systematic approach to determine the optimal
architecture of a neural network which 
parametrizes a function $F$ is related to 
how many terms are needed in an expansion of
$F$ in terms of the activation function $g$.
Starting from a large neural network, we can reduce
the architecture up to the optimal size with the
{\it weight decay} approach, in which weights which are
rarely updated are allowed to decay according to
\be
\delta_{ij}^{(l)}=-\eta \frac{\partial \chi^2}{\partial \omega_{ij}^{(l)}}
-\epsilon \omega_{ij}^{(l)} \ ,
\ee
where $\epsilon$ is the decay parameter, typically a  small number.
This corresponds to adding an extra complexity term to the
error function \cite{decayweight}, 
\be
\chi^2\to \chi^2+\frac{\epsilon}{2\eta}\sum_{i,j,l}\omega_{ij}^{(l)2} \ ,
\ee
that is, larger weights lead to larger contributions in the
error function. A more advanced complexity terms is
\be
\chi^2\to \chi^2+\lambda \sum_{i,j,l}
\frac{\omega_{ij}^{(l)2}/\omega_0^2}{1+\omega_{ij}^{(l)2}/\omega_0^2} \ ,
\ee
which again penalizes those weights whose absolute value
is larger. With this technique, the network gets forced
to only contain the weights that are really needed to
represent the problem under consideration.

For each training, the parameters that
define the behavior of the neural networks,
its weights and thresholds, are initialized at random.
To explore in a more efficient way the space of minima,
it has been checked to be specially useful to
initialize randomly not only the values
of the neural network parameters  but also the same range in
which these parameters are initialized.
That is, the parameter 
initialization range is determined at random
between
\be
\lc -\la \omega \ra-\sigma_{\omega}, \la \omega \ra+\sigma_{\omega}\rc \ ,
\ee
and
\be
\lc -\la \omega \ra+\sigma_{\omega}, \la \omega \ra-\sigma_{\omega}\rc \ ,
\ee
where $\la \omega \ra$ is the average value of the
neural network parameters and $\sigma_{\omega}$ is the
associated variance.

It has to be taken into account that
a minimization strategy is defined not only by an algorithm
to vary parameters so that a given quantity is
minimized, but also by a convergence condition
that determines when the minimization is stopped.
Several stopping criteria, each with its own
advantages and drawbacks, have been considered during the
course of this thesis.

A first criterion is the dynamical stopping of the training.
With this criterion, for each replica, we 
stop the training either when the condition $E^{(k)}\le
\chi^2_{\mathrm{stop}}$
is satisfied or, if the previous condition
cannot the fulfilled, when the maximum number of iterations
of the minimization algorithm $N_{\max}$ is
reached. That is, the length of the neural network
training is different for each replica. 
The value of the $\chi^2_{\mathrm{stop}}$ parameter
is determined by the value of the total $\chi^2$, Eq. \ref{chi2tot},
that defines the parametrization.
The dynamical stopping of the training allows to avoid
both overlearning and insufficient training, as
discussed in detail below.

 The method to define the optimal value
for $N_{\max}$ is the following: first perform
a training with a very large value for $N_{\max}$. 
For $\overline{N}_{\rep}$ replicas, the condition
 $E^{(k)}\le
\chi^2_{\mathrm{stop}}$ will not be satisfied due
to statistical fluctuations in the
generation of the data replicas. 
If $E^{(k)}_{N_{\max}}$ is the final value of the error
function at the end of the
training of the k-th replica, which has not reached the
dynamical stopping criterion,
then determine for each replica 
which was the iteration $\mathrm{it}$ for which
the condition
\be
E^{(k)}_{N^{(k)}_{\mathrm{it}}}\le \lp 1 +r_{\chi^2}\rp
E^{(k)}_{N_{\max}} , \qquad N^{(k)}_{\mathrm{it}}\le 
N_{\max} \ ,
\ee
was verified, were $r_{\chi^2}$ is the required tolerance. 
The new and optimal value for
$N_{\max}$ is determined then
as the average value of such iterations.
\be
N_{\max}=\frac{1}{\overline{N}_{\rep}}\sum_{k=1}^{\overline{N}_{\rep}}
  N^{(k)}_{\mathrm{it}} \ ,
\ee
where the sum is over those replicas which have
not been dynamically stopped.
With dynamical stopping of the training, the final distribution
of error functions $E^{(k)}$ is even and there is no problem
of overlearning, as will be discussed in brief.

An alternative criterion to stop the neural network 
minimization is
fixing the maximum number of iterations of the minimization
algorithm to a value large enough so that once is sure that
within a given tolerance the fit has converged. However, this
approach does not take into account that the length of the training of each
replica depends on each precise replica, so it is not
computationally efficient and might lead to overlearning.

The rigorous statistical method to determine the
value of $\chi^2_{\stopp}$ with dynamical stopping
of the training is the so called {\it overlearning} criterion.
Since the number of parameters of the neural network parametrization is
large, in principle, without the presence of inconsistent data, the
final $\chi^2$ could be lowered to arbitrarily low values for
a large enough neural network.
The overlearning criterion to determine the length of the training
states that the training should be stopped when the
neural network begins to overlearn, that is, it begins to follow
the statistical fluctuations of the experimental data rather
than the underlying law. The onset of overlearning can
be determined by separating the data set into two
disjoint sets, called the {\it training} set and
the {\it validation} set. One then minimizes the
error function, Eq. \ref{er3}, computed only with the
data points of the training set, and analyzes the dependence
of the error function  of the validation set
as a function of the number of iterations.

Then one computes the total $\chi^2$, Eq. \ref{chi2tot}, 
for both the training, $\chi^2_{\mathrm{tr}}$, 
and validation, $\chi^2_{\mathrm{val}}$, subsets. It might 
turn out that 
statistical fluctuations are large, and one has to average over
a large enough number of partitions to obtain stable
results. The onset of overlearning is determined as
the point in the neural network
 training such that the $\chi^2_{\mathrm{val}}$ of the
validation set saturates or even rises while the $\chi^2_{\mathrm{tr}}$
of the training set is still decreasing. This implies that
the neural network is learning only statistical fluctuations,
and signals the point where the training should be stopped.

A drawback of this approach is that one needs to assume that
the training subset  reproduces the main features
of the full set of data points. While this is the case in global
fits of parton distributions, where in each kinematical region there
are several overlapping measurements, in other physical
situations experimental data
is much more scarce and the overlearning criterion cannot be used to
determine the length of the training.

\begin{figure}[ht]
\begin{center}
\epsfig{width=0.62\textwidth,figure=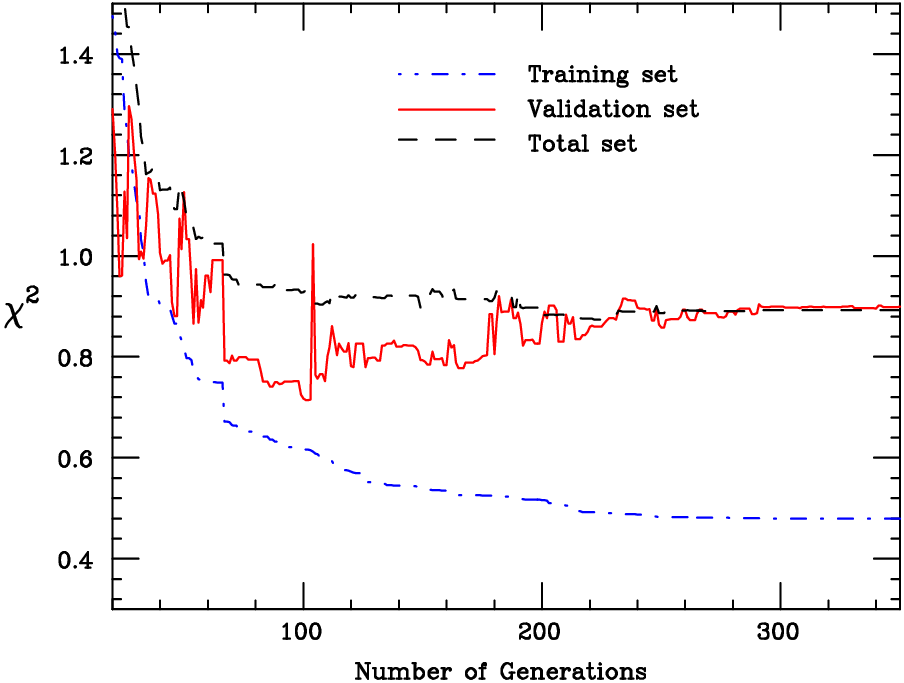}  
\end{center}
\caption{}{\small An example of a neural network learning
process with overlearning. The point where the $\chi^2$
as computed with the validation set begins to rise
while the $\chi^2$ of the training set is still
decreasing is the sign of this overlearning.}
\label{thesis_overlearn}
\end{figure}

In Fig. \ref{thesis_overlearn} we show the expected behavior
for the onset of overlearning. One can see that while
the $\chi^2$ of the validation set begins to increase,
the  $\chi^2$ of the training set is still decreasing.
This point signals the onset of overlearning, that is, the fact
that the neural network is learning statistical
fluctuations rather than the underlying law in the
experimental data. Note also that for some problems, like
for example those in which the data points are
very dense, overlearning could not be possible.

An alternative criterion to determine the optimal
$\chi^2$ which defines the neural network training is the
so-called {\it leave-one-out} strategy \cite{leaveoneout}.
In this strategy, out of the total $N_{\dat}$ data points,
one selects one data point at random and leaves it out of
the training of all the remaining points. One computes then
the prediction of the neural network for this point that has
been left out. This procedure is repeated over
all the data points, and the total $\chi^2$ as computed
with the predictions for those points that have
been left out is the average $\chi^2$ for a pure
neural network prediction for a point in the
same range. If the underlying law
followed by the data points is clear,
then the $\chi^2$ of a prediction can be
as good as the total $\chi^2$ of the trained data points, and
allows to determine which is the optimal value
of the total $\chi^2$ to aim for in the neural network training.
The conditions for the {\it leave-one-out} strategy can be
modified to take into account different types of correlations
between the data points, for example choosing
sets of independent points at random instead
of single data points.

\subsection{Implementation of theoretical constraints}
\label{constraints}
In addition to experimental measurements, in general we have 
additional theoretical information concerning the function we want
to parametrize, like for example kinematical constraints. Since 
this theoretical information is usually essential
to determine the shape of the function that is
going to be parametrized, it
 must be incorporated in our fit. 
In this section we discuss
the different approaches to incorporate theoretical
information in our neural network fits. These theoretical
constraints are specially useful in general to
constrain the shape of the parametrization in regions where there
is no experimental data.
The methods used in the present thesis have been the following:
\begin{enumerate}
\item The Lagrange multipliers method:

 If the function to be parametrized, $F$ has
to verify a set of $N_{\con}$ theoretical
constraints $\mathcal{F}_i\lc F_i\rc=
\mathcal{F}_i^{(\theo)}$,
one can use the Lagrange multiplier technique, which consist on
minimizing for each replica the standard error function while constraining the
function to satisfy the theoretical requirements,
\be
\chi^{2(k)}_{\tot}=\chi^{2(k)}_{\dat}+\sum_{i=1}^{N_{\con}}\lambda_i \lc
\mathcal{F}_i\lc F_i^{(\net)(k)}\rc-\mathcal{F}_i^{(\theo)}\rc^2 \ ,
\ee
where $\chi^{2(k)}_{\dat}$ is the contribution from the
experimental errors, Eqns. \ref{er2} or \ref{er3}, and
where the Lagrange multipliers $\lambda_i$ are computed via
a stability analysis. They have to be such that the
constraints are imposed with the required accuracy but at the
same time the space of minima of $\chi^{2(k)}_{\tot}$
has to be close to that of the data error function,
$\chi^{2(k)}_{\dat}$.

\item Pseudo-data:

 If the constraints to be imposed of the
function $F$ are local, they can be implemented as if they were
data points. That is, if the theoretical constraint is
\be
F\lp x_1,x_2,\ldots,\tilde{x}_j,\ldots\rp=b_j \ ,
\ee
then the minimized quantity should be
\be
\chi^{2(k)}_{\tot}=\chi^{2(k)}_{\dat}
+\sum_{j=1}^{N_{\con}}\frac{1}{\sigma_j^2} \lc
F^{(\net)(k)}\lp x_1,x_2,\ldots,\tilde{x}_j,\ldots\rp-b_j\rc^2 \ ,
\ee
where $\sigma_j$ is the accuracy with which the corresponding constraint
must be satisfied. This approach differs from the
Lagrange multiplier approach in that the accuracy to which the
constraints are to be satisfied is 
fixed by several requirements, for example,
by requiring the error $\sigma_i$ to be a fixed fraction of the
average error of the experimental data points.
Note that this is equivalent to adding to the experimental data set
artificial data points with central value 
$F\lp x_1,x_2,\ldots,\tilde{x}_j,\ldots\rp=b_j$
and total uncorrelated error $\sigma_j$. The addition of artificial
data points is helpful to constrain the shape of the
parametrized function in regions without experimental data, for 
example near kinematical thresholds.

\item Hard-wired parametrization: 

Another method
is to hard-wire the constraint in the neural
network parametrization. This method is specially useful
to implement the vanishing of the function for some
kinematical region. If we have
\be
F\lp x_1,x_2,\ldots,\tilde{x}_j,\ldots\rp=0 \ ,
\ee
then we redefine the function to be parametrized to be
\be
F^{(\net)(k)}\lp x_1,x_2,\ldots\rp\equiv\prod_{j=1}^{N_{\con}}\lp x_j-
\tilde{x}_j\rp^{n_j}\overline{F}^{(\net)(k)}\lp x_1,x_2,\ldots\rp \ ,
\ee
where now is the function $\overline{F}^{(\net)}$ the one that is
going to be parametrized with a neural network.
The introduction of a partial functional form 
dependence in the parametrization does not mean
that there is any functional form bias, since one can check the the results
of the fit do not depend on the values of $n_j$, provided
they verify $n_j>0$. The default values
of the $n_j$ parameters have to be determined from a stability
analysis to assess which is the optimal neural
network training {\it preprocessing}.

\end{enumerate}

All of the above techniques can be combined for a given
parametrization.
Using any of these techniques, it has to be shown that in each
case the result of the fit is modified as expected by the
implementation of the kinematical constraints. For example,
the dominant contribution to $\chi^2_{\tot}$ has to
be always that of the experimental data points.

\section{Validation of the results}
The third step of our strategy consists on the validation of the
results of the neural network training using suitable
statistical estimators. These results are defined by the sample
of trained neural networks which constitutes the sought-for
probability measure for the observable $F$, $\mathcal{P}\lc F\rc$, from
which  expectation values for an arbitrary functional of
it, $\mathcal{F}[F]$, can be computed using
\be
\la \mathcal{F}[F]\ra=\int \mathcal{D}F \mathcal{P}\lc F\rc \mathcal{F}[F]
=\frac{1}{N_{\rep}}\sum_{k=1}^{N_{\rep}}\mathcal{F}[F^{(\net)(k)}]
 \ ,
\ee
just as with standard probability measures.
We define now such estimators, that can be
divided into two categories: statistical estimators
to assess how well the sample of trained neural networks 
reproduce the features of the experimental data, 
and statistical estimators which assess the stability of a given
fit with respect some parameters, for example the
number of trained neural networks. 

\label{validation}

\subsection{Statistical estimators: probability measure 
vs. experimental data}
\label{estimators}
\label{est_fit}
Now we describe the statistical estimators that we
use to assess the quality of
the constructed probability measure in the
space of the observable $F$ by comparing with
the experimental data.
The goodness of the final fit
is measured with the total $\chi^2$, as constructed from the full experimental
covariance matrix
\be
\label{chi2tot}
\chi^2=\frac{1}{N_{\dat}}\sum_{i,j=1}^{N_{\dat}}\lp \la
F_i^{(\net)}\ra_{\rep}-
F_i^{(\mrexp)}\rp \lp \mathrm{cov}^{-1}\rp_{ij} \lp\la
F_j^{(\net)}\ra_{\rep}-
F_j^{(\mrexp)}\rp \ ,
\ee
where now the total experimental covariance matrix
Eq. \ref{covmat} includes the contribution from the
normalization uncertainties.
If experimental errors were correctly estimated and there were
no incompatibilities between measurements, on statistical
grounds one would expect $\chi^2\sim 1$, with deviations from
this value scaling with the number of data points as $1/\sqrt{N_{\dat}}$.
To be precise, instead of $N_{\dat}$ one should have the
number of degrees of freedom $N_{\mathrm{dof}}$, the
difference between the number of data points and the
number of free parameters of the theoretical model
$N_{\parr}$, and the standard deviation of the $\chi^2$
distribution is given by $\sigma_{\chi^2}=\sqrt{2/N_{\mathrm{dof}}}$.

Another important estimator is the
average error,
\be
\label{avchi2}
\la E\ra=\frac{1}{N_{\rep}}\sum_{k=1}^{N_{\rep}}E^{(k)} \ ,
\ee
where $E^{(k)}$ is either Eq. \ref{er2} or Eq. \ref{er3},
depending on the estimator which has been used in the
neural network training.
It is instructive to estimate the typical values that the
average error Eq. \ref{avchi2} can take. We will compute
now Eq. \ref{avchi2} in a simplified
model, in which correlated systematics are neglected. 
Let us consider a set of measurement $m_i$ of $F$. Then if
one assumes that experimental measurements are
distributed gaussianly around the true value $t_i$
of the observable, one has
\be
m_i=t_i+\sigma_is_i \ ,
\ee
where $\sigma_i$ is the total error and 
$s_i$ a univariate zero mean gaussian number.
The k-th replica of generated artificial data $g^{(k)}$ is as in
Eq. \ref{gen1} without correlated uncertainties, 
\be
g_i^{(k)}=m_i+r_i^{(k)}\sigma_i=t_i+(s_i+r_i^{(k)})\sigma_i \ ,
\ee
where $r_i^k$ is another univariate zero mean gaussian random 
number. Now let us assume that the best fit neural networks are
distributed around the true values of the
observable $F$ with error $\hat{\sigma}_i$.
For the k-th neural network $n^{(k)}$ we will have
\be
n_i^{(k)}=t_i+l_i^{(k)} \hat{\sigma}_i \ ,
\ee
where $l_i^{(k)}$ are highly correlated to both
$r_i^{(k)}$ and $t_i$.
For a large enough set of measurements, and a large enough
number of generated replicas, so that
correlations between $s_i$ and $r_i^{(k)}$ can be
neglected, it can be seen that the
average error is given by,
\be
\la E \ra_{\rep}=2+\la \frac{\hat{\sigma}^2}{\sigma^2}\ra_{\dat}
-2\la\la rl\ra_{\rep} \frac{\hat{\sigma}}{\sigma} \ra_{\dat} \ ,
\ee
where the error function is the diagonal error, Eq. \ref{er2}.
Note also that within the same model, the error function of the
k-th net as compared to the experimental measurement, defined
as
\be
\label{ertilde}
\widetilde{E}^{(k)}_2=\frac{1}{N_{\dat}}\sum_{i=1}^{N_{\dat}}
\frac{\lp m_i-n_i^{(k)}\rp^2}{\sigma^2_{\stat,i}} \ ,
\ee
has as average
\be
\la \widetilde{E} 
\ra_{\rep}=1+\la \frac{\hat{\sigma}^2}{\sigma^2}\ra_{\dat}\ge 1
\ ,
\ee
and finally   in the case that the neural network
prediction coincides with the true value of the function, $\hat{\sigma}_i=0$,
the average error is $\la \tilde{E}\ra_{\rep}=1$, just as expected from
textbook statistics.

Now we define the estimators which are used to assess how well
the sample of trained neural networks reproduce the sample
of experimental data. These estimators are
\begin{itemize}
\item Average  for each experimental point
\be
\la
 F_i^{(\net)}\ra_{\rep}=\frac{1}{N_{\rep}}\sum_{k=1}^{N_{\rep}}
F_i^{(\net)(k)}\ .
\ee 
\item Associated variance
\be
\sigma_i^{(\net)}=\sqrt{\la\lp F_i^{(\net)}\rp^2\ra_{\rep}-
\la F_i^{(\net)}\ra^2_{\rep}} \  .
\ee
\item Associated covariance
\be
\rho_{ij}^{(\net)}=\frac{\la F_i^{(\net)}F_j^{(\net)}\ra_{\rep}-
\la F_i^{(\net)}\ra_{\rep}\la F_j^{(\net)}\ra_{\rep}}{\sigma_i^{(\net)}
\sigma_j^{(\net)}} \ .
\ee
\be
\mathrm{cov}_{ij}^{(\net)}=\rho_{ij}^{(\net)}\sigma_i^{(\net)}
\sigma_j^{(\net)} \ .
\ee
As in the case of the artificial replicas, the 
three above quantities provide the estimators of the
experimental central values, errors and correlations
as computed from the sample of trained neural networks.
\item Mean variance and percentage error on central values over the
  $N_{\dat}$ data points.
\be
\la V\lc\la F^{(\net)}\ra_{\rep}\rc\ra_{\dat}=
\frac{1}{N_{\dat}}\sum_{i=1}^{N_{\dat}}\lp \la F_i^{(\net)}\ra_{\rep}-
F_i^{(\mrexp)}\rp^2 \ ,
\label{vnets}
\ee
\be
\la PE\lc\la F^{(\net)}\ra_{\rep}\rc\ra_{\dat}=
\frac{1}{N_{\dat}}\sum_{i=1}^{N_{\dat}}\lc\frac{ \la F_i^{(\net)}\ra_{\rep}-
F_i^{(\mrexp)}}{F_i^{(\mrexp)}}\rc \ .
\label{penets}
\ee
We define analogously the mean variance and the 
percentage error for errors, correlations and
covariances. 
\item Scatter correlation
\be
r\lc F^{(\net)}\rc=\frac{\la F^{(\mrexp)}\la F^{(\net)}
\ra_{\rep}\ra_{\dat}-\la F^{(\mrexp)}\ra_{\dat}\la\la F^{(\net)}
\ra_{\rep}\ra_{\dat}}{\sigma_s^{(\mrexp)}\sigma_s^{(\net)}} \ ,
\label{sccnets}
\ee 
where the scatter variance $\sigma_s^{(\net)}$
associated to the neural network prediction is defined as
\be
\sigma_s^{(\net)}=\sqrt{\la \lp \la F^{(\net)}\ra_{\rep}\rp^2\ra_{\dat}-
\lp \la  \la F^{(\net)}\ra_{\rep} \ra_{\dat}\rp^2} \ .
\ee 
We define analogously $r[\sigma^{(\net)}]$, 
$r[\rho^{(\net)}]$ and $r[\mathrm{cov}^{(\net)}]$.
\item  ${\cal R}$-ratio
\be
{\cal R}=\frac{\langle \tilde E \rangle}{\langle  E \rangle},
\label{rratdef}
\ee
where ${\langle  E \rangle}$ is given by Eq.~\ref{avchi2},
with covariance matrix error, and
for $\tilde E^{(k)}$ one uses the analog of Eq. \ref{ertilde}
including correlated errors, 
\be
 \la \widetilde{E}
\ra=\frac{1}{N_{\rep}}\sum_{k=1}^{N_{\rep}}{\widetilde{E}}^{(k)}
\label{avchi2til} \ ,
\ee
\be
\label{tilchicov}
{\widetilde{E}}^{(k)}=\frac{1}{N_{\dat}}\sum_{i,j=1}^{N_{\dat}}\lp
F_i^{(\net)(k)}-F_i^{(\exp)}\rp{\overline\mathrm{cov}^{(k)}}^{-1}_{ij}
\lp
F_j^{(\net)(k)}-F_j^{(\exp)}\rp \ .
\ee
The $\mathcal{R}-$ratio is interesting since it provides an estimator
which is capable to determine whether or not when
error reduction is observed it is due
to the fact that the neural network by combining data points
has found the underlying law or whether in the other
hand it is due to artificial smoothing.
To verify this property, assume that correlated systematic
uncertainties can be neglected, as we have
done in the example above, then the estimator reads
\be
\mathcal{R} = \frac{1+\la \frac{\hat{\sigma}^2}{\sigma^2}\ra_{\dat}}{2+
\la \frac{\hat{\sigma}^2}{\sigma^2}\ra_{\dat}
-2\la\la r l \ra_{\rep} \frac{\hat{\sigma}}{\sigma} \ra_{\dat}} \ ,
\ee
so that if error reduction is due to the fact that the neural network
has found the underlying law that describes the experimental data, 
one will have $\hat{\sigma}\ll \sigma$ and therefore the
$\mathcal{R}$-ratio will satisfy $\mathcal{R}\sim 1/2$.
\end{itemize}

Note that even if the covariance matrix is determined
from the correlation matrix and the total error
from Eq. \ref{cormat}, the reproduction of
covariances by the probability measure 
has to be validated independently
of correlations and error. This is so because
even if it is clear that when correlations and
errors are correctly reproduced, also covariances
will be reproduced, the inverse is not necessarily
true.

\subsection{Statistical estimators: stability of the probability measure}
\label{stabest}

Let us define the statistical estimators that are used
to analyze the compatibility and stability of the
constructed probability measure with respect of some of the
parameters that define the fit.
These estimators are also
 useful in order to perform a quantitative comparison
between fits, that is, between probability measures.
For example, we know how the different estimators depend
on the size of the sample $N_{\rep}$ in the replica generation,
but we do not know if the same behavior holds for the sample
of trained networks.
The estimators in the above section were used to compare the
results of either the replica generation or the
neural network fit to {\ experimental data}.
In this section we want to define statistical
estimators which allow us to compare
a given fit with another fit.
Since in practical applications we will have a reference fit,
and compare other fits with it, let us label
 the first
fit as the {\it reference fit} and is denoted by the
superscript (ref) and the second fit as the {\it current fit}, denoted by the
superscript (fit).

These estimators are divided in estimators for central
values, standard deviations and correlations. 
The estimators are computed for a set of $\widetilde{N}_{\dat}$ 
points  which
need not to be the same points where there is experimental data.
In particular one can compute these estimators in the extrapolation region,
to check the stability of the fit also in the region where there
is no experimental data. These estimators are given by:
\begin{itemize}
\item Central values:
\begin{itemize}
\item Relative error:
\be
\la \mathrm{RE} \lc F\rc\ra_{\dat}=
\frac{1}{\widetilde{N}_{\dat}}\sum_{i=1}^{
\widetilde{N}_{\dat}}\Bigg|2\lp
\frac{ \la F_i^{(\reff)}\ra-
  \la F_i^{(\fit)}\ra }{ \sqrt{V\lc F_i^{(\reff)}\rc}+
\sqrt{V\lc F_i^{(\fit)}\rc}}\rp\Bigg|\equiv \la 
\Bigg|2
\frac{ \la F_i^{(\reff)}\ra-
  \la F_i^{(\fit)}\ra}{ \sqrt{V\lc F_i^{(\reff)}\rc}+
\sqrt{V\lc F_i^{(\fit)}\rc} }\Bigg| \ra_{\dat} \ ,
\ee
\item Scatter correlation:
\be
r\lc \la F \ra \rc=\frac{\la \la F^{(\fit)}F^{(\reff)}\ra
\ra_{\dat}-\la 
\la F^{(\fit)}\ra \ra_{\dat} \la \la F^{(\reff)}\ra \ra_{\dat}}{
\sqrt{ \la 
\la F^{(\fit)2}\ra \ra_{\dat}-\la 
\la F^{(\fit)}\ra \ra_{\dat}^2  }\sqrt{
 \la 
\la F^{(\reff)2}\ra \ra_{\dat}-\la 
\la F^{(\reff)}\ra \ra_{\dat}^2  }} \ .
\ee
\end{itemize}
\item Standard deviations:\\

\begin{itemize}
\item Relative error:
\be
\la \mathrm{RE} \lc \sigma^2 \rc\ra=\frac{1}{\widetilde{N}_{\dat}}
\sum_{i=1}^{\widetilde{N}_{\dat}}\Bigg|
2\frac{\sigma_i^{(\reff)2}-\sigma_i^{(\fit)2}}{
\sqrt{V\lc \sigma_i^{(\fit)}\rc} +
\sqrt{V\lc \sigma_i^{(\reff)}\rc}} \Bigg|=
\la 2\Bigg|
\frac{\sigma_i^{(\reff)2}-\sigma_i^{(\fit)2}}{
\sqrt{V\lc \sigma_i^{(\fit)}\rc} +
\sqrt{V\lc \sigma_i^{(\reff)}\rc}} \Bigg|
\ra_{\dat} \ ,
\ee
\item Scatter correlation:
\be
r\lc \sigma \rc=\frac{\la  \la \sigma^{(\fit)}
 \sigma^{(\reff)}\ra\ra_{\dat}-\la 
\sigma^{(\fit)} \ra_{\dat} \la \sigma^{(\reff)} \ra_{\dat}}{
\sqrt{ \la 
\sigma^{(\fit)2} \ra_{\dat}-\la 
\sigma^{(\fit)} \ra_{\dat}^2  }\sqrt{ \la 
\sigma^{(\reff)2} \ra_{\dat}-\la 
\sigma^{(\reff)} \ra_{\dat}^2  }} \ .
\ee
\end{itemize}
\item Correlations: \\
\begin{itemize}
\item Relative error:
\bea
\la \mathrm{RE} \lc \rho \rc\ra= \frac{1}{\widetilde{N}_{\dat}\lp
\widetilde{N}_{\dat}+1\rp/2}\sum_{i=1}^{\widetilde{N}_{\dat}}
\sum_{j=i}^{\widetilde{N}_{\dat}}\Bigg|
2\frac{\rho_{ij}^{(\reff)}-\rho_{ij}^{(\fit)}}{
\sqrt{V\lc \rho_{ij}^{(\fit)}\rc} +
\sqrt{V\lc \rho_{ij}^{(\reff)}\rc}
} \Bigg|
\equiv \nonumber \\\la \Bigg|
2\frac{\rho_{ij}^{(\reff)}-\rho_{ij}^{(\fit)}}{
\sqrt{V\lc \rho_{ij}^{(\fit)}\rc} +
\sqrt{V\lc \rho_{ij}^{(\reff)}\rc}} \Bigg|
\ra_{\dat}\qquad\qquad\qquad \ ,
\eea
\item Scatter correlation:
\be
r\lc \rho \rc=\frac{\la  \la\rho^{(\fit)}
 \rho^{(\reff)}\ra\ra_{\dat}-\la 
\rho^{(\fit)} \ra_{\dat} \la \rho^{(\reff)} \ra_{\dat}}{
\sqrt{ \la 
\rho^{(\fit)2} \ra_{\dat}-\la 
\rho^{(\fit)} \ra_{\dat}^2  }\sqrt{ \la 
\rho^{(\reff)2} \ra_{\dat}-\la 
\rho^{(\reff)} \ra_{\dat}^2  }} \ .
\ee
\end{itemize}
\end{itemize}
were averages and variances of central values, errors and
correlations can be computed using the expressions in
Appendix \ref{probrev}.
Two probability measures are said to be equivalent to
each other if 
 the conditions $\la \mathrm{RE}\lc F\rc\ra_{\dat}  \lsim 1$,
$\la \mathrm{RE}\lc \sigma^2\rc\ra_{\dat}  \lsim 1$ and
$\la \mathrm{RE}\lc \rho\rc\ra_{\dat}  \lsim 1$
are fulfilled. For the scatter correlations
one expects for two compatible
probability measures the condition $r \lsim 1$ to hold.
Note that the $\widetilde{N}_{\dat}$ points where the
stability estimators defined above are computed
do not need to be those were there is experimental data, for
instance, these estimators
can be used to compute the stability of a probability measure
also in the extrapolation region. 
Since fluctuations might be large, it might be needed to average
over different partitions of the sets of trained 
neural networks to obtain stable results.

Another relevant estimator is the so called pull. 
The average pull of two different fits is defined as
\be
\la P \ra_{\dat}= \frac{1}{N_{\dat}}=\sum_{i=1}^{N_{\dat}}P_i
=\sum_{i=1}^{N_{\dat}} \frac{ \la F_i^{(\reff)}\ra
-\la F_i^{(\fit)}\ra}{\sqrt{\sigma_i^{(\fit)2}+
\sigma_i^{(\reff)2}}} \ ,
\ee
For two fits to be consistent within the respective uncertainties
the condition $P_i \lsim 1$ should be
satisfied. Note that the condition $P_i\le 1$ is necessary
for two fits to be compatible within errors, but it
is not sufficient for two fits to define the
same probability measure, since in 
particular, if either $\sigma^{(\fit)}$ or $\sigma^{(\reff)}$
is much larger than the other error, then the two fits will
be very different yet the pull will still satisfy $P_i\le 1$.

Finally, there are  other conditions that must be verified
for the sample of trained neural networks for a
given fit to be considered as satisfactory.
These conditions are related to the criteria used
to
stop the minimization of the individual replicas.
If we use dynamical stopping of the training, as
described in Section \ref{trainstrat}, the
 distribution of errors $E^{(k)}$ 
at the end of the training over the trained replica sample
 must be peaked around the average result $\la E\ra$,
Eq. \ref{avchi2}, because
 the opposite case would mean that the averaged result is obtained combining
 good fits with bad fits (in the sense of fits with  large values of
$E^{(k)}$). Another estimator of the consistency of
the results is the  
 distribution of training length, where the training length
measures the length of the minimization procedure, cannot
be peaked near $N_{\max}$, the maximum number of iterations,
because if it is this means that the dynamical stopping
of the minimization is not being effective, and one
has instead fixed training length minimization.
In Section \ref{bdecay_appl} we show how these two conditions
are satisfied in a particular example.

\clearpage

\newpage
~
\newpage

\chapter{The neural network approach: Applications}
\label{appl}

In this Chapter we review four applications of the
general strategy to parametrize experimental data that
has been described in the previous Chapter.
First we describe a parametrization of the vector-axial vector
 spectral function
$\rho_{V-A}(s)$
from the hadronic decays of the tau lepton
which
incorporates theoretical information in the form of sum rules, 
with the motivation
of extracting from this parametrization the values 
of the nonperturbative vacuum condensates.
Then we present a parametrization of the
proton structure function $F_2^p(x,Q^2)$, which updates the
results of Ref. \cite{f2ns} by including the data from the
HERA experiments. This parametrization is not only interesting
as a new application of the general technique,
but also it allows us to develop the necessary techniques
to apply our general strategy to problems with data coming
from many different experiments.
The third application is a parametrization of the
lepton energy distribution
$d\Gamma(E_l)/dE_l$ in the semileptonic decays of the
B meson. As a byproduct of this parametrization we will
provide a determination of the b quark mass $\overline{m}_b
\lp \overline{m}_b\rp$.

We devote special attention to the last and most important
application, which is the main motivation for the
development of the techniques described in
 Chapter \ref{general}: the neural network
parametrization of parton distributions. 
To this purpose a new strategy to solve the DGLAP
evolution equations is introduced. We then review the
parametrization of the non-singlet parton distribution 
$q_{NS}(x,Q_0^2)$ from experimental data
on the non-singlet structure function $F_2^{NS}(x,Q^2)$, and we
devote special attention to the comparison of our results
with those of the standard approach to parton distributions
described in Section \ref{globalfits}.

\section{Spectral functions in hadronic tau decays}

\label{tau_appl}

In this Section, we describe the first application of the
technique introduced in Chapter \ref{general}
to parametrize experimental data that was
implemented during the present thesis. We
construct a parametrization of the vector minus axial-vector
 spectral function
$\rho_{V-A}(s)$
from the hadronic decays of the tau
lepton. As has been 
described in Section \ref{tau_theo}, this spectral function 
is determined from purely nonperturbative dynamics,
so it is specially suited for the determination
of non perturbative parameters like the chiral vacuum
condensates.
The  
motivation to determine a parametrization of the
spectral function $\rho_{V-A}(s)$ is thus to
provide an extraction
of the QCD vacuum condensates $\la \mathcal{O}_6 \ra$, $\la \mathcal{O}_8
 \ra$ and
higher
dimensional condensates from experimental data.
A more detailed description of this work can be found in 
Ref. \cite{condensates}.

Following the strategy described in the
previous Chapter, the parametrization of the 
spectral function $\rho_{V-A}(s)$ will
combine all available experimental information, that is, central values,
errors and correlations, together with information
from theoretical constraints: the
chiral sum rules, Eqns. \ref{dmo}-\ref{empionmass} 
and the asymptotic vanishing of the
spectral function in the perturbative $s\to \infty$ limit,
Eq. \ref{specpert}.
This way we will obtain a determination of the QCD vacuum condensates 
which is unbiased with respect to the parametrization of the
spectral function and with faithful estimation and
propagation of the experimental uncertainties.
All sources of
uncertainty related to the method of analysis are kept under control,
and their contribution to the total error in the
extraction of the condensates is estimated,
as  discussed in \cite{condensates}.

Since the relevant spectral function for the determination of the
condensates is the $\rho_{V-A}(s)$ spectral function, we need a
simultaneous measurement of the vector and axial-vector spectral
functions from the
hadronic decays of the tau lepton. Data from the ALEPH 
Collaboration \cite{exp1}
 and from the OPAL Collaboration \cite{expopal} will be used\footnote{
In this analysis we used the ALEPH data as presented in 
Ref. \cite{exp1}.
Recently, the  ALEPH collaboration released
their final results on hadronic spectral functions
and branching fractions of the hadronic tau decays \cite{finalaleph},
which have reduced uncertainties due to the larger statistics,
specially in the large $s$ region. It would be interesting to
repeat the present analysis with this updated experimental data
on the $\rho_{V-A}(s)$ spectral function,
as has been done with other physical applications \cite{physicsaleph}.}, 
 which provide a simultaneous determination of the vector
and axial vector spectral functions in the same kinematic region and
also provide the full set of correlated uncertainties for these measurements.
There exists additional data on spectral functions coming from
electron-positron annihilation, but their quality is lower than the
data from hadronic tau decays and will not be
incorporated to the present analysis.

Experimental data does not consist on the
spectral function directly, but rather  on the invariant mass-squared
spectra for both the vector and axial-vector
components, that are related to the spectral functions
by a kinematic factor and a branching ratio
normalization,
\be
\label{invmass}
\rho_{V/A}(s)\equiv\frac{M^2_{\tau}}{6|V_{ud}|^2S_{EW}}\frac{B(\tau^-\to
V^-/A^-\nu_{\tau} )}{B(\tau^-\to e^-\bar{\nu}_e\nu_{\tau}
)} \frac{1}{N_{V/A}}\frac{dN_{V/A(s)}}{ds}\left[\left(1-\frac{s}{M_{\tau}^2}
\right)\left(1-\frac{2s}{M_{\tau}^2}\right)\right]^{-1} \ , \quad s\le 
M_{\tau}^2 \ .
\ee  
In the above equation,
\be
\frac{1}{N_{V/A}}\frac{dN_{V/A}(s)}{ds} \ ,
\ee
is the normalized invariant mass distribution,
 $M_{\tau}$ is the
tau lepton mass, $s$ is the invariant mass
of the hadronic final state and $S_{EW}$ are the electroweak radiative
corrections.

\begin{figure}[ht]
\begin{center}
\includegraphics[scale=0.28,angle=-90]{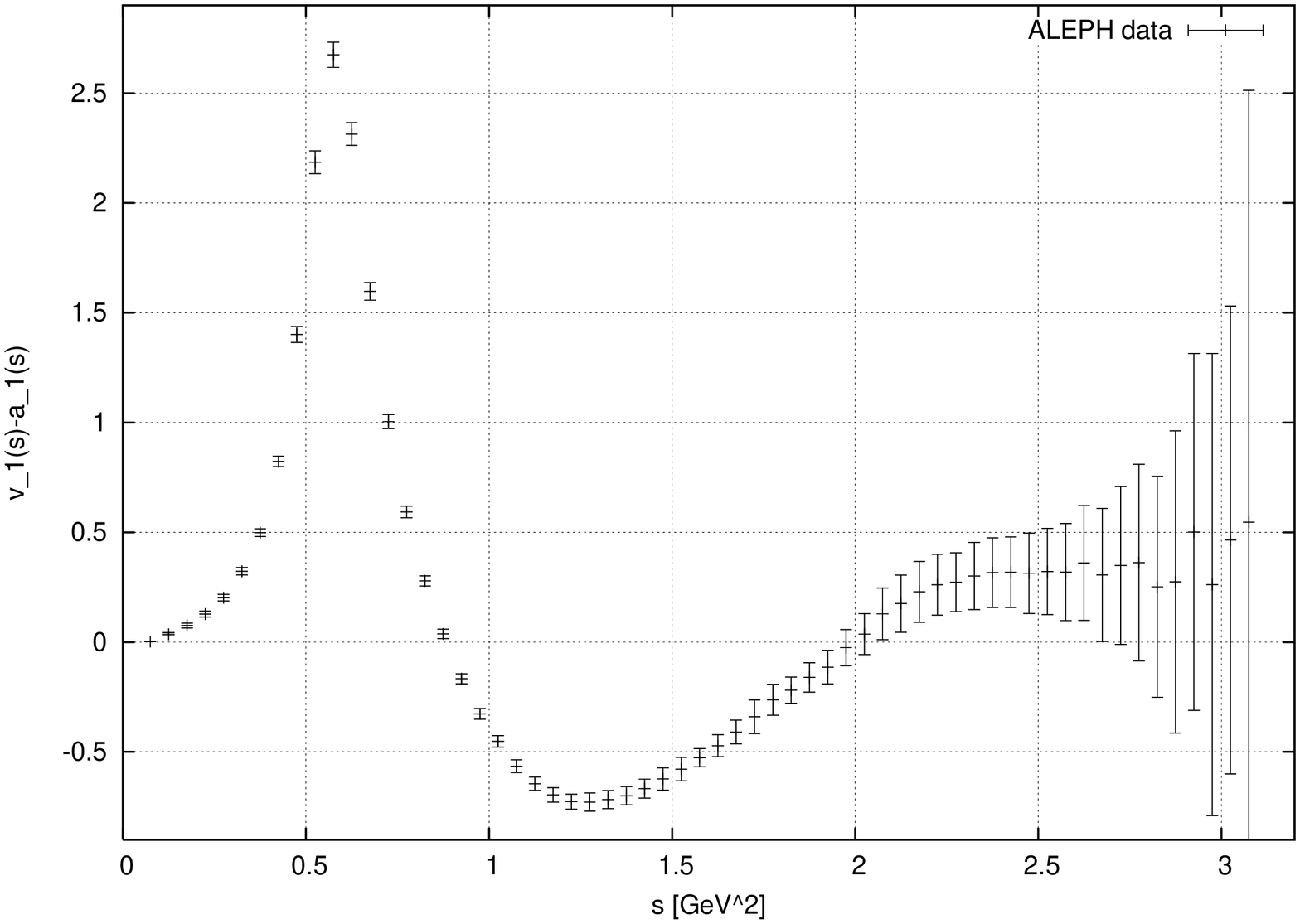}
\includegraphics[scale=0.28,angle=-90]{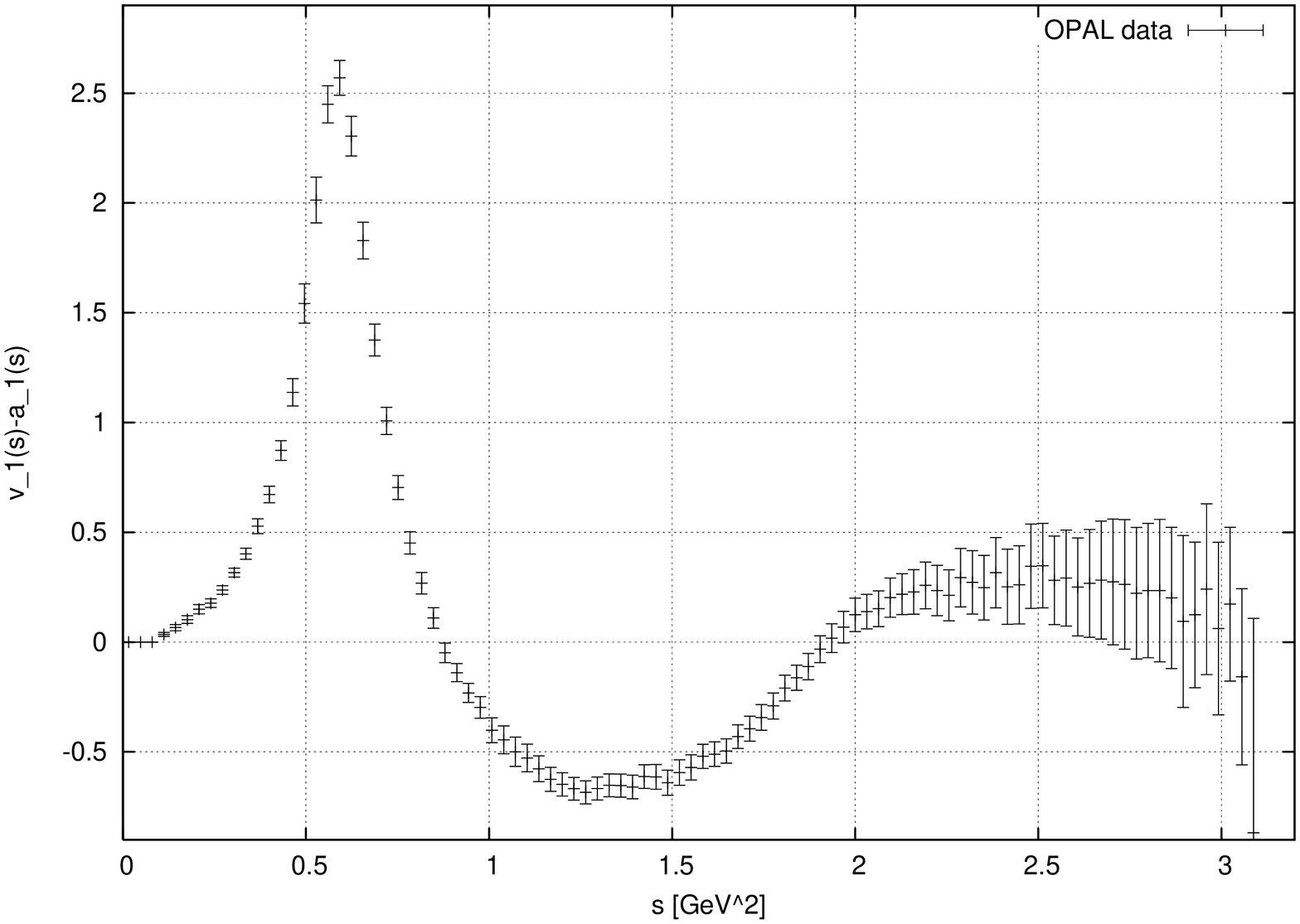}
\caption{}{\small Experimental data for $\rho_{V-A}(s)$ spectral function from
  the ALEPH (left) and OPAL (right) collaborations. Note that
errors increase as we approach the kinematic
threshold at $s=M_{\tau}^2$.
}
\label{specdata}
\end{center}
\end{figure}

In the following
$\rho^{(\exp)}_{V-A,i}$ will denote the $i$-th data point,
\be
\rho^{(\exp)}_{V-A,i}=\rho_{V-A}(s_i)\equiv
\rho_V(s_i)-\rho_A(s_i) \ , \qquad i=1,\ldots,N_{\dat} \ ,
\ee
where $N_{\dat}$  is the number of available data points.
 Fig. \ref{specdata} 
shows the experimental data
used in the present analysis from the two LEP experiments.
Note that errors are  small in the low and middle $s$
regions and that they become larger as we approach the tau mass threshold.
The last points are almost zero in the invariant mass spectrum, 
since near threshold the reduced phase space implies
a lack of statistics, and
are only enhanced in the spectral functions due to the large
kinematic factor for $s$ near $M_{\tau}^2$ (the last term
in Eq. \ref{invmass}), so  special
care must be taken with the physical significance of these points.

It is clear that the asymptotic vanishing of the spectral function,
\be
\lim_{s\to\infty}\rho_{V-A}(s) \to 0 \ ,
\ee
implied by perturbative QCD
is not reached for $s\le M_{\tau}^2$, at least for the
central experimental values, and therefore should be enforced
artificially on the parametrization of the $\rho_{V-A}(s)$
spectral function that we will construct.
The method that we use
takes advantage of the smooth, unbiased interpolation
capability of  neural networks: artificial points are added to
the data set with adjusted errors in a region where 
$s$ is high enough that the $\rho_{V-A}(s)$ spectral function
should vanish, as discussed in Section
\ref{constraints}. That is, we add to the
experimental data a set of artificial  points,
\be
\rho_{V-A}(s_i)= \ 0,  \qquad  i=N_{\dat}+1,\ldots,N_{\tot} \ ,
\qquad s_i\ge M_{\tau}^2  \ ,
\ee
with error $\sigma_i$, and where $N_{\tot}$ is the total
number of points (data and artificial).
 Once these artificial points are
included,
 the neural network will smoothly interpolate between the
real and artificial data points, also taking into account
 the constraints of the sum  rules, as explained below.

\begin{table}[t]  
\begin{center}  
\begin{tabular}{|c|c|c|c|} 
\multicolumn{4}{c}{$\rho_{V-A}(s)$}\\   
\hline
$N_{\rep}$ & 10 & 100 & 1000 \\
\hline  
$r [\rho_{V-A}(s)^{(\art)}]$ 
  & 0.98  & 0.99 & 0.99 \\
\hline  
$r [\sigma^{(\art)} ]$ 
  & 0.98 & 0.99 & 0.99 \\
\hline
$r [\rho^{(\art)}]$ 
  & 0.61 & 0.95 & 0.99 \\
\hline
\end{tabular}
\end{center}
\caption{}{\small Comparison between experimental data the
Monte Carlo sample of artifical replicas for
the $\rho_{V-A}(s)$ spectral function.}
\label{reptesttable}
\end{table}

As has been described in Chapter \ref{general}, the first step
in the parametrization of the spectral function $\rho_{V-A}(s)$
is the generation of a Monte Carlo sample of replicas of the
experimental data.
The experimental data for the invariant hadronic mass
spectrum consist on the central values, the total error
and the correlations between different invariant mass
bins. Therefore,
to generate replicas of the experimental data, we use Eq. \ref{gen2},
which in the present case reads
\be
\label{gen2tau}
\rho_{V-A,i}^{(\art)(k)}=\rho_{V-A,i}^{(\exp)}+r_i^{(k)}\sigma_{\tot,i} \ ,
\qquad k=1,\ldots,N_{\rep} \ ,
\ee
where the gaussian random numbers $r_i^{(k)}$
 have the same correlation as experimental data.
The replica generation has the  statistical estimators
that can be seen in Table \ref{reptesttable}. 
It can be seen that a Monte Carlo sample with $N_{\rep}=1000$ is required
to have scatter correlations larger than 0.99 for both
central values, errors and correlations.

Once the Monte Carlo sample of replicas has been generated, 
following the method described above, one has to train
a neural network in each replica. 
The estimator which will be minimized
in the present case is the diagonal error Eq. \ref{er2},
which  reads
\be
E_2^{(k)}=\frac{1}{N_{\tot}}\sum_{k=1}^{N_{\tot}} \frac{
\lp \rho_{V-A,i}^{(\art)(k)}-
\rho_{V-A,i}^{(\net)(k)}\rp^2}{\sigma_{\stat,i}^2} \ ,
\quad k=1,\ldots,N_{\rep} \ ,
\ee
where $\rho_{V-A,i}^{(\net)(k)}$ is the k-th neural network, trained
on the k-th replica of the experimental data. Note that
as explained in \cite{f2ns}, correlations are correctly incorporated
in the parametrization of $\rho_{V-A}(s)$ through the
Monte Carlo pseudo-data generation, Eq. \ref{gen2tau}.
As has been described in Section \ref{tau_theo}, the parametrization
of the $\rho_{V-A}(s)$
spectral function has to satisfy several theoretical
constraints.
These theoretical constraints, the chiral sum rules, are
implemented using the Lagrange Multipliers technique, as described in
Section \ref{constraints}.
Therefore the total error function to be minimized will be
\be
\label{etot}
E_{\tot}^{(k)}=E_2^{(k)}+\sum_{i=1}^{N_{\con}}\lambda_i
\lp \mathcal{F}_i\lp \rho_{V-A}^{(\net)(k)}(s)\rp
-\mathcal{F}_i^{(\theo)} \rp^2 \ ,
\ee
where for example, for the first theoretical constraint, the 
DMO sum rule, Eq. \ref{dmo}, one has
\be
\mathcal{F}_1\lc \rho_{V-A}^{(\net)(k)}(s) \rc=
\frac{1}{4\pi^2}\int_0^{\infty} ds\frac{\rho_{V-A}^{(\net)(k)}(s)}{s}, 
\qquad \mathcal{F}_i^{(\theo)}=\frac{f_{\pi}\la r_{\pi}^2\ra}{3}-F_A \ ,
\ee
and similarly for the remaining chiral sum rules, up to $N_{\con}=4$.
These sum rules act as constraints on the neural network output, that is,
the main contribution to the error function $E_{\tot}^{(k)}$
(which determines the 
learning of the network) still comes from the experimental errors, 
that is, the $E_2^{(k)}$ term, and the
sum rules are only relevant in the region where the errors are larger. 
The relative weights of the chiral sum rules $\lambda_i$
are determined 
according to a stability 
analysis, as discussed in \cite{condensates}.

The length of the training is fixed by studying the behavior of the 
error function $E_{\tot}^{(0)}$ for the neural
net fitted to the central experimental values, and asking that 
$E_{\tot}^{(0)}$ stabilizes to a value close to one, which on 
statistical grounds in the value expected for a correct fit.
The minimization algorithm that is used for the neural
network training is Genetic Algorithms, introduced in
Section \ref{minimstratt}, which is required since
the total error function to be minimized, Eq. \ref{etot},
depends non-linearly with the $\rho_{V-A}(s)$ spectral function
through the convolutions of the chiral sum rules.

With the strategy discussed above, $N_{\rep}$ neural networks
are trained on the $N_{\rep}$ Monte Carlo replicas of the
experimental data for the spectral function $\rho_{V-A}(s)$.
As has been described in Chapter \ref{general},
once the probability measure in the space of spectral functions
$\mathcal{P}\lc \rho_{V-A}\rc$
has been constructed, it is crucial to validate
the results with suitable statistical estimators.
A number of checks is then performed in order to be sure that an
unbiased representation of the probability density has been obtained. 
The values for the scatter correlations for central
values, errors and correlations 
are presented in Table \ref{reptraintesttable}. 
It is seen that the central values and
the correlations are  well reproduced, whereas this is not the
case for the total errors.
The average standard deviation for each data point
computed from the Monte Carlo sample of nets is substantially smaller than
the experimental error. This is due to the fact that the network is
combining the information from different data points by capturing and 
underlying law. 
This effect is enhanced by the inclusion of sum rules constraints. 
All networks have to fulfill these constraints which forces
the fit to behave smoothly in a region where experimental data
errors
are very large. This should be understood as a success
of the fitting procedure.

\begin{table}[t]  
\begin{center}  
\begin{tabular}{|c|c|c|} 
\multicolumn{3}{c}{$\rho_{V-A}(s)$}\\   
\hline
$N_{rep}$ & 10 & 100  \\
\hline  
$r [\rho_{V-A}(s)^{(\net)}]$ 
  & 0.98  & 0.98  \\
\hline  
$r [\sigma^{(\net)} ]$ 
  & -0.21 & -0.20  \\
\hline
$r [\rho^{(\net)}]$ 
  & 0.80 & 0.85  \\
\hline
\end{tabular}
\end{center}
\caption{}{\small Comparison between experimental data and the averages
computed from the sample of trained neural networks for
  the $\rho_{V-A}(s)$ spectral function.}
\label{reptraintesttable}
\end{table}

\begin{figure}[ht]
\begin{center}
\epsfig{width=0.32\textwidth,figure=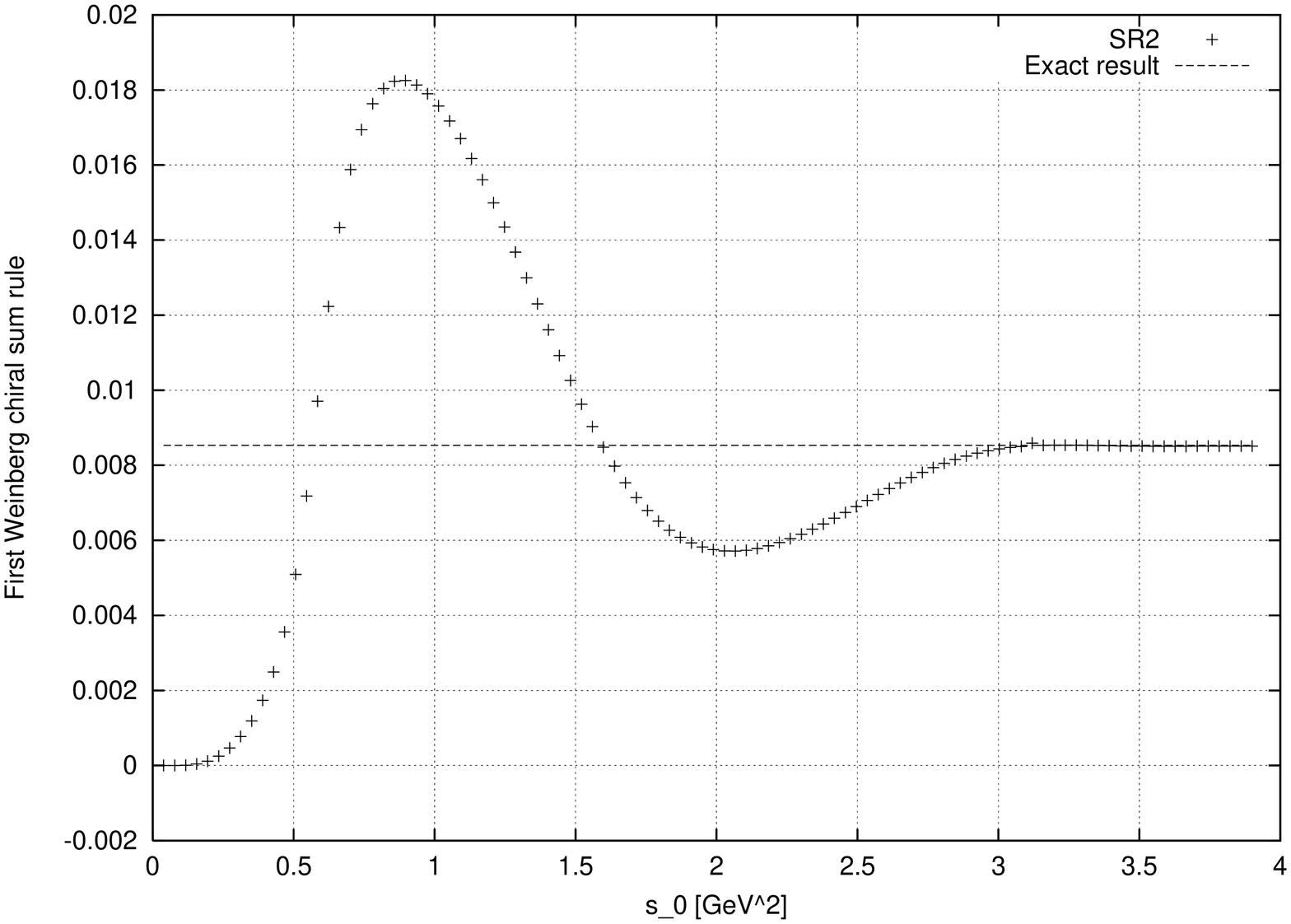,angle=-90}  
\epsfig{width=0.32\textwidth,figure=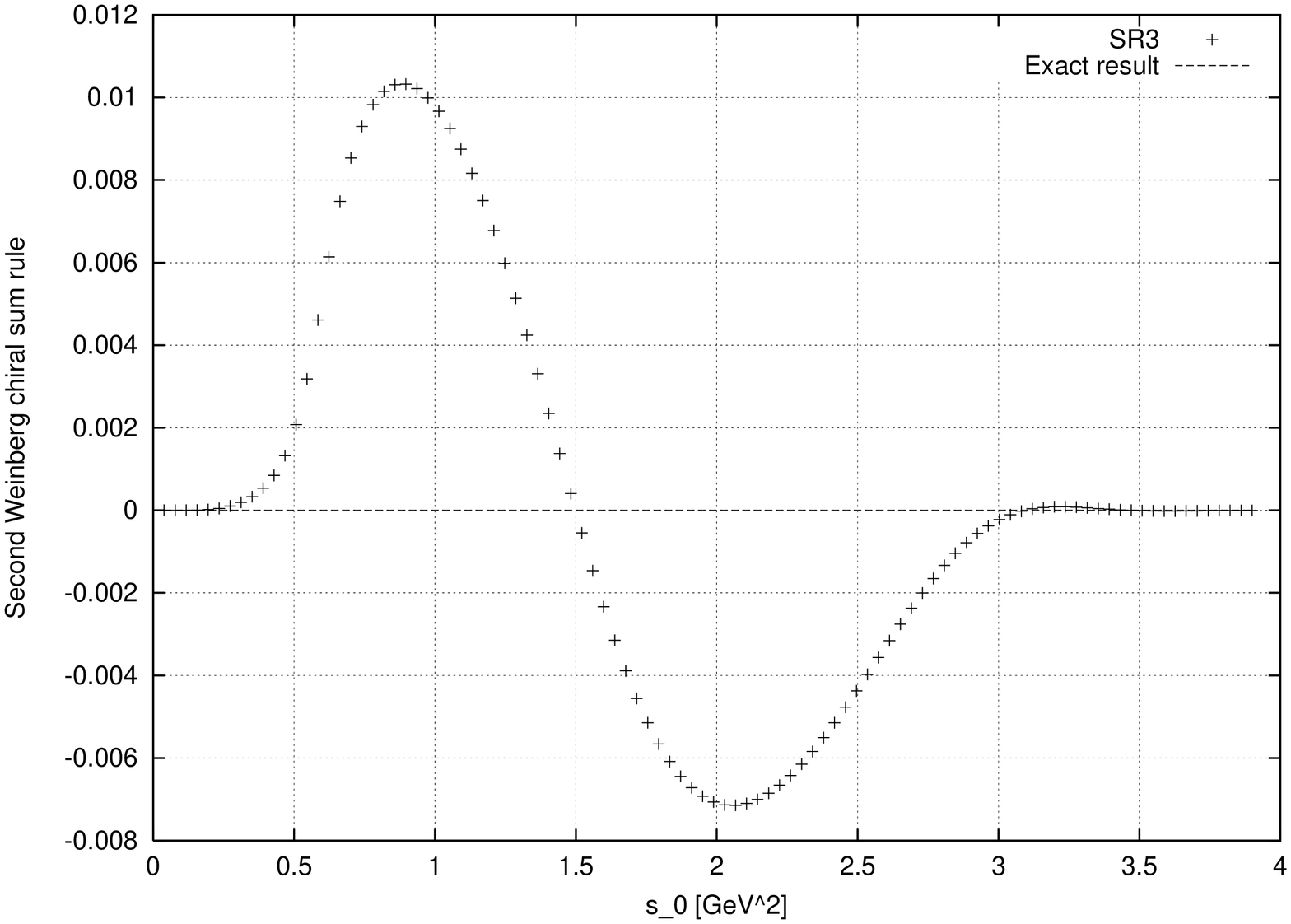,angle=-90}  
\end{center}
\caption{}{\small The two 
Weinberg chiral sum rules, Eqs. \ref{wsr1} and \ref{wsr2},
evaluated from the neural network
 parametrization
of $\rho_{V-A}(s)$. Only central values are shown.}
\label{srstab}
\end{figure}

The constructed probability measure for the 
$\rho_{V-A}(s)$ spectral function has built-in
the theoretical constraints for the chiral sum rules.
For example, it can be checked that the two Weinberg chiral sum rules are
well
verified by our neural network parametrization, and thus have been
incorporated to the information contained on the 
experimental data. This fact will be  crucial  because 
different extraction methods, differing in combinations
of these chiral sum rules, can be shown to be equivalent in the 
asymptotic region $s_0\to\infty$. In Fig. \ref{srstab} the 
two Weinberg sum rules, Eqs. \ref{wsr1} and \ref{wsr2} evaluated with
the neural network parametrization of the spectral function 
$\rho_{V-A}(s)$ are represented. Both chiral sum rules are well
verified
in the asymptotic region, beyond the range of available 
experimental data. This also will ensure the stability of the
evaluation of the condensates with respect to the
specific value of $s_0$ chosen as long as it stays in the
asymptotic region.

Using the neural network parametrization of the
$\rho_{V-A}$ spectral function, we can compute
 any given sum rule with associated
uncertainties. Because the neural 
parametrization retains all the experimental information, we can view
values coming from the neural networks
 as direct experimental determinations of convolutions
of the spectral function $\rho_{V-A}(s)$.
The value of the  condensates $\la\mo_6\ra,\la\mo_8\ra$  and higher 
dimensional condensates are then extracted
from the value of an appropriate sum rule, to be
discussed in brief.
The method we will follow is the evaluation of the vacuum condensates
as a function of the upper limit of integration for each replica
and compute the mean and standard deviation. As has been
explained before, it is crucial to represent the value of the
different sum rules as a function of the upper limit 
of integration, to check both its convergence and its stability.

Once the neural network parametrization of the 
$\rho_{V-A}(s)$ spectral function has been constructed and
validated, we can use it to determine the
nonperturbative chiral vacuum condensates.
These condensates can be determined by virtue of the dispersion
relation from another sum rule, that is, a
 convolution of the $\rho_{V-A}(s)$ spectral function 
with an appropriate weight function. As discussed in
Section \ref{tau_theo}, we define the operator product
expansion of the chiral correlator in the following way
\be
\label{chiralope2}
\Pi(Q^2)|_{V-A}=
\sum_{n=1}^{\infty}\frac{1}{Q^{2n+4}}C_{2n+4}(Q^2,\mu^2)
\left\langle \overline{\mathcal{O}}_{2n+4}(\mu^2)\right\rangle
\equiv \sum_{n=1}^{\infty}\frac{1}{Q^{2n+4}}
\left\langle \mathcal{O}_{2n+4}\right\rangle \ .
\ee
The Wilson coefficients, including radiative corrections, are
absorbed into the nonperturbative vacuum expectation values, to
facilitate comparison with the current literature.  
As has been explained in Sect. \ref{tau_theo}
the analytic structure of the 
chiral correlator implies that it has
to obey the  dispersion relation,
\be
\label{chiraldisp}
\Pi_{V-A}(Q^2)=\int_0^{\infty}\mathrm{d}s\frac{1}{s+Q^2}\frac{1}{\pi}
\mathrm{Im}\Pi_{V-A}(s)=\frac{1}{\pi}
\sum_{n=0}^{\infty}
\int_0^{\infty} ds
\frac{s^n}{Q^{2n+2}}\mathrm{Im}\Pi_{V-A}(s) \ .
\ee
Recalling that the imaginary part of the chiral correlator
is proportional to the spectral function, 
\be
\mathrm{Im}\Pi_{V-A}(s)=\frac{1}{2\pi}\rho_{V-A}(s)\ ,
\ee
comparing terms
of the same order in the $1/Q^ 2$
expansion in Eq. \ref{chiralope2} and in 
Eq. \ref{chiraldisp}, it can
be seen that
condensates of arbitrary dimension are given by
\be
\label{condenconv}
\left\langle\mathcal{O}_{2n+2}\right\rangle= 
(-1)^n\int_0^{s_0}dss^2\frac{1}{2\pi^2}\rho_{V-A}(s) \ ,\quad
n\ge 2
\ ,
\ee
which, if the asymptotic regime has reached, should be independent
of the upper integration limit for large enough $s_0$. 

As long as all previous integrals have to be cut at some finite 
energy $s_0\le M_{\tau}^2$, since no experimental information
on the $\rho_{V-A}(s)$
spectral function is available above $M_{\tau}^2$, 
a truncation of the integration should be
performed that competes with all other sources of statistical and 
systematic errors, introducing a theoretical bias which is difficult
to estimate. Many techniques have been developed to deal
with this finite energy integrals, leading to the so-called 
Finite Energy Sum Rules (FESR). A detailed analysis
of some alternative techniques and methods
to extract the condensates can be found in the
original work \cite{condensates}.

\begin{figure}
\begin{center}
\includegraphics[angle=-90,scale=0.28]{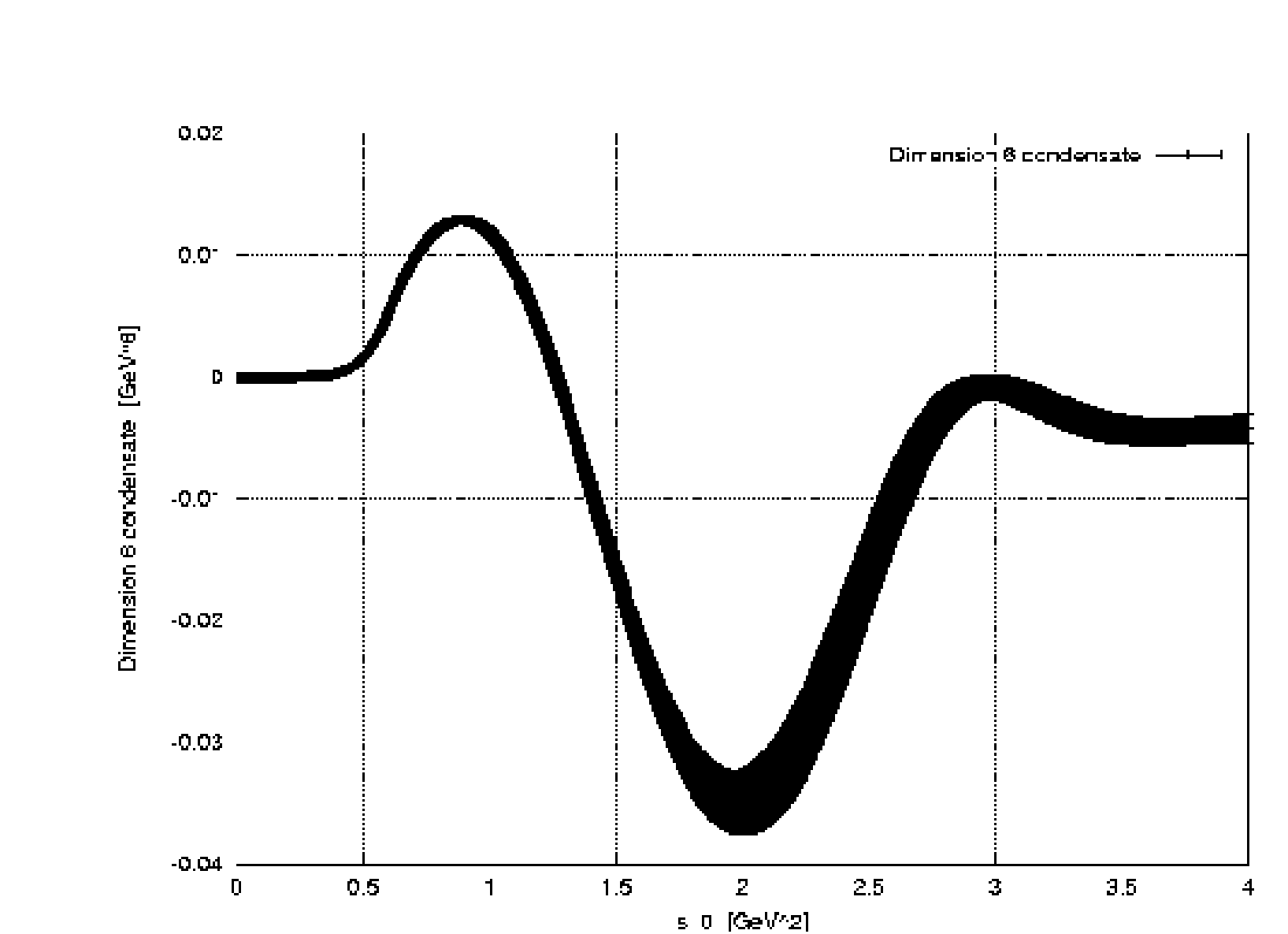}
\includegraphics[angle=-90,scale=0.28]{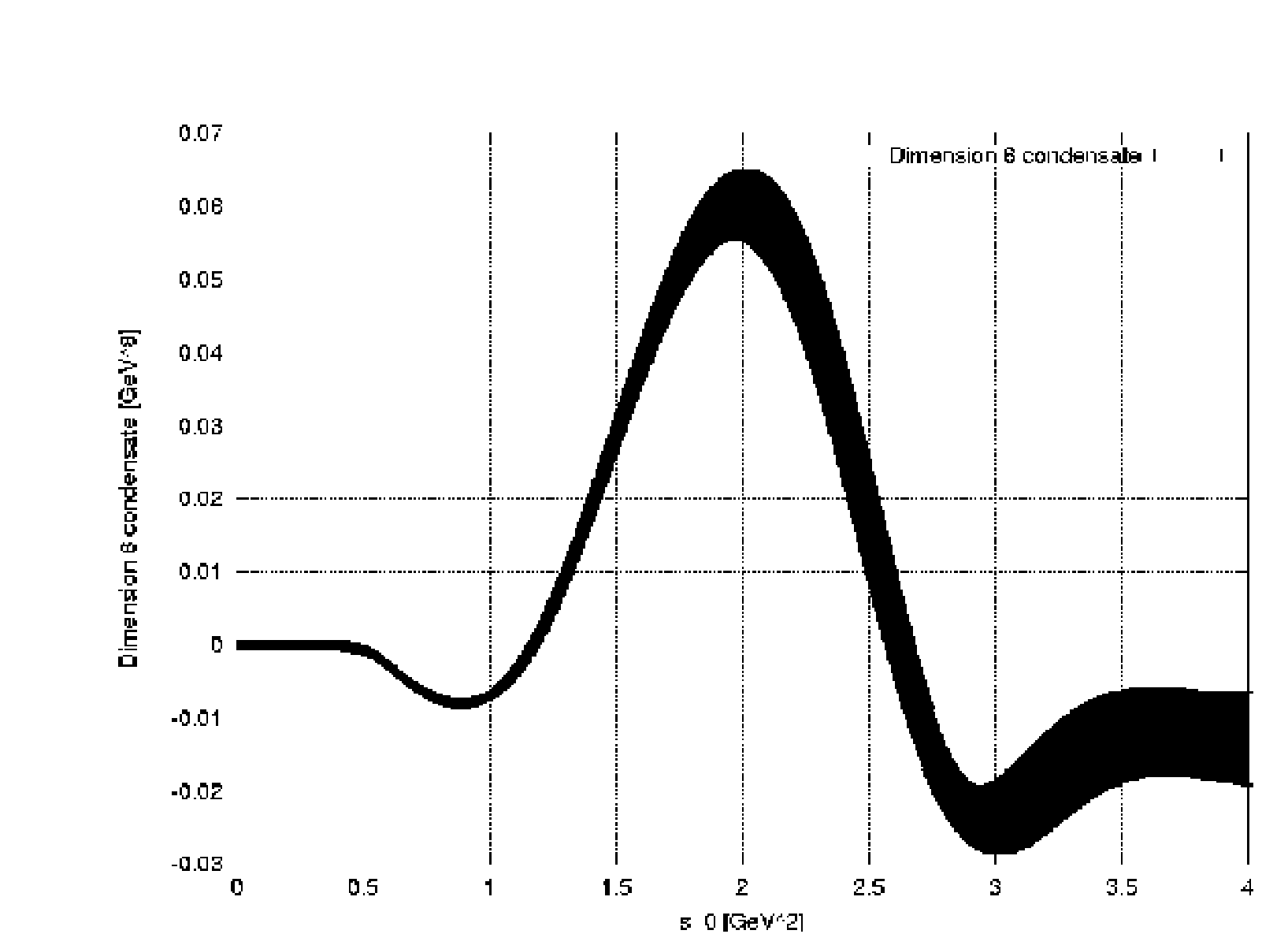}
\caption{}{\small 
Condensates $\la \mo_6 \ra$ and  $\la \mo_8 \ra$ as a function of
  $s_0$. The error bands include  the propagation of experimental
uncertainties. \label{condenextracterr}}
\end{center}
\end{figure}

Using Eq. \ref{condenconv}, we can extract from our
neural network
parametrization the values of the
nonperturbative condensates.
To be explicit, one would compute the
dimension 6 condensate from the sample
of trained neural networks in the following
way,
\be
\la \mo_6 \ra=\frac{1}{N_{\rep}}
\sum_{k=1}^{N_{\rep}} \la \mo_6^{(k)}\ra_{\rep}
=\frac{1}{N_{\rep}}
\sum_{k=1}^{N_{\rep}} \int_0^{s_0} ds s^2
\frac{1}{2\pi^2}\rho_{V-A}^{(k)(\net)}(s) \ .
\ee
Stable results are obtained for the dimension six condensate $\la \mo_6 \ra$ 
whereas higher condensates {\it e. g.}  $\la \mo_8 \ra$ are less stable. 
Fig. \ref{condenextracterr}
shows the outcome for $\la \mo_6\ra$ and
$\la \mo_8 \ra$ including the propagation of
statistical errors. It is clearly seen that convergence
in the limit of integration $s_0$ is obtained
due to the addition of sum rules and endpoints in the
learning procedure. The central values for the
condensates in the asymptotic limit, that is,
in the limit in which $s_0\to\infty$, come out to be:
\begin{eqnarray}
 \la \mo_6 \ra= -4.2~10^{-3}~\mathrm{GeV}^6 \ ,
\nonumber
\\
\la \mo_8 \ra= -12.7~10^{-3}~\mathrm{GeV}^8 \ .
\end{eqnarray}
The
value of the $\la \mo_6 \ra$ is a cross-check of the validity of our 
treatment: not only there are strong theoretical arguments that
support
the fact that $ \la \mo_6 \ra$ is negative
\cite{witten,latorre} 
but also all previous determinations with different techniques yield
negative results, being the majority of them compatible with
ours within errors. 

\begin{figure}[t]
\begin{center}
\epsfig{width=0.33\textwidth,figure=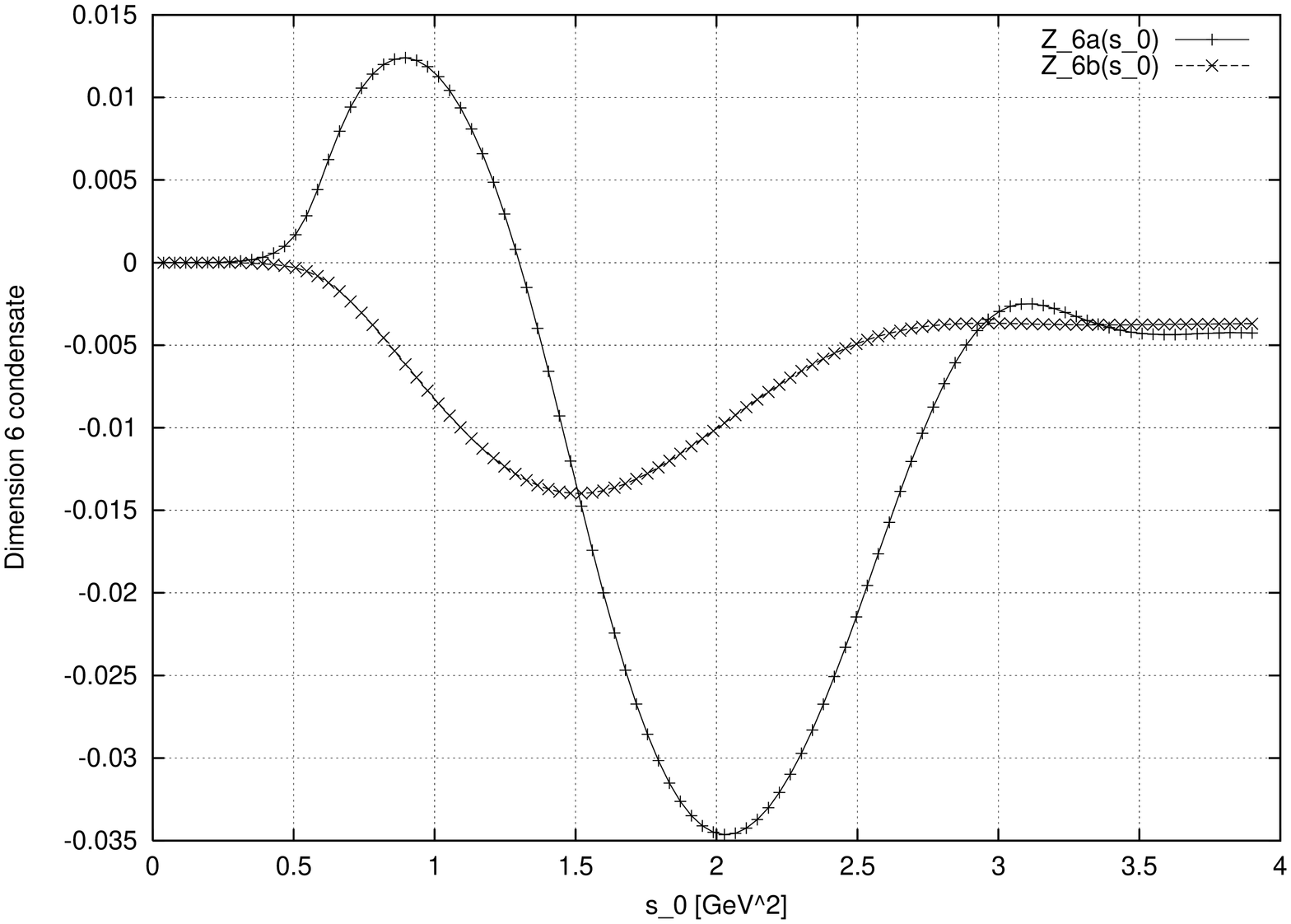,angle=-90} 
\epsfig{width=0.33\textwidth,figure=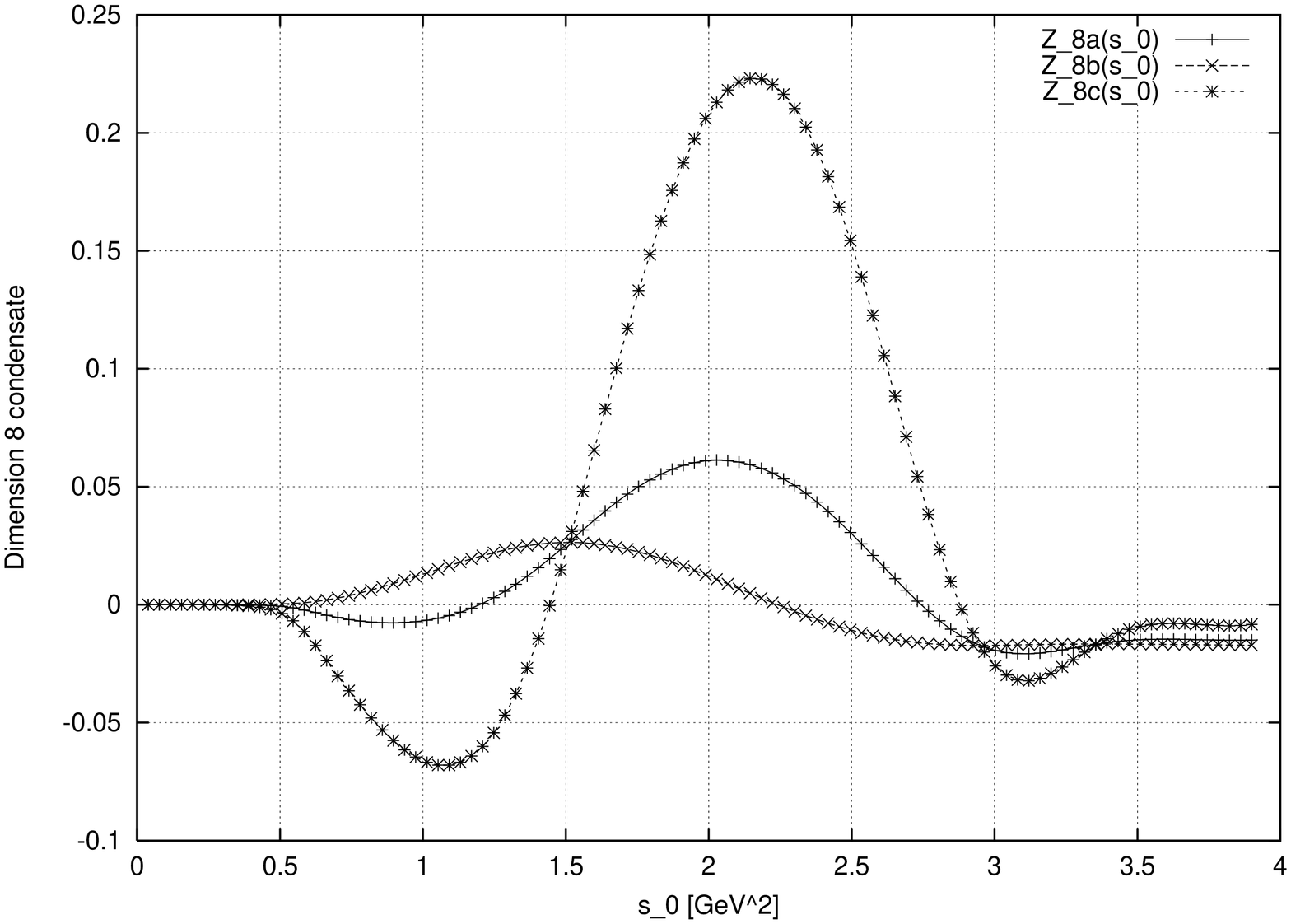,angle=-90}    
\end{center}
\caption{}{\small Determination of $\la \mo_6 \ra$ and $\la \mo_8 \ra$ 
 using different finite energy sum rules, as described
in Ref. \cite{condensates}.}
\label{fesr}
\end{figure}

We note that our evaluation of the condensates is compatible
with some of our previous evaluations and has a similar error.
This is though misleading as the error quoted here is only
statistical and a discussion on systematic errors is needed.
The discussion of the
various
sources of errors is  crucial  to our treatment.
The first criterion to judge the reliability of a QCD sum rule
is its independence, at large values of $s$ from the value of the upper
integration limit, that its, its saturation. We then need
to explore the values for the final condensates which are
stable against the limit of integration of the sum rule. This
stability criterion is completed with demanding independence
of the results on the specific polynomial entering the sum rule.
Further criteria are stability with respect to the 
artificial endpoints added to the data and with respect to
the relative weights in the error function used to train
the neural networks. 
A detailed analysis of the contributions of the different sources 
of uncertainty to the values of the
condensates can be found in \cite{condensates}.
These uncertainties include
the statistical error propagation from the experimental covariance
  matrix, which 
is  the best understood and treated error source
in our analysis, the 
choice of the finite energy sum rule, the dependence on the
implementation of the asymptotic vanishing of the spectral functions
and the dependence on the implementation of the
chiral sum rules.
For example, in Fig. \ref{fesr} we show that the final 
values of the condensates do not depend on the precise
finite energy sum rule used in their extraction.

Our  final determination of the nonperturbative condensates
including all relevant sources of uncertainties is 
\bea
 \la \mo_6 \ra=\lp -4.0 ~\pm 2.0\rp~10^{-3}~\mathrm{GeV}^6 \ ,
\nonumber \\
\la \mo_8 \ra=\lp -12 ~\epm{7}{11} \rp 10^{-3}~\mathrm{GeV}^8 \ ,
\\
 \la \mo_{10} \ra= \lp 7.8\pm 2.4\rp~10^{-2}~\mathrm{GeV}^{10} \ ,
\nonumber
\\
\la \mo_{12} \ra=\lp-2.6\pm 0.8 \rp~10^{-1}~\mathrm{GeV}^{12} \ .
\label{finalresults}
\eea

The values of the 
QCD nonperturbative condensates have been previously 
extracted from the experimental data, with different techniques and
different results, as summarized in Table \ref{lit}.
We include also the results of works which were published
after the original publication of Ref. \cite{condensates}.
Note that our results agree, at least on the sign, with that
of Refs. \cite{prades,cirigliano,sumrules,cirigliano2,valencia}.
See \cite{friot,narisoncon} for a detailed comparison of 
the different methods for the extraction of the condensates.

\begin{table}[t]  
\begin{center}  
\begin{tabular}{|c|c|c|}  
\hline
 Reference & $ \la \mo_6 \ra \times 10^{3}~ \mathrm{GeV}^6  $ 
&  $ \la \mo_8 \ra \times 10^{3} ~\mathrm{GeV}^8  $  \\
\hline  
 
 Ref. \cite{davier} 
  & $-6.4 \pm 1.6$  & $ 8.7\pm 2.4 $\\
\hline
 Ref. \cite{ioffe} 
  & $-6.8\pm 2.1$ & $7\pm 4$ \\
\hline
 Ref. \cite{prades} 
  & $-3.2\pm 2.0 $ & $ -12.4\pm 9.0$ \\
\hline
 Ref. \cite{periscond} 
  & $-9.5\pm 3 $ & $16.2\pm 5$ \\
\hline
Ref. \cite{cirigliano} 
  & $-4.45\pm 0.7$ & $-6.2\pm 3.2$  \\
\hline  
 Ref. \cite{sumrules} 
  & $-4\pm 1 $ & $-1.2\pm 6$ \\
\hline
 Ref. \cite{condensates} (This work)
  & $-4\pm 2 $ & $ -12 ~\epm{7}{11} $ \\
\hline
Ref. \cite{zyabluk}
  &  $-7.2\pm 1.2$ & $7.8\pm 2.5$ \\
\hline
Ref. \cite{friot}
  &  $-7.9\pm 1.6$ & $ 11.7\pm 2.7$  \\
\hline
Ref. \cite{narisoncon}
  &  $-8.7 \pm 2.3$ & $15.6 \pm 4.0 $  \\
\hline
Ref. \cite{valencia}
  & $-2.3 \pm  0.5$ & $-5 \pm 3$ \\
\hline
\end{tabular}
\end{center}
\caption{}{\small Summary of different extractions of the QCD vacuum
condensates.
Appropriate rescaling have been performed to allow the comparison of
different determinations. Note that some of the above
determinations appeared after the original publication of
Ref. \cite{condensates}.}
\label{lit}
\end{table}

Since the work presented in Ref. \cite{condensates} was published, there
have appeared several studies which also  determine the values
of the higher dimensional condensates from experimental data using a wide
variety of methods and techniques 
\cite{zyabluk,friot,narisoncon,ioffe,cata,physicstau,valencia,string}, from
large $N_C$ methods to new sum rule approaches and even
a determination inspired in higher dimensional
string theories. The spread of the results obtained
for the higher dimensional 
vacuum condensates using different techniques
show that their determination from experimental data is
still an open issue.

Summarizing, in this part of the thesis 
we have presented a determination
a bias-free neural network parametrization
of the $\rho_{V-A}(s)$ spectral function, inferred from the data, which
retains all the information on experimental errors and correlations, and
is supplemented with the additional theoretical input of the chiral sum rules.
As a byproduct of this analysis, we have
performed an extraction of the nonperturbative 
vacuum condensates $\la\mo_6\ra$ and $\la\mo_8\ra$
 aimed at minimizing the sources of
theoretical bias which might be cause of concern in existing determinations
of these condensates from spectral functions. Our final results give 
negative central values for  the dimension 6 and 8 condensates. 
These results take into account the propagation of statistical
errors and their correlations. Higher dimension condensates 
carry larger errors, although the sign of the condensates
seem to remain unaltered.

\clearpage

\section{Structure functions in deep inelastic scattering}
\label{dis_appl}

As has been discussed in the Introduction and in Section 
\ref{dis_theo},
the requirements of precision physics at hadron colliders have recently led
to a rapid improvement in the techniques for the determination of
parton distributions of the nucleon, 
which are mostly extracted from deep-inelastic
structure functions~\cite{pdfs}. Specifically, it is now mandatory to
determine accurately the uncertainty on these quantities. The main
problem to be faced here is that one is trying to determine the
uncertainty on a function, i.e., a probability measure on a space of
functions, and to extract it from a finite set of experimental data.
In Ref. 
\cite{f2ns} this problem was studied in a simpler context, namely, the
determination of a structure function and associate error from the
pertinent data. This sidesteps the technical complication of
extracting parton distributions from structure functions, but it does
tackle the main issue, namely the determination of an error on a
function.
Furthermore, the determination of a structure function and associate
error might be useful for a variety of applications, such as precision
tests of QCD (determination of $\alpha_s$~\cite{alphas}, tests of
sum rules) or the determination of polarized structure functions from
asymmetry data~\cite{smc}.

In this part of the thesis  we
extend the results of Ref.~\cite{f2ns} by constructing a parametrization
of the proton $F_2(x,Q^2)$ structure function which includes all available
data, in particular the HERA collider data. Besides the obvious
motivation of having state-of-the art results for this quantity, the
main aim of this work is to develop a set of techniques which are
required for the application of the method of Ref.~\cite{f2ns} to cases
where the handling of a large number of disparate data sets is
required. A more detailed description of this
part of the thesis can be found in the original work,
Ref. \cite{f2nnp}.
\begin{figure}[t]
\begin{center}
\epsfig{width=0.5\textwidth,figure=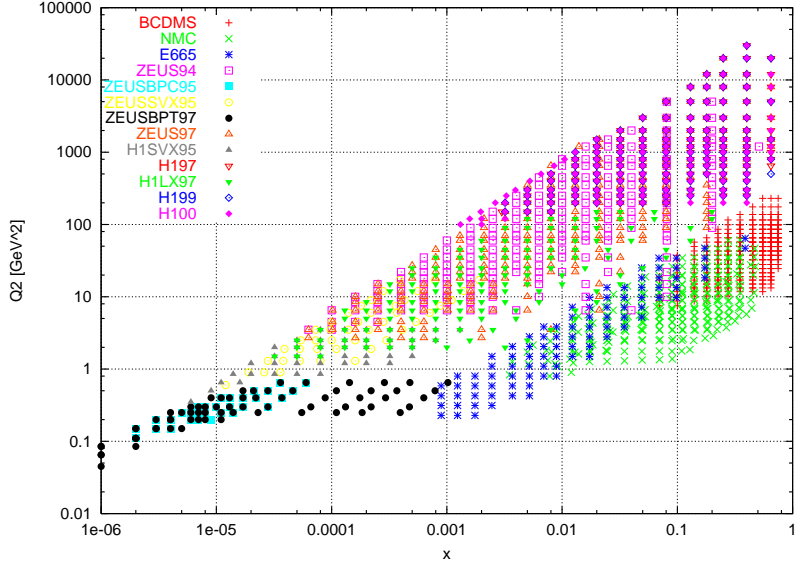,angle=-90}
\end{center}
\caption{\small Kinematic range of the available experimental data
 for the proton structure function
$F_2^p(x,Q^2)$.}
\label{kinall}
\end{figure}

Therefore, we construct a parametrization of 
the $F_2(x,Q^2)$ structure function
 based on all available unpolarized
charged lepton-proton deep-inelastic scattering
data. However, we do not include
early SLAC data, for which the covariance matrix is not available, since
they do not provide any extra kinematic coverage, and are anyway less precise
than later data.
This leaves a total of 13  experiments, listed in Table
\ref{exp}, along with
their main features. Note that the averages
of the different types of
uncertainties are given in percentage.
 The coverage of the $(x,Q^2)$ kinematic plane
afforded by these data is shown in Fig. \ref{kinall}.
Note that the kinematical coverage of the experimental data
included in the present parametrization
span 6 orders of magnitude in both $x$ and $Q^2$.

As has been discussed in Section \ref{dis_theo},
deep-inelastic
structure functions are defined by parametrizing the deep-inelastic
neutral current scattering cross section as
\be
\frac{d^2\sigma}{dxdQ^2}=\frac{4\pi \alpha^2}{xQ^4}
\lc xy^2F_1(x,Q^2)+(1-y)F_2(x,Q^2)+y\lp1-\frac{y}{2}\rp F_3(x,Q^2)\rc \ .
\label{f2}
\ee
We will construct a parametrization of the structure function
$F_2(x,Q^2)$, which provides the bulk of the contribution to
Eq.~\ref{f2}. In fact, a large number of experiments
present their results in terms of the
reduced cross section,
\be
\label{reduced}
\tilde{\sigma}(x,Q^2)\equiv \frac{xQ^4}{4\pi\alpha^2\lp 1+(1-y)^2\rp}
\frac{d^2\sigma(x,Q^2)}{dxdQ^2} \ ,
\ee
which is equal to $F_2(x,Q^2)$ in most of the $(x,Q^2)$
kinematical range.
For all experiments the longitudinal structure function $F_L(x,Q^2)$
contribution, defined as
\be
F_L(x,Q^2)=\lp 1+\frac{4M^2x^2}{Q^2}\rp F_2(x,Q^2)-2xF_1(x,Q^2) \ ,
\ee
 to the cross section  has already been subtracted by the
experimental collaborations, except for ZEUSBPC95, where
we subtracted it using the values published by the same experiment.
Note that the structure function $F_2$ receives contributions
from both $\gamma$ and $Z$ exchange, though the $Z$ contribution is
only non-negligible for the high $Q^2$ datasets
ZEUS94, H197,  H199 and H100. We will construct 
a parametrization of the structure
function $F_2$ defined in Eq. \ref{f2}, i.e. containing all
contributions, since it is closer
to the quantity which is experimentally measured,
the reduced cross section Eq. \ref{reduced}. 
When the experimental collaborations provide separately
the contributions to $F_2$ due to $\gamma$ or $Z$ exchange we have
recombined them in order to
get the full $F_2$ Eq. \ref{f2}.

\begin{table}[t]  
\begin{center}  
\tiny
\begin{tabular}{|cc|cc|c|ccccc|cc|} 
\hline
 Experiment & Ref. & $x$ range & $Q^2$ range & $N_{\dat}$   
& $\la\sigma_{\stat}\ra$& $\la\sigma_{\sys}\ra$& 
$\la\frac{\sigma_{\sys}}{\sigma_{\stat}}\ra$ &
$\la \sigma_{N}\ra$   & $\la\sigma_{tot}\ra$ 
&$\la\rho\ra$& $\la\mathrm{cov}\ra$ 
 \\
\hline
NMC & \cite{nmc} &$2.0~10^{-3}~-~6.0~10^{-1}$ & $0.5-75$ & 
 288 & 3.7 & 2.3 & 0.76  &2.0 &5.0 & 0.17 
& 3.8 
\\
\hline  
BCDMS & \cite{bcdms1,bcdms2} 
& $6.0~10^{-2}~-~8.0~10^{-1}$ & $7-260$ &  351 & 3.2 & 2.0 
& 0.56 &3.0 & 5.4 & 0.52 & 13.1 
\\
\hline
E665 & \cite{e665} &$8.0~10^{-4}~-~6.0~10^{-1}$ & $0.2-75$ & 
 91 & 8.7 & 5.2 & 0.67
&2.0 & 11.0 &0.21 &
   21.7  
 \\
\hline
ZEUS94 & \cite{ZEUS94} &$6.3~10^{-5}~-~5.0~10^{-1}$ & $3.5-5000$ &
 188 &7.9 & 3.5 & 1.04 &  2.0
& 10.2  &0.12 & 6.4  
\\
ZEUSBPC95 & \cite{ZEUSBPC95} & $2.0~10^{-6}~-~6.0~10^{-5}$ & $0.11-0.65$ & 34 
& 2.9 & 6.6 & 2.38 &2.0 & 7.6 & 0.61 & 34.1  
\\
ZEUSSVX95 & \cite{ZEUSSVX95} & $1.2~10^{-5}~-~1.9~10^{-3}$ & $0.6-17$ 
&  44 & 3.8& 5.7 & 1.53 &1.0 & 7.1  & 0.10 &  4.1  
\\
ZEUS97 & \cite{ZEUS97} & $6.0~10^{-5}~-~6.5~10^{-1}$ & $2.7-30000$ &  240 & 
5.0 & 3.1 & 0.93 & 3.0 & 6.7 & 0.29& 7.0 
\\ZEUSBPT97 & \cite{ZEUSBPT97} &$6.0~10^{-7}~-~1.3~10^{-3}$ & $0.045-0.65$ &
  70 
& 2.6 & 3.6 & 1.40 & 1.8 & 4.9  & 0.41 & 8.8
\\
\hline
H1SVX95 & \cite{H1SVX95} &$6.0~10^{-6}~-~1.3~10^{-3}$ & $0.35-3.5$ &  44 &  
 6.7& 4.6& 0.74 & 3.0 & 8.9 &0.36& 28.1 
\\
H197 & \cite{H197} & $3.2~10^{-3}~-~6.5~10^{-1}$& $150-30000$ & 130 & 
 12.5 & 3.2 & 0.31 &1.5 & 13.3 & 0.06 & 10.9 
\\
H1LX97 & \cite{H1LX97} & $3.0~10^{-5}~-~2.0~10^{-1}$ & $1.5-150$ &  133 &
2.6&2.2& 0.87 &1.7 & 3.9 & 0.30 & 3.9 
\\
H199 & \cite{H199} &$2.0~10^{-3}~-~6.5~10^{-1}$ & $150-30000$ &  126 &
 14.7& 2.8 &0.24 &1.8 & 15.2 & 0.05 & 11.0 
\\
H100 & \cite{H100} & $1.3~10^{-3}~-~6.5~10^{-1}$ & $100-30000$ &  147 &
9.4 & 3.2 & 0.42 &1.8 & 10.4 & 0.09 & 8.6 
\\
\hline

\end{tabular}
\caption{\small Experiments included in this analysis. All values of
  $\sigma$ and cov are given as percentages.}
\label{exp}
\end{center} 
\end{table}
\normalsize

Experimental data on deep-inelastic structure functions
consist on central values, statistical errors, and the
contributions from the different sources of correlated and
uncorrelated uncertainties.
Uncorrelated systematic errors are added in quadrature to the
statistical errors to construct the total uncorrelated
uncertainty.
On top of this, for some experiments, in particular for
the ZEUS94,
ZEUSSVX95 and ZEUSBPT97 experiments some uncertainties are
asymmetric. 
For the treatment of asymmetric uncertainties we follow
the prescription discussed in Section \ref{asymm}.

\begin{figure}[t]
\begin{center}
\epsfig{width=0.32\textwidth,figure=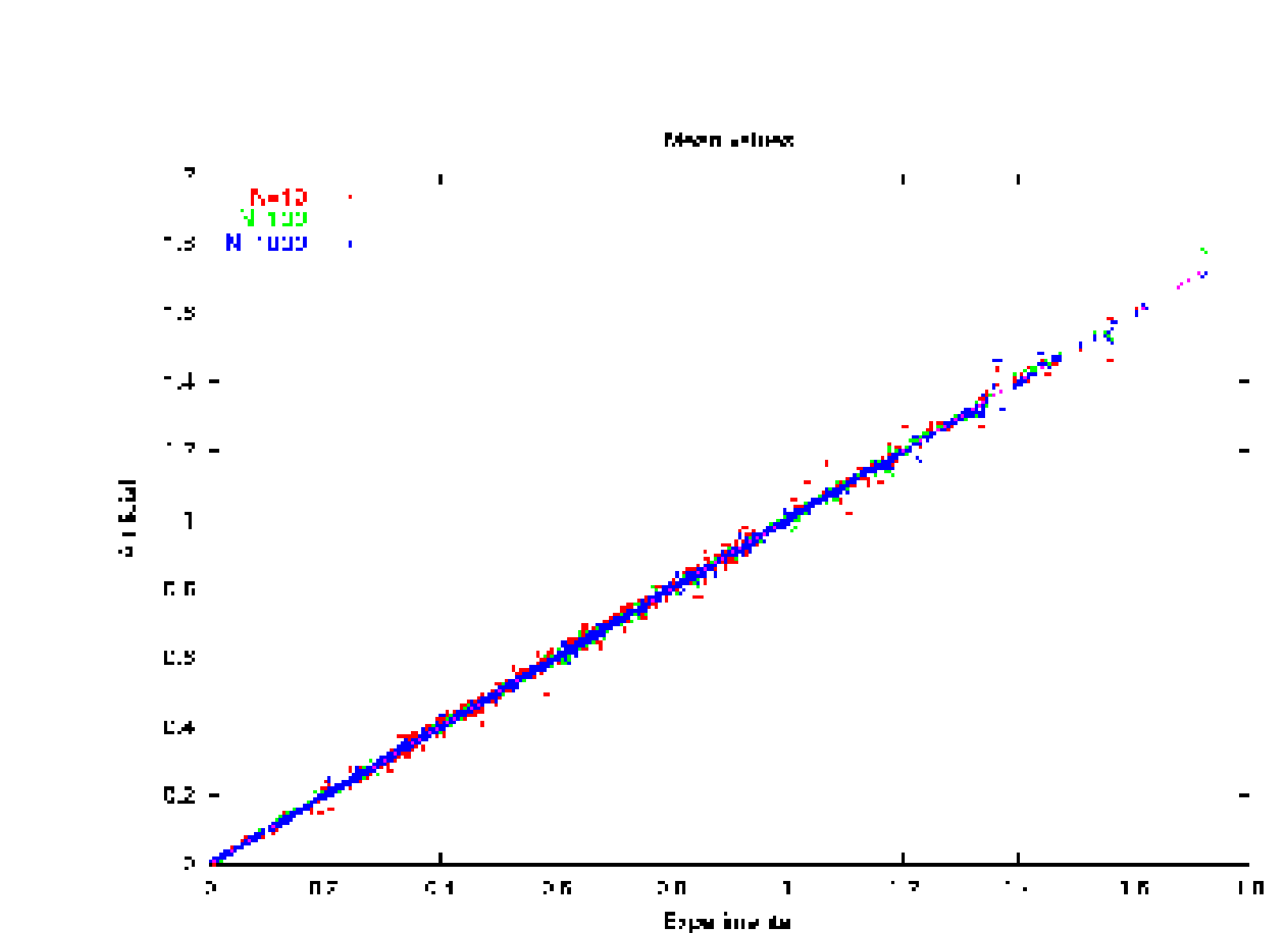,angle=-90}
\epsfig{width=0.32\textwidth,figure=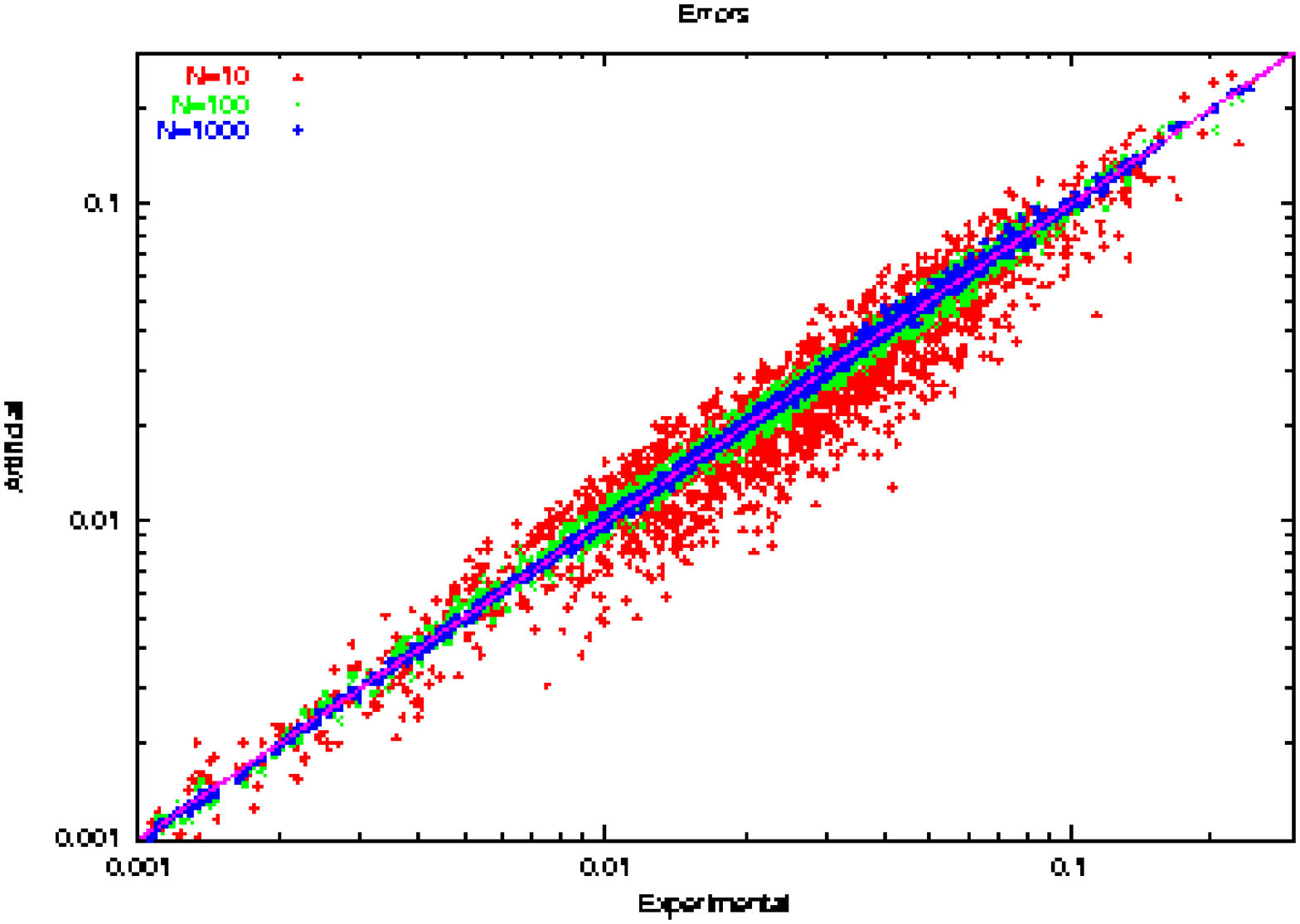,angle=-90}
\caption{\small
Scatter plot of experimental data versus
 Monte Carlo artifical data for both central values and errors.}
\label{genplots}
\end{center}
\end{figure}

The construction of a parametrization of $F_2(x,Q^2)$ according to
general strategy described in
Chapter \ref{general} consists in three steps: generation of a
set of Monte Carlo replicas of the original 
experimental data,  training of a
neural network to each replica and finally statistical validation
of the constructed probability measure. 
The Monte Carlo replicas of the original experiment are generated as a
multi-gaussian distribution: each replica is given, following
Eq. \ref{gen1}, by a set of values
\be
\label{replicas}
F_i^{(\art)(k)}=\lp 1+r_{N}^{(k)}\sigma_N\rp\lp F_i^{\rm (\mrexp)}+
 \sum_{l=1}^{N_{\sys}}r_{\sys,li}^{(k)}\sigma_{\sys,li}+r_{t,i}^{(k)}
\sigma_{t,i}\rp
 \ , \qquad k=1,\ldots,N_{\rep} \ ,
\ee
where  the various errors are 
defined in Eqns. \ref{totuncorr}-\ref{toterr}.
As has been discussed before,
the value of $N_{\rep}$ is determined in such a way that
the Monte Carlo set of replicas models faithfully the probability
distribution of the data in the  original  set.
A comparison of
expectation values, variances and correlations of the Monte Carlo set
with the corresponding input experimental values as a function of the
number of replicas is shown in Fig.~\ref{genplots},
where we display scatter plots of the central values and errors 
for samples of 10, 100 and 1000 replicas. The corresponding plot for
correlations is essentially the same as  that shown in Ref.~\cite{f2ns}.
A more quantitative comparison 
is performed using the statistical estimators as
defined in Section \ref{mcerr}. 
The results for these estimators  are presented in
Table \ref{gendata_fd}. Note in particular the scatter correlations $r$
for central values, errors and correlations, which
indicate
the size of the spread of data around a straight line. The table shows
that a sample of  1000 replicas is sufficient to ensure  average 
scatter correlations of 99\% and accuracies
of a few percent on structure functions, errors and correlations. 

\begin{table}[t]  
\begin{center}  
\begin{tabular}{|c|ccc|} 
\multicolumn{4}{c}{
$F_2^p(x,Q^2)$}\\   
\hline
 $N_{\rep}$ & 10 & 100 & 1000 \\
\hline
$\la PE\lc\la F^{(\art)}\ra_{\rep}\rc\ra$ & 1.88\% 
& 0.64\%  &  0.20\%    \\
$r\lc F^{(\art)} \rc$ & 0.99919 & 0.99992 &  0.99999 \\
\hline
$\la V\lc \sigma^{(\art)}\rc\ra_{\dat}$ & $6.7\times 10^{-4}$ &
$2.0\times 10^{-4}$ & $6.9\times 10^{-5}$  \\
$\la PE\lc \sigma^{(\art)}\rc\ra_{\dat}$ & 37.21\%
& 11.77\% & 3.43\%     \\

$\la \sigma^{(\art)}\ra_{\dat}$ & 0.0292 & 0.0317& 0.0316\\
$r\lc \sigma^{(\art)}\rc$ & 0.945 &  0.995 & 0.999 \\
\hline
$\la V\lc \rho^{(\art)}\rc\ra_{\dat}$ & $8.1\times 10^{-2}$  
&  $7.8\times 10^{-3}$ &  $7.3\times 10^{-4}$ \\

$\la \rho^{(\art)}\ra_{\dat}$ & 0.3048 & 0.3115 & 0.2920\\
$r\lc \rho^{(\art)}\rc$ & 0.696 &  0.951 & 0.995 \\
\hline
$\la V\lc \mathrm{cov}^{(\art)}\rc\ra_{\dat}$ & $5.2\times 10^{-7}$  
&  $6.8\times 10^{-8}$ &  $6.9\times 10^{-9}$ \\

$\la \mathrm{cov}^{(\art)}\ra_{\dat}$ & 0.00013 & 0.00018 & 0.00015\\
$r\lc \mathrm{cov}^{(\art)}\rc$ & 0.687 &  0.941 & 0.994 \\
\hline

\end{tabular}
\caption{\small Comparison between experimental and 
Monte Carlo data.\hfill\break
The experimental data have
$\la \sigma^{(\mrexp)}\ra_{\dat}=0.0311$, $\la \rho^{(\mrexp)}\ra_{\dat}=
0.2914$ and $\la \mathrm{cov}^{(\mrexp)}\ra_{\dat}=0.00015$. All
statistical indicators are defined in Section
\ref{mcerr}.}
\label{gendata_fd}
\end{center} 
\end{table}

$N_{\rep}$ neural networks are then trained on the Monte
Carlo replicas, by training each
neural network on all the $F_i^{(k)(\art)}$ data points in the
  $k$-th replica. The architecture of the networks is the same as in
Ref.~\cite{f2ns}. The training is
subdivided in three epochs, each
based on the minimization of a different error function,
as described in Section \ref{minimstratt}.
 First one minimizes
$E_1^{(k)}$, then $E_2^{(k)}$ and finally correlated
systematics are incorporated in the 
minimization of $E_3^{(k)}$.
The rationale behind this three-step procedure is that the true minimum which
the fitting procedure must determine is that of the 
error function with correlated systematics $E_3^{(k)}$
Eq.~\ref{er3}. However, this is nonlocal and
time consuming to compute. It is therefore advantageous to 
look for its rough features at first, 
then refine its search, and finally determine its actual location.

The minimum during the first two epochs is found using
back-propagation (BP), discussed in
Section \ref{minimstratt}. This method is  not
suitable for the minimization of the nonlocal function
Eq.~\ref{er3}. 
In Ref.~\cite{f2ns} BP was used throughout, and the
third epoch was omitted. This is acceptable
provided the total systematics is small in comparison
to the statistical errors, and indeed it was verified that a
good approximation to the minimum of Eq.~\ref{er3} could be
obtained from the ensemble of neural networks. This is no longer the
case for the present extended data set, as we shall see explicitly in
brief.
Therefore, the full $E_3^{(k)}$ Eq.~\ref{er3} is
minimized in the third training
epoch by means of genetic algorithms (GA), 
also discussed in Section \ref{minimstratt}. 

At the end of the GA training we are left with a sample of $N_{\rep}$
neural networks, from which e.g. the value of the structure function
at $(x,Q^2)$ can be computed as
\be
F_2(x,Q^2)=\frac{1}{N_{\rep}}\sum_{k=1}^{N_{\rep}} F^{(\net)(k)} (x,Q^2)
 \ .
\label{avf2}
\ee
The goodness of fit of the final set is thus measured by the $\chi^2$
per data point, Eq. \ref{chi2tot},
 which, given the large number of data points is
essentially identical to the $\chi^2$ per degree of freedom.
        
In order to apply the general strategy to the
present problem several points must
be considered: the choice
of training parameters and training length, the choice of the actual
data set, and the choice of theoretical constraints.
We now address these issues in turn.
The parameters and length
for the first two training epochs have been determined by
inspection of the fit of a neural network to the central experimental
values. Clearly, this choice is less critical, in that it is
only important in order for the
 training to be reasonably fast, but it does not impact on final result.

After these first two  training epochs, 
the diagonal error function $E_2^{(0)}$,  Eq.~\ref{er2},
for a training on central values 
is of
order two for the central data set, with a similar length
of the training than that that
 was required to reach $E_2^{(0)}\approx 1.3$ for
the smaller data set of Ref.~\cite{f2ns}.
The value of $E_3^{(0)}$, Eq.~\ref{er3}, which is always bounded by it, 
$E_3 \le E_2 $ is accordingly smaller (see Table \ref{chis}).
The training algorithm then switches to GA minimization of
the $E_3$, Eq.~\ref{er3}. The determination of the length of this
training epoch is critical, in that it controls the form of the final
fit. This can only be done by looking at the features of the full
Monte Carlo sample of neural networks.

\begin{table}[ht]  
\begin{center}  
\begin{tabular}{|c||cc|cc|}
\multicolumn{5}{c}{TABLE 3} \\
\hline
 & \multicolumn{2}{c|}{A} & \multicolumn{2}{c|}{B} \\
\hline
 Experiment & $E_2$  &  $E_3$ 
& $E_2$  &  $E_3$ \\
\hline
Total & 2.05 & 1.54 & 2.03 & 1.36 \\
\hline
NMC       &  1.97 & 1.56 & 1.74  & 1.54 \\
BCDMS     &  1.93 & 1.66 & 1.32  & 1.26 \\
E665      &  1.64 & 1.37 & 1.83  & 1.38 \\
ZEUS94    &  3.15 & 2.26 & 3.01  & 2.21 \\
ZEUSBPC95 &  4.18 & 1.32 & 5.18  & 1.24 \\
ZEUSSVX95 &  3.37 & 1.88 & 5.68  & 2.11 \\
ZEUS97    &  2.33 & 1.54 & 3.02  & 1.37 \\
ZEUSBPT97 &  2.82 & 1.97 & 2.08  & 1.22 \\
H1SVX95   &  3.21 & 0.96 & 4.74  & 1.09  \\
H197      &  0.86 & 0.76 & 1.08  & 0.87  \\
H1LX97    &  1.96 & 1.46 & 1.50  & 1.18 \\
H199      &  1.15 & 1.07 & 1.10  & 1.01 \\
H100      &  1.59 & 1.50 & 1.48  & 1.26 \\
\hline
\end{tabular}
\caption{\small The uncorrelated 
error function, $E_2^{(0)}$, Eq.~\ref{er2}, and  the 
correlated one, $E_3^{(0)}$,  
Eq.~\ref{er3}, for
  the fit to the central data points: (A)
after the back-propagation training epoch  and (B)
after the final genetic algorithms training epoch.}
\label{chis}
\end{center} 
\end{table}

Before addressing this issue, however, it turns out to be necessary to
consider the possibility of introducing cuts in the data set. Indeed,
consider the results
obtained after a GA training of $ 4\times 10^4$
generation  to the central data set,
displayed in
Table \ref{chis}. This is a rather long training: indeed, 
in each  GA generation all the data are shown
to the nets. Hence  in $ 4\times 10^4$ GA generations the
data are shown to the nets $0.7\times10^8$ times, comparable to the
number of times they are shown to the nets during BP training.
It is apparent that whereas $E_3\sim1$ for most
experiments, it remains abnormally high for NMC and especially ZEUS94
and ZEUSSVX95. Because of the weighted training which has been
adopted, this is unlikely to be due to insufficient training of these
data sets, and is more likely related to problems of these data sets. 

Whereas ZEUSSVX95 only contains a small number of
data points, NMC and ZEUS94 account each for more than 10\% of the
total number of data points, and thus they can bias final results
considerably.
 The case of NMC was discussed in detail in
Ref.~\cite{f2ns}.
This data set is the only one to cover the
medium-$x$, medium-small $Q^2$ region (see Fig. \ref{kinall}) 
and thus it
cannot be excluded from the fit. As discussed in Ref.~\cite{f2ns}, 
 the relatively large value of $E_3$ for this
experiment is a consequence of internal inconsistencies within
the data set. A
value of $E_3\approx1.5$ indicates that the neural nets do not
reproduce the subset of data which are incompatible with the
bulk, as it should be, whereas  a 
value $E_3\approx 1$ could only be obtained by
overlearning, i.e. essentially by fitting irrelevant fluctuations (see
Ref.~\cite{f2ns}). 

Let us now consider the case of ZEUS94. The kinematic region of
this experiment is entirely covered by the ZEUS97, H197, H199 and H100
experiments. We
can therefore study the impact of excluding this experiment from the
global fit, without information loss. The results obtained in such
case are displayed in Table \ref{zeus94inc}: 
when the experiment is not fitted the
$E_3$ value for all experiments with which it overlaps improves
and so does the global $E_3$, whereas $E_3$
for ZEUS94 itself only deteriorates by a comparable amount, despite the fact
that the experiment is now not fitted at all. We conclude that the
experiment should be excluded from the fit, since its inclusion results
in a deterioration of the fit quality, whereas its exclusion does not entail
information loss. Difficulties in the inclusion of this experiment in
global fits were already pointed out in Refs.~\cite{GKK,alekhin},
where it was suggested that they may be due to underestimated
or non-gaussian uncertainties. 
It is likely that  ZEUSSVX95 has similar problems. However, its
inclusion 
in the fit is no
reason of concern, even if its high $E_3$ value were due to incompatibility
of this experiment with the others or underestimate of its
experimental uncertainties, because of the small number of
data points. It is therefore retained in the data set. Our final data
set thus includes all experiments in Table \ref{exp}, except ZEUS94.
We are thus left
with $N_{\dat}=$1698 data points.

\begin{table}[ht]  
\begin{center}  
\begin{tabular}{|cc|}
\multicolumn{2}{c}{TABLE 4} \\
\hline
 Experiment & $E_3$  \\
\hline
Total     & 1.25    \\
\hline
NMC       & 1.51    \\
BCDMS     & 1.24    \\
E665      & 1.23    \\
ZEUS94    & 2.28    \\
ZEUSBPC95 & 1.16    \\
ZEUSSVX95 & 2.08    \\
ZEUS97    & 1.37    \\
ZEUSBPT97 & 1.00    \\
H1SVX95   & 1.04    \\
H197      & 0.84    \\
H1LX97    & 1.19    \\
H199      & 1.00    \\
H100      & 1.24    \\
\hline
\end{tabular}
\caption{\small The same fit as the last column of 
Table \ref{chis} if the ZEUS94
  data are excluded from the fit. }
\label{zeus94inc}
\end{center} 
\end{table}

For the sake of future applications, it is interesting to ask how the
procedure of selecting experiments in the data set can be
automatized. This can be done in an iterative way as follows: first, 
a neural net (or  sample of neural nets) is trained 
on only one experiment; then, 
the total $E_3$ for the full
data set is computed using this neural net (or sample of nets); the
procedure is then repeated for all experiments, and 
the experiment which leads to the smallest total $E_3$ is selected. In
the second iteration, the net (or sample of nets) is trained on the
experiment selected in the first iteration plus any of the other
experiments, thereby leading to the selection of a second experiment
to be added to that selected previously, and so on. The process can be
terminated before all experiments are selected, for instance if it is
seen that the addition of a new experiment does not lead to a
significant improvement in $E_3$ for a large
enough length of training.

\begin{figure}[t]
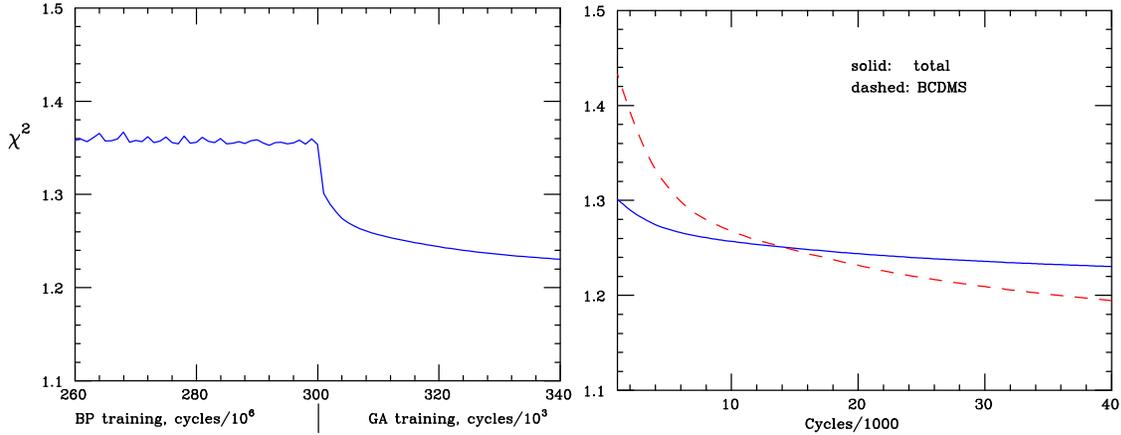

\begin{center}
\epsfig{width=0.495\textwidth,figure=chi2cme.ps}
\epsfig{width=0.47\textwidth,figure=chi2cme_bcd.ps}
\caption{\small Dependence of the total
$\chi^2$ Eq.~\ref{chi2tot} on the length of
  training: (left) total training (right) detail of the GA training.}
\label{trlength}
\end{center}
\end{figure}

We now proceed to discuss the length of training for our final
 data set.
The total $\chi^2$, Eq.~\ref{chi2tot}, as
computed from averages over the trained
network sample, is shown in Fig. \ref{trlength} 
as a function of the number of GA
 generations.
The $\chi^2$ decreases very rapidly during
the first few hundreds of training generations, a typical
feature of genetic algorithms minimization.
After about 5000 training generations, the $\chi^2$ as a function of the
 training length
essentially flattens for all experiments but BCDMS. The further
decrease of
 the total $\chi^2$ is then due essentially to the decrease  of the
 contribution from BCDMS. A
 training length of $ 4\times 10^4$ GA generations
 is necessary in order for the
 $\chi^2$ of BCDMS to flatten out at $\chi^2\sim 1.2$. 
As discussed in Ref.~\cite{f2ns}, the
 BCDMS data can only be learnt with a  longer training because they have  high
 precision while being located in the intermediate $x$ (valence)
 region, where the parton distributions display significant variation.

 The $E_3^{(0)}$ values for the fit of a neural net to the
 central data with this training is given in Table
\ref{chis}.  It
 shows that all experiments are well reproduced with the exceptions
discussed above. It is interesting to observe that while  $E_3^{(0)}$
 Eq.~\ref{er3} decreases significantly during the GA training,
 the uncorrelated  $E_2^{(0)}$, Eq.~\ref{er2}, decreases
 marginally, and in fact it actually increases for several HERA 
 experiments. This
 shows that correlations are sizable for the HERA experiments, so
 that the approach of Ref.~\cite{f2ns}, based on the minimization of
 $E_2$,
is not adequate in this case. GA minimization appears to be very
 efficient in reducing the $E_3$ value relatively fast.

\begin{table}[ht]  
\begin{center}  
\begin{tabular}{|cc|} 
\multicolumn{2}{c}{
$F_2^p(x,Q^2)$}\\   
\hline
 $N_{\rep}$ & 1000 \\
\hline
$\chi^2$  & 1.18 \\
$\la E \ra$  & 2.52 \\
$r\lc F^{(\net)}\rc$ & 0.99  \\
${\cal R}$& 0.54  \\
\hline
$\la V\lc \sigma^{(\net)}\rc\ra_{\dat}$ & $1.2~10^{-3}$  \\
$\la PE\lc \sigma^{(\net)}\rc\ra_{\dat}$ &  $80\%$ \\

$\la \sigma^{(\mrexp)}\ra_{\dat}$ & 0.027   \\
$\la \sigma^{(\net)}\ra_{\dat}$ & 0.008   \\

$r\lc \sigma^{(\net)}\rc$ &  0.73   \\
\hline
$\la V\lc \rho^{(\net)}\rc\ra_{\dat}$ & 0.20 \\
$\la \rho^{(\mrexp)}\ra_{\dat}$ &  0.31\\
$\la \rho^{(\net)}\ra_{\dat}$ & 0.67 \\
$r\lc \rho^{(\net)}\rc$ &  0.54 \\
\hline
$\la V\lc \mathrm{cov}^{(\art)}\rc\ra_{\dat}$ & $3.3~10^{-7}$ \\
$\la \mathrm{cov}^{(\mrexp)}\ra_{\dat}$ & $1.3~10^{-4}$ \\
$\la \mathrm{cov}^{(\net)}\ra_{\dat}$ & $3.6~10^{-5}$ \\
$r\lc \mathrm{cov}^{(\net)}\rc$ & 0.49 \\
\hline
\end{tabular}
\caption{}{\small Estimators of the final  results for the
constructed probability measure of $F_2(x,Q^2)$.}
\label{results_new}
\end{center} 
\end{table}

\begin{table}[ht]  
\begin{center}
\tiny
\begin{tabular}{|c|ccc|}    
\hline
 Experiment & NMC & BCDMS & E665 \\
\hline
$\chi^2$  & 1.47  & 1.19 & 1.20 \\
$\la E \ra $ & 2.69  & 3.17 & 2.29 \\
$r\lc F^{(\net)}\rc$ &  0.96  & 0.99 & 0.91 \\
${\cal R}$&  0.59 & 0.50 & 0.56 \\
\hline
$\la V\lc \sigma^{(\net)}\rc\ra_{\dat}$ & $0.002$ & $1.9~10^{-5}$ 
& 0.0013 \\
$\la PE\lc \sigma^{(\net)}\rc\ra_{\dat}$ & 0.63 & 0.56 & 0.89 \\

$\la \sigma^{(\mrexp)}\ra_{\dat}$ & 0.017 & 0.007 & 0.032 \\
$\la \sigma^{(\net)}\ra_{\dat}$ & 0.008  & 0.005 & 0.008 \\

$r\lc \sigma^{(\net)}\rc$ &  0.23   & 0.98 & 0.17 \\
\hline
$\la V\lc \rho^{(\net)}\rc\ra_{\dat}$ & 0.51  & 0.69 & 0.29 \\

$\la \rho^{(\mrexp)}\ra_{\dat}$ & 0.17  & 0.52 & 0.20\\
$\la \rho^{(\net)}\ra_{\dat}$ & 0.84  & 0.86 & 0.60 \\
$r\lc \rho^{(\net)}\rc$ & 0.08 & 0.73 & 0.05 \\
\hline
$\la V\lc \mathrm{cov}^{(\art)}\rc\ra_{\dat}$ & $2.4~10^{-9}$ &
$1.8~10^{-9}$  &  $4.5~10^{-9}$\\
$\la \mathrm{cov}^{(\mrexp)}\ra_{\dat}$ &  $4.4~10^{-5}$ & 
$3.8~10^{-5}$&  $1.7~10^{-4}$ \\
$\la \mathrm{cov}^{(\net)}\ra_{\dat}$  &  $ 5.2~10^{-5}$ &
 $2.3~10^{-5}$&  $3.3~10^{-5}$ \\
$r\lc \mathrm{cov}^{(\net)}\rc$ & -0.03 & 0.98 & 0.16 \\
\hline
\end{tabular}\\\bigskip
\begin{tabular}{|c|ccccccccc|}    
\hline
 Experiment &  ZEUSBPC95 & ZEUSSVX95 &
ZEUS97 & ZEUSBPT97 & H1SVX95 & H197 & H1LX97 & H199 & H100 \\
\hline
$\chi^2$  & 1.02 & 2.08 & 1.35 & 
0.86& 0.67 & 0.71& 1.07 & 0.90 & 1.11 \\
$\la E \ra $ & 2.07 & 2.03 & 2.24 
& 2.08 & 2.03 & 1.91 & 2.41 & 1.93 & 2.11 \\
$r\lc F^{(\net)}\rc$ & 0.98 & 0.96 & 0.99& 0.99&
0.97 & 0.99 & 0.99& 0.98 & 0.99 \\
${\cal R}$& 0.51 & 0.66 & 0.55& 0.55&
0.44 & 0.46 & 0.53& 0.48 & 0.54 \\
\hline
$\la V\lc \sigma^{(\net)}\rc\ra_{\dat}$ 
& $4.3~10^{-4}$ & 0.0035 & 0.0010 &$1.3~10^{-4}$& 0.0043& 
0.0030 & 0.0005& 0.003& 0.0013 \\

$\la PE\lc \sigma^{(\net)}\rc\ra_{\dat}$ & 0.91 &
0.94&0.87& 0.72& 0.96& 0.95 & 0.75 & 0.96 & 0.93\\

$\la \sigma^{(\mrexp)}\ra_{\dat}$ & 0.022 & 0.061 &
0.037& 0.012 & 0.063& 0.040 & 0.027 & 0.051 & 0.030 \\
$\la \sigma^{(\net)}\ra_{\dat}$ & 0.006 & 0.013
& 0.011 & 0.006 & 0.011 & 0.008& 0.008& 0.008 & 0.009 \\

$r\lc \sigma^{(\net)}\rc$ & 0.85 &0.72 & 0.86&
0.73 & 0.84 & 0.87 & 0.42 & 0.82 & 0.89\\
\hline
$\la V\lc \rho^{(\net)}\rc\ra_{\dat}$ & 0.09 &
0.30 & 0.12 & 0.14& 0.118 & 0.14 & 0.31& 0.16 & 0.14\\

$\la \rho^{(\mrexp)}\ra_{\dat}$ & 0.61 & 0.24 &
0.28 & 0.40 & 0.36 & 0.06 & 0.29& 0.05 & 0.09\\
$\la \rho^{(\net)}\ra_{\dat}$ & 0.77 & 0.64 &
0.39& 0.63 & 0.57 & 0.27 &0.58& 0.29 & 0.26 \\
$r\lc \rho^{(\net)}\rc$ & 0.53 & 0.40 &
0.66& 0.60 & 0.48 & 0.51 &0.69 & 0.37 & 0.55 \\
\hline
$\la V\lc \mathrm{cov}^{(\art)}\rc\ra_{\dat}$ & $6.4~10^{-8}$& $1.9~10^{-6}$ &
$3.4~10^{-7}$& $1.4~10^{-9}$& $3.0~10^{-6}$&
$3.8~10^{-7}$& $3.8~10^{-8}$& $2.7~10^{-7}$& $1.7~10^{-7}$\\

$\la \mathrm{cov}^{(\mrexp)}\ra_{\dat}$ & $2.8~10^{-4}$& $8.5~10^{-4}$&
$3.7~10^{-4}$& $5.8~10^{-5}$ &
0.0014 & $1.0~10^{-4}$& $2.1~10^{-4}$ & 
$1.4~10^{-4}$& $9.6~10^{-5}$\\

$\la \mathrm{cov}^{(\net)}\ra_{\dat}$  & $2.8~10^{-5}$ & $1.2~10^{-4}$&
$3.2~10^{-5}$& $2.3~10^{-5}$ & $7.0~10^{-5}$&
$1.510^{-5}$& $6.9~10^{-5}$ & $1.6~10^{-5}$& 
$2.2~10^{-5}$\\

$r\lc \mathrm{cov}^{(\net)}\rc$ & 0.69 & 0.48 & 
0.77 & 0.65 & 0.53 & 0.61& 0.57 & 0.54 & 0.58\\
\hline
\end{tabular}
\caption{\small Final results for the individual 
experiments: fixed target (top) and HERA (bottom)
\label{resultsexp_new} }
\end{center} 
\end{table}

We finally turn to the issue of theoretical constraints. The
only theoretical assumption on the shape of $\fd$ 
 is that it satisfies the kinematic constraint $F_2(1,Q^2)=0$
for all $Q^2$. As this constraint is local, its implementation
is straightforward: it can be enforced by including in the data set a
number
of artificial data points which satisfy the constraint with a suitably 
tuned error. In the present fit we have checked that
the best choice is to add a number of artificial  points at $x=1$,
as discussed in Section \ref{constraints},
equal to
2\% of the experimental trained points (33 points with ZEUS94 excluded
from the fits), and with error 
equal to one tenth of the mean statistical error of  the
trained points. These points are equally spaced in  $\ln Q^2$,
within the  range covered by the 
experimental data. 

The result of the minimization of a single neural net
to the central data points is shown
in Table \ref{zeus94inc}. 
The results for the final set of 1000 neural networks are  
displayed in Table \ref{results_new}, while in
Table \ref{resultsexp_new} 
we give the details of results for each experiment. Note that
the figure of merit for the minimization $E_3$, Eq.~\ref{er3}, and
its average, defined by Eq. \ref{avchi2}, differs from
the full $\chi^2$, Eq.~\ref{chi2tot}, not only because the latter is
computed from the structure function averaged over nets
Eq.~\ref{avf2}, but also because of the different treatment of
normalization errors in the respective covariance matrices,
Eq.~\ref{covmatnn} and Eq.~\ref{covmat}.
Besides the $\chi^2$ we
also list the values of various quantities, defined in 
Section \ref{validation},
which can be used to assess the goodness of fit.

The quality of the final fit is somewhat better than that of the fit
to the central data points shown in Table
\ref{zeus94inc}. In particular, with the
exception of NMC (which is likely to have internal 
inconsistencies~\cite{f2ns}) and
ZEUSSVX95 (which is likely to have the same problems as those of ZEUS94) 
the $\chi^2$ per degree of freedom is of order
1 for all experiments. It is interesting to note that the $\chi^2$ for
the neural network average is rather better than the average
$\langle E_3 \rangle$.  The (scatter)
correlation between experimental data and the neural network
prediction equals one to about 1\% accuracy, with the exception of NMC,
ZEUSSVX95 (which have the aforementioned problems) and E665. The
E665 kinematic region
overlaps almost entirely (apart from very small $Q^2<1$~GeV$^2$) with
that of NMC and BCDMS, while having lower accuracy (this is why the
experiment was not included in the fits of Ref.~\cite{f2ns}). The data
points corresponding to this experiment are therefore essentially
predicted by the fit to other experiments, thus explaining the
somewhat smaller scatter correlation.

The  average neural network variance is in general substantially smaller than
the average experimental error, typically by a factor $3-4$. This is
the reason why $\la E \ra > \chi^2$: the
neural nets fluctuate less about central experimental values than the
Monte Carlo replicas. In the presence of substantial
error reduction, the (scatter) correlation between
network covariance and experimental error is generally not very high,
and can take low values when a small number of data points from one
experiment is enough to determine the outcome of the fit, such as in
the case of the NMC experiment, even more so for E665~\cite{f2ns}.

\begin{figure}[ht]
\begin{center}
\epsfig{width=0.55\textwidth,figure=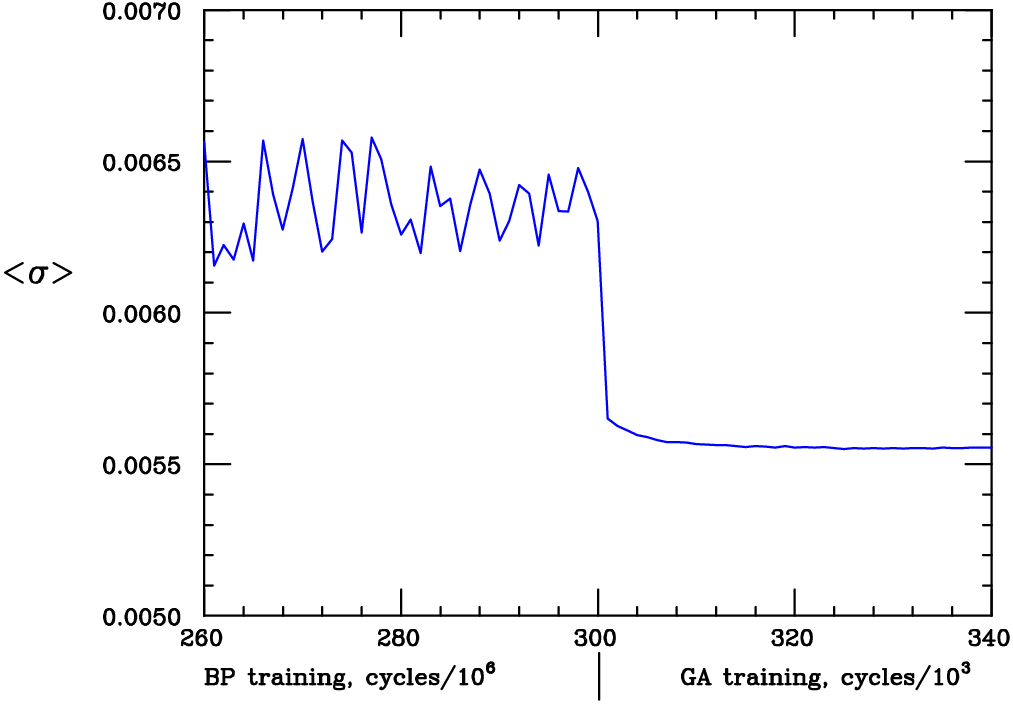}
\caption{}{\small Dependence of $\la\sigma^{(\net)}\ra_{\dat}$ on the length of
  training for the BCDMS experiment.}
\label{bcdsig}
\end{center}
\end{figure}

\begin{figure}[ht]
\begin{center}
\epsfig{width=0.60\textwidth,figure=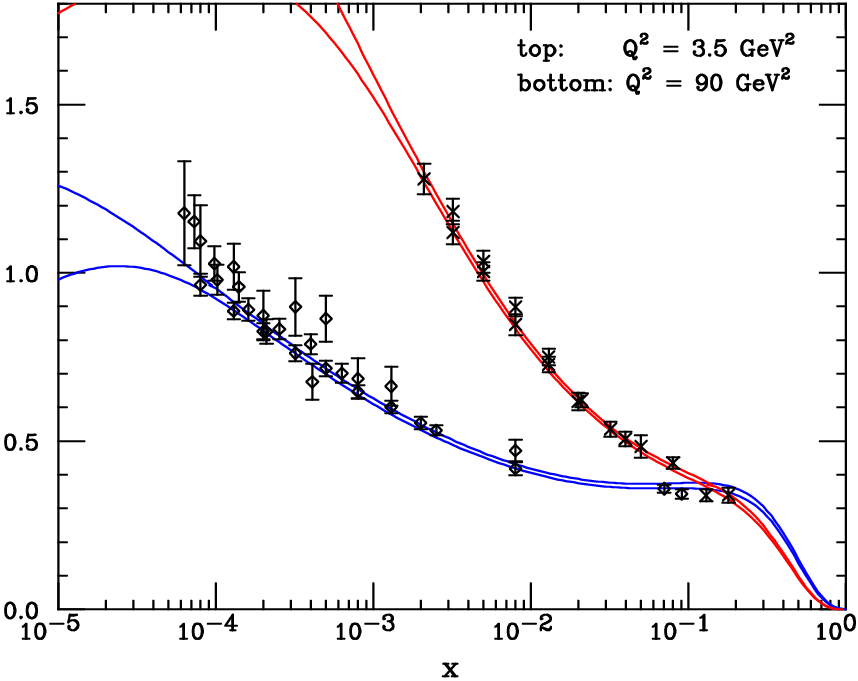}
\caption{}{\small Final results for $\fd$ compared to 
experimental data. For the
 neural network result, the
 one-$\sigma$ error band is shown.}
\label{f2plot}
\end{center}
\end{figure}
 
As discussed extensively in Ref.~\cite{f2ns} it is important to make sure
that this is due to the fact that information from individual
data points is combined through an underlying law by the neural
networks, and not due to parametrization bias. To this purpose, the ${\cal
  R}$-estimator has been introduced in
Section \ref{validation}, where it was 
shown that in the presence of substantial
error reduction ${\cal
  R}\gsim 1$ if there is parametrization bias, whereas ${\cal
  R}\approx 0.5$ in the absence of parametrization 
bias. It is apparent
from Tables \ref{results_new} and
\ref{resultsexp_new} that indeed ${\cal
  R}\approx 0.5$ for all experiments. Note that, contrary to what was
found in ref.~\cite{f2ns}, there is now some error reduction also for the
BCDMS experiment, though by a somewhat smaller  amount than for other
experiments. We will come back to this issue  when comparing
results to those of Ref.~\cite{f2ns}. 

Further evidence that the error reduction is not due to
parametrization bias can be obtained by studying the dependence of
$\la\sigma^{(\net)}\ra_{\dat}$ on the length of training. This
dependence 
is shown in Fig.~\ref{bcdsig} for the BCDMS experiment. It is apparent that the
error reduction is correlated with the goodness of fit displayed in
Fig. \ref{trlength}, and it occurs during the GA training, 
thereby suggesting that
error reduction occurs when the neural networks start reproducing an
underlying law. If error reduction were
due to parametrization bias it would be essentially independent of the
length of training. 

The point-to-point correlation $\rho$ of the neural nets is somewhat
larger than that of the data, as one might expect as a consequence of
an underlying law which is being learnt by the neural networks. 
In fact, for the NMC experiment the increase in correlation
essentially compensates the reduction in error, in such a way that the
average covariance of the nets and the data are essentially the
same. This again shows that in the case of the NMC
experiment a small number of points is sufficient to predict the
remaining ones.
 For  all other experiments, however, the
covariance of the nets is substantially smaller than that of the
data. As a consequence the (scatter) correlation of covariance
remains relatively high for all experiments, except NMC, and
especially E665 whose points are essentially predicted by the fit to
other experiments.

\begin{figure}[ht]
\begin{center}
\epsfig{width=0.535\textwidth,figure=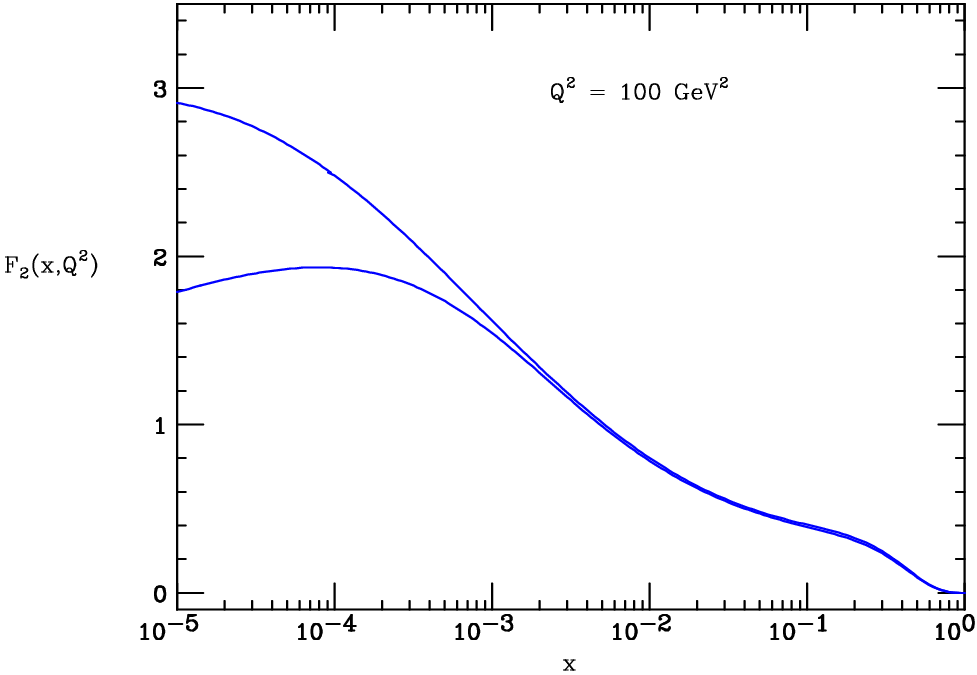}
\epsfig{width=0.45\textwidth,figure=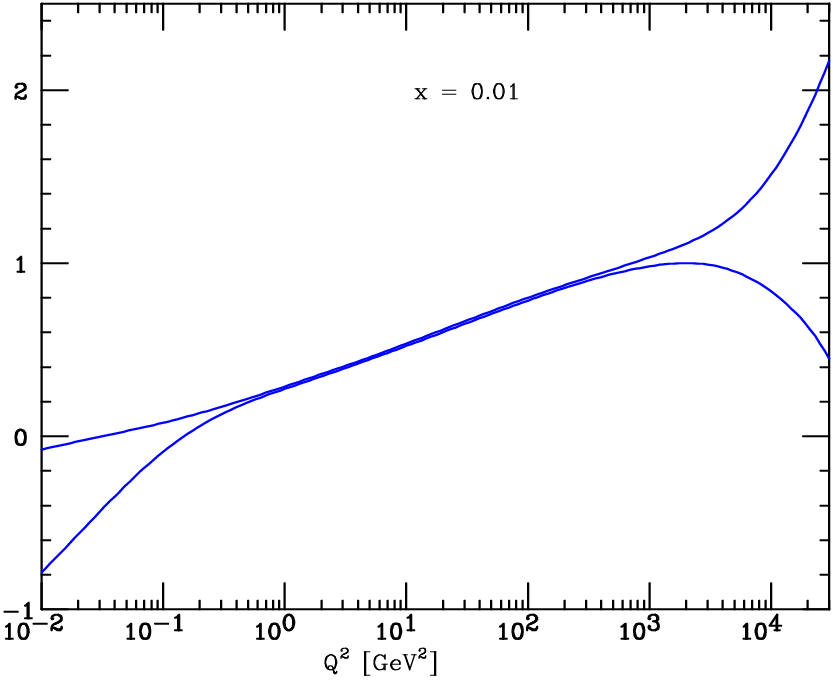}
\caption{}{\small One-$\sigma$ error band for the structure function
  $F_2(x,Q^2)$ 
computed from neural
  nets. Note the different scale on the $y$ axis in the two plots.}
\label{f2extra}
\end{center}
\end{figure}

The structure function and associated one-$\sigma$ error band is
compared to the data as a function of $x$ for a pair of typical
values of $Q^2$ in Fig.~\ref{f2plot}. In Fig.~\ref{f2extra} the behavior 
of the structure
function as a function of $x$ at fixed $Q^2$ and as a function of
$Q^2$ at fixed $x$ is also shown. It is apparent that in the data
region the error on the neural nets is rather smaller than that on the
data used to train them. The error however grows rapidly as soon as
the nets are extrapolated outside the region of the data, specially
in the small-$x$ region and in the large-$Q^2$ region. At large
$x$, however, the
extrapolation is kept under control by the kinematic constraint
$F_2(1,Q^2)=0$. Note that the increase of the uncertainties in
the extrapolation region, as opposed to the case for fits
with functional forms, is due to the fact that the behavior
of neural networks in this region is not determined
by the corresponding behavior in the data (interpolation) region.

\begin{figure}[ht]
\begin{center}
\epsfig{width=0.65\textwidth,figure=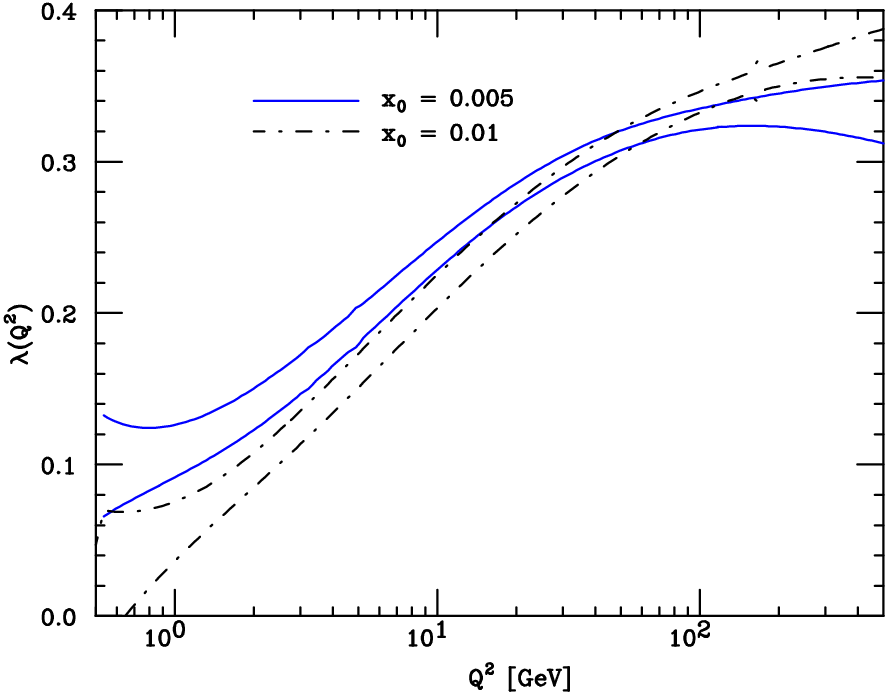}
\caption{}{\small The logarithmic derivative of $F_2(x,Q^2)$,
$\lambda(Q^2)$, as determined from the neural network
parametrization in the transition region between perturbative
and nonperturbative regimes.}
\label{tranplot}
\end{center}
\end{figure}

\begin{figure}[ht]
\begin{center}
\epsfig{width=0.49\textwidth,figure=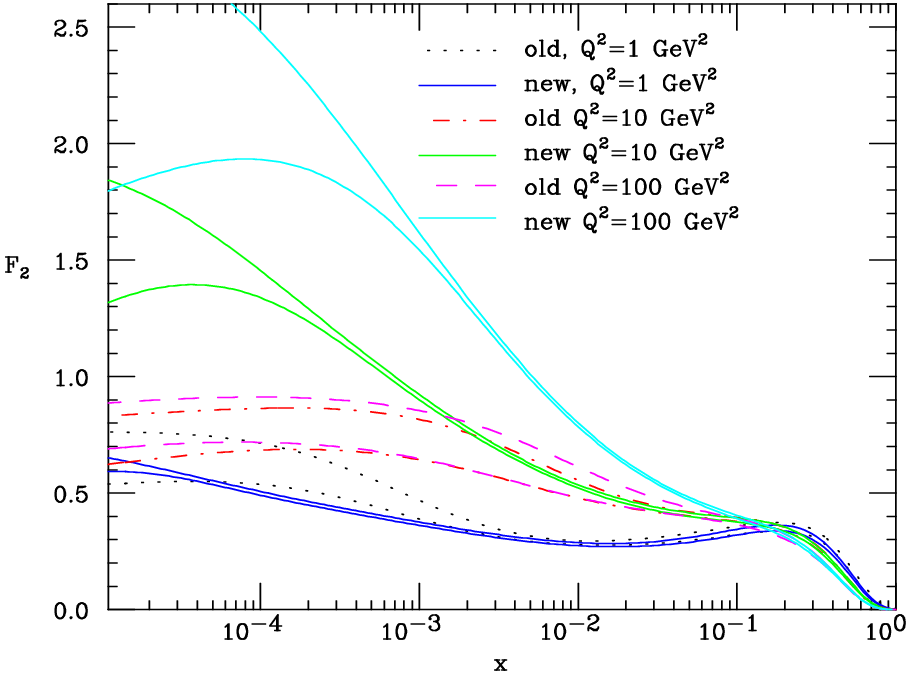}
\epsfig{width=0.49\textwidth,figure=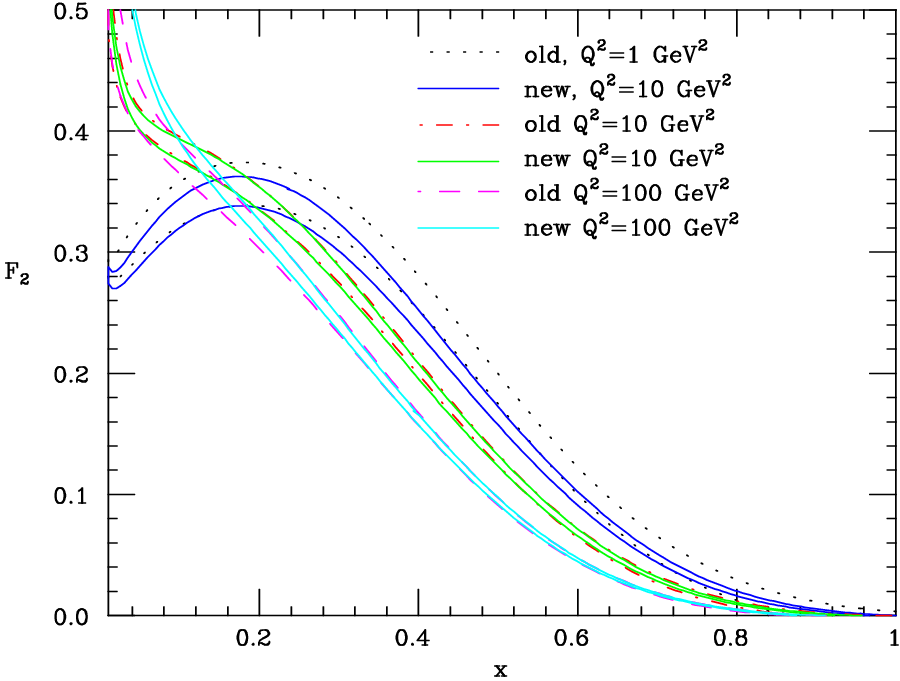}
\caption{}{\small Comparison of the parametrization of $F_2(x,Q^2)$ of
  ref.~\cite{f2ns} (old) with that 
described in \cite{f2nnp} (new). The pairs
  of curves correspond to a one-$\sigma$ error band.}
\label{f2comp}
\end{center}
\end{figure}

The number of possible phenomenological applications of 
the neural network parametrization of the proton
structure function $F_2^p(x,Q^2)$ is  rather large,
from determinations of the strong coupling $\aq$, comparison
of different theoretical predictions at low $x$, checks of sum
rules, or its effects of the extraction of polarized structure
functions and polarized parton distributions.
Another possibility is a detailed quantitative study
of $F_2^p(x,Q^2)$ in the transition region between
perturbative and nonperturbative regimes around $Q^2=1$ GeV$^2$.
Our parametrization is specially suited for this purpose since
it incorporates all the information from experimental data without
introducing any bias from the functional form
behavior of the transition region. In Fig. \ref{tranplot}
we show the logarithmic derivative of the structure function,
defined
as
\be
\lambda(Q^2,x_0)\equiv \frac{\ln F_2(x,Q^2)}{d\ln x}\Bigg|_{x=x_0} \ ,
\ee
for two different values of $x_0$. Note that at very low 
$Q^2$ expectations are that the Pomeron exponent, $\lambda(Q^2=0)=0.08$
is recovered.

\begin{figure}[ht]
\begin{center}
\epsfig{width=0.55\textwidth,figure=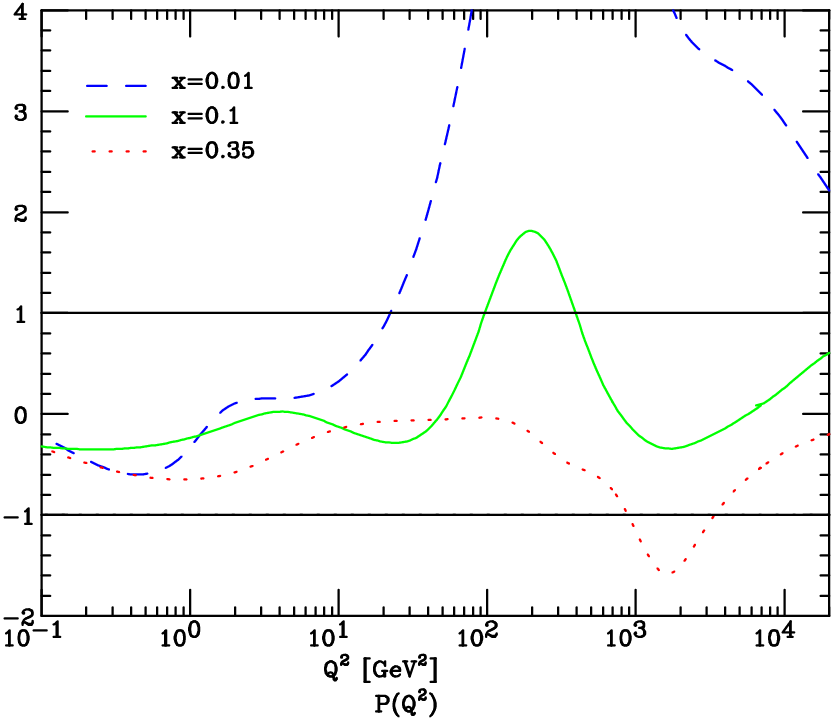}
\caption{}{\small The relative pull, Eq.~\ref{pull}, of the new and old
 $F_2(x,Q^2)$
  parametrizations. The one-$\sigma$ band is also shown. Note that
in the kinematical region where experimental data
for the two versions of the fit overlaps,
the pull always verifies $\Big| P(x,Q^2)\Big|\le 1$, 
as expected from consistency arguments.} 
\label{pullfig}
\end{center}
\end{figure}

Let us finally compare the determination of $F_2(x,Q^2)$ presented
here with that of Ref.~\cite{f2ns}, which was based on pre-HERA data. 
In Fig.~\ref{f2comp} one-$\sigma$ error bands for the two
parametrizations are compared, whereas in Fig.~\ref{pullfig} we
display the relative pull of the two parametrizations,
introduced in Section \ref{validation}, which in the
present situation is given by 
\be
P(x,Q^2)\equiv \frac{F_2^{\mathrm{new}}(x,Q^2)-F_2^{\mathrm{old}}(x,Q^2)}
{\sqrt{\sigma_{\mathrm{new}}^2(x,Q^2)+\sigma_{\mathrm{old}}^2(x,Q^2)}}
\ ,
\label{pull}
\ee
where $\sigma(x,Q^2)$ is the error on the structure function
determined as the variance of the neural network sample.
In view of the fact that the old fit only included BCDMS and NMC data,
it is interesting to consider four regions:  (a)
 the BCDMS region (large $x$, intermediate $Q^2$, e.g. $x=0.3$,
 $Q^2=20$~GeV$^2$); (b)
the NMC region (intermediate $x$, not too large $Q^2$, e.g. $x=0.1$,
$Q^2=2$~GeV$^2$); (c) the HERA region  
(small $x$ and large $Q^2$, e.g. $0.01$ and $Q^2>10$~GeV$^2$); (d) the region
where neither the old nor the new fit had data (very large or very
small $Q^2$). In region (a) the new fit is rather more precise than
the old one, for reasons to be discussed shortly, while central values
agree, with $P\lsim 1$). In region (b) the new fit is significantly more
precise than the old one, while central values agree to about one
sigma. In region (c) the new fit is rather accurate while
the old fit had large errors, but $P\gg1$ nevertheless, 
because the HERA rise of $F_2$ is outside the error bands extrapolated
from NMC. This shows that even though errors on  extrapolated data
grow rapidly they become unreliable when extrapolating far from the data.
Finally in region (d) all errors are very large and $P$ is
consequently small, except at small $x$ and large $Q^2$, where the new
fits extrapolate the rise in the HERA data, which is missing
altogether in the old fits.

Let us finally come to the issue of the BCDMS error, which, as
already mentioned, is reduced somewhat in the current
fit in comparison to the data and the previous fit. This may appear 
surprising, in that the new fit does not contain any new data in
the BCDMS region.
However,
 as is apparent
from Fig.~\ref{bcdsig}, this error reduction takes place in the GA
training stage, when $E_3$  is
minimized. Furthermore, we have verified that if the uncorrelated
$E_2$  is minimized during the GA training
no error reduction is observed for BCDMS. 
Hence, we conclude that the reason why error reduction for BCDMS  was
not found in
Ref.~\cite{f2ns} is that in that reference neural networks were trained by 
minimizing $E_2$. In fact, as discussed above, 
the BCDMS experiment turns out to require the longest time to learn, 
especially after inclusion of the HERA data. Error reduction only
obtains after this lengthy minimization process.

In summary, we have presented a determination of the probability density in the
space of structure functions for the structure function $\fd$ 
for the proton, based on all available data from the  NMC,
BCDMS,
E665, ZEUS and H1 collaborations. Our results take the form of a Monte
Carlo sample of 1000 neural networks, each of which provides a
determination of the structure function for all $(x,Q^2)$. The
structure function and its  statistical moments 
(errors, correlations and so on) can be determined by
averaging over this sample. The results of this
part of the thesis  are made available as a
FORTRAN routine which gives $F_2(x,Q^2)$, determined by a  set of
parameters, and 1000 sets of parameters corresponding to the Monte
Carlo sample of structure functions. They can be downloaded
from the web site {\tt
http://sophia.ecm.ub.es/f2neural/}. 

This works updates and upgrades that of Ref.~\cite{f2ns}, where similar
results were obtained from the BCDMS and NMC
data only. The main improvements in the present work are related to the
need of handling a large number of experimental data, 
affected by large correlated systematics. 
Apart from many
smaller technical aspects, the main improvement introduced here is the
use of genetic algorithms to train neural networks on top of
back-propagation. This has allowed for a more accurate handling of
correlated systematics.

Whereas the results of this part of the
thesis are of direct practical use for
any application where an accurate determination of $F_2^p(x,Q^2)$ and
its associate error are necessary, its main motivation is the
development of a set of techniques which will be required for the
construction of a full set of parton distributions with faithful
uncertainty estimation based on the same method.
The application of this strategy to the
parametrization of parton distribution
functions is discussed in Section \ref{nnqns}.

\clearpage

\section{Lepton energy spectrum in B meson decays}

\label{bdecay_appl}

The third application of the general strategy defined in
Chapter \ref{general} is the parametrization of the
lepton energy distribution from
semileptonic decays of B mesons. The motivation 
for such a parametrization is the application
of the general strategy of Chapter \ref{general}
to the construction of the probability density of an observable
when only its moments are available experimentally.
As a byproduct of this parametrization, we will provide
a determination of the b quark mass $\overline{m}_b\lp m_b\rp$.

In the last decade the field of B meson physics has been the object
of a wealth of studies (see Ref. \cite{superB} and references therein),
motivated by the high precision measurements
coming from the B factories Belle and Babar. In particular
the inclusive semileptonic decays $B\to X l\nu$, where $X$ stands for
a hadronic system, have received a lot of
attention, both in the theoretical 
and in the experimental sides, due
to its paramount importance for the determination of elements
of the CKM matrix, and also since they provide us with important
information on the underlying strong interaction dynamics.

Therefore,
our purpose in this part of the thesis
is to allow a more general comparison of the
theoretical predictions with the experimental data, constructing
a parametrization of the lepton energy spectrum from available
experimental information on its moments, supplemented by 
constraints from the kinematics of the process. 
This goal is part of a more general motivation, namely the
development of a suitable general strategy
to determine  a function with uncertainties if the only available 
experimental information is given in terms of its convolutions
and  additional constraints (like kinematical cuts).

The success of previous applications of the technique
introduced in Chapter \ref{general}
 to other problems motivated us to implement this technique
in the context of B physics. Therefore, in this work we
construct a unbiased parametrization of the lepton energy spectrum in
semileptonic B meson decays with a faithful estimation
of the uncertainties.

The semileptonic decays of the B meson have been introduced
in Section \ref{bmeson_theo}.
We will consider therefore experimental data for 
leptonic moments with kinematical cuts
of the lepton energy  distribution, Eq. \ref{specdef}, 
in  semileptonic
B meson decays to charmed final states $B\to X_c l\nu$.
These moments have recently been measured with great accuracy at the 
B-factories, Babar \cite{Aubert:2004td} and Belle \cite{elmombelle}, as well as
by the Cleo \cite{cleoleptonmom} detector. We will therefore
use in the present analysis the latest data from these
three experiments. We do not use data from CDF \cite{cdfhadmom} since it is
restricted to hadronic moments and leptonic moments are not available.
The features of the experimental data of these three experiments
can be seen in Table \ref{datafeat1}. Note that for all experiments
correlations are rather large, so it is compulsory to incorporate them
in a consistent way in the statistical analysis of the data.

\begin{table}[ht]
\begin{center}
\begin{tabular}{|c|c|cc|cc|c|}  
\hline
Experiment & $N_{\dat}$ &  $n$ &  $E_0$ (GeV) & 
$\la \sigma_{\stat}\ra$ &    $\la \sigma_{\tot}\ra$ &
$\la \rho\ra$ \\
\hline
Babar \cite{Aubert:2004td} & 20 & 0~-~3 & 0.6~-~1.5& 0.06 & 0.08 & 0.50   \\
Belle \cite{elmombelle} & 18 & 1~-~3 & 0.4~-~1.5& 0.15 & 0.16 & 0.34  \\
Cleo\cite{cleoleptonmom} & 20 & 1~-~2 & 0.6~-~1.5& 0.008 & 0.0013 & 0.65  \\
\hline
\end{tabular}
\caption{\small Features of experimental data on lepton
moments $M_n(E_0)$,Eqns. \ref{momexp1}-\ref{momexp2}, 
where $n$ stands for the order
of the moment and $E_0$ the lower cut in the
lepton energy, see Eq. \ref{leptonmom}. Experimental
errors are given as percentages. 
\label{datafeat1}}
\end{center}
\end{table}

The final published observables for the
semileptonic decays of the B meson are the  moments of the lepton
energy spectrum, Eq. \ref{leptonmom}, 
with different cuts in the lepton energy, rather than the
full differential spectrum itself.
The data that we use for the present 
parametrization of the lepton energy  spectrum
is the following:
the Babar Collaboration \cite{Aubert:2004td}
provides  the partial branching fractions,
\be
\label{momexp1}
M_0(E_0)=\tau_BL_0(E_0,0)=\tau_B\int_{E_0}^{E_{\max}}
\frac{d\Gamma}{dE_l}(E_l)~dE_l \ ,
\ee
where $\tau_B$ is the average B meson lifetime \cite{hfag},
the first moment,
\be
M_1(E_0)=\frac{L_1(E_0,0)}{L_0(E_0,0)} \ ,
\label{babmom}
\ee
and the central moments,
\be
\label{momexp2}
M_n(E_0)=\frac{L_n(E_0,M_1(E_0))}{L_0(E_0,0)}, \qquad n=2,3 \ ,
\ee
for five different values of $E_0$ from 0.6 to 1.5 GeV,
which makes a total of 20 data points.
The leptonic moments have been defined in Eq. \ref{leptonmom}.

The Belle Collaboration \cite{elmombelle} provides the
same moments, $M_n(E_0)$  for $n=1$, $n=2$ and $n=3$.
For example, they define $M_1=\la E_l\ra$, which if one
takes into account that the corresponding normalized
probability density is given by
\be
\mathcal{P}(E_l)=\lp\frac{1}{\int_{E_0}^{E_{\max}}\frac{d\Gamma}{dE_l}
dE_l}\rp\frac{d\Gamma}{dE_l}(E_l),\qquad E_0 \le E_l \le E_{\max} \ ,
\ee
one ends up with Eq. \ref{babmom}, and
similarly for the remaining moments.
The difference with the Babar data is that the partial branching
fraction Eq. \ref{momexp1} is not measured, and that the
Belle data cover a somewhat larger lepton energy range. since the
lowest value of $E_0$ of their data set is $E_0=0.4$ GeV.
These moments, for six different values of $E_0$ from 0.4 to 1.5 GeV,
make up a total of 18 data points. 
Finally the Cleo Collaboration \cite{cleoleptonmom} provides the moments
$M_n(E_0)$ for
$n=1,2$, for energies between 0.6 to 1.5 GeV, for a total of
20 data points (10 for $n=1$ and 10 for $n=2$). The correlations
for this experiment are larger since measurements of the same
moment at different energies $E_0$ are highly correlated.
The three collaborations provide also the total 
and statistical errors, as well as the
correlation between different measurements. These characteristics
are summarized in Table \ref{datafeat1}.

We have noticed that the experimental correlation
matrices, $\rho_{ij}^{(\mrexp)}$, 
as presented with the published data of the
three experiments \cite{elmombelle,Aubert:2004td,cleoleptonmom},
are not positive definite (see also \cite{bauerglobalfit}). 
The source of this problem can be traced to the
fact that
off-diagonal elements of the experimental correlation matrix
are large, as expected since moments with similar energy cut
contain almost the same amount of information and are therefore
highly correlated. Then one can check that 
some eigenvalues are negative and small, 
and 
this is pointing that the source of the problem is
an insufficient accuracy in the computation of the
elements of the correlation matrix.

However, whatever is the
original source of the problem, the fact that the
experimental correlation matrices
are not positive definite has an important consequence:  the
 technique introduced in
Chapter \ref{general} for the generation of a sample of replicas of the
experimental data taking into account correlations 
 relies on the existence of a positive definite
correlation matrix, and therefore if the experimental 
correlation matrix is not positive definite, our
strategy cannot be applied. 

A method to overcome these difficulties while keeping the maximum
amount on information on experimental correlations as
possible consists on removing those data points for which
the experimental correlations are larger 
than a maximum correlation, $\rho_{ij}^{(\exp)}\ge \rho^{\max}$.
The value of $\rho^{\max}$ is determined separately for each
experiment as the maximum value for which the resulting
correlation matrix is positive definite. In Table
\ref{datafeat2} we show the value of $\rho^{\max}$ for each
experiment, together with the features of the remaining experimental data
after cutting those data points with too large correlations.
Similar considerations about the need to take a restricted
subset of data points due to the large
point to point correlations have been discussed in the
context of global fits
in B physics, see for example \cite{fitbabar,fitskin}. 

\begin{table}[ht]
\begin{center}
\begin{tabular}{|c|c|cc|cc|cc|}  
\hline
Experiment & $N_{\dat}$ &  $n$ &  $E_0$ (GeV) & 
$\la \sigma_{\stat}\ra$ &    $\la \sigma_{\tot}\ra$ &
$\rho^{\max}$ & $\la \rho\ra$ \\
\hline
Babar \cite{Aubert:2004td} & 16 & 0~-~3 & 0.6~-~1.5& 0.04 & 0.05 & 
0.97 & 0.49   \\
Belle \cite{elmombelle} & 15 & 1~-~3 & 0.4~-~1.5& 0.18 & 0.19 & 
0.88 & 0.31  \\
Cleo \cite{cleoleptonmom} & 10 & 1~-~2 & 0.6~-~1.5& 
0.005 & 0.009 &  0.95 &0.69  \\
\hline
\end{tabular}
\caption{\small Features of experimental data that is
included in the fit, after cutting data points
with too large correlations. Experimental
error are given as percentages. } 
\label{datafeat2}
\end{center}
\end{table}

As has been discussed in Chapter \ref{general}, the first step of our
strategy to parametrize experimental data 
is the generation of an ensemble of replicas of the original
measurements, which in the present case consists
 in the measured moments, which we
will denote by
\be
M_i^{(\mrexp)}, \quad  i=1,\ldots,N_{\dat} \ ,
\ee
where $M_i^{(\mrexp)}$ stands for any of Eqns. 
\ref{momexp1}-\ref{momexp2}, and $N_{\dat}$ is the 
total number of experimental data points,
together with the total error and the correlation matrix.

To generate replicas we proceed in the 
usual way. Since experimental data
consists on central values for the
moments, together with its total error
and the associated correlations, the k-th replica of the
experimental data is constructed, following
Eq. \ref{gen2}, as
\be
M_j^{(\art)(k)}=M_j^{(\exp)}+r^{(k)}_j\sigma_{\tot,j}  \ ,
\quad
j=1,\ldots,N_{\dat}, \quad, k=1,\ldots,N_{\rep}\ .
\ee
One can check that the sample of replicas is able to reproduce
the errors and  the correlations of the experimental data. 
In Table \ref{gendata} we have
the statistical estimators for the
replica generation.
One can observe that
to reach the desired accuracy of a few percent
and to have scatter correlations $r\ge 0.99$
for central values, errors and correlations,
 a sample of 1000 replicas is 
needed.

\begin{table}
\begin{center}
\begin{tabular}{|c|ccc|}  
\hline
 $N_{\rep}$ & 10 & 100 & 1000\\
\hline
$\la PE\lc\la M \ra_{\rep}\rc\ra$ & 2.47\%
& 0.40\% & 0.24\%   \\
\hline
$\la PE\lc \sigma^{(\art)}\rc\ra_{\dat}$ & 32.4\% 
& 13.8\% & 3.4\% \\
$\la \sigma^{(\art)}\ra_{\dat}$ &  0.00265 & 0.00277 & 0.00268\\
$r\lc \sigma^{(\art)}\rc$ &  0.95 & 0.99 & 0.99 \\
\hline
$\la PE \lc \rho^{(\art)} \rc \ra_{\dat}$ & 60.1\%
& 19.6\% & 6.7\%   \\
$\la \rho^{(\art)}\ra_{\dat}$ & 0.132 & 0.138 & 0.155 \\
$r\lc \rho^{(\art)}\rc$ &  0.75 & 0.96 & 0.99  \\
\hline
$\la \mathrm{cov}^{(\art)}\ra_{\dat}$ & $1.1~10^{-6}$ 
& $1.4~10^{-6}$  & $1.3~10^{-6}$ \\
$r\lc \mathrm{cov}^{(\art)}\rc$ & 0.86 & 0.98 & 0.99 \\
\hline
\end{tabular}
\caption{\small Comparison between experimental and 
Monte Carlo data.\hfill\break
The experimental data  have
$\la \sigma^{(\mrexp)}\ra_{\dat}=0.00267$ ,
 $\la \rho^{(\mrexp)}\ra_{\dat}=
0.166$ and $\la \mathrm{cov}^{(\mrexp)}\ra_{\dat}=1.4~10^{-6}$, for a total of
41 data points.
\label{gendata}}
\end{center}
\end{table}

The next step is to train a neural network parametrizing
the lepton spectrum for each replica of the experimental data.
Therefore we parametrize the lepton energy spectrum, Eq.
\ref{specdef}, 
\be
\lp \frac{d\Gamma}{dE_l}\rp^{(\net)(k)}(E_l), \qquad k=1,\ldots,N_{\rep} \ ,
\ee 
where $E_l$ is the lepton energy,
with a neural network. From this neural network parametrization
one can compute the corresponding moments, to compare with
experimental data. For example, the leptonic moment $L_1(E_0,0)$
is computed for the k-th neural network as
\be
\label{leptondis}
L_1^{(k)}(E_0,0)=\int_{E_0}^{E_{\max}} E_l
\lp \frac{d\Gamma}{dE_l}\rp^{(\net)(k)}(E_l) dE_l \ .
\ee

Now we describe the details of the neural network
training.
Concerning the topology of the
neural network, in this case we find
that an acceptable compromise is given by an architecture 1-4-3-1.
Concerning the training strategy, 
in the present situation 
we will have a single training epoch minimizing 
the 
diagonal error defined in Eq. \ref{er2} but with the total error
$\sigma^{(\mrexp)}_{i,\mathrm{tot}}$ 
instead of only the statistical error with
dynamical stopping of the training, as discussed in 
Section \ref{minimstratt}.
The minimization technique that we will use for the
neural network training is Genetic
Algorithms, which is suitable for minimization of highly
nonlinear error functions, as in the present case. 
We also use weighted training to obtain a more even
$\chi^2$ distribution between the different experiments.
See the
original work \cite{bmeson} for additional details
on the neural network training.

\begin{figure}[ht]
\begin{center}
\epsfig{width=0.50\textwidth,figure=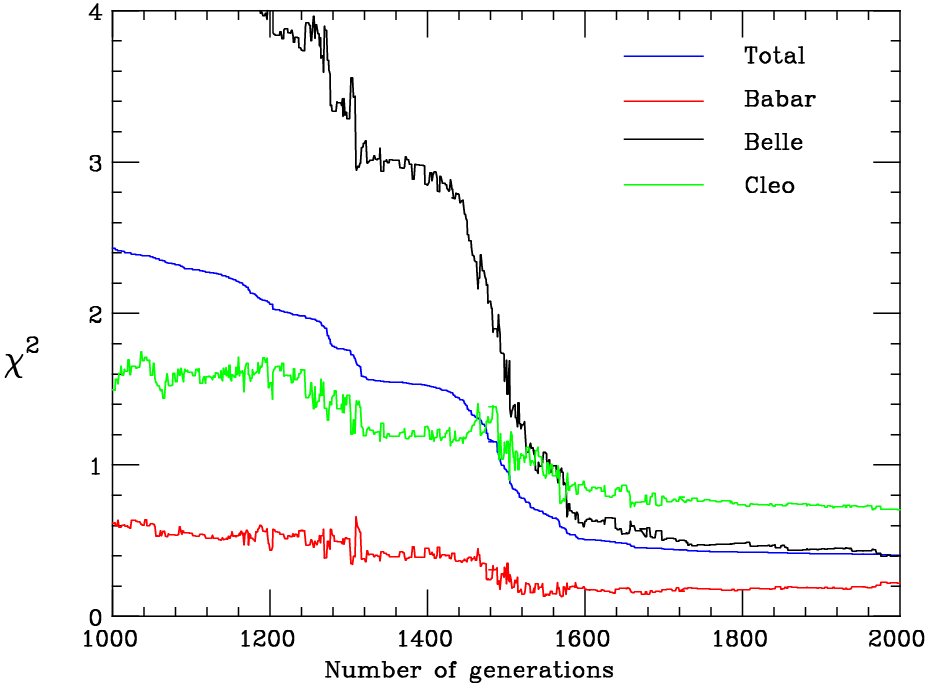}
\end{center}
\caption{}{\small Dependence of the error function $E_2^{(0)}$
for a training of all three experiments on central
experimental data. Note that at the end of the
minimization $E_2^{(0)}\ll 1$ for all experiments.}
\label{comp_bmeson}
\end{figure}

In situations in which experimental data consists of
different experiments, as it is the case now (with Babar, Belle and Cleo),
one has also to address the issue of 
possible inconsistency between different experiments, that is,
the possibility that a subset of points from the
two experiments in the same kinematical region
do not agree with each other within 
experimental errors. 
This issue has been discussed in Section \ref{dis_appl},
regarding the possible inconsistency of the ZEUS94 experiment
with the other experimental data sets in the
parametrization of the structure function
$F_2^p(x,Q^2)$.
This issue
 is of paramount importance in the context of global 
parton distribution fits, see for example \cite{incon}.
As has been discussed in detail in Ref. \cite{bmeson},
in this case the three experiments are fully compatible, as
can be seen in Fig. \ref{comp_bmeson} for a training to the
experimental data. 

The lepton energy spectrum, Eq. \ref{leptondis}, as
parametrized with a neural network, has to satisfy three
constraints independently of the dynamics of the process. 
First of all, it is 
zero outside the region where it has kinematical support, in particular
it has to vanish at the kinematical endpoints, $E_l=0$ and
$E_l=E_{\max}$.
Second, the spectrum is a positive definite quantity (since any
integral over it is an observable, a partial branching 
ratio), therefore it must satisfy a local positivity condition. 

As we have discussed in 
Section \ref{constraints}, there are several methods in our strategy to
introduce kinematical constraints in an unbiased way. 
We have found that for the present application, the optimal method 
to implement the kinematical constraints that the spectrum
should vanish at the endpoints is hard-wiring them
 into the parametrization, that is,
the lepton energy spectrum parametrized by a neural net will be
given by
\be
\label{gammacon}
\lp \frac{d\Gamma}{dE_l}\rp^{(\net)(k)}
(E_l)=E_l^{n_1}\xi_1^{(L)}(E_l)(E_{\max}-E_l)^{n_2} \ ,
\ee
with $n_1,n_2$ positive numbers, and $\xi_1^{(L)}$ is the output
of the neural network for a given value of
$E_l$. Assuming this functional behavior at the
endpoints of the spectrum introduces no bias since it can be
 shown  that our results
do not depend on the value of $n_1$ and $n_2$. For the reference training
we have verified that $n_1=1$ and $n_2=1$ are suitable
values for these polynomial exponents.

We will impose the remaining kinematical constraint, the
positivity constraint, as a
Lagrange multiplier in the total error. That is, the total
quantity to be minimized is the sum of two terms,
\be
\label{chi2totcon}
E_{\tot}^{(k)}=E_2^{(k)}+\lambda_p P\lc
\lp \frac{d\Gamma}{dE_l}
\rp^{(\net)} \rc \ ,
\ee
where $E_2^{(k)}$ is 
Eq. \ref{er2}  and the
 the positivity condition is 
implemented penalizing those configurations
in which a region of the spectrum is negative, 
\be
 P\lc
\lp \frac{d\Gamma}{dE_l}
\rp^{(\net)} \rc=-\int_0^{E_{\max}} dE_l \lp \frac{d\Gamma}{dE_l}
\rp^{(\net)}(E_l) \theta \lp -
 \lp \frac{d\Gamma}{dE_l}
\rp^{(\net)}(E_l) \rp,
\ee
since P is zero for a positive spectrum.
The relative weight $\lambda_p$  is determined via a
stability analysis, requiring that $\lambda_p$ is large
enough so that the constraint is verified, but small
enough so that experimental data can still be learned in
an efficient way. 
As we will show in brief,
the implementation of the kinematical constraints plays an essential
role in the parametrization of the lepton spectrum. In particular,
if kinematic
constraints are not included in the fit the error at small $E_l$
grows very large and the extrapolation to $E_0=0$ becomes
completely unreliable.

\begin{figure}[ht]
\begin{center}
\epsfig{width=0.70\textwidth,figure=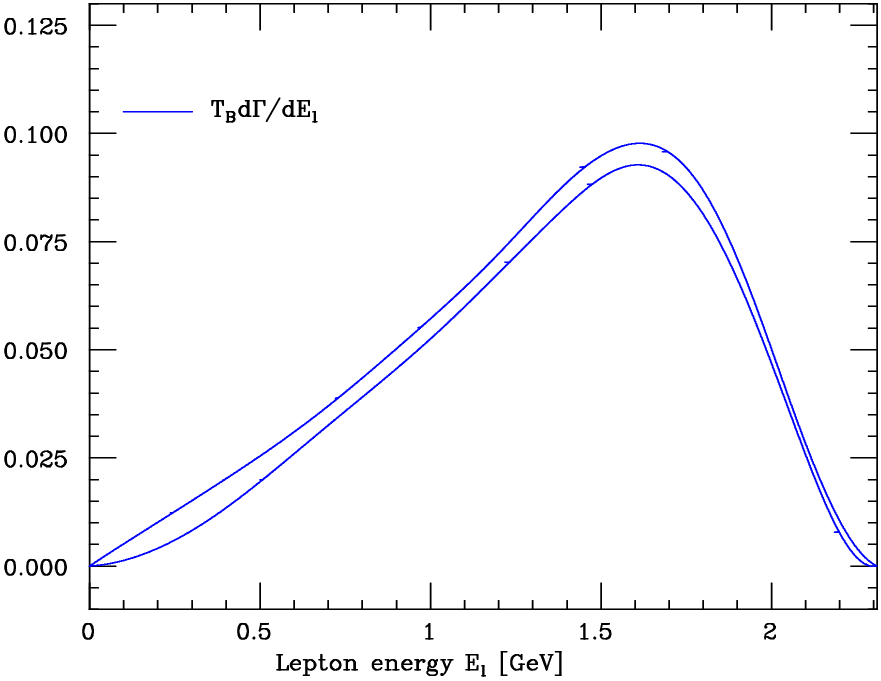}
\end{center}
\caption{}{\small Lepton energy spectrum, Eq. \ref{specdef},
as parametrized by the Monte Carlo ensemble of neural networks.
The bands represent the 1-$\sigma$ uncertainty region.}
\label{specav2}
\end{figure}

\begin{figure}[ht]
\begin{center}
\epsfig{width=0.49\textwidth,figure=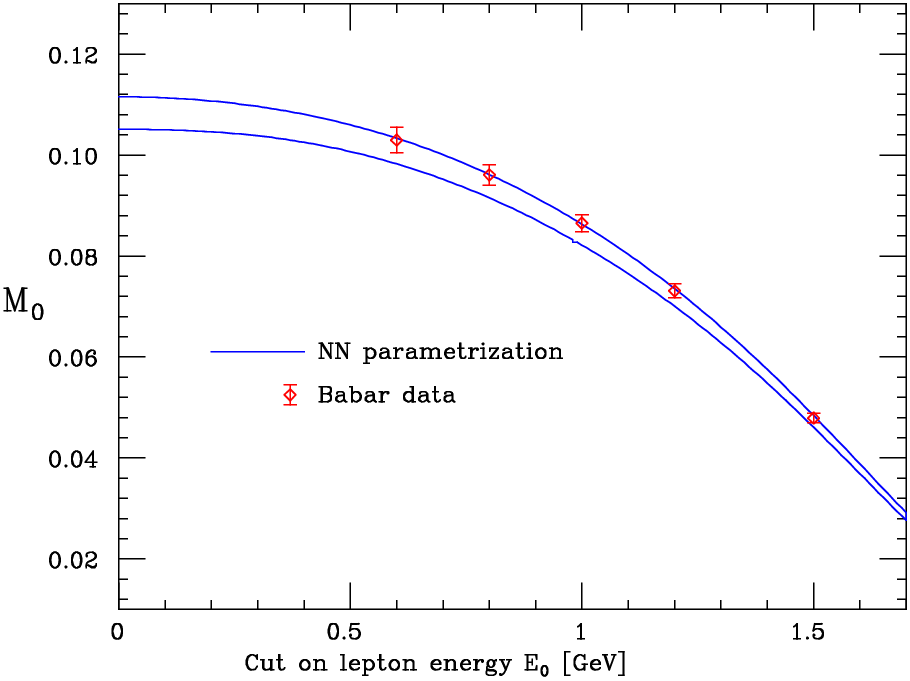}
\epsfig{width=0.49\textwidth,figure=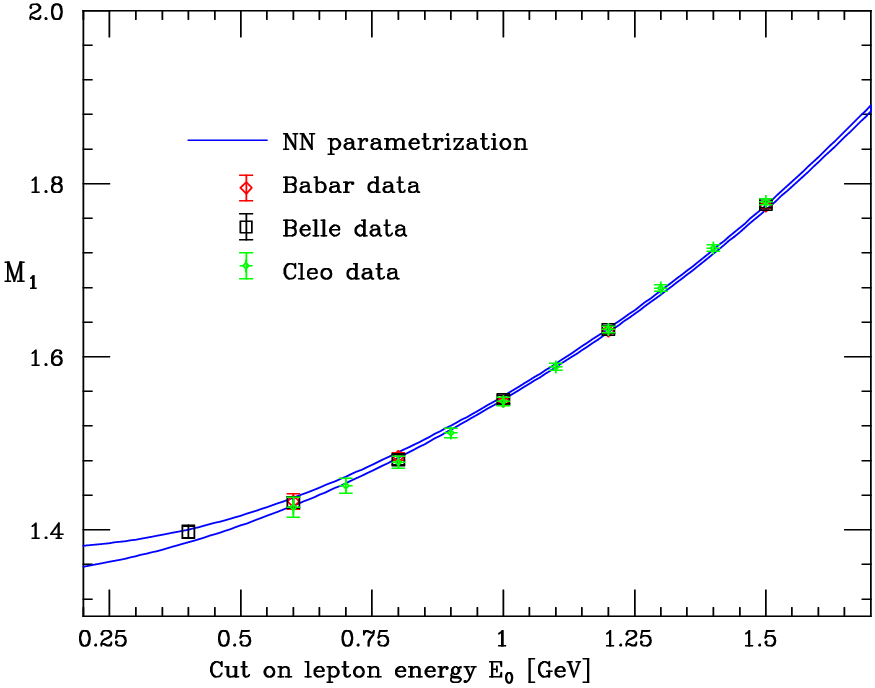}
\epsfig{width=0.49\textwidth,figure=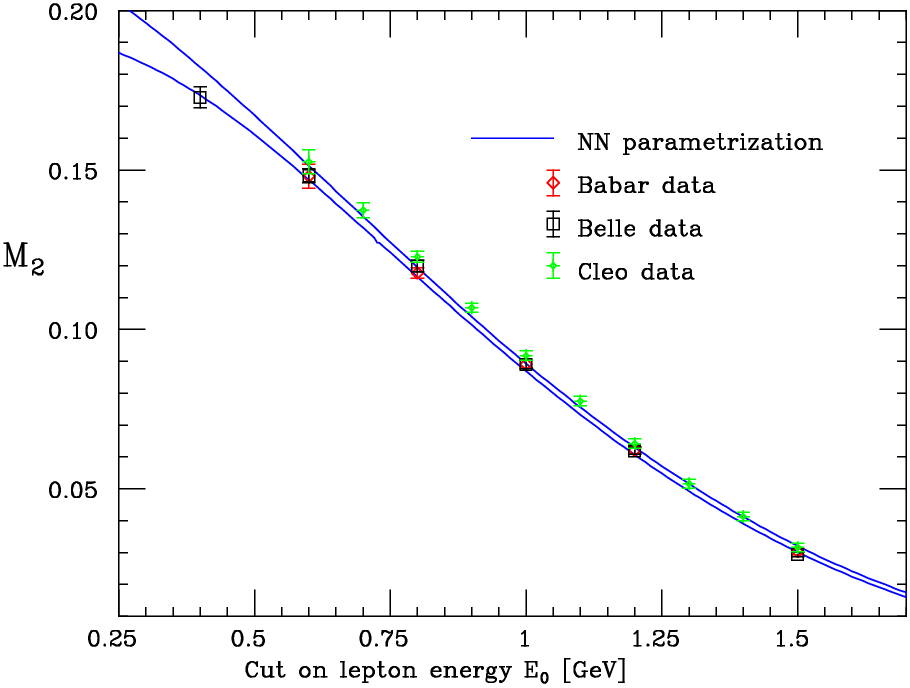}
\epsfig{width=0.49\textwidth,figure=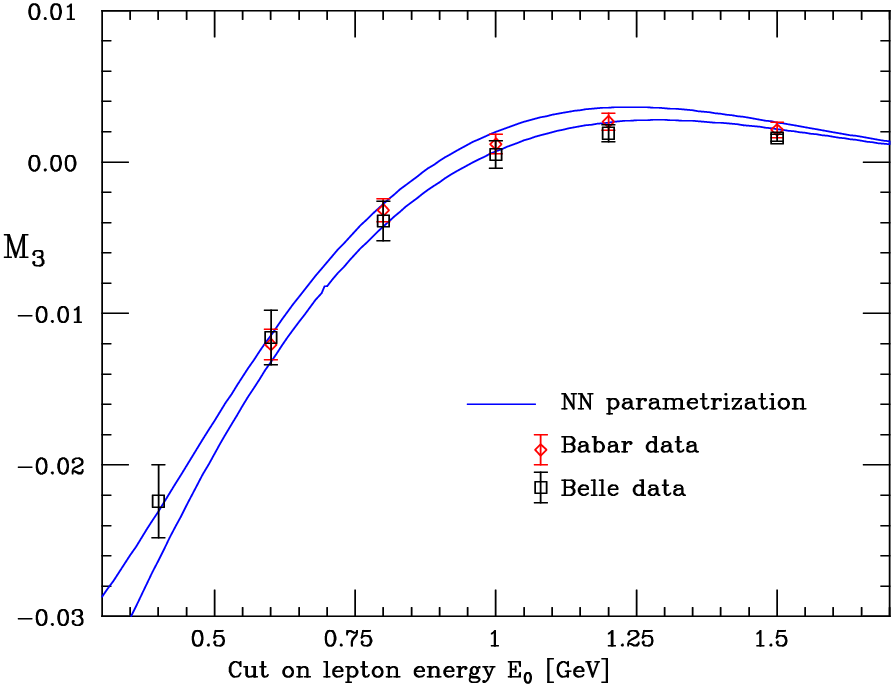}
\end{center}
\caption{}{\small Comparison of the
different experimental moments, Eqns. \ref{momexp1}-\ref{momexp2}
as obtained from our parametrization
with the original measurements, as a function of the lower
cut on the lepton energy $E_0$.}
\label{momav0}
\end{figure}

The set of neural networks parametrizing the
lepton energy distribution 
 trained on the Monte Carlo sample
of replicas of the experimental data defines the probability 
measure in the space of lepton energy spectra. 
From this probability measure, as explained in Chapter
\ref{general}, expectation values of
functionals of the lepton spectrum can be computed as
\be
\la \mathcal{F}\lc \frac{d\Gamma}{dE_l}\rc\ra\equiv\int 
\mathcal{D}\frac{d\Gamma}{dE_l} \mathcal{P}\lc \frac{d\Gamma}{dE_l}\rc 
\mathcal{F}\lc \frac{d\Gamma}{dE_l}\rc=
\frac{1}{N_{\rep}}\sum_{k=1}^{N_{\rep}}
\mathcal{F} \lc\lp\frac{d\Gamma}{dE_l}\rp^{(\net)(k)}\rc \ .
\ee
 We now present the results for this parametrization, 
 from which averages and moments can be computed with
the corresponding uncertainties. 
In Fig. \ref{specav2} we plot the resulting lepton energy spectrum
with uncertainties.
In Fig. \ref{momav0} we compare the computation of the
moments of the lepton energy spectrum 
using our parametrization to the experimental results from
Babar, Belle and Cleo. We observe good agreement for all the
data points. 
As it has been explained in Section \ref{validation}, it is crucial to
validate the results of the parametrization using suitable 
statistical estimators. In Table \ref{resdata1} we have
the most relevant statistical estimators for all the
data points, and in Table \ref{resdata2} the same estimators
for the different experiments included in the fit.

\begin{figure}[ht]
\begin{center}
\epsfig{width=0.49\textwidth,figure=specav_ndep.ps}
\epsfig{width=0.49\textwidth,figure=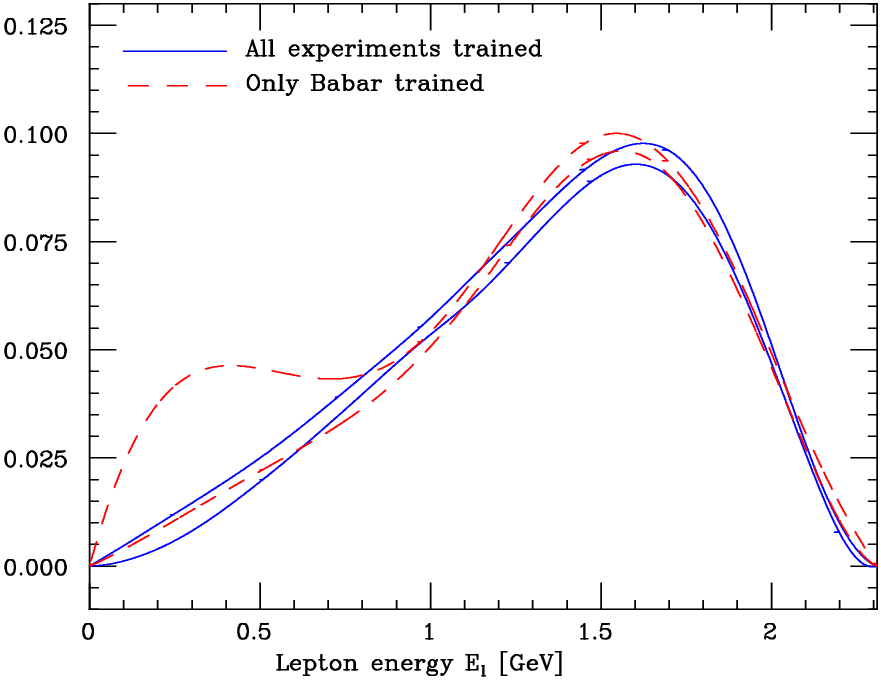}
\end{center}
\caption{}{\small Comparison of the lepton energy spectrum when
the parameters $n_1$ and $n_2$ are changed (left).
Comparison of the lepton energy spectrum when
only Babar data is incorporated in the fit with the result when
all experiments are incorporated in the fit (right).\label{specav_babar}
\label{specav_ndep}}
\end{figure}

With the results described in this section the 
total branching ratio can be computed, even
if experimental information was restricted to a finite 
value of $E_0$. This is possible because
the continuity condition implicit in the neural
network definition together with  the kinematic
constraints  allow for  an accurate
extrapolation from the
experimental data with lowest $E_l=0.4$ GeV to the
kinematic endpoint $E_l=0$. 
Note that this is not true
if the Belle data is not included in the fit,
see Fig. \ref{specav_babar}. 
The result that is obtained for the
partial decay rate,
 computed from the neural network
sample,
\be
\la \mathcal{B}\lp B\to X_c l\nu\rp \ra= 
\la M_0(E_l=0)\ra=\tau_B\frac{1}{N_{\rep}}\sum_{k=1}^{N_{\rep}}
\int_0^{E_{\max}} dE_l \lp \frac{d\Gamma}{dE_l}\rp^{(\net)(k)} \lp E_l
\rp\ ,
\ee
is the total branching ratio,
\be
\mathcal{B}\lp B\to X_c l\nu\rp=\lp 10.8 \pm 0.4\rp~10^{-2} \ ,
\ee
which is to be compared with the PDG result \cite{pdg},
\be
\mathcal{B}\lp B\to X_c l\nu\rp_{\mathrm{PDG}}=\lp 10.2 \pm 0.9\rp~10^{-2} \ ,
\ee
and with the
direct Delphi measurement of the total branching ratio \cite{delphifit},
which is measured without restrictions on the lepton energy,
which yields
\be
\mathcal{B}\lp B\to X_c l\nu\rp_{\mathrm{Delphi}}=
\lp 10.5 \pm 0.2\rp~10^{-2} \ .
\ee
Is is observed that the three results are compatible, even
if our determination is somewhat closer, both in the central
value and in the size of the uncertainty, to the Delphi
measurement.
The small error in our
determination of $\mathcal{B}\lp B\to X_c l\nu\rp$ shows that the
technique discussed in this work can be used also to extrapolate
in a faithful way into regions where there is no
experimental data available.

\begin{table}[ht]
\begin{center}
\begin{tabular}{|c|ccc|} 
\hline
 $\qquad$ $\qquad$$\qquad$
$\qquad$  &$\qquad$  10  $\qquad$ & $\qquad$ 100  $\qquad$ & 1000\\
\hline
$\chi^2_{\tot}$  & 1.31  & 1.11 & 1.11 \\
$\la \chi^2\ra $   & 2.50      & 2.23 & 2.21  \\
\hline
$\la PE\lc\la M \ra_{\rep}\rc\ra$ &  9\% & 8\% & 5\%\\
$r\lc M \rc$ & 0.999 & 0.999 & 0.999 \\
\hline
$\la PE\lc \sigma^{(\net)}\rc\ra_{\dat}$ &  67\% & 58\% & 10\%\\
$\la \sigma^{(\exp)}\ra_{\dat}$ &  0.00267 & 0.00267 & 0.00267\\
$\la \sigma^{(\net)}\ra_{\dat}$ & 0.00180 & 0.00168 & 0.00168 \\
$r\lc \sigma^{(\net)}\rc$ &  0.77 &  0.85 &  0.85\\
\hline
$\la \rho^{(\exp)}\ra_{\dat}$ &  0.166 & 0.166 &  0.166\\
$\la \rho^{(\net)}\ra_{\dat}$ &  0.32 & 0.245  & 0.245\\
$r\lc \rho^{(\net)}\rc$ & 0.35 & 0.38  & 0.38\\
\hline
$\la \mathrm{cov}^{(\exp)}\ra_{\dat}$ &
$1.4~10^{-6}$ & $1.4~10^{-6}$ & $1.4~10^{-6}$\\
$\la \mathrm{cov}^{(\net)}\ra_{\dat}$ &  $7.8~10^{-7}$ & $6.7~10^{-7}$   &
$6.7~10^{-7}$\\
$r\lc \mathrm{cov}^{(\net)}\rc$ & 0.49  & 0.53  & 0.53 \\
\hline
\end{tabular}
\caption{\small Statistical estimators for the
ensemble of trained neural networks, for 10, 100 and 1000
trained replicas.}
\label{resdata1}
\end{center}
\end{table}

\begin{table}[ht]
\begin{center}
\begin{tabular}{|c|ccc|}
\hline
  &  Babar  & Belle & Cleo  \\
\hline
$\chi^2_{\tot}$  & 0.42 & 1.81 & 1.14\\
$\la \chi^2\ra $   & 1.75    & 2.75  & 2.21  \\
\hline
$\la PE\lc\la M \ra_{\rep}\rc\ra$ & 0.78\% & 18\% & 0.6\%\\
$r\lc M \rc$ & 0.999 & 0.999& 0.999\\
\hline
$\la PE\lc \sigma^{(\net)}\rc\ra_{\dat}$ & 47\% & 60\%& 71\% \\
$\la \sigma^{(\exp)}\ra_{\dat}$ & 0.0023 & 0.0021& 0.0041\\
$\la \sigma^{(\net)}\ra_{\dat}$ & 0.0016& 0.0015 & 0.0020 \\
$r\lc \sigma^{(\net)}\rc$ &  0.91 & 0.89 & 0.96\\
\hline
$\la \rho^{(\exp)}\ra_{\dat}$ & 0.16 & 0.40 & 0.31\\
$\la \rho^{(\net)}\ra_{\dat}$ & 0.17 & 0.21 & 0.33\\
$r\lc \rho^{(\net)}\rc$ & 0.87 & 0.19 & 0.34 \\
\hline
$\la \mathrm{cov}^{(\exp)}\ra_{\dat}$ & $6.9~10^{-6}$& $1.5~10^{-6}$ &
$6.5~10^{-7}$\\
$\la \mathrm{cov}^{(\net)}\ra_{\dat}$ & $1.9~10^{-5}$ &$1.4~10^{-6}$ &
$2.6~10^{-7}$\\
$r\lc \mathrm{cov}^{(\net)}\rc$ &  0.98 & 0.47 & -0.04 \\
\hline
\end{tabular}
\caption{\small Statistical estimators for the
ensemble of trained neural networks, for those experiments
included in the fit. The replica sample consists of 
$N_{\rep}=$1000 neural networks.}
\label{resdata2}
\end{center}
\end{table}

\begin{figure}[ht]
\begin{center}
\epsfig{width=0.60\textwidth,figure=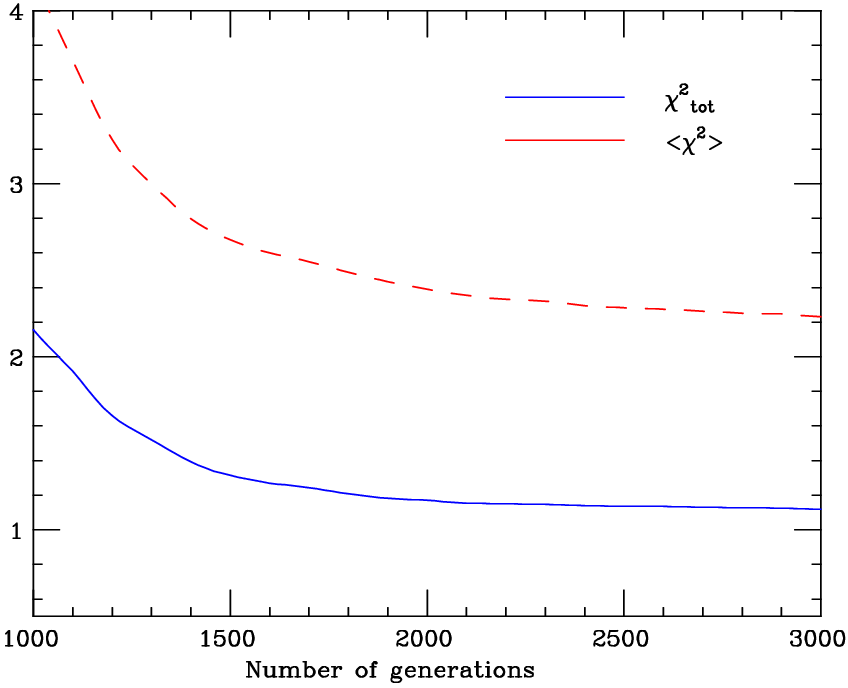}
\end{center}
\caption{}{\small Total $\chi^{2}_{\tot}$, Eq. \ref{chi2tot}, of the
replica sample, compared 
with average error, $\la \chi^2\ra$.}
\label{averr}
\end{figure}

\begin{figure}[ht]
\begin{center}
\epsfig{width=0.49\textwidth,figure=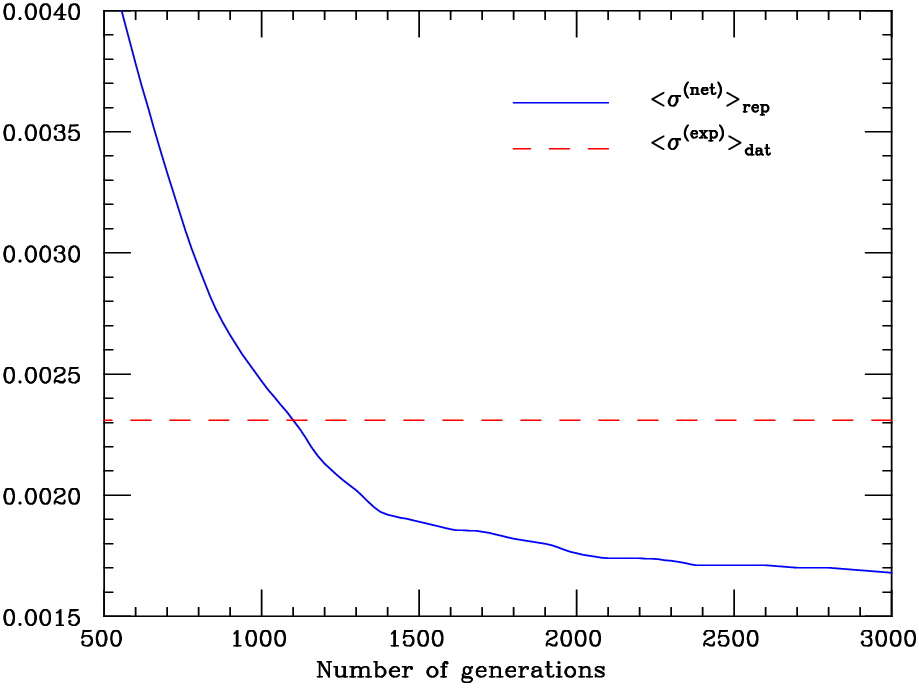}
\epsfig{width=0.49\textwidth,figure=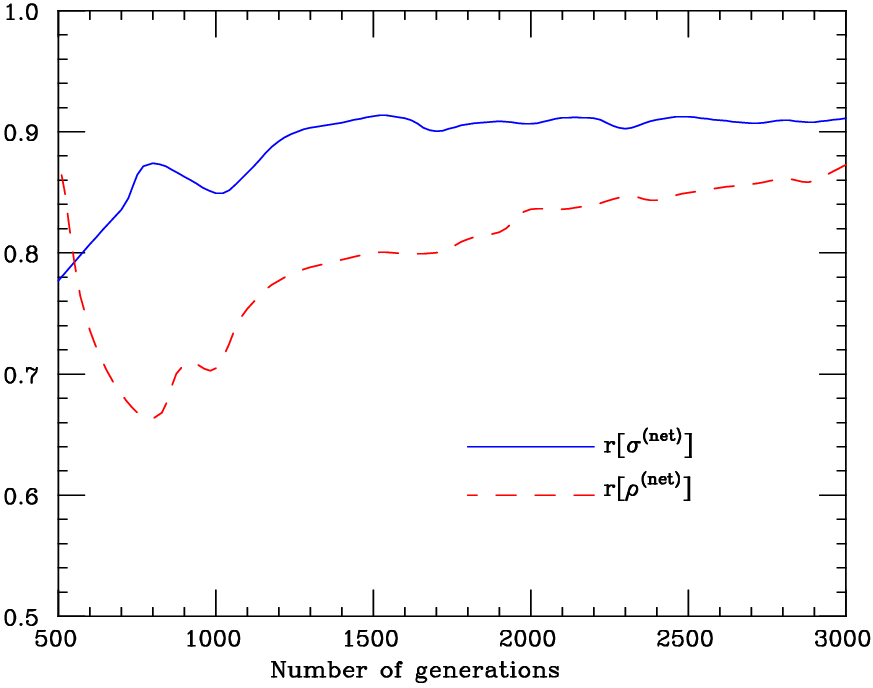}
\end{center}
\caption{}{\small Average error of the data points
as computed from the neural network sample, Eq. \ref{var},
as compared with the experimental value (left). 
Dependence during the training of the values of the scatter correlations,
          Eq. \ref{sccnets}, during the training, for the
Babar experimental data (right).\label{averr3}}
\label{averr2}
\end{figure}

\begin{figure}[ht]
\begin{center}
\epsfig{width=0.49\textwidth,figure=hist.ps}
\epsfig{width=0.49\textwidth,figure=histtl.ps}
\end{center}
\caption{}{\small Distribution 
of error functions, $E_2^{(k)}$, over the sample
        of trained replicas (left) and distribution
of training lenghts in GA generations (right). }
\label{hist}
\end{figure}

\begin{figure}[ht]
\begin{center}
\epsfig{width=0.55\textwidth,figure=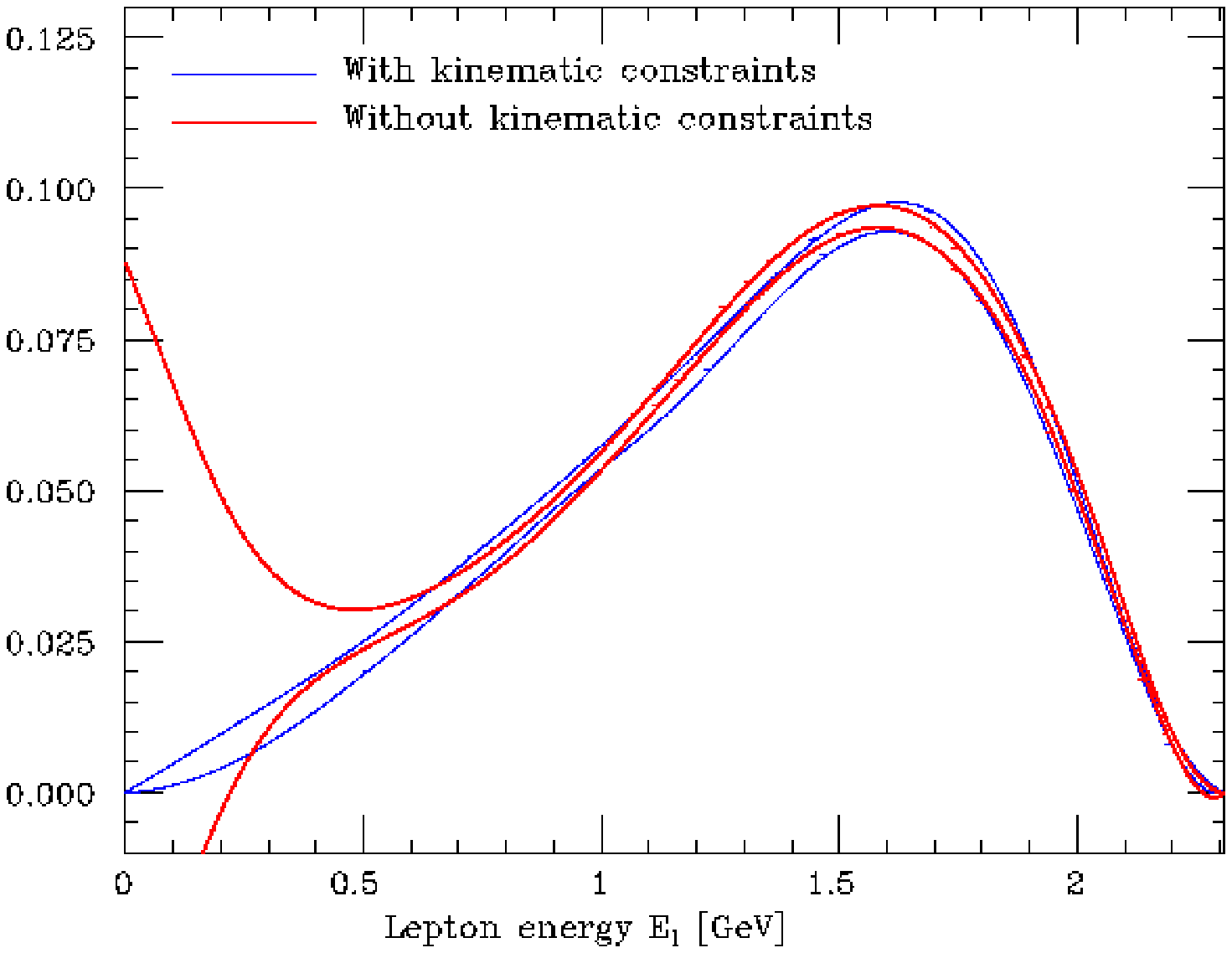}
\end{center}
\caption{}{\small Comparison of the lepton energy spectrum when
no kinematic constraints are incorporated.\label{specav_nokincos}}
\end{figure} 

Important information on the quality of the fit can be
obtained from the analysis of the dependence of 
the different statistical estimators
with respect to the number of genetic algorithms generations, like
the total $\chi^2$ or
the average error. This dependence is represented in
 Fig. \ref{averr}.
 Note that at the end
of the training $\chi^{2}_{\tot}\sim 1$ and $\la \chi^2\ra \sim 2$,
as expected. Note also that the fit has reached convergence 
since  the
$\chi^2_{\tot}$ profile is very flat for a large number of generations. 

This can be repeated for other
statistical estimators, like for example
the average spread of the data points
 as computed from the neural network
ensemble, defined
in Section \ref{estimators},
which is  compared with the same quantity computed from
the experimental data, $\sigma_i^{(\mrexp)}$ in 
Fig. \ref{averr2} as a function of the number
of genetic algorithms generations. 
The fact that one has error reduction, as has been explained in
\cite{f2ns},
is the sign that the network has found an underlying law
to the experimental data, in this case the lepton
energy spectrum.

Another relevant estimator is the 
 scatter correlations of the spread of the points
(see Fig. \ref{averr3}).
One can define similarly a scatter correlation for the
net correlation $\rho_{ij}^{(\net)}$, also represented in Fig. 
\ref{averr3} for the Babar experimental data. We observe that when
the training ends both values of $r$ are close to 1, a sign
that errors and correlations are being faithfully reproduced.
Another relevant estimator of the goodness of the fit is the distribution 
of both $E_2^{(k)}$ and of the training lengths over the replica sample,
see Fig.  \ref{hist}. We have checked that these
two distributions satisfy the requirements described
in Section \ref{validation}.

We can compare also  fits with and without the inclusion of kinematical
constraints, to see which is the effect of their
implementation.
 The endpoint constraint at $E_l=E_{\max}$ is crucial
to obtain convergent results. In Fig. \ref{specav_nokincos}
one can observe that when the endpoint
constraint at $E_l=0$ and the positivity constraint are
removed the error becomes very large at small $E_l$, and
on top of that sightly negative at large $E_l$ near the endpoint. Note that
the physical value for the spectrum at the endpoint, 
$d\Gamma/dE_l(E_l=0)=0$,
is contained within the small-$E_l$ error bars, as expected.

In Fig.~\ref{specav_babar}  we show a comparison of 
a fit of a  single experiment, Babar. We observe that when only the 
Babar data is incorporated  in the fit, the error at small
values of $E_l$ is much larger. This is so because, as discussed 
above, the Belle data, which extends to lower values of $E_l$, is crucial
to determine the low $E_l$ behavior of the lepton spectrum.
Finally, in Fig. \ref{specav_ndep} we show that our results
are independent of the precise choice of $n_1$ and $n_2$ in
Eq. \ref{gammacon}. In particular using $n_1=1.5$ and $n_2=1.5$
results in the same fit as when using the reference values, 
$n_1=1$ and $n_2=2$.

As one example of the  applications of the 
present parametrization of the lepton energy spectrum,
in this section our results are 
compared with the
theoretical calculation of Ref. \cite{agru} (AGRU). Their
formalism allows the computation of moments of
arbitrary differential distributions from
semileptonic B meson decays, with arbitrary kinematical cuts,
like the lepton energy spectrum in charmed decays
that is analyzed in the present work. 
In particular  their computation of
the lepton energy spectrum will be studied, which they define as
\be
N_k\equiv\frac{1}{\Gamma_{\mathrm{LO}}}\int_{E_0}^{E_{\max}} 
dE_ldq^2dr \tilde{E}_l^k\frac{d^3\Gamma}{dE_ldrdq^2}\ ,
\ee
with $\tilde{E}\equiv E/m_b$, and 
where the leading order partonic semileptonic decay rate 
$\Gamma_{\mathrm{LO}}$ is
given by 
\be
\Gamma_{\mathrm{LO}}=\Gamma_0|V_{cb}|^2 
z_0(\rho) \ ,
\ee
where $\Gamma_0$ is defined in Eq. 
\ref{gammalo}, $\rho\equiv m_c^2/m_b^2$ and the phase space factor is
defined in Eq. \ref{zo}.
These moments can be related to the 
experimentally measured moments, defined in Eqns. \ref{momexp1}-
\ref{momexp2}, in a straightforward way, for example for the
first two moments one has
\be
M_0=\tau_B\Gamma_0N_0, \qquad M_1=m_b\frac{N_1}{N_0} \ ,
\ee
and similarly for the other moments.

\begin{figure}[t]
\begin{center}
\epsfig{width=0.45\textwidth,figure=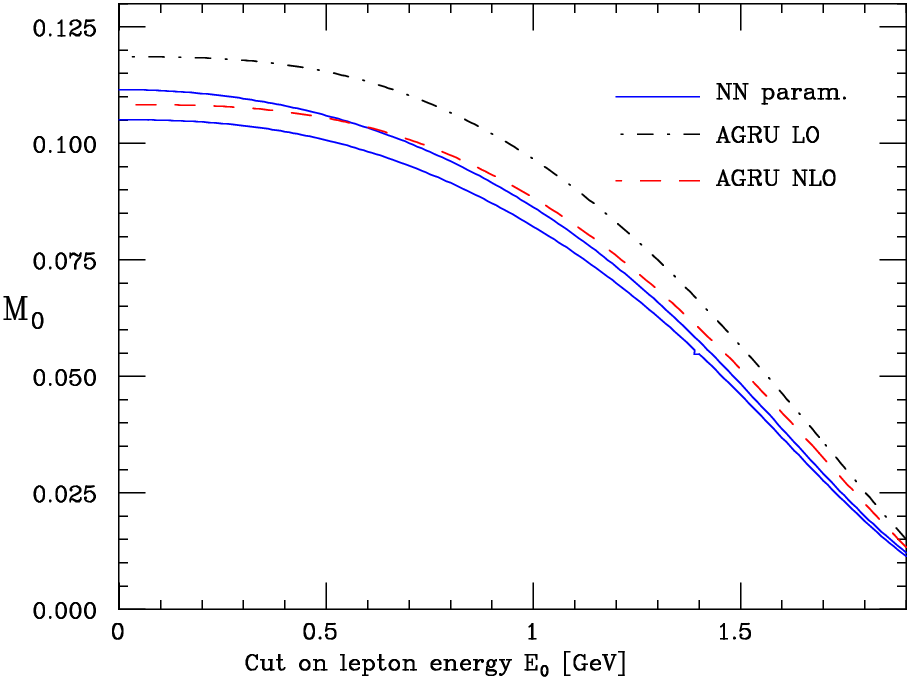}
\epsfig{width=0.45\textwidth,figure=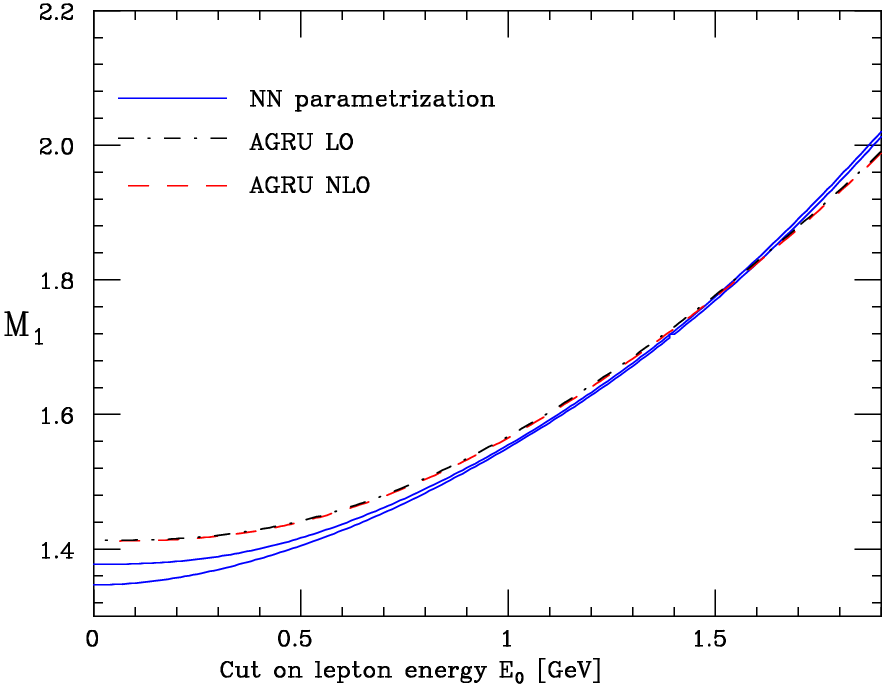}
\caption{}{\small Comparison of the results of Ref. \cite{agru}
on the partial branching ratio Eq. \ref{momexp1} both
at leading order (LO) and at next-to-leading
order (NLO) with the
same quantity computed from our parametrization. }
\label{agruplots}
\end{center}
\end{figure}

\begin{figure}[t]
\begin{center}
\epsfig{width=0.6\textwidth,figure=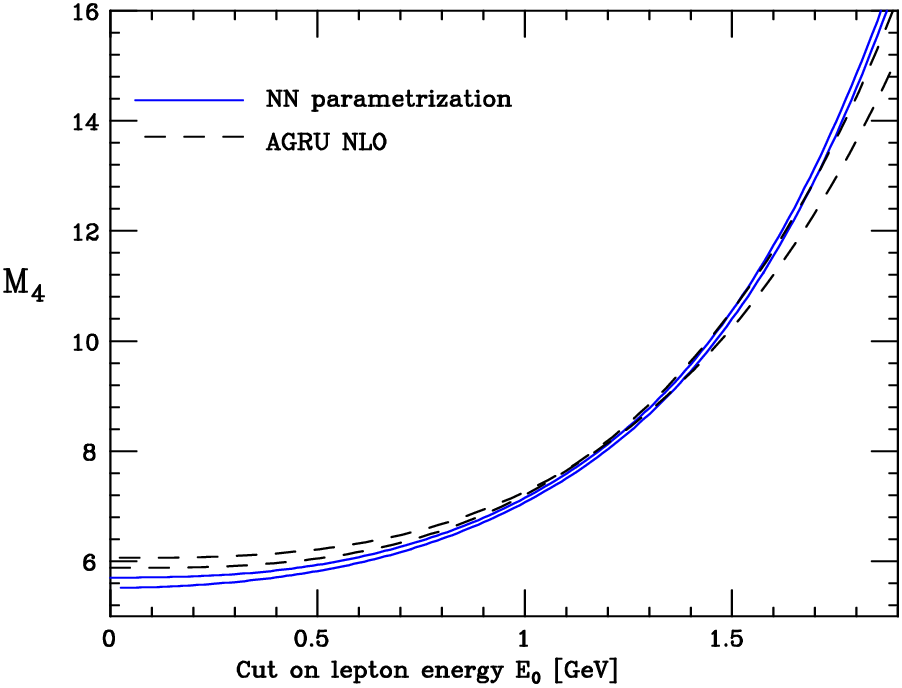}
\caption{}{\small Comparison of the results of Ref. \cite{agru}
on the  $n=4$ moment and at next-to-leading
order (NLO), with the corresponding
theoretical uncertainties, with the
same quantity computed from the neural network
parametrization parametrization. }
\label{agruplots2}
\end{center}
\end{figure}

 In Figs. \ref{agruplots}
 the results of \cite{agru} both at leading order (LO)
and at next-to-leading order (NLO) are
compared with the
moments obtained from our parametrization as a function of the lower
cut in the lepton energy $E_0$. Comparing
results at different perturbative orders is interesting
to asses the behavior of the perturbative expansion.
 One should
take into account in this comparison that the results of
\cite{agru} are purely perturbative, therefore the 
difference between the two results could be an indicator of the
size of the missing nonperturbative corrections. 
Another interesting feature is that while for $M_0$, the partial
branching fraction, the NLO corrections are sizable and
bring the theoretical prediction in better agreement with
the experimental measurement, for $M_1$ (which is the
ratio of two perturbative expansions) the size
of the perturbative corrections is much smaller.
In addition, we show in Fig. \ref{agruplots2}
the comparison for the $n=4$ moment (not measured experimentally)
with an estimation of the uncertainties in the theoretical prediction,
as described in \cite{bmeson}.

A more general comparison with theoretical predictions should include
also the known nonperturbative power corrections up to
order $\mathcal{O}(1/m_b^3)$ to the
expressions for the moments of the spectrum, since in this case
the difference of the theoretical results from our
parametrization would indicate the size of the missing 
unknown corrections,
both perturbative and nonperturbative. A more
detailed study of this point, together with an analysis of possible
violations of local quark-hadron duality \cite{shifman}, 
is left for future work. 

As another
application of our parametrization of the lepton
energy spectrum,
it will be used to determine the
b quark mass $m_b^{1S}$ from the
experimental data using a novel
strategy. To this purpose the
technique of Ref. \cite{bauertrott} will be used, which
consists on
the minimization of the size of higher order corrections
to obtain sets of moments of the
lepton energy spectrum which have reduced theoretical
uncertainty
for the extraction of  nonperturbative parameters like 
$\bar{\Lambda}_{1S}$ and $
\lambda_1$.  In Ref. \cite{bmeson}
we describe in detail the method we use to
determine the b quark mass from our neural network
parametrization of the leptonic spectrum, which
is summarized here.

The moments that minimize the impact of
the higher order nonperturbative corrections are given by
\be
R_1\equiv\frac{
\int_{1.3}^{E_{\max}}E_l^{1.4}
 \frac{d\Gamma}{dE_l}dE_l}{\int_1^{E_{\max}}E_l \frac{d\Gamma}{dE_l}dE_l}\ ,
\ee
and
\be
R_2\equiv\frac{
\int_{1.4}^{E_{\max}}E_l^{1.7}
 \frac{d\Gamma}{dE_l}dE_l}{\int_{0.8}^{E_{\max}}E_l^{1.2} 
\frac{d\Gamma}{dE_l}dE_l} \ .
\ee
The full expression for this moments in terms of
heavy-quark non-perturbative 
parameters can be found in Ref. \cite{bauertrott}.
These leptonic moments $R_1$  and $R_2$ depend on 9 nonperturbative
parameters, up to $\mathcal{O}(1/m_b^{3})$: $\bar{\Lambda}_{1S}$,
$\lambda_1$ and $\lambda_2$, and six matrix elements,
$\rho_1,\rho_2,\tau_1,\tau_2,\tau_3$ and $\tau_4$, that
contribute at order $1/m_b^3$ in the heavy quark
expansion, of which not all of them are
independent  \cite{gremm,shapevar}.

The most relevant feature of these leptonic moments $R_1$ and $R_2$ 
is that they have non-integer powers and to the
best of our knowledge have not been yet
experimentally measured, at least in a published
form. Therefore,  
 the values of $R_1$ and $R_2$ that will be
used in this analysis are extracted from 
our neural network parametrization of the
lepton spectrum, which allows the
computation of arbitrary moments, together  with their
associated error and correlation. Let us recall that the central values
are determined as
\be
\la R_1^{(\net)}\ra=\frac{1}{N_{\rep}}\sum_{k=1}^{N_{\rep}}
R_1^{(\net)(k)} , \qquad R_1^{(\net)(k)}=
\frac{
\int_{1.3}^{E_{\max}}E_l^{1.4}
\frac{d\Gamma^{(\net)(k)}}{dE_l}(E_l)dE_l}{\int_1^{E_{\max}}
E_l\frac{d\Gamma^{(\net)(k)}}{dE_l} (E_l)dE_l} \ ,
\ee
and similarly for $R_2$,
and the error and the correlation of the moments $R_1$ and 
$R_2$ are computed in the
standard way. The following values for the moments with
the associated errors and their
correlation are obtained,
\be
R_1^{(\net)}=1.017\pm 0.003,  \quad R_2^{(\net)}=0.938\pm 0.004, \quad
\rho_{12}=0.94 \ ,
\ee
that as expected are highly correlated.
Then to determine the nonperturbative parameters $\Lambda_{1S}$
and $\lambda_1$ the associated 
$\chi^2_{\mathrm{fit}}$ is minimized,
\be
\label{chi2fit}
\chi^2_{\mathrm{fit}} = \sum_{i,j=1}^2 \lp R_i^{(\net)}-R_i^{(\tth)}\rp \lp
\mathrm{cov}^{-1}\rp_{ij} 
\lp R_j^{(\net)}-R_j^{(\tth)}\rp \ ,
\ee
where $\mathrm{cov}^{-1}_{ij}$ is the inverse of the covariance matrix
associated to the two moments $R_1^{(\net)}$ and $R_2^{(\net)}$, 
and $R_i^{(\tth)}\lp
\bar{\Lambda}_{1S},\lambda_1\rp$
is the theoretical prediction for these moments as a function of the
two nonperturbative parameters \cite{bauertrott}. 

Once the values of $\bar{\Lambda}_{1S}$ and $\lambda_1$ have
been determined from the minimization of Eq. \ref{chi2fit},
one obtains for the b quark mass $m_b^{1S}$ mass in the 1S
scheme the following
value:
\be
m_b^{1S}=\bar{m}_{B}-\bar{\Lambda}_{1S}=\lp 4.84 \pm 0.16^{\mathrm{exp}}
\pm 0.05^{\mathrm{th}}\rp ~\mathrm{GeV}=\lp 4.84 \pm 0.17^{\mathrm{tot}}
\rp~\mathrm{GeV} \ ,
\ee
From the above results one observes
 that the dominant source of uncertainty is the
experimental uncertainty, that is, the uncertainty associated
to our parametrization of the lepton energy spectrum. This determination of
the b quark mass is consistent with determinations
from other analysis. The b quark mass
has been determined using different techniques, like the sum rule
approach, using either non-relativistic \cite{hoang,
benekesum,pinedasumrule}
or relativistic \cite{kuhnbmass,corcella} sum rules, 
 global fits of moments of
differential distributions in
 B decays, \cite{bauerglobalfit,delphifit,fitskin},
 the renormalon analysis of Ref. \cite{pineda}, and several other
methods related to heavy-quarkonium physics \cite{quarkonium1,
quarkonium2}
(see \cite{quarkonium3} for a review).
 To compare our results with some of the above references, it is 
useful to 
 relate the $m^{1S}_b$ mass to the MS-bar $\bar{m}_b \lp \bar{m}_b\rp$
mass \cite{hoangfull}, and once the conversion
is performed the value
\be
\bar{m}_b\lp \bar{m}_b \rp=\lp 4.31 \pm 0.17^{\tot} \rp~\mathrm{GeV} \ ,
\ee
is obtained,
where we have used $\alpha_s(M_Z^2)=0.1182$ and included 
perturbative corrections up to two loops.
It turns out that our determination of 
$m^{1S}_b$ is not competitive with respect
to other determinations due to the large
 uncertainties associated to the fact that only
two moments we use in the determination of these parameters.
The inclusion of additional moments in the fit would
constrain more the nonperturbative parameters and
reduce the experimental uncertainty associated to them.

For the nonperturbative parameter $\lambda_1$  the
following value is obtained
\be
\lambda_{1}=\lp -0.17 \pm 0.15^{\mathrm{exp}}  \pm 0.05^{\mathrm{th}}\rp~
\mathrm{GeV^2} =\lp -0.17 \pm 0.16^{\mathrm{tot}}\rp ~\mathrm{GeV^2} 
\ .
\ee
As in the determination of $\bar{\Lambda}_{1S}$ 
it can be seen that the theoretical uncertainties are
smaller than the experimental ones, which are the dominant ones.
Our result for the parameter $\lambda_1$ is consistent with other 
extractions in the context of  global fits of
B decay data \cite{bauerglobalfit,delphifit}, but again not competitive
due to the rather large  uncertainties.

To summarize, in this part of the thesis
we have presented a determination of the probability density in the
space of the lepton energy spectrum from semileptonic B meson decays, based
on the latest available data from the Babar, Belle and
Cleo collaborations.
In addition, this application shows the implementation of
a  well defined strategy to reconstruct 
functions with uncertainties
 when the only available experimental information comes 
through convolutions of these functions.
As a byproduct of our analysis, using 
our parametrization of the lepton spectrum, we have
extracted the nonperturbative parameters $\bar{\Lambda}_{1S}$
and $\lambda_1$, with a method that reduces
the contributions from the theoretical uncertainties.

The number of possible applications of this strategy to other 
problems in B physics is rather large,
and is discussed in some detail in Ref. \cite{bmeson}.
The most promising
application is to use our neural network strategy 
to construct a parametrization of the shape function $S(k)$ 
of the B meson, a universal characteristic
of the B meson governing inclusive decay spectra
in processes with massless partons in the
final state, as extracted from the
$B\to X_s\gamma$ and $B\to X_ul\nu$ decay modes.
 In this case we have additional theoretical
information on its behaviour. For example,
at tree level  its moments 
\be
A_n\equiv \int dk k^n S(k) \ ,
\ee
have to satisfy $A_0=1,A_1=0$ and $A_2=\mu_{\pi}^2/3$, 
where $\mu_{\pi}$ is a nonperturbative parameter of the
heavy quark expansion.
At higher orders these relations are theoretically more
controversial \cite{neubertfactorization,shapemanohar}.
Since the uncertainty from the extraction of $S(k)$
 is the dominant source of theoretical 
uncertainty in some CKM matrix
elements extractions, it would therefore be interesting to 
estimate again this uncertainty using the technique presented in this
work, since in the current approach \cite{shape2} 
the shape function uncertainties are estimated in
a rather crude way, just trying different functional
forms compatible with the theoretical constraints.

\clearpage

\section{Parton distribution functions}

\label{nnqns}

In this Section we describe the most important
application of the general strategy introduced in 
Chapter~\ref{general}. Indeed, it was the 
determination of the probability measure in the
space of parton distributions the main motivation
for the development of the technique that is
the main subject of this thesis.
First we describe an alternative approach to evolve 
parton distributions, required by the use
of neural networks as interpolants for parton
distributions, and then we present results
on the neural network parametrization of the 
nonsinglet parton distribution $q_{NS}(x,Q_0^2)$ from
experimental data on the nonsinglet structure function $F_2^{NS}(x,Q^2)$.

\subsection{A new approach to parton evolution}
\label{evolform}

In order to
construct the probability measure in the space
of  nonsinglet parton distributions, 
for reasons to be described in the following, we need
a dedicated parton evolution formalism.
In this Section the strategy that will be used
for the evolution
of parton distributions is introduced. 
For the present application only
nonsinglet parton distributions will be considered, and
therefore
 the description of this
parton evolution technique  will be
restricted to the nonsinglet sector. This discussion
can be generalized straightforwardly to the singlet sector and will
be throughly described in Ref. \cite{nnsinglet}.

There are two classes of methods for solving the perturbative QCD
DGLAP evolution equations, introduced in
Section \ref{dis_theo}: the N-space methods (like
the one described in \cite{pegasus}) which 
solve analytically the DGLAP evolution equations in 
Mellin space and then  compute the inverse Mellin transformation
to x-space,
and the x-space methods (like \cite{qcdnum}), which
perform a direct numerical integration of the
integro-differential DGLAP equations.
Both methods have different advantages and drawbacks, but in
principle they lead to a similar accuracy in the evolution
of parton distributions, as shown in the dedicated analysis
of Ref. \cite{lh}.
However,  since we are going to parametrize parton distributions not
with simple functional forms but with neural networks, none of the
standard techniques can be applied to our case.
For example, in standard N-space evolution codes one needs to
perform the Mellin transform of the parton distribution and extend
it to complex values of N, but in our case this is not allowed
since  an analytical
expansion of neural networks to complex values of the inputs
and the outputs is mathematically ill-defined.

Hence we need to use a hybrid strategy: we will solve the
DGLAP 
evolution equations \cite{gl,ap,dok} in N-space,
introduced in Section \ref{dis_theo}, and then we will Mellin invert
only the evolution factor $\Gamma(N)$ in Eq. \ref{evfactn}, 
so that the evolved x-space parton
distribution $q(x,Q^2)$ can
be written as the convolution of the  x-space 
parton distribution at the
initial evolution scale $q(x,Q^2_0)$ and
an x-space evolution factor $\Gamma(x)$.
Schematically, we will have
\be
q(x,Q^2)=\Gamma(x)\otimes q(x,Q_0^2), \qquad \Gamma(N)\equiv\int_0^1dxx^{N-1}
\Gamma(x) \ .
\ee 
Then we will interpolate the x-space
evolution factor $\Gamma(x)$,
so that the heavy numerical task of its computation
is decoupled from the determination
of the parameters describing the parton
distribution from experimental data. With all these
considerations taken into account one ends
up with a fast and efficient evolution code, which will described in this
Section.

First of all we define the notation that will be used
for the strong coupling constant.
The convention that we use is
\be
Q^2\frac{da_s(Q^2)}{Q^2}=-\sum_{k=0}^{\infty} \beta_k a_s(Q^2)^{k+1} , \qquad
a_s(Q^2)=\frac{\alpha_s(Q^2)}{4\pi} \ ,
\ee
where the beta function coefficients are given by
\be
\beta_0=11-\frac{2N_f}{3}, \qquad \beta_1=102-\frac{38}{3}N_f \ ,
\ee
the scheme-dependent coefficent $\beta_2$ is given in
\cite{pegasus}, and $N_f$ is the number of active quark flavors.

The explicit solution for the above equation at the NNLO 
accuracy can be
written in terms of a boundary condition $\alpha_s\lp M_Z^2\rp$
and is given by
\be
\aq_{\NNLO}=\aq_{\LO}\lp 1+\aq_{\LO}\lp \aq_{\LO}- \alpha_s\lp M_Z^2\rp \rp
(b_2-b_1^2)+\aq_{\NLO}b_1 \ln\frac{\aq_{\NLO}}{\alpha_s\lp M_Z^2\rp}\rp  \ ,
\ee
where we have defined in a recursive way,
\be
\aq_{\NLO}=\aq_{\LO}\lp 1-b_1\aq_{\LO}\ln\lp 1+\beta_0\alpha_s\lp M_Z^2\rp
\ln\frac{Q^2}{M_Z^2}\rp\rp  \ ,
\ee
\be
\aq_{\LO}=\frac{\alpha_s\lp M_Z^2\rp}{ 1+\beta_0\alpha_s\lp M_Z^2\rp
\ln\frac{Q^2}{M_Z^2}} \ .
\ee
and we have defined $b_k=\beta_k/\beta_0$.

The evolution equations for nonsinglet combinations
of parton distributions, as we have seen in Section
\ref{dis_theo},
read in Mellin space
\be
\label{evolN}
\frac{d}{d\ln Q^2}q^i_{NS}(N,Q^2)=\frac{\aq}{2\pi}\gamma_{NS}^i
(N,\aq)q_{NS}^i(N,Q^2) \ ,
\ee
where $i=\pm,v$ stand for the three different  
nonsinglet combinations that evolve independently at NNLO, given by
\be
q_{NS,ik}^{\pm}=q_i\pm \bar{q}_i-\lp q_k \pm \bar{q}_k\rp, \qquad
q_{NS}^v=\sum_{r=1}^{N_f}\lp q_r-\bar{q}_r\rp \ .
\ee
In the following discussion we assume that in each case
we use the appropriate anomalous dimension for 
each type of nonsinglet parton distribution, and
since we are restricted now to the
analysis of nonsinglet parton distributions,  we will assume also
that all quantities
(parton distributions or anomalous dimensions) correspond
to this nonsinglet sector.
The nonsinglet anomalous dimension has been computed in powers of $\aq$
up to NNLO,
\be
\gamma\lp N,\aq\rp=\gamma^{(0)}(N)+\frac{\aq}{4\pi}\gamma^{(1)}(N)
+\lp \frac{\aq}{4\pi}\rp^2\gamma^{(2)}(N) \ ,
\ee
where the  N-space anomalous dimension was
computed at LO in Ref. \cite{ap},
at NLO in Ref. \cite{petronzio}, and recently the
full NNLO contribution was computed in Ref. \cite{mvvns}.
The leading order nonsinglet anomalous dimension has the explicit form
\be
\gamma^{(0)}(N)=C_F\lc 3-4\sum_{k=1}^N\frac{1}{k}+\frac{2}{N(N+1)}\rc \ ,
\qquad C_F=\frac{4}{3} \ ,
\ee
which has the following large-N limit,
\be
\label{gammalargen}
\lim_{N\to\infty}\gamma^{(0)}(N)=-4C_F\ln N\ +\mathcal{O}\lp N^0\rp, 
\ee
which will be needed for the computation of the
large-x limit of the x-space evolution factor $\Gamma(x)$. 

The solution to the  nonsinglet evolution equation, Eq.
\ref{evolN},  reads at NNLO accuracy
\be
\label{evoln}
q(N,Q^2)=\Gamma(N,\aq,\aqq)q(N,Q^2_0) \ ,
\ee
where the N-space evolution factor 
$\Gamma(N)$
is given by
\bea
\label{gammaN}
\Gamma(N,\aq,\aqq)=\lp \frac{\aq}{\aqq}\rp^{-\gamma^{(0)}(N)/\beta_0}
\Bigg( 1 -\lp \aq-\aqq \rp U_{NS}^{(1)}(N)
\nonumber
\\
+\lp \aq^2-\aqq^2\rp U_{NS}^{(2)}(N)
-\aq\aqq \lp U_{NS}^{(1)}(N)\rp^2
\Bigg) \ ,
\eea
where we have defined the nonsinglet evolution coefficients as
\be
U_{NS}^{(1)}(N)=-\frac{1}{\beta_0}\lp \gamma^{(1)}(N)-
b_1\gamma^{(0)}(N)\rp \ ,
\ee
\be
U_{NS}^{(2)}(N)= -\frac{1}{2}\lp \frac{1}{\beta_0}\gamma^{(2)}(N)+
b_1U_{NS}^{(1)}(N)-\frac{b_2}{\beta_0}\gamma^{(0)}(N) \rp \ .
\ee

Once we have solved the DGLAP equations in Mellin space, Eq. \ref{evoln}, 
we Mellin invert
the evolution factor in Eq. \ref{gammaN} to perform parton evolution in
x-space.
Using the convolution theorem one can check that the x-space
evolved parton distribution is given by
\be
\label{evolx}
q(x,Q^2)=\int_x^1 \frac{dy}{y}\Gamma(y,\aq,\aqq)q\lp \frac{x}{y},Q_0^2\rp \ ,
\ee
where the evolution kernel $\Gamma(x)$ is the Mellin inverse of
Eq. \ref{gammaN},
\be
\label{gammaX}
\Gamma(x,\aq,\aqq)=\int_{c-i\infty}^{c+i\infty}
\frac{dN}{2\pi i}x^{-N}\Gamma(N,\aq,\aqq) \ ,
\ee
where $c$ lies to the right of the rightmost singularity of
$\Gamma(N)$. That is, one can check that the Mellin transform
of the LHS of Eq. \ref{evolx} is equal to the RHS of
Eq. \ref{evoln},
\be
\int_0^1 dx x^{N-1}q(x,Q^2)=\Gamma\lp N,\aq,\aqq\rp q(N,Q_0^2) \ .
\ee
The complicated numerical task in this evolution formalism is
the numerical computation of the Mellin inverse transformation,
Eq. \ref{gammaX}. However note that as opposite to standard
N-space parton evolution methods, the evolution
of the parton distribution can be decoupled from the Mellin
inversion of the evolution factor, which is the
 most time consuming task.

We will use the Fixed
Talbot algorithm to perform the numerical inversion of the Mellin
transform Eq. \ref{gammaX}. For a detailed description
of this algorithm and its efficiency see Refs. \cite{ft,ft2}.
The Fixed Talbot algorithm for the numerical inversion
of Mellin-Laplace transforms is specially accurate for the following reason:
the numerical computation of inverse Mellin transforms is in general
difficult since the integrand is highly oscillatory since 
the integration path
 moves through the complex plane. The Fixed Talbot algorithm
 bypasses this problem by choosing a path in the
complex plane which minimizes the imaginary part of the
integrand and therefore minimizes the impact of these oscillations.
In the Fixed Talbot algorithm a
 generic inverse Mellin transform is computed as
\be
\label{invmellin}
f(x)=\frac{1}{2\pi i} \int_C x^{-N}f(N)dN \ ,
 \ee
where the Talbot path $C$ is defined by the condition
\be
N(\theta)=r\theta \lp 1/\tan \theta +i\rp, \quad  -\pi \le \theta \le \pi \ ,
\ee
\be
r\equiv\frac{2M}{5\ln \frac{1}{x}} \ ,
\ee
where M is the number of required precision digits.
The Talbot path minimizes the contribution from oscillatory terms
to the inverse Mellin Eq. \ref{invmellin}, and therefore
avoids the associated numerical instabilities.
The resulting integral, Eq. \ref{invmellin}, can be
computed with adaptative gaussian quadratures to any desired
accuracy.

\begin{figure}[t]
\begin{center}
\epsfig{width=0.48\textwidth,figure=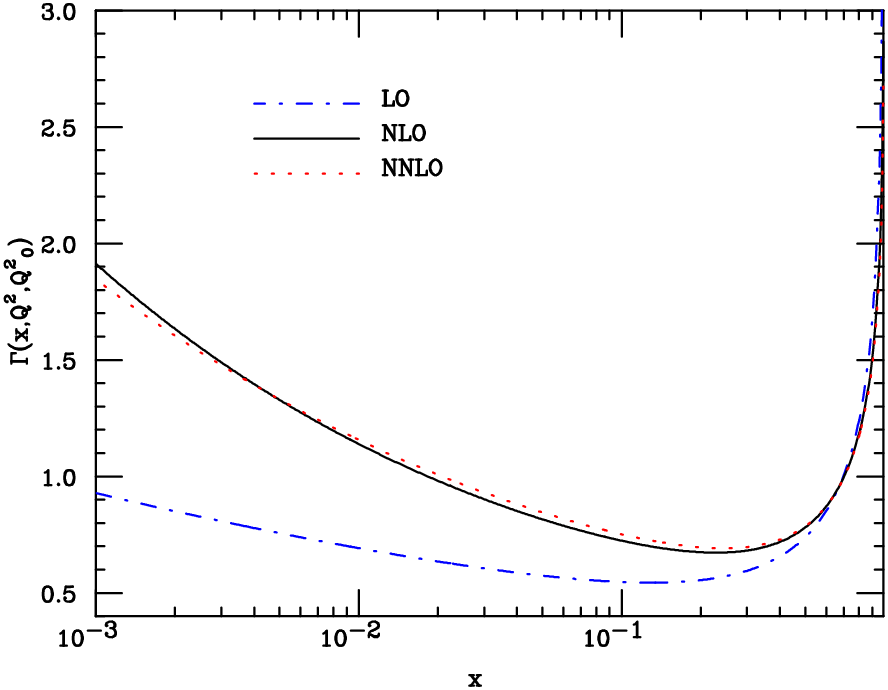}
\epsfig{width=0.48\textwidth,figure=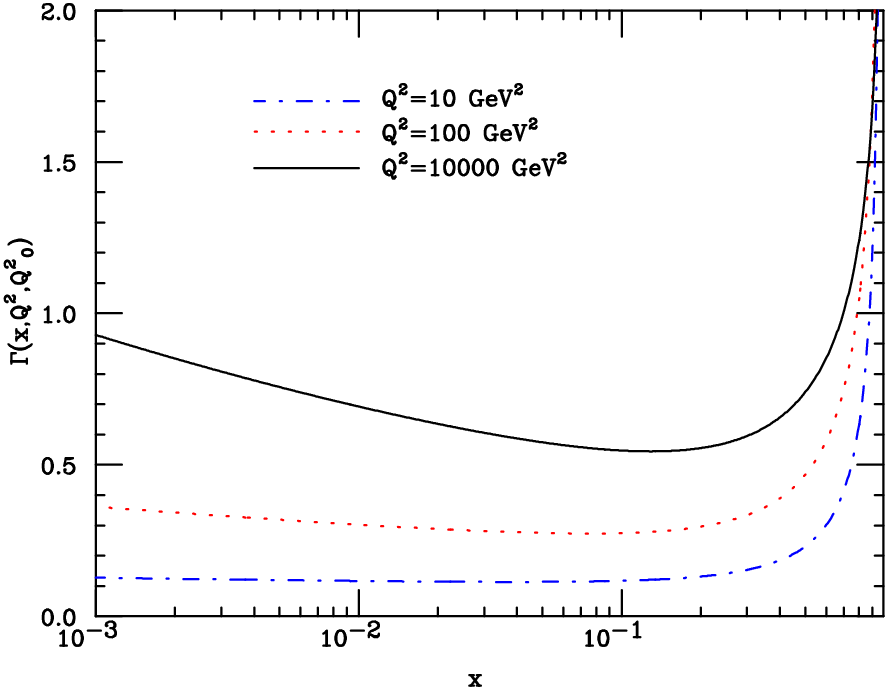}
\caption{}{\small The x-space evolution factor $\Gamma\lp x,\aq,\aqq\rp$
for different perturbative orders at $Q^2=10^4$
GeV$^2$ (left) and for different
evolution lenghts at leading order (right). In all cases
the starting evolution scale is $Q_0^2=2$   GeV$^2$.
Note the sharp rise of the evolution factor at large values
of $x$.}
\label{evolfactxplots}
\end{center}
\end{figure}

The x-space evolution factor
$\Gamma\lp x,\aq,\aqq\rp$, Eq. \ref{gammaX},
 has a flat behavior at intermediate $x$
together with a growing at both large and small $x$, where the
behavior in both limits can be computed analytically.
Now the explicit analytic expressions for
the evolution factor $\Gamma(x)$ at both large $x$ 
and small $x$ will be computed. These results are interesting both for
themselves, to obtain a more quantitative understanding of our
evolution technique, and also for practical purposes, since they
allow to perform a more efficient interpolation of $\Gamma(x)$.
In Fig. \ref{evolfactxplots} we represent the x-space
evolution factor  $\Gamma(x)$ for different perturbative
orders and different evolution lenghts. Note that in the
$x$ range relevant for nonsinglet evolution, $x\ge 5~10^{-3}$, 
the perturbative
expansion for $\Gamma(x)$ appears to be near to convergence at the NNLO
level. Note also that the small-x behaviour is very smooth in the
relevant x range, unlike the large-x one.

First of all, the large $x$ limit of the evolution
kernel will be computed in two equivalent ways.
At leading order in $\aq$, in the
large $x$ limit, the dominant contribution to the
evolution kernel comes from the large $N$ limit of the
LO anomalous dimension, Eq. \ref{gammalargen}, 
and therefore one has
\be
\label{largeN}
\Gamma_{x\to 1}(x)=\int_{-i\infty}^{+i\infty}\frac{dN}{2\pi i}x^{-N}
\lp \frac{\aq}{\aqq}\rp^{2C_F\ln N /b_0} \ ,
\ee
then one has to use the Mellin transform
\be
\int_0^1x^{N-1}\ln^{\eta-1}\frac{1}{x}=\frac{\Gamma(\eta)}{N^{\eta}},\qquad
\eta \ge 0 \ ,
\ee
and the final result is
\be
\label{largex}
\Gamma_{x\to 1}\lp x,\aq,\aqq\rp=
\frac{1}{\Gamma\lp \frac{2C_F}{b_0}\ln\frac{\aqq}{\aq}\rp}
\lp \ln\frac{1}{x}\rp^{-1+ \frac{2C_F}{b_0}\ln\frac{\aqq}{\aq}} \ ,
\ee
so that at large $x$ one has 
\be
\Gamma_{x\to 1}\lp x,\aq,\aqq\rp\sim (1-x)^{\lambda(Q^2,Q_0^2)} \ ,
\qquad 
\lambda(Q^2,Q_0^2)=-1+ \frac{2C_F}{b_0}\ln\frac{\aqq}{\aq} \ ,
\ee
therefore, the growth of $\Gamma(x)$ at large $x$ depends only mildly on the
evolution lenght.

A second related method makes use of the analytic
results for the all logarithmic orders Mellin inversions
of Ref. \cite{resum}.
The large $x$ limit of the evolution kernel is given by Eq. \ref{largeN}.
Its inverse Mellin transformation can be performed analytically
using the formulas of Ref. \cite{resum}, which yield
\bea
\Gamma(x)=\sum_{n=1}^{\infty} \frac{\Delta^{(n-1)}(1)}{(n-1)!}
\lc \frac{1}{1-x} \frac{d^n}{d\ln^n(1-x)} \lp\frac{\aq}{\aqq}
\rp^{2C_F\ln(1-x)/b_0}\rc_++\mathcal{O}((1-x)^0,\alpha_s)=
\nonumber
\\
\sum_{n=1}^{\infty} \frac{\Delta^{(n-1)}(1)}{(n-1)!}\lc
\frac{2C_F}{b_0}\ln \lp\frac{\aq}{\aqq}
\rp\rc^n
\lc \frac{1}{1-x}  \lp\frac{\aq}{\aqq}
\rp^{2C_F\ln(1-x)/b_0}\rc_++\mathcal{O}((1-x)^0,\alpha_s) \ ,
\nonumber
\eea
\be
\Gamma_{x\to 1}\lp x,\aq,\aqq\rp=\Delta\lp \frac{2C_F}{b_0}
\ln \lp\frac{\aq}{\aqq}
\rp \rp \lc (1-x)^{-1+
\frac{2C_F}{b_0}
\ln \lp\frac{\aq}{\aqq}
\rp }\rc_+ \ .
\label{gammadis}
\ee
where $\Delta(\eta)=1/\Gamma(\eta)$,
which coincides with Eq. \ref{largex} once one takes into
account that for large x one has that $\ln (1/x)\sim (1-x)$. 
On top of the computation of the large-x limit of
$\Gamma(x)$, the above derivation shows that x-space
evolution factor can be defined as a distribution, just as
standard splitting functions.

We can compute also analytically the small $x$ behavior.
In the nonsinglet sector $\Gamma(x)$ grows at 
small $x$ as dictated by Double Asymptotic Scaling \cite{dasns}.
To see this, recall that $\Gamma(x)$ at low $x$ is given by
Eq. \ref{gammaX} expanding the LO anomalous dimension around
its rightmost singularity, in this case the $N=0$ pole, so that
one has
\be
\Gamma_{x\to 0}\lp 
x,\aq,\aqq\rp=\int_{-i\infty}^{+i\infty}\frac{dN}{2\pi i}
\exp\lp N\ln\frac{1}{x}+\frac{2C_F}{\beta_0}\ln\lp
\frac{\aqq}{\aq} \rp\lc \frac{1}{N}+\frac{1}{2}\rc\rp \ .
\ee
If in the expression above we perform a saddle point
integration, one obtains that the
leading small$-x$ behavior of the nonsinglet
x-space evolution kernel is given by
\be
\label{gammaxdas}
\Gamma_{x\to 0}\lp x,\aq,\aqq\rp=\mathcal{N}\exp\lp
2\sqrt{\frac{2C_F}{\beta_0}\ln\frac{1}{x}
\ln\lp
\frac{\aqq}{\aq} \rp}\rp \ ,
\ee
where $\mathcal{N}$ is a normalization factor.
The growing of $\Gamma(x)$
at low $x$ is more important for larger values of $Q^2$,
 as can be seen in Fig. \ref{evolfactxplots}.
 In the
singlet sector, the leading behavior of $\Gamma(x)$ at low
$x$ can also be computed exactly and is given again by 
Double Asymptotic Scaling \cite{das},
\be
\Gamma_{x\to 0}\lp x,\aq,\aqq \rp=\widetilde{\mathcal{N}}
\frac{1}{x}\exp\sqrt{\frac{2C_F}{\beta_0}
\ln\frac{\aqq}{\aq}\ln\frac{1}{x}} \ ,
\ee
which is much larger at small $x$ than the
corresponding non-singlet result, Eq. \ref{gammaxdas}. 
In the singlet case therefore, one would need to
substract the effects of the small-x growth of the evolution
factor, in a similar way that is done now with the large-x
growth in the interpolation of  $\Gamma(x)$, to
be discussed in the following.

\begin{figure}[t]
\begin{center}
\epsfig{width=0.60\textwidth,figure=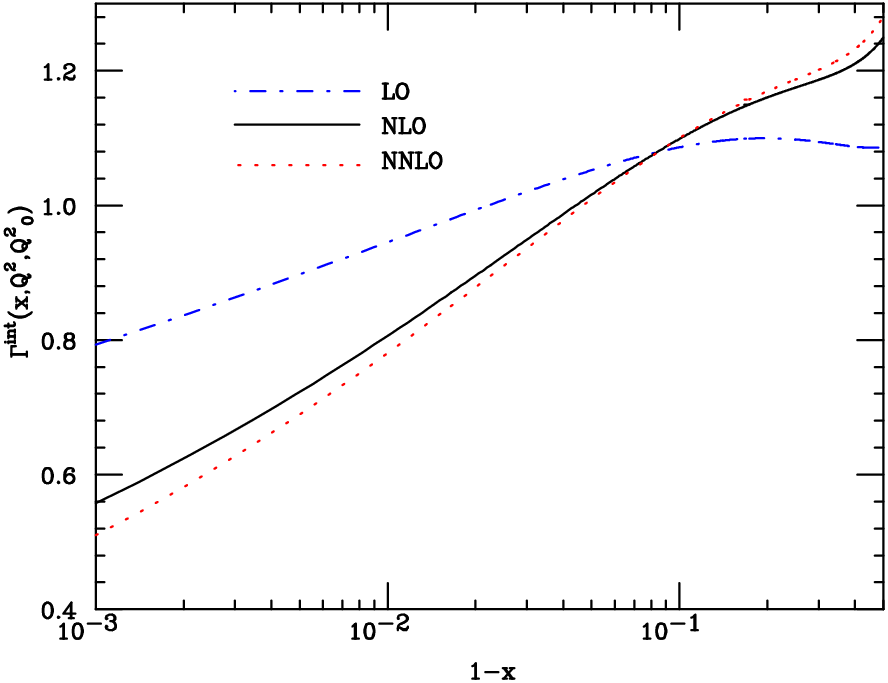}
\caption{}{\small The x-space evolution factor $\Gamma(x)$
once the leading large-x behaviour has been substracted,
Eq. \ref{gammaint}, for different perturbative orders
for $Q^2=10^4$ GeV$^2$. Note that the resulting function
is much more smooth, and therefore much more efficient to
interpolate.}
\label{evolfactxplots2}
\end{center}
\end{figure}

It is known that all splitting functions $P_{ij}(x,\aq)$, except the
non diagonal entries of the singlet matrix, diverge when $x=1$.
Therefore the nonsinglet evolution
factor $\Gamma(x)$, Eq. \ref{gammaX}, will likewise be divergent at $x=1$. 
In a similar way as what it is done
to regularize splitting functions, also evolution factors
$\Gamma(x)$ can be defined as distributions, as we have
seen explicitely in Eq. \ref{gammadis}. This divergent yet
integrable behavior poses also numerical problems that will be
 analyzed now.

Now let us consider the evolution of a generic
parton distribution,
\be
\label{evol1}
f(x,Q^2)
=\int_x^1 \frac{dy}{y}\Gamma(y)f\lp \frac{x}{y},Q_0^2\rp \ ,
\ee 
and now let us add and subtract a delta function to
the evolution factor $\Gamma(y)$,
\be
\Gamma(y)=\Gamma(y)+\gamma \delta(1-y)-\gamma \delta(1-y) \ ,\qquad 
\gamma\equiv \int_0^1\Gamma(z)dz \ .
\ee
Inserting this decomposition in Eq. \ref{evol1} one finds that
it can be written as
\be
\label{evol2}
f(x,Q^2)=\int_x^1 \frac{dy}{y}\Gamma(y)\lp f\lp \frac{x}{y},Q_0^2\rp
-y f\lp x,Q_0^2\rp \rp+ f\lp x,Q_0^2\rp\int_x^1 \Gamma(y) dy \ ,
\ee
so that now thanks to the subtraction that we have performed all the
integrals in Eq. \ref{evol2} converge and thus can be computed
numerically. Note that this is equivalent to the definition of
the x-space evolution factor in terms of the plus distribution prescription,
\be
f(x,Q^2)=\int_x^1 \frac{dy}{y}\Gamma(y)_+ f\lp \frac{x}{y},Q_0^2\rp
+ f\lp x,Q_0^2\rp\int_x^1 \Gamma(y) dy \ , 
\ee
where the plus distribution is defined in the standard way \cite{collider}.

Therefore, we will use to evolve our parton distributions Eq.
\ref{evol2}, which can be written as
\be
\label{evolfull}
q(x,Q^2)=q(x,Q_0^2)\gamma(x,\aq,\aqq)+\int_x^1 \frac{dy}{y}
\Gamma(y,\aq,\aqq)\lp q\lp\frac{x}{y},Q_0^2\rp- yq(x,Q_0^2)\rp  \ ,
\ee
where we have defined
\be
\gamma\lp x,\aq,\aqq \rp=\int_x^1 dy \Gamma(y,\aq,\aqq)\ ,
\qquad \gamma(0)\equiv
\gamma \ .
\ee
Note that $\gamma$ can be written as
\be
\gamma=\int_0^1 dx \int_{-i\infty}^{i\infty} \frac{dN}{2\pi i}
x^{-N}\Gamma(N)=\int_{-i\infty}^{i\infty} \frac{dN}{2\pi i}
\frac{\Gamma(N)}{1-N}=\Gamma(N=1) \ .
\ee
At leading order the anomalous dimension satisfies 
$\gamma^{(0)}(N=1)=0$ and therefore it follows that
$\gamma=1$. This result follows from momentum conservation.
At higher orders this result applies only to certain combinations
of nonsinglet parton distributions. 
For practical implementations we use the equality
\be
\gamma\lp x,\aq,\aqq \rp=\Gamma(N=1)-\int_0^x 
dy \Gamma(y,\aq,\aqq) \ ,
\ee
since in the nonsinglet sector the integral in the above equation is
very fast to compute. Note that the above property
does not hold in the singlet sector since
the gluon-gluon and gluon-quark anomalous
dimensions have a pole at $N=1$.

We have benchmarked our evolution formalism with the 
benchmark evolution tables first presented in Ref. \cite{lh}
and recently updated including the full NNLO anomalous
dimension in Ref. \cite{heralhc}. The results and the
accuracy of this benchmark can be seen in Table \ref{lh}, where
we use exactly the same parameters than in \cite{lh,heralhc} for
parton evolution in the Fixed Flavor Number (FFN) scheme.
More details about this benchmarking procedure of QCD
evolution codes can be found in Ref \cite{lh}.
 We have checked that the accuracy is always of the order $\mathcal{O}
\lp10^{-5}\rp$,
which is the required accuracy on 
modern QCD evolution codes. Similar checks have been performed
for the evolution of other parton distributions as well
as for evolution in the Variable Flavor Number (VFN) scheme.

\begin{table}[ht]
\begin{center}
\begin{tabular}{|c|c|cc|} 
\hline
x & $xu_v(x,Q^2_0)$ (LH) &  $xu_v(x,Q^2_0)$ (FT) & Rel. error\\
\hline
\multicolumn{4}{c}{Leading order}\\
\hline
$10^{-7}$   &   $5.7722~10^{-5}$    & $5.7722~10^{-5}$  & $3.3760~10^{-6}$  \\
$10^{-6}$   &   $3.3373~10^{-4}$    & $3.3373~10^{-4}$&$1.6880~10^{-6}$    \\
$10^{-5}$   &   $1.8724~10^{-3}$    & $1.8724~10^{-3}$&$1.9212~10^{-6}$ \\
$10^{-4}$   &   $1.0057~10^{-2}$    & $1.0057~10^{-2}$&$   
1.4095~10^{-6}$    \\
$10^{-3}$   &   $5.0392~10^{-2}$    & $5.0392~10^{-2} $&$  2.6145~10^{-6} $  
   \\
$10^{-2}$   &   $2.1955~10^{-1}$    &  $2.1955~10^{-1} $&$3.1065~10^{-6} $   
  \\
$0.1$       &   $5.7267~10^{-1}$    & $5.7267~10^{-1} $&$  6.4524~10^{-6}   $ 
 \\
$0.3$       &   $3.7925~10^{-1}$    & $3.7925~10^{-1}$&$   9.2674~10^{-6}    $
   \\
$0.5$       &   $1.3476~10^{-1}$    &$1.3476~10^{-1} $&$  1.1307~10^{-5}$\\
$0.7$       &   $2.3123~10^{-2}$    & $2.3122~10^{-2} $&$  2.1165~10^{-5}$    
 \\
$0.9$       &   $4.3443~10^{-4}$ &$4.3440~10^{-4} $&$6.3630~10^{-5}$       \\
\hline
\multicolumn{4}{c}{Next-to-Leading order}\\
\hline
$10^{-7}$   & $1.0616~10^{-4}$  &  $1.0616~10^{-4}$&$   2.1462~10^{-6}$    \\
$10^{-6}$   & $5.4177~10^{-4}$ &  $5.4177~10^{-4} $&$  8.7799~10^{-6}$      \\
$10^{-5}$   & $2.6870~10^{-3}$ &  $2.6870~10^{-3}$&$   9.7796~10^{-6}$      \\
$10^{-4}$   & $1.2841~10^{-2}$ &  $1.2841~10^{-2}$&$   1.3380~10^{-5}$      \\
$10^{-3}$   & $5.7926~10^{-2}$&  $5.7926~10^{-2}$&$   8.5063~10^{-6}$        \\
$10^{-2}$   & $2.3026~10^{-1}$&  $2.3026~10^{-1}$&$   3.0757~10^{-7}$       \\
$0.1$       & $5.5452~10^{-1}$&  $5.5452~10^{-1}$&$   7.6419~10^{-7}$     \\
$0.3$       & $3.5393~10^{-1}$&  $3.5393~10^{-1}$&$   2.6979~10^{-6}$        \\
$0.5$       & $1.2271~10^{-1}$&$1.2271~10^{-1}$&$   2.4466~10^{-5}$        \\
$0.7$       & $2.0429~10^{-2}$ &  $2.0429~10^{-2}$&$   1.4810~10^{-5}$       \\
$0.9$       & $3.6096~10^{-4}$ &  $3.6094~10^{-4}$&$   6.0762~10^{-5}$       \\
\hline
\multicolumn{4}{c}{Next-to-Next-to-Leading order}      \\
\hline
$10^{-7}$   & $1.5287~10^{-4}$&   $1.5287~10^{-4}$&$   1.5497~10^{-5}$    \\
$10^{-6}$   & $6.9176~10^{-4}$  & $6.9176~10^{-4}$&$   5.0711~10^{-6}$    \\
$10^{-5}$   & $3.0981~10^{-3}$  & $3.0981~10^{-3}$&$   9.5455~10^{-6}$    \\
$10^{-4}$   & $1.3722~10^{-2}$  & $1.3722~10^{-2}$&$   1.8022~10^{-5}$   \\
$10^{-3}$   & $5.9160~10^{-2}$ & $5.9160~10^{-2}$&$   5.0631~10^{-6}$      \\
$10^{-2}$   & $2.3078~10^{-1}$  & $2.3078~10^{-1}$&$   2.4853~10^{-6}$    \\
$0.1$       & $5.5177~10^{-1}$   & $5.5177~10^{-1}$&$   2.4747~10^{-6}$    \\
$0.3$       & $3.5071~10^{-1}$  & $3.5071~10^{-1}$&$   2.8430~10^{-7}$     \\
$0.5$       & $1.2117~10^{-1}$   & $1.2117~10^{-1}$&$   3.5893~10^{-5}$    \\
$0.7$       & $2.0077~10^{-2}$  & $2.0077~10^{-2}$&$   5.5823~10^{-6}$    \\
$0.9$       & $3.5111~10^{-4}$  & $3.5109~10^{-4}$&$   5.8172~10^{-5}$    \\
\hline
\end{tabular}
\end{center}
\caption{}{\small \label{lh} Benchmark tables for the evolution
formalism
described in this Section in the Fixed
Flavor Number Scheme with $N_f=4$ active flavors. 
The procedure for the
benchmarking is described in detail in 
Section of 1.3 of Ref. \cite{lh} and in Section 4.4 
of \cite{heralhc}. We use
the same notation as in the above references.}
\end{table}

The experimental observable that determines the nonsinglet parton
distribution is the nonsinglet structure function, defined
as the difference between structure functions in the
proton and in the deuteron,
\be
\label{f2pd}
F_2^{NS}(x,Q^2)\equiv F_2^p(x,Q^2)-F_2^d(x,Q^2) \ ,
\ee
which is related to
the nonsinglet parton distribution
 via a perturbative coefficient function,
\be
\label{coef}
\label{nsst}
F_2^{NS}(x,Q^2)=
x\int_x^1 \frac{dy}{y}C_{NS}(y,\aq)q_{NS}\lp
\frac{x}{y},Q^2\rp \ ,
\ee
where the coefficient function has the following expansion
up to NNLO in perturbation theory,
\be
C_{NS}(x,\aq)=\delta(1-x)+\frac{\aq}{4\pi}C_{NS}^{(1)}(x)+
\lp \frac{\aq}{4\pi}\rp^2 C_{NS}^{(2)}(x) \ .
\ee
The NLO coefficient $C^{(1)}(x)$ 
was computed in Ref. \cite{petronzio}, while the
 NNLO nonsinglet coefficient function was first computed
in Ref. \cite{neerven}.
However, for the NNLO  coefficient function
we do not use the exact result but rather the
 N-space parametrization of Ref. \cite{coefnnlo}, which is fast and
accurate enough for our purposes. 

The way that we incorporate the coefficient functions into
our evolution formalism is through a redefinition of the
x-space evolution kernel,
\be
\label{gammaX2}
\widetilde{\Gamma}(x,\aq,\aqq)=\int_{c-i\infty}^{c+i\infty}
\frac{dN}{2\pi i}x^{-N}C\lp N,\aq\rp\Gamma(N,\aq,\aqq) \ ,
\ee
so that now the nonsinglet structure function can be written 
in terms of the parton distribution at the initial evolution scale as
\be
\label{f2qns}
F_2^{NS}(x,Q^2)=x\int_x^1 \frac{dy}{y}
\widetilde{\Gamma}(y,\aq,\aqq)q\lp \frac{x}{y},Q_0^2\rp \ .
\ee
The rationale behind this procedure is to improve the 
speed of the evolution code, that is, if coefficient
functions where introduced as in Eq. \ref{coef}, we would need
to perform an additional convolution integral each time a 
structure function was computed. The only drawback of this
method
is that the evolution factor becomes process-dependent, while
the bare evolution factor $\Gamma(x)$, Eq. \ref{gammaX},
is process independent and indeed it could be used by itself as an
alternative procedure for evolution of standard parton dstributions.

We use a Variable Flavor Number (VFN) scheme with zero mass
partons to incorporate the effects of heavy quark thresholds in the
evolution.
At NNLO one has to take into account that both the strong coupling and the
parton distributions are discontinuous when crossing the
heavy quark thresholds. 
We compute $\aq$ in the Variable Flavor Number scheme, taking
into account the discontinuity at NNLO at heavy quark 
thresholds,
\be
\alpha_{s,f+1}(m_{f}^2)=\alpha_{s,f}(m_{f}^2)+\lp \frac{C_2}{4\pi} \rp^2
\alpha_{s,f}(m_{f}^2)^3 \ ,
\ee
where $\alpha_{s,f}$ is the strong coupling in the effective theory
with $N_f$ active light quarl flavors, $m_f^2$ is the position
of the heavy quarl threshold, and
 $C_2=14/3$ was computed in \cite{coefalphannlo}. 
For the parton distribution at heavy quark
thresholds, the corresponding N-space matching condition is
given by
\be
q_{NS}^{(n_f+1)}(N,m_h^2)=q_{NS}^{(n_f)}(N,m_h^2)\lp1+\lp 
\frac{\alpha_{s,f}(m_h^2)}{4\pi}\rp^2
A_{qq}^{NS,(2)}(N))\rp \ ,
\ee
where the  NNLO matching coefficients are determined
in Ref. \cite{pdfnnlo}. A more refined treatment
of heavy quark mass effects \cite{acot,rt} is postponed
to the case of singlet evolution, since it is known
that the influcence of heavy quark mass effects in the
nonsinglet sector is rather small.

The evolution approach described above
is very accurate but also very CPU time consuming. This is
so since one has to compute both the N-space evolution
factor and its Mellin inverse each time one wants
to evolve a parton distribution. This is 
specially a problem in our approach, where we are parametrizing
our parton distributions with neural networks,
with a very large parameter space and thus
the minimization routine requires more time than 
in the standard approach.
What we do then is to interpolate the evolution kernel $\Gamma(x)$
so that the  the evolution of
parton distributions, Eq. \ref{evolfull}, is much faster.

The only problem is since $\Gamma(x)$ grows heavily at large $x$,
it is numerically difficult to interpolate it. A way to overcome
this difficulty is
to substract to the exact result for $\Gamma(x)$
the large-x behavior Eq. \ref{largex} so that
the resulting function to be interpolated is a smooth one.
We interpolate  the subtracted x-space evolution
kernel, defined as
\be
\label{gammaint}
\widetilde{\Gamma}^{(\mathrm{int})}(x,\aq,\aqq)=
\frac{\widetilde{\Gamma}(x,\aq,\aqq)}{
\Gamma_{x\to 1}(x,\aq,\aqq)} \ ,
\ee
where $\Gamma_{x\to 1}(x)$ is given by Eq. \ref{largex} and the
inclusion of the coefficient function does not affect the 
leading large $x$ behavior.
We also interpolate $\gamma(x)$ in Eq. \ref{evolfull}. 
In Fig \ref{evolfactxplots2} we represent the behaviour
of the substracted evolution factor 
$\tilde{\Gamma}^{(\mathrm{int})}(x)$, it is clear
that this functional dependence is much more
efficient to interpolate than that of the
bare evolution factor, represented in Fig. \ref{evolfactxplots2}.
Therefore, the nonsinglet structure function will be
given in terms of an interpolated evolution factor as
\be
F_2^{NS}(x,Q^2)=x\int_x^1 \frac{dy}{y}
\widetilde{\Gamma}^{(\mathrm{int})}(x,\aq,\aqq)
\Gamma_{x\to 1}(x,\aq,\aqq)
q\lp \frac{x}{y},Q_0^2\rp \ ,
\ee
instead of with Eq. \ref{f2qns}.

We use
Chebishev interpolation in $x$ for all the values of $Q^2$ where
there is experimental data.
The rationale for using Chebishev polynomials for the interpolation is that
the Chebishev approximation for a given function is very close
to the minimax polynomial, defined as the approximating polynomial
with the smallest maximum deviation from the exact function. On top
of this property, Chebishev interpolation is extremely stable, and the
increase in accuracy as we increase the order of the interpolation can
be controlled in a stringent way.
We require an accuracy in the interpolation $\mathcal{O}\lp 10^{-5}\rp$
for all the $(x,Q^2)$ range covered by experimental data. This accuracy
is enough for our present purposes since this is the accuracy to which
the exact evolution formalism has been benchmarked.
Thanks to Eq. \ref{gammaint} this accuracy is achieved with a relatively
small number of Chebyshev polynomials, and therefore
the interpolated evolution kernel is rather fast to be
evaluated for an arbitrary value of $x$, and for those
values of $Q^2$ where there is experimental data.

In order to incorporate data in our fit with small $Q^2$, we must
take into account the target mass corrections to the nonsinglet
structure function \cite{tmc}.
These target mass corrections are incorporated into our analysis
using the following standard relations,
\be
F_2^{NS,LT,TMC}(x,Q^2)=
\frac{x^2}{\tau^{3/2}}\frac{F_2^{NS,LT}(\xi_{TMC},Q^2)}{\xi^2_{TMC}}
+6\frac{M^2}{Q^2}\frac{x^3}{\tau^2}I_2(\xi_{TMC},Q^2) \ ,
\label{tmc}
\ee
where we have defined
\be
\label{tmc2}
I_2(\xi_{TMC},Q^2)=\int_{\xi_{TMC}}^1
dz\frac{F_2^{NS,LT}(z,Q^2)}{z^2} \ ,
\ee
\be
\xi_{TMC}=\frac{2x}{1+\sqrt{\tau}},\qquad \tau=1+\frac{4M^2x^2}{Q^2} \ ,
\ee
where $M$ is the mass of an isoscalar nucleus.
This way one separates kinematical target mass corrections
from genuine dynamical target mass corrections.

In summary, in this part of the thesis we have described an alternative
parton evolution formalism which combines the advantages
of both x-space and N-space evolution codes.
This formalism has been described for the nonsinglet sector,
and its extension to the singlet sector
will be discussed in Ref. \cite{nnsinglet}.
In the next Section we will use this evolution formalism
to construct a neural network parametrization of the
nonsinglet parton distribution.

\clearpage

\subsection{The non-singlet parton distribution}
\label{nnqns_appl}

\begin{figure}[ht]
\begin{center}
\epsfig{width=0.84\textwidth,figure=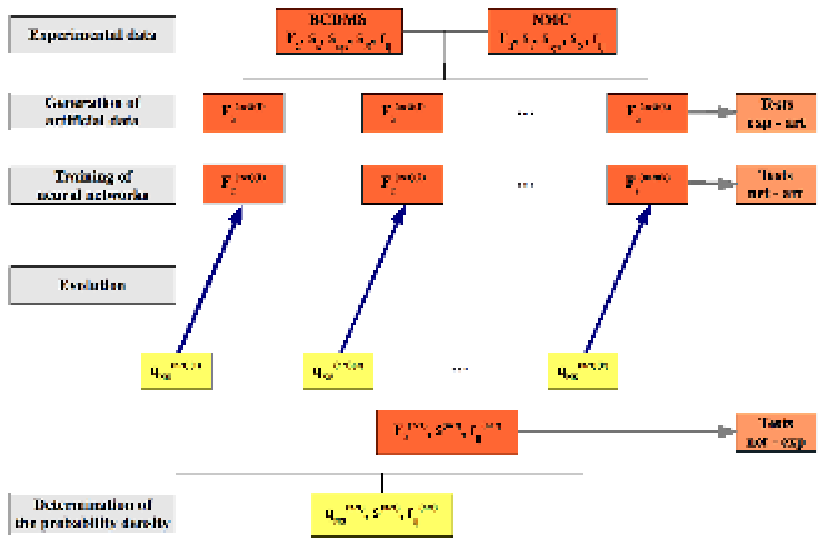}
\caption{}{\small General strategy for the
parametrization of the nonsinglet parton
distribution $q_{NS}(x,Q^2_0)$ from experimental data on the
nonsinglet structure function $F_2^{NS}(x,Q^2)$.}
\label{genstrat}
\end{center}
\end{figure}

Once the formalism that will be used for the
evolution of the nonsinglet parton distribution has
been introduced in the previous Section, we now turn
to the first application of the general technique
described in  Chapter \ref{general} to the parametrization
of parton distributions functions. We will consider  the parametrization
of the nonsinglet parton distribution $q_{NS}(x,Q^2_0)$ from
experimental data on the nonsinglet structure function, 
$F_2^{NS}(x,Q^2)$, defined in Eq. \ref{nsst}.
The restriction to the nonsinglet sector makes the problem
technically simpler, since nonsinglet evolution
requires a single parton distribution
parametrized with a neural network. Hence,
nonsinglet evolution does not involve the complications from
the training of several neural networks.
In the general case of  singlet evolution,
one requires the simultaneous minimization
of additional neural networks which parametrize different independent
combinations of quark parton distributions as well as the 
 gluon distribution, and this issue  will be
considered in Ref. \cite{nnsinglet}.
In Fig. \ref{genstrat} we show a diagram which
summarizes our approach to
parametrize the nonsinglet parton distribution 
$q_{NS}(x,Q_0^2)$. Note that the main difference with respect
Ref. \cite{f2ns} is that the effects of DGLAP parton evolution
reduce the dimensionality of the quantity to be parametrized
from $d=2$ for $F_2^{NS}(x,Q^2)$ to $d=1$ for the
parton distribution $q_{NS}(x)$.

\begin{figure}[ht]
\begin{center}
\epsfig{width=0.6\textwidth,figure=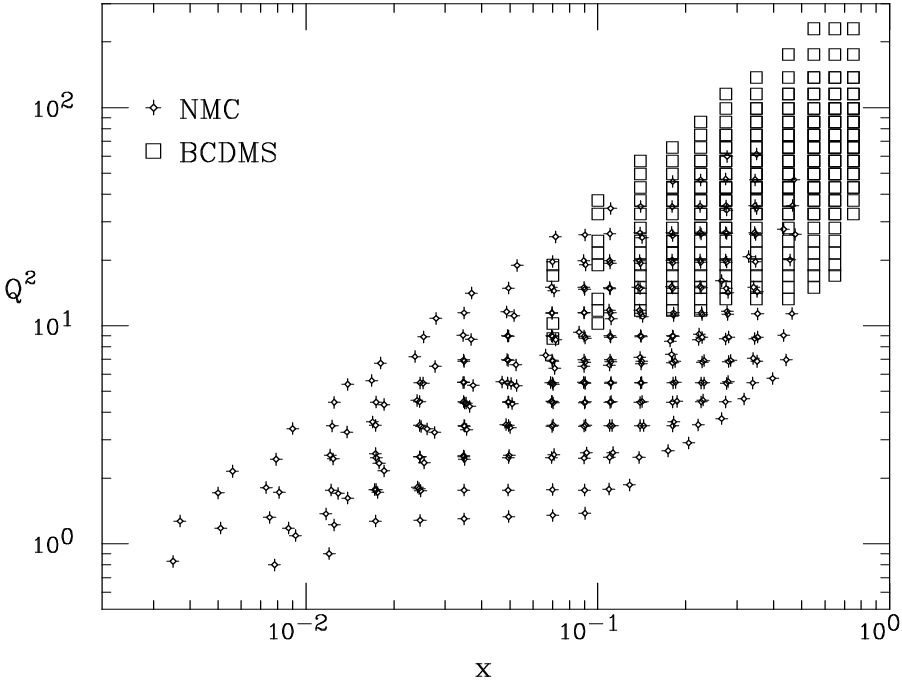}
\caption{}{\small Kinematical coverage of the
available experimental data on the $F_2^{NS}(x,Q^2)$
structure function in the $\lp x,Q^2\rp$ plane. Note that
fixed target scattering geometry implies that large-$Q^2$
data is at large-$x$ and conversely.}
\label{kincov}
\end{center}
\end{figure}

For the parametrization of the
nonsinglet parton distribution $q_{NS}(x,Q^2_0)$,  the same data on the
nonsinglet structure function $F_2^{NS}(x,Q^2)$
that was used in Ref. \cite{f2ns}
from the NMC \cite{nmc} and the BCDMS
\cite{bcdms1,bcdms2}
collaborations will be used in the present analysis.
 The main features of this experimental data
from   was 
discussed in Ref. \cite{f2ns}. While BCDMS covers the
large-x, large-$Q^2$ kinematical region, NMC covers
the complementary small-$x$, small $Q^2$ region, with
some overlap in the intermediate regions. BCDMS data is rather precise,
while the small-$x$ NMC data has larger uncertainties. 
The kinematical coverage of the $F_2^{NS}$ experimental data available
from these two experiments can be seen in Fig. \ref{kincov}.
The requirement of a simultanous measurement of the proton
$F_2^p$ and deuteron $F_2^d$ structure functions for the
determination of the nonsinglet combination, Eq. \ref{f2pd},
restricts sizeably this kinematical coverage as
compared with the one for $F_2^p$ only, see Fig. \ref{kinall}.
However, there are proposals \cite{erhic,lhec} 
for future deep-inelastic
scattering facilities which should extend this
kinematical range in the small-x region and in addition reduce the
associated statistical and systematic
uncertainties. The inclusion of additional high
statistics data from JLAB
\cite{jlab1,jlab2} at the largest values of $x$
would require the use of resummed parton
evolution.

The main difference with respect the dataset
used in Ref. \cite{f2ns} is that now
we have applied to this dataset the kinematical
cuts $Q^2\ge 3$ GeV$^2$ and $W^2\ge 6.25$ GeV$^2$.
The motivation for the two kinematical
cuts is to remove those data points for which the
application of perturbation theory is questionable, as 
has been discussed in Section \ref{dis_theo}.
The cut in $Q^2$ is required to remove experimental data for
which higher twist corrections might be sizable, and the
cut in $W^2$ removes those data points at very large x
for which the application of unresummed perturbation
theory is not reliable. After these kinematical cuts, we are
left with a total of $N_{\dat}=483$ data points.

In Table \ref{datafeat}  the features of the experimental data
that is included in the present analysis are summarized. Note that the
average error is substantially larger for the NMC experiment than for
the more precise BCDMS measurements.
All the experiments included in our analysis provide full correlated
systematics, as well as normalization errors. 
The description of the different uncertainties of the
experimental data that will be used
can be found in Ref. \cite{f2ns}.

\begin{table}[ht]
\begin{center}
\begin{tabular}{|cc|cc|c|c|cc|} 
\hline
 Experiment & Ref. & $x$ range & $Q^2$ range (GeV$^2$) & $N_{\dat}$   
& $\la\sigma_{\tot}\ra$ 
&$\la\rho\ra$& $\la\mathrm{cov}\ra$ 
 \\
\hline
NMC & \cite{nmc} &$9.0~10^{-3}~-~4.7~10^{-1}$ & $3.2-61$ & 
 229   & 102.0  & 0.034  &  0.200
\\
\hline  
BCDMS & \cite{bcdms1,bcdms2} & $7.0~10^{-2}~-~7.5~10^{-1}$ & $8.7-230$ & 
 254 & 24.6 &  0.163 &  0.007\\
\hline
\end{tabular}
\caption{\small Experiments included in this analysis. Note that the
 values of
  $\sigma$ and cov are given as percentages. The data from the
two experiments partially overlaps in the medium-$x$, medium-$Q^2$
region.}
\label{datafeat}
\end{center}
\end{table}

\begin{table}[ht]
\begin{center}
\begin{tabular}{|c|ccc|} 
\multicolumn{4}{c}{
$F_2^{NS}(x,Q^2)$}\\   
\hline
 $N_{\rep}$ & 10 & 100 & 1000 \\
\hline
$\la PE\lc\la F^{(\art)}\ra_{\rep}\rc\ra$ & 
20\% & 6.4\% & 1.3\%    \\
$r\lc F^{(\art)} \rc$ & 0.97 &  0.997& 0.999 \\
\hline
$\la V\lc \sigma^{(\art)}\rc\ra_{\dat}$ & $6.1~10^{-5}$ & $1.9~10^{-5}$
 & $6.7~10^{-6}$  \\
$\la PE\lc \sigma^{(\art)}\rc\ra_{\dat}$ & 33\%
& 11\% &  3\%   \\
$\la \sigma^{(\art)}\ra_{\dat}$ & 0.011 & 0.011 & 0.011 \\
$r\lc \sigma^{(\art)}\rc$ & 0.94 & 0.994  & 0.999 \\
\hline
$\la V\lc \rho^{(\art)}\rc\ra_{\dat}$ & $0.10$
&  $9.4~10^{-3}$ &  $1.0~10^{-3}$ \\
$\la \rho^{(\art)}\ra_{\dat}$ & 0.182 & 0.097 & 0.100\\
$r\lc \rho^{(\art)}\rc$ &  0.47&  0.79 & 0.97  \\
\hline
$\la V\lc \mathrm{cov}^{(\art)}\rc\ra_{\dat}$ & $ 5.5~10^{-9}$ 
&  $1.7~10^{-10}$  &  $5.7~10^{-11}$ \\
$\la \mathrm{cov}^{(\art)}\ra_{\dat}$ & 
$1.3~10^{-5}$ & $7.6~10^{-6}$ & $8.1~10^{-6}$ \\
$r\lc \mathrm{cov}^{(\art)}\rc$ & 0.41 &  0.81 & 0.975\\
\hline
\end{tabular}
\caption{\small Comparison between experimental and 
Monte Carlo data.\hfill\break
The experimental data have
$\la \sigma^{(\mrexp)}\ra_{\dat}=0.011$, $\la \rho^{(\mrexp)}\ra_{\dat}=
0.107$ and $\la \mathrm{cov}^{(\mrexp)}\ra_{\dat}=8.63~10^{-6}$. }
\label{gendata_pdf}
\end{center}
\end{table}

The first step of our strategy, as discussed in
Chapter \ref{general}, is to construct a Monte Carlo
sample of replicas of the experimental data.
Note that as in the case of the parametrization of the
lepton energy spectra discussed in Section \ref{bdecay_appl},
 the quantity for which the sample of replicas is generated,
$F_2^{NS}(x,Q^2)$, 
does not coincide with the quantity that is parametrized with
neural networks, the parton distribution $q_{NS}(x,Q_0^2)$. 
Since in this case the experimental data is the same as in
Ref. \cite{f2ns}, we use the same relations to generate the
Monte Carlo
sample, namely Eq. \ref{gen1}, which for the present situation reads,
\be
F_i^{NS(\art)(k)}=\lp 1 +r_N^{(k)}\sigma_N\rp\lc F_i^{NS(\exp)}
+ r_{t,i}^{(k)}\sigma_{t,i}
+\sum_{j=1}^{N_{\sys}} r^{(k)}_{\sys,j}\sigma_{\sys,ji}\rc,
\quad i=1,\ldots,N_{\dat} \ , 
\quad k=1,\ldots,N_{\rep} \ .
\ee
In Table \ref{gendata_pdf} the statistical estimators
from the sample of generated replicas are presented, 
for the data set that is used
in the fit, where the statistical estimators have been defined
in Section \ref{mcerr}. 
This table shows
that a sample of  1000 replicas is sufficient to ensure  average 
scatter correlations of 99\% and accuracies
of a few percent on structure functions, errors and correlations. 

Once the Monte Carlo sample of replicas of the
structure function $F_2^{NS}(x,Q^2)$
has been generated,
the second step is to train a neural network
on each replica of the experimental data.
However the situation is now more complicated than in the
structure function case \cite{f2ns}. Now, while the Monte Carlo sample
is constructed from the experimental data on the nonsinglet structure
function, the neural networks  parametrize the
nonsinglet parton distribution.
Therefore, following Eqns. \ref{evolfull} and \ref{nsst}, the
k-th neural network which parametrizes a parton distribution 
at the starting evolution scale $Q_0^2$ is related
to the k-th replica of the experimental data on the
nonsinglet structure function at $(x,Q^2)$ as
\be
\label{nsnet}
F_2^{NS(\net)(k)}\lp x,Q^2\rp
=x\int_x^1 \frac{dy}{y}\widetilde{\Gamma}
\lp x,\aq,\aqq \rp q_{NS}^{(\net)(k)}\lp \frac{x}{y},Q_0^2\rp \ ,\qquad 
k=1,\ldots,N_{\rep} \ ,
\ee 
where the evolution factor $\widetilde{\Gamma}$ which includes
the effect of the perturbative coefficient function $C_{NS}\lp x,\aq\rp$
has been defined in Eq. \ref{gammaX2}.
  
In the present analysis the value of the
strong coupling $\aq$ will be kept fixed
at the current world average
 \cite{pdg}  given by
\be
\label{aq} 
\alpha_s(M_Z^2)=0.118 \pm 0.002 \ ,
\ee
and our fits will be repeated for the extreme values
of the strong coupling allowed by the errors in 
Eq. \ref{aq}, to estimate the theoretical uncertainties
associated to $q_{NS}(x,Q_0^2)$ from the 
finite precision with which the strong coupling $\aq$
is determined from data. 
The determination of $\aqz$ altogether with the nonsinglet parton
distribution from experimental data is possible in principle, but it has
been decided to have $\aqz$ as an external input since first the
opposite complicates considerably the practical implementation of the
evolution formalism discussed in Section \ref{evolform}, and second,
the strong coupling in much better determined from other
processes involving the strong interaction \cite{bethke}.
The interplay between the strong coupling and global
parton distributions fits has been recently discussed in Ref. \cite{ascteq}.

As been been discussed in detail in Section \ref{minimstratt},
there exist several techniques to implement the training of
neural networks.
In the present case the optimal training strategy has been found to be
a single training epoch, in which the covariance matrix error,
$E_3^{(k)}$, Eq. \ref{er3}, is minimized for each replica. 
In the present case one has
\be
\label{er3pdf}
E_3^{(k)}=\frac{1}{N_{\dat}-N_{\mathrm{par}}}
\sum_{i,j=1}^{N_{\dat}} 
\lp F^{NS(\net)(k)}_i-F^{NS(\art)(k)}_i\rp 
\lp\lp\overline{\mathrm{cov}}^{(k)}\rp^{-1}\rp_{ij}\lp F^{NS(\net)(k)}_j-
F^{NS(\art)(k)}_j\rp \ ,
\ee
with the covariance matrix defined in Eq. \ref{covmatnn}.
Note that the error function is normalized to the
number of degrees of freedom \cite{cowan}.
The minimization algorithm used is
genetic algorithms with dynamical stopping of the training.
Weighted training will also be used in order to guarantee that
at the end of the training the total $\chi^2$, Eq. \ref{chi2tot}, which
now reads
\be
\label{chi2totpdf}
\chi^2=\frac{1}{N_{\dat}-N_{\mathrm{par}}}
\sum_{i,j=1}^{N_{\dat}} 
\lp \la F^{NS(\net)}_i\ra_{\rep}-F^{NS(\exp)}_i\rp 
\lp{\mathrm{cov}}^{-1}\rp_{ij}\lp \la F^{NS(\net)}_j\ra_{\rep}-
F^{NS(\exp)}_j\rp \ ,
\ee
as computed for the two experiments has a similar value.

The architecture of the neural network that parametrizes
$q_{NS}(x,Q_0^2)$ must be determined as
discussed in Section \ref{minimstratt}. 
This optimal architecture is obtained by the requirements that
it must be complex enough to reproduce the experimental data
patterns and 
that the results are independent of the precise number
of neurons. In particular it has been checked that the results are
stable for architectures with one more or one less neuron than the
reference architecture, 2-5-3-1. This ensures that the network
architecture is redundant for the problem under consideration.

To define the optimal training strategy we should determine
which is the suitable value of the $\chi^2_{\mathrm{stop}}$
parameter that defines the dynamical stopping of the
training, as discussed in Section \ref{minimstratt}.
We will use the overlearning criterion, introduced in the same
Section, to determine its value.
The overlearning criterion to determine the length of the training
states that the training should be stopped when the
neural network begins to overlearn, that is, it begins to follow
the statistical fluctuations of the experimental data rather
than the underlying physical law. The onset of overlearning can
be determined by separating the data set into two
disjoint sets, called the {\it training} set and
the {\it validation} set. One then minimizes the
error function, Eq. \ref{er3}, computed only with the
data points of the training set, and analyzes the dependence
of the error function Eq. \ref{er3} of the validation set
as a function of the number of generations.

Then one computes the total $\chi^2$,
for both the training and validation subsets. It turns
out that in the present case fluctuations in the data set
turn out to be very large, and one has to average over
a large enough number of partitions to obtain stable
results. The onset of overlearning is determined as
the length of the training such that the $\chi^2$ of the
validation set saturates or even rises while the $\chi^2$
of the training set is still decreasing.

\begin{figure}[ht]
\begin{center}
\epsfig{width=0.78\textwidth,figure=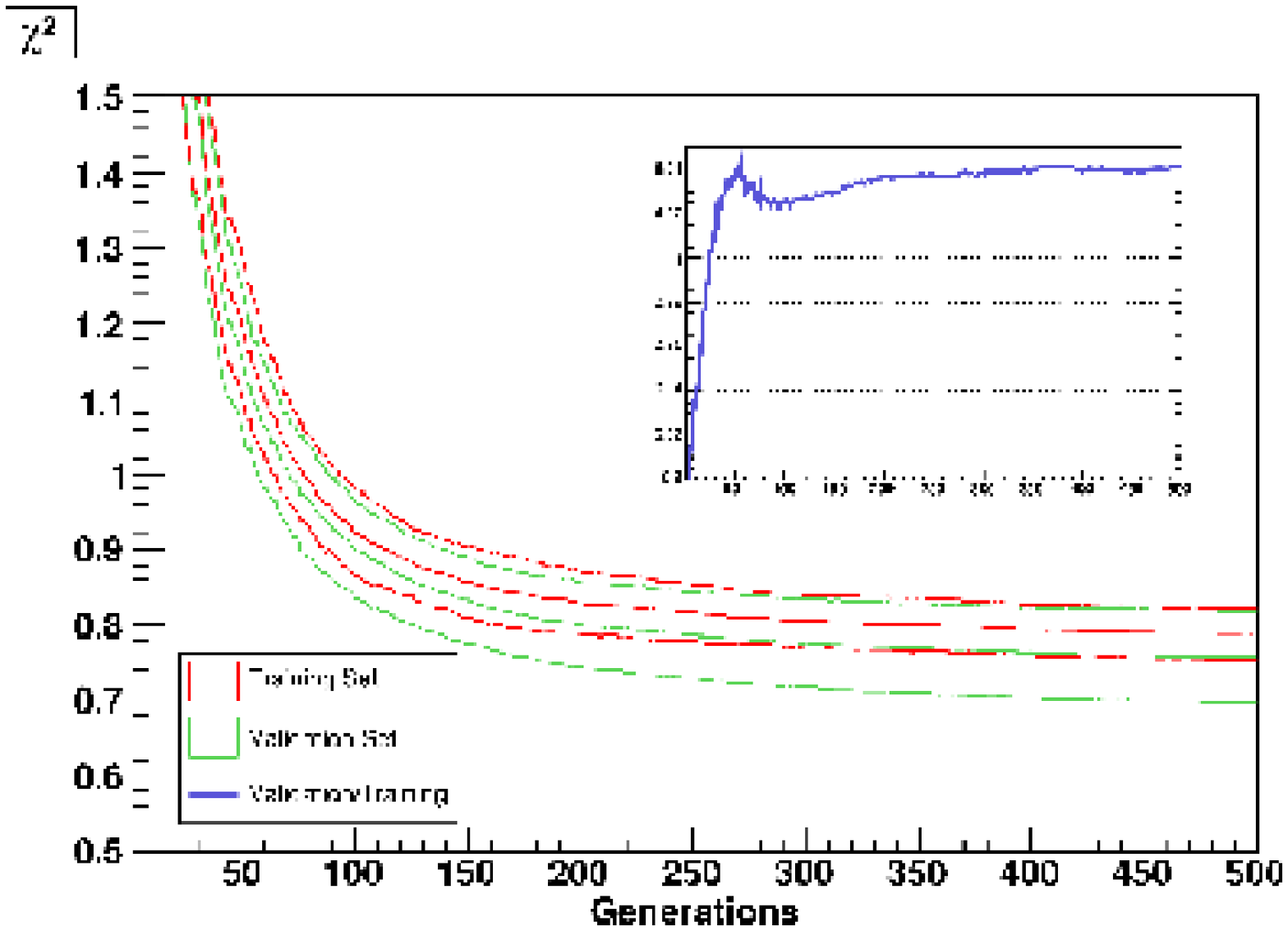}
\caption{}{\small The average $\la \chi^2\ra_{\mathrm{part}}$ over partitions
as a function of the number of 
genetic algorithms generations. The bands
show also the associated variance $\sigma_{\chi^2}$. The up-right
plot shows the ratio $\chi^2_{\val}/\chi^2_{\tr}$.
\label{chi2_overlearn}}
\end{center}
\end{figure}

\begin{figure}[ht]
\begin{center}
\epsfig{width=0.48\textwidth,figure=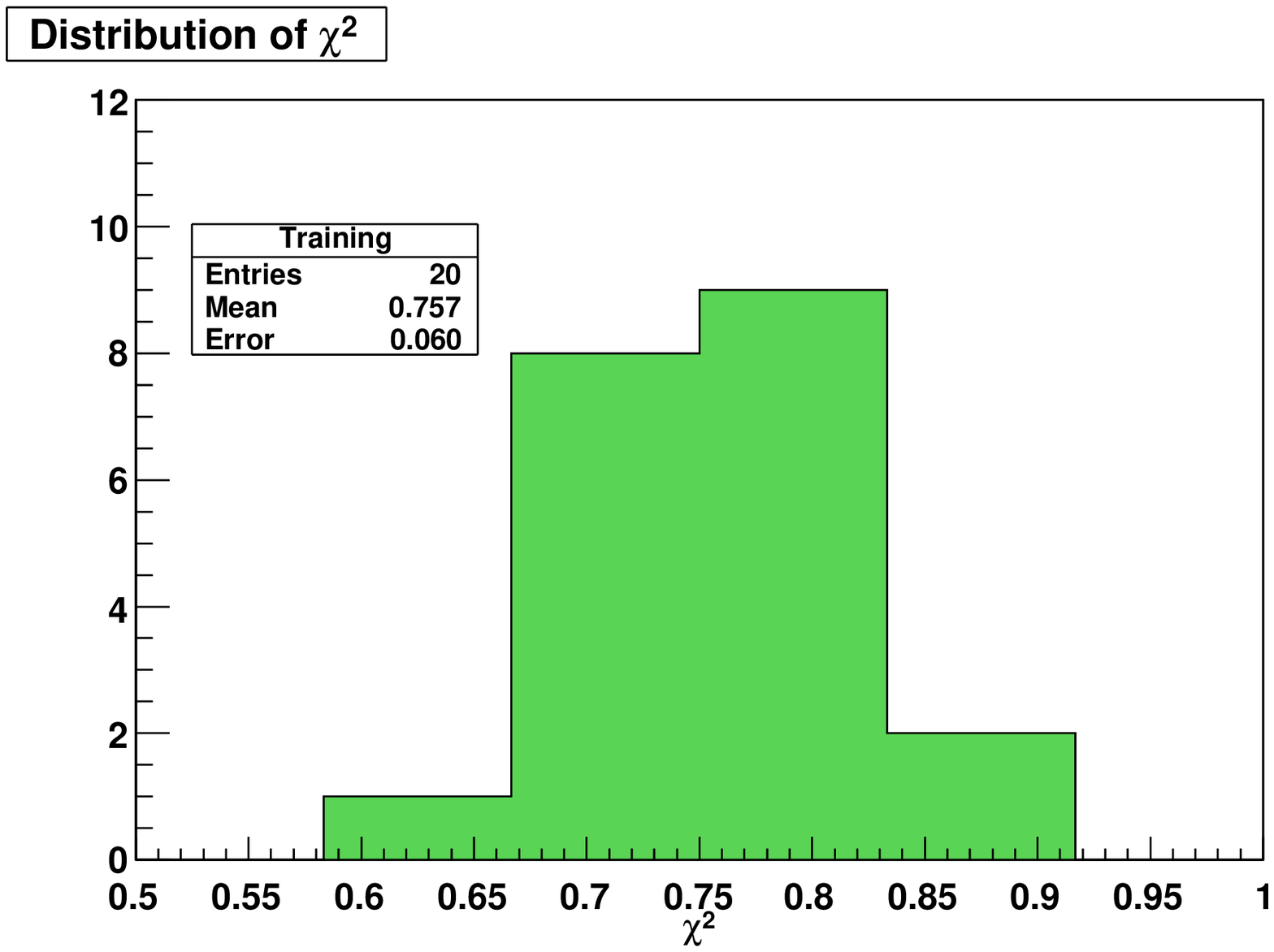}
\epsfig{width=0.48\textwidth,figure=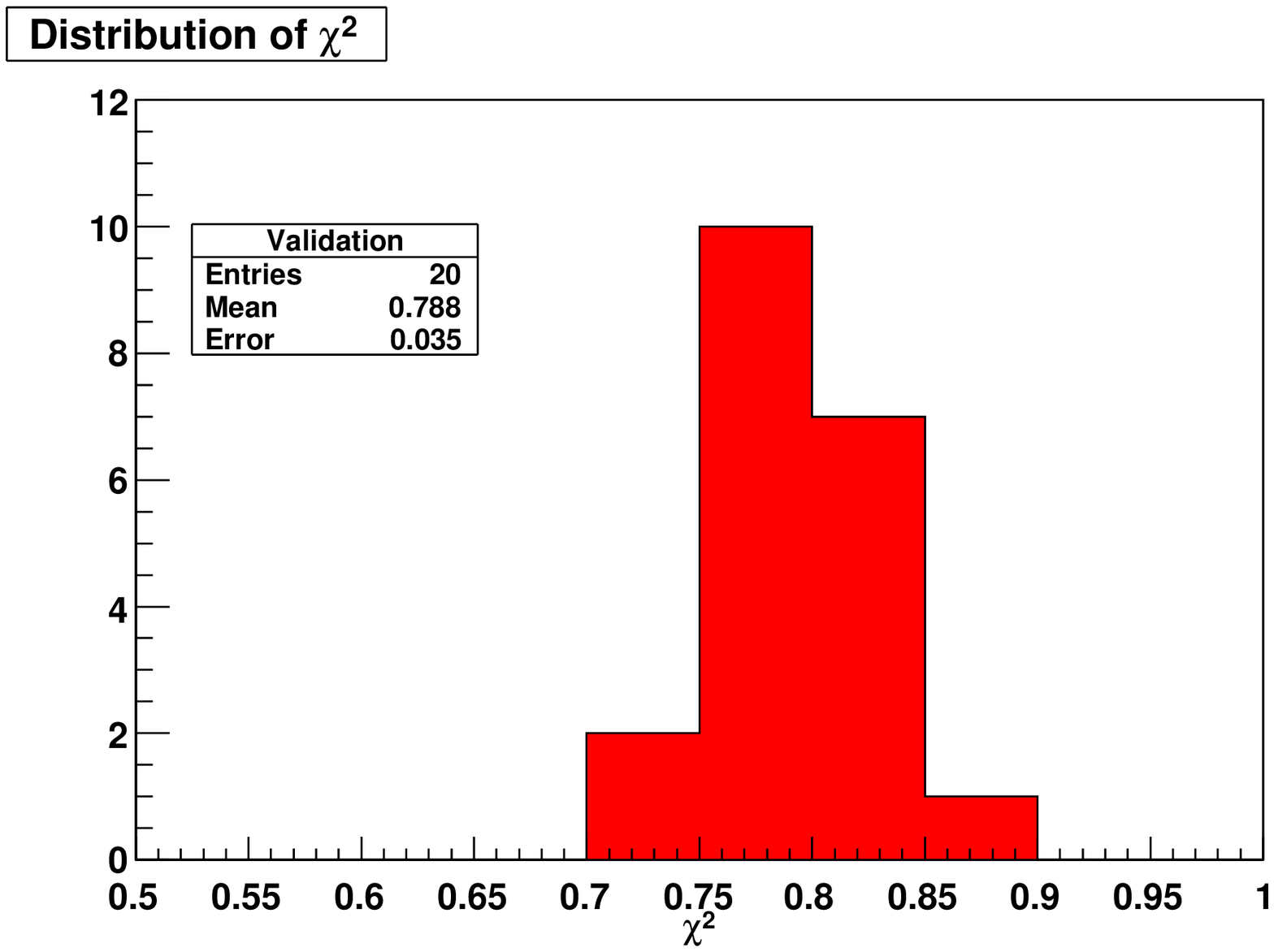}
\caption{}{\small \label{histtr} Distribution of $\chi^2_l$ for 
the different partitions
used in the overlearning test, for both the training (left)
and the validation (right) subset of points.}
\end{center}
\end{figure}

\begin{figure}[ht]
\begin{center}
\epsfig{width=0.6\textwidth,figure=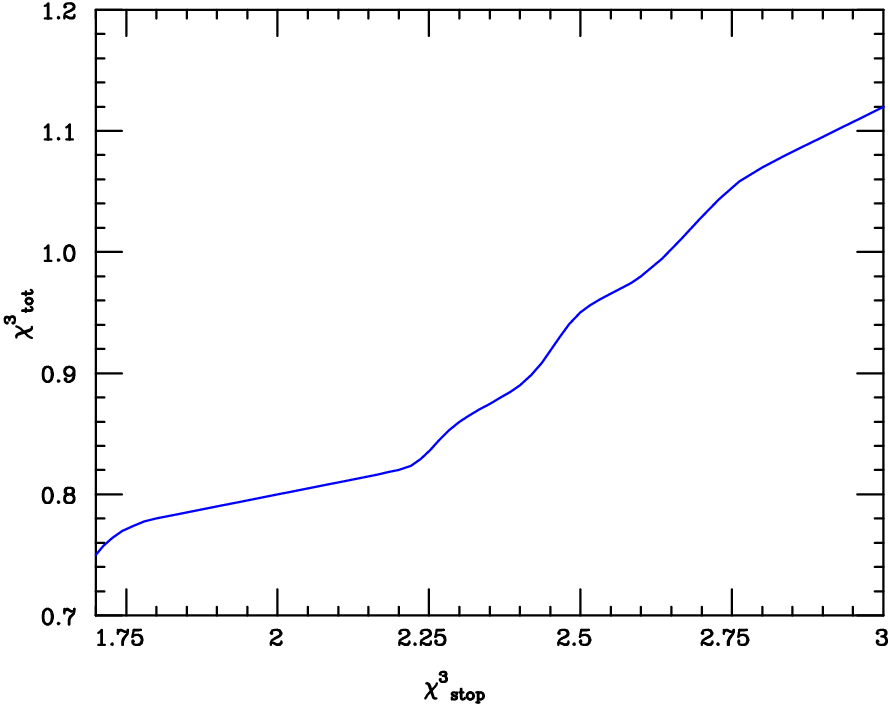}
\caption{}{\small \label{chi2stop} The total $\chi^2$, as computed in
Eq. \ref{chi2tot} from averages over replicas, as a function
of the $\chi^2_{\stopp}$ used in the dynamical stopping
of the training.}
\end{center}
\end{figure}

In Fig \ref{chi2_overlearn} we show the $\chi^2$ 
for both the training and validation subsets as a function
of genetic algorithms generations, averaged over 
a large enough number of partitions. 
The training partition contains the 30\% of all the data points,
selected at random, while the validation partition includes
the complementary 70\% data points.
If $N_{\mathrm{part}}$ is the number of different partitions
used to determine the onset of overlearning, in Fig. 
\ref{chi2_overlearn} we show both the average value
of the total $\la \chi^2\ra_{\mathrm{part}}$ 
and its variance $\sigma_{\chi^2}$,
defined as
\be
\la \chi^2\ra_{\mathrm{part}}=\frac{1}{N_{\partt}}\sum_{l=1}^{N_{\partt}}
\chi^2_l \ ,
\ee
\be
\label{stchi2part}
\sigma_{\chi^2}^2=\frac{1}{N_{\partt}}\sum_{l=1}^{N_{\partt}}
\lp \chi^2_l\rp^2-\lp \la \chi^2\ra_{\mathrm{part}}\rp^2 \ ,
\ee
where $\chi^2_l$ is the value of the error
function for the l-th partition.
The number of required partitions $N_{\mathrm{part}}$ has
to be large enough so that the resulting distribution
is gaussian and therefore the standard deviation, 
Eq. \ref{stchi2part}, has the standard statistical interpretation.
In Fig. \ref{histtr} we show the
histograms for the distributions of $\chi^2_{\tr,l}$ and 
$\chi^2_{\val,l}$ over partitions. One observes that in the
present case  $N_{\mathrm{part}}=20$ is enough to achieve convergence.
In Fig. \ref{chi2stop} we show the relation between 
the total $\chi^2$ and 
 the $\chi^2_{\stopp}$ used in the dynamical stopping
of the training. This relation is needed in the determination
of the appropiate value of $\chi^2_{\mathrm{stop}}$ in the dynamical
stopping of the training to achieve a given value
of the total $\chi^2$.
The overlearning anaysis points out to the fact that the final
total $\chi^2$ should be around  $\chi^2 \sim 0.8$, which
from Fig. \ref{chi2stop} implies a value $\chi^2_{\mathrm{stop}}\sim
2.0$ for the dynamical stopping of the training.

With the stability estimators defined in Section \ref{stabest}, one
can assess which is the required number of trained replicas $N_{\rep}$
to obtain stability of the results. To this purpose,
one compares in Table \ref{stabtablenrep} the results
for the probability measure of $q_{NS}(x,Q_0^2)$ for different
values of $N_{\rep}$. That is, one
compares the probability measure as
obtained from training a sample of
neural networks on $N_{\rep}$ Monte
Carlo replicas of the experimental data with
another probability measure, constructed from
a different set of $N_{\rep}$ Monte
Carlo replicas. One expects that the differences
in this comparison are reduced as the value of 
$N_{\rep}$ is increased.
For these stability comparisons, we use
in the data regions $\widetilde{N}_{\dat}=14$ points linearly
spaced between $x=0.04$ and $x=0.75$, and the same number
of points in the extrapolation region logarithmically spaced
between $x=0.001$ and $x=0.01$.
It can be observed that the required stability is obtained for
$N_{\rep}=500$, as expected.

\begin{table}[ht]
\begin{center}
\begin{tabular}{|c|c|c|c|}
\hline 
$N_{\rep}$&  $\quad$ 10 $\quad$  &  
 $\quad$ 100 $\quad$ &   $\quad$ 500 $\quad$ \\
\hline
$\la RE\lc q \rc\ra_{\dat}$ & 1.26 & 1.12
  &  1.10 \\
$\la RE\lc q \rc\ra_{\extra}$ & 1.29 & 1.15 & 1.16  \\
\hline
$\la RE\lc \sigma_q \rc\ra_{\dat}$ & 1.48 & 1.15 & 1.14  \\
$\la RE\lc \sigma_q \rc\ra_{\extra}$ & 1.48 & 1.32 & 1.21 \\
\hline
\end{tabular}
\caption{\small  Stability estimators, defined in Section
\ref{stabest} for the
probability measure of $q_{NS}(x,Q_0^2)$ as a function of the
number of trained replicas $N_{\rep}$, both in the
data and in the extrapolation region. }
\label{stabtablenrep}
\end{center}
\end{table}

\begin{figure}[ht]
\begin{center}
\epsfig{width=0.48\textwidth,figure=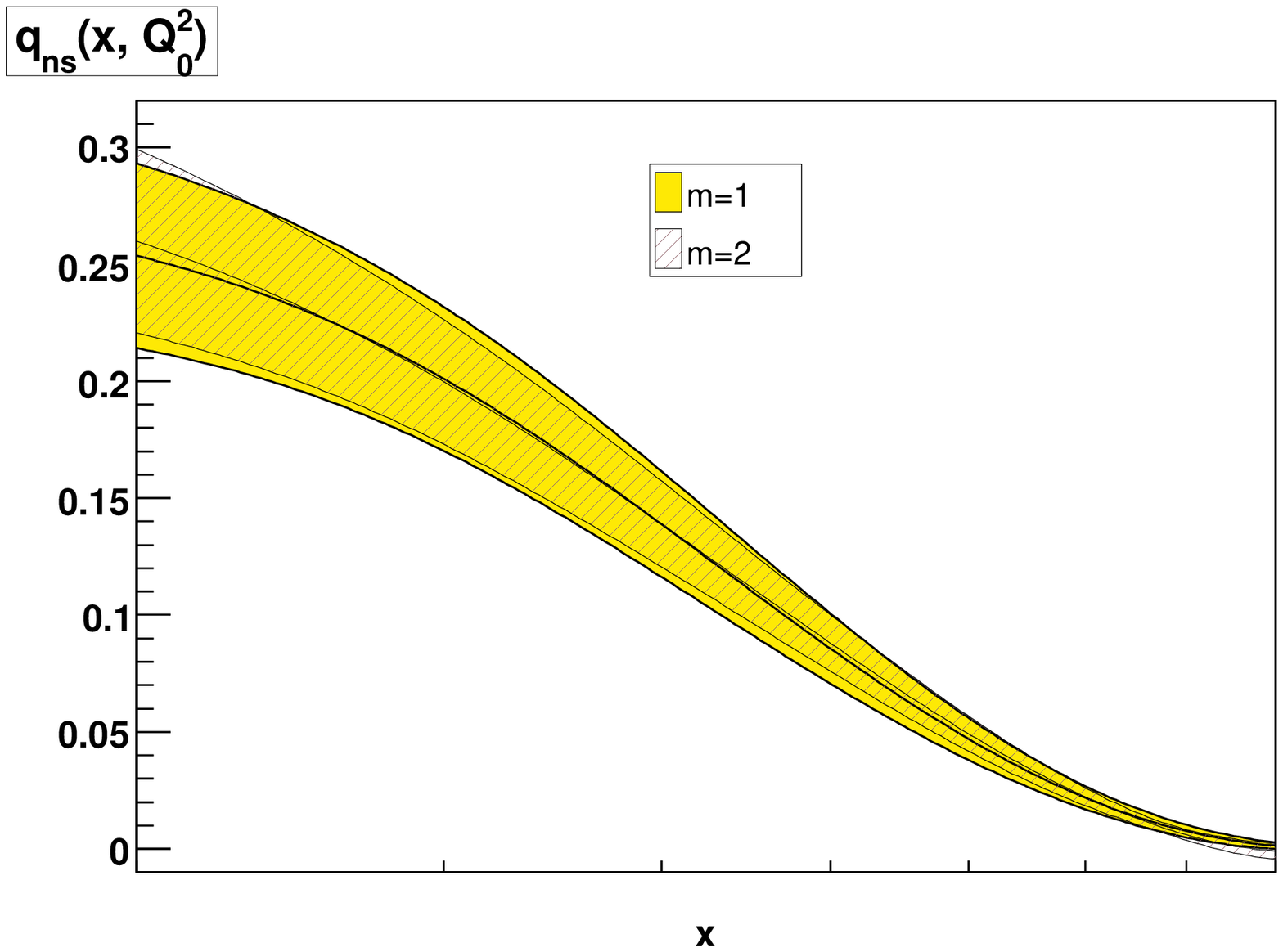}
\epsfig{width=0.48\textwidth,figure=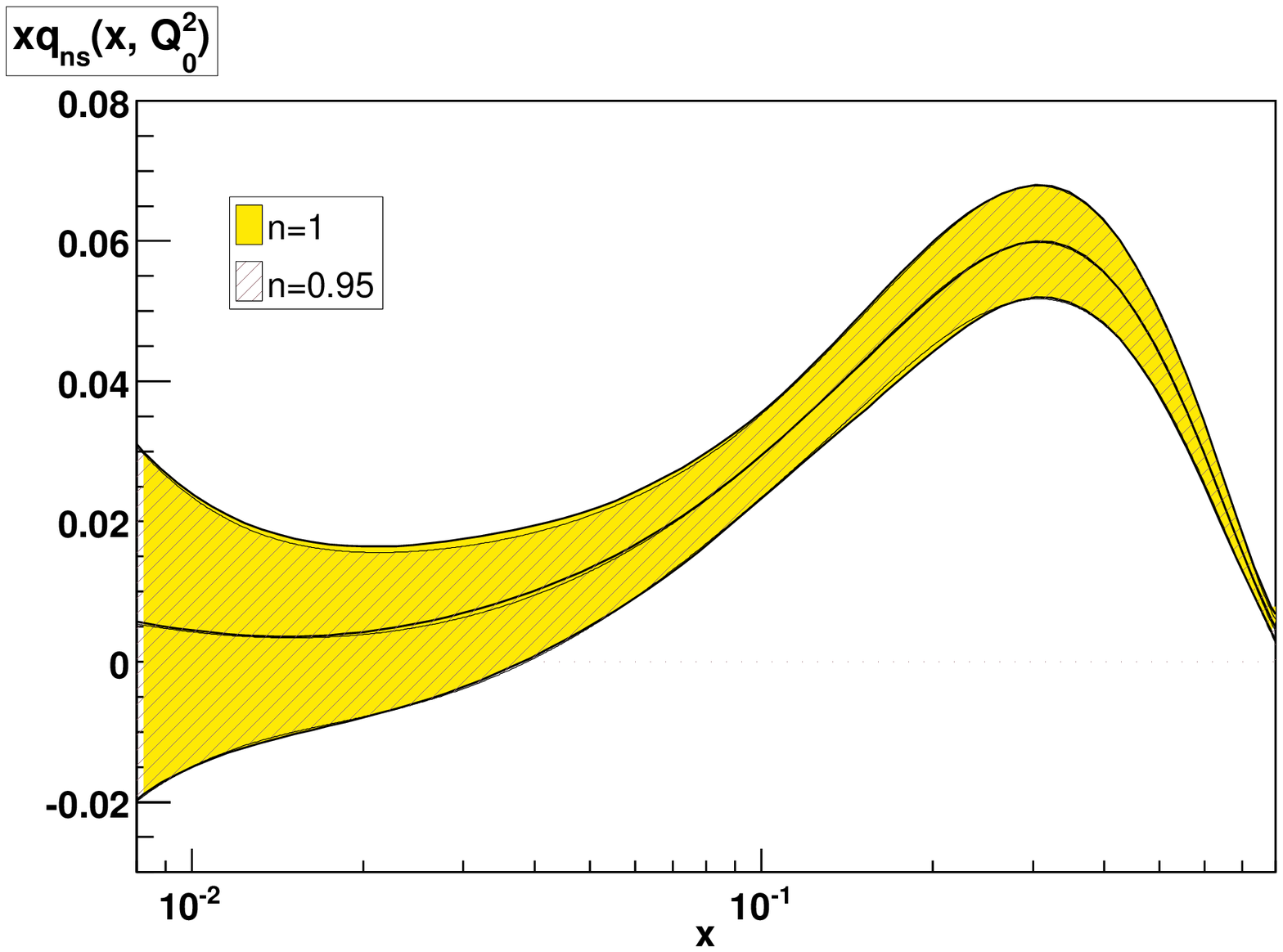}
\caption{}{\small The nonsinglet parton distribution
$q_{NS}(x,Q_0^2)$ for different values of the large$-x$
exponent (right) and of the small-$x$ exponent (left).
The polynomial exponents $m$ and $n$ are defined in
Eq. \ref{polyform}.}
\label{pdfexp}
\end{center}
\end{figure}

As  discussed in  detail in Ref. \cite{nnqns}, the most
unbiased fitting strategy is to parametrize with a neural
network the quantity $xq_{NS}(x,Q_0^2)$.
On top of that, the nonsinglet parton distribution has
to satisy the kinematical constraint that 
\be
q_{NS}(x=1,Q_0^2)=0 \ , {\mathrm{for~all~}} Q_0^2 \ ,
\ee
that is, parton distributions vanish in the elastic limit.
This constraint will be implemented with one of the techniques
discussed in Section \ref{constraints}, the hard-wiring
of the kinematical constraint in the neural network parametrization.
With this two considerations, we write the non-singlet parton
distribution in Eq. \ref{nsnet} as
\be
q_{NS}^{(\net)(k)}(x,Q_0^2)=\lp 1-x \rp\frac{
\widetilde{q}^{(\net)(k)}_{NS}(x,Q_0^2)}{x} \ ,
\ee
where now it is the quantity $\widetilde{q}^{(\net)(k)}_{NS}(x,Q_0^2)$ the
one that is parametrized with a neural network. It can be checked
that the neural network complemented with some simple
functional form dependence is still an unbiased approximant to the
true value of the nonsinglet parton distribution. In particular
we will show that the results of the fit are not affected if
in the expression
\be
\label{polyform}
q_{NS,mn}^{(\net)(k)}(x,Q_0^2)=\lp 1-x \rp^m\frac{
\widetilde{q}^{(\net)(k)}_{NS}(x,Q_0^2)}{x^n} \ ,
\ee
the values of the parameters $m$ and $n$ are modified
from their default values, $m=1$ and $n=1$. The
stability of the results with respect to reasonable variations
of the $m,n$ exponents can be made quantitative 
 by means of the stability statistical estimators
introduced in Section \ref{stabest}. Fig. \ref{pdfexp}
show in two particular cases how the results of the fit are stable
againts different choices of the polynomial exponents $m$ 
and $n$ in Eq. \ref{polyform}. This comparison is made more
quantitative with the stability estimators, as can be
seen in Table \ref{stabtable}.

\begin{table}[ht]
\begin{center}
\begin{tabular}{|c|c|c|}
\hline 
& Small-$x$ exponent $n$  &   Large-$x$ exponent $m$   \\
\hline
$\la RE\lc q \rc\ra_{\dat}$ & 0.48  & 1.92
  \\
$\la RE\lc q \rc\ra_{\extra}$ & 0.39 &  0.38 \\
\hline
$\la RE\lc \sigma_q \rc\ra_{\dat}$ & 0.81 & 1.80  \\
$\la RE\lc \sigma_q \rc\ra_{\extra}$ & 0.92 & 2.27   \\
\hline
\end{tabular}
\caption{\small  The stability estimators defined in
Section \ref{stabest} for the comparison of fits with
different polynomial exponents, see Fig. \ref{pdfexp}.
The method used to compute these estimators is the same
that has been used to assess the stability with respect
$N_{\rep}$.}
\label{stabtable}
\end{center}
\end{table}

Now we discuss how the results for the parametrization of
$q_{NS}(x,Q_0^2)$ depend on the kinematical cut in $Q^2$.
The kinematical cut in $Q^2$ has been chosen rather low
($Q^2\ge 3$ GeV$^2$) since
first of all Target Mass Corrections taken
into account in the theoretical
expression for the nonsinglet structure function
$F_2^{NS}$, Eq. \ref{tmc}, and
second we have not observed evidence, within experimental
uncertainties, of a dynamical higher twist correction.
High quality data at large-$x$ would allow to extract
the higher twist contribution $\mathrm{HT}(x)$ to
the nonsinglet structure function with higher
accuracy
with a variety of techniques \cite{hightwist,hightwist2}, even
if it is known that the magnitude of the extracted HT
correction is reduced if evolution is performed
at the NNLO level, as it is in our case.
The kinematical cut in the invariant mass of the
final hadronic state $W^2\sim Q^2(1-x)\ge 6.25$ GeV$^2$
removes those points at the largest values of $x$ for which
a Sudakov resummed evolution \cite{corcellaresum,
stermanresum} would be needed.

\begin{figure}[ht]
\begin{center}
\epsfig{width=0.63\textwidth,figure=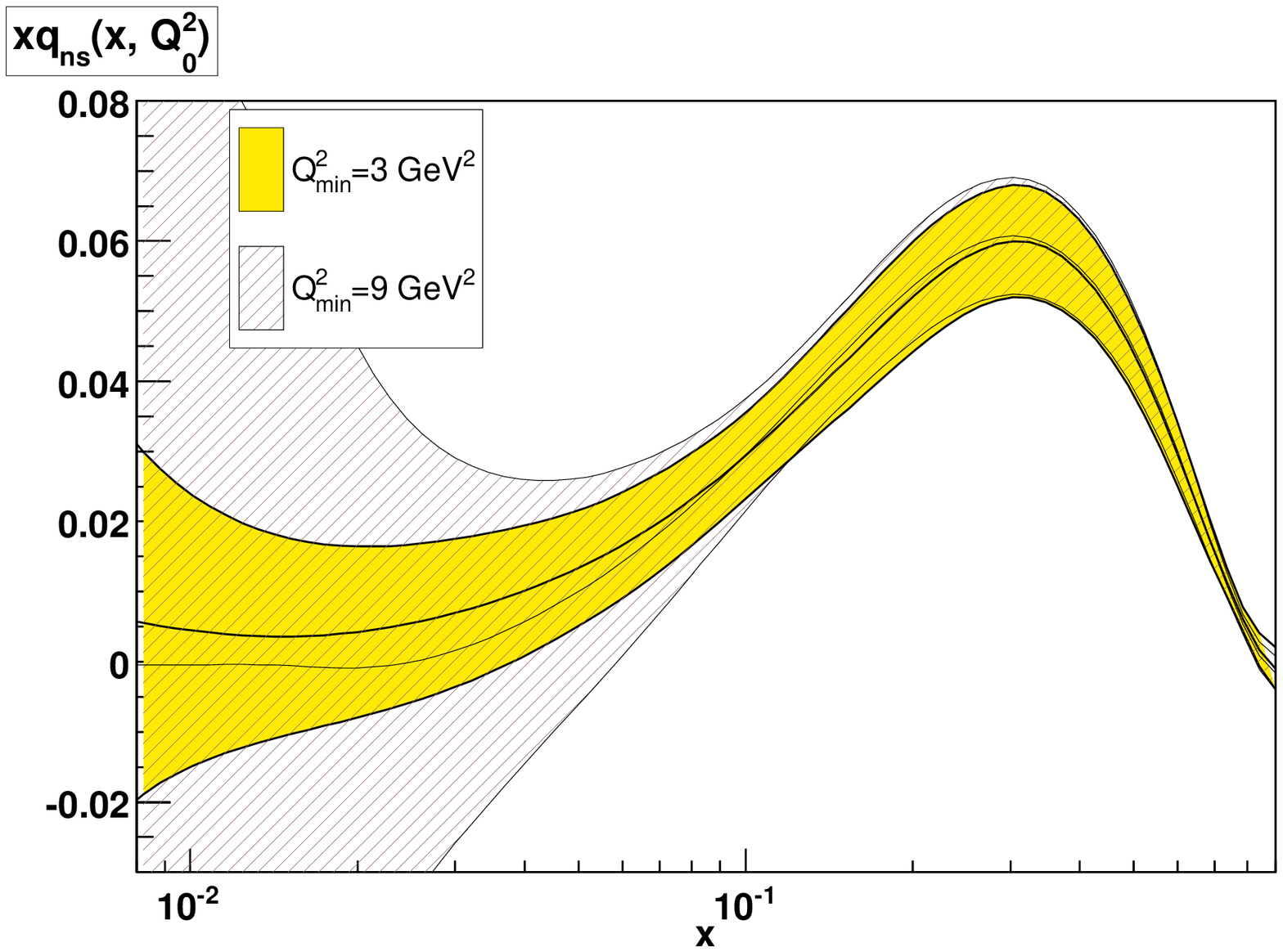}
 \caption{}{\small Comparison of the results for the
nonsinglet parton distribution $xq_{NS}(x,Q_0^2)$ 
for two different kinematical cuts: the one
for the reference fit $Q^2\ge 3$ GeV$^2$ and another with a more
conservative one $Q^2\ge 9$ GeV$^2$. }
  \label{kincuts}
\end{center}
\end{figure}

In Fig. \ref{kincuts} we compare the results of the reference final
fit, with the standard kinematical cut in $Q^2\ge 3$ GeV$^2$
with another fit with exactly the same training parameters but with
kinematics restricted to  $Q^2\ge 9$ GeV$^2$. It is clear that
the uncertainties in the parametrization of $q_{NS}(x,Q_0^2)$
grow sizeably at medium and small $x$ once the subset of data points
with $3 \le Q^2 \le 9$ GeV$^2$ is removed from the fit.
Note that the size this subset of points is rather large, $\sim$ 150
data points, the bulk of the NMC experimental data. 

\begin{figure}[ht]
\begin{center}
\epsfig{width=0.48\textwidth,figure=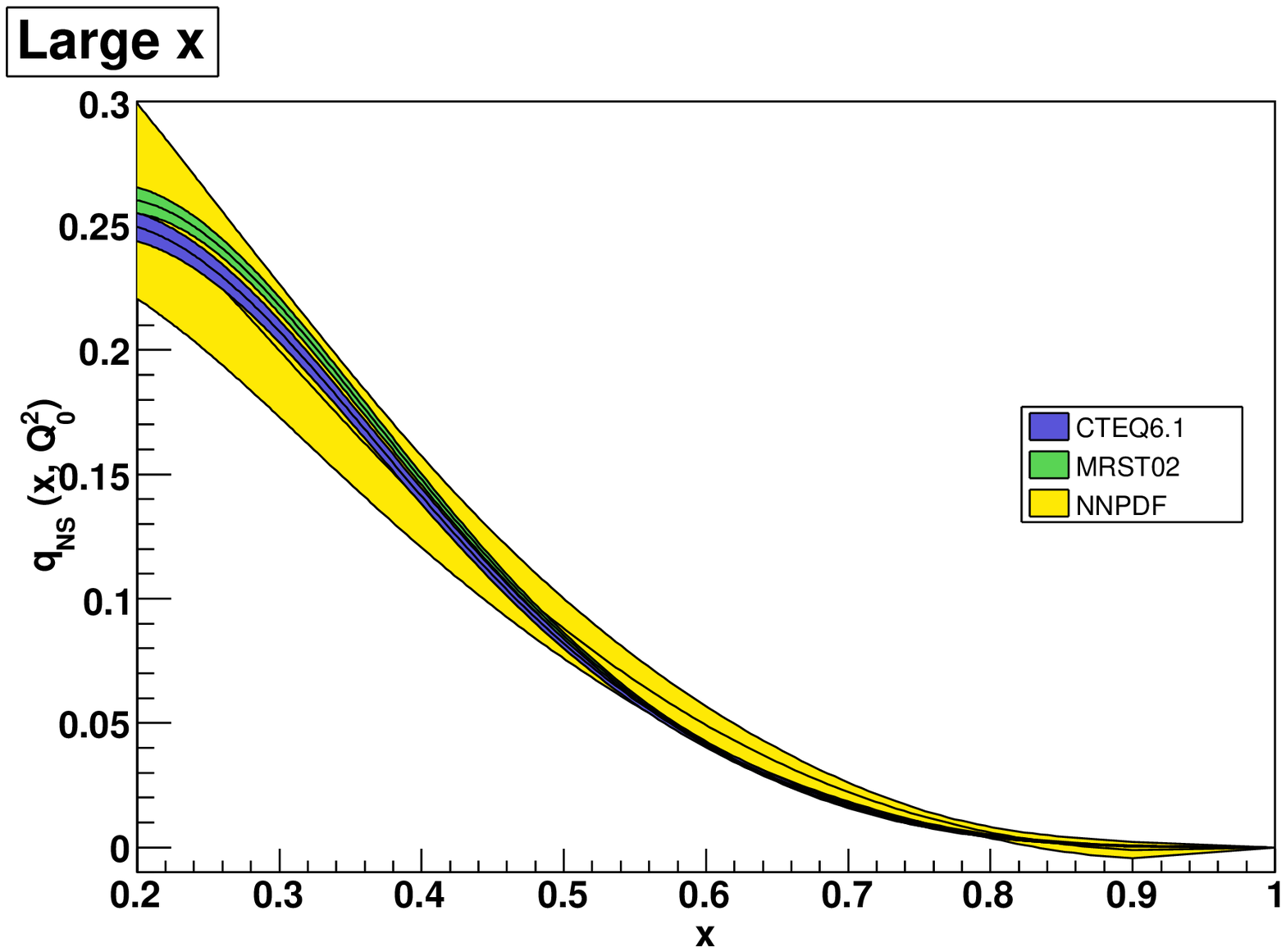}
\epsfig{width=0.48\textwidth,figure=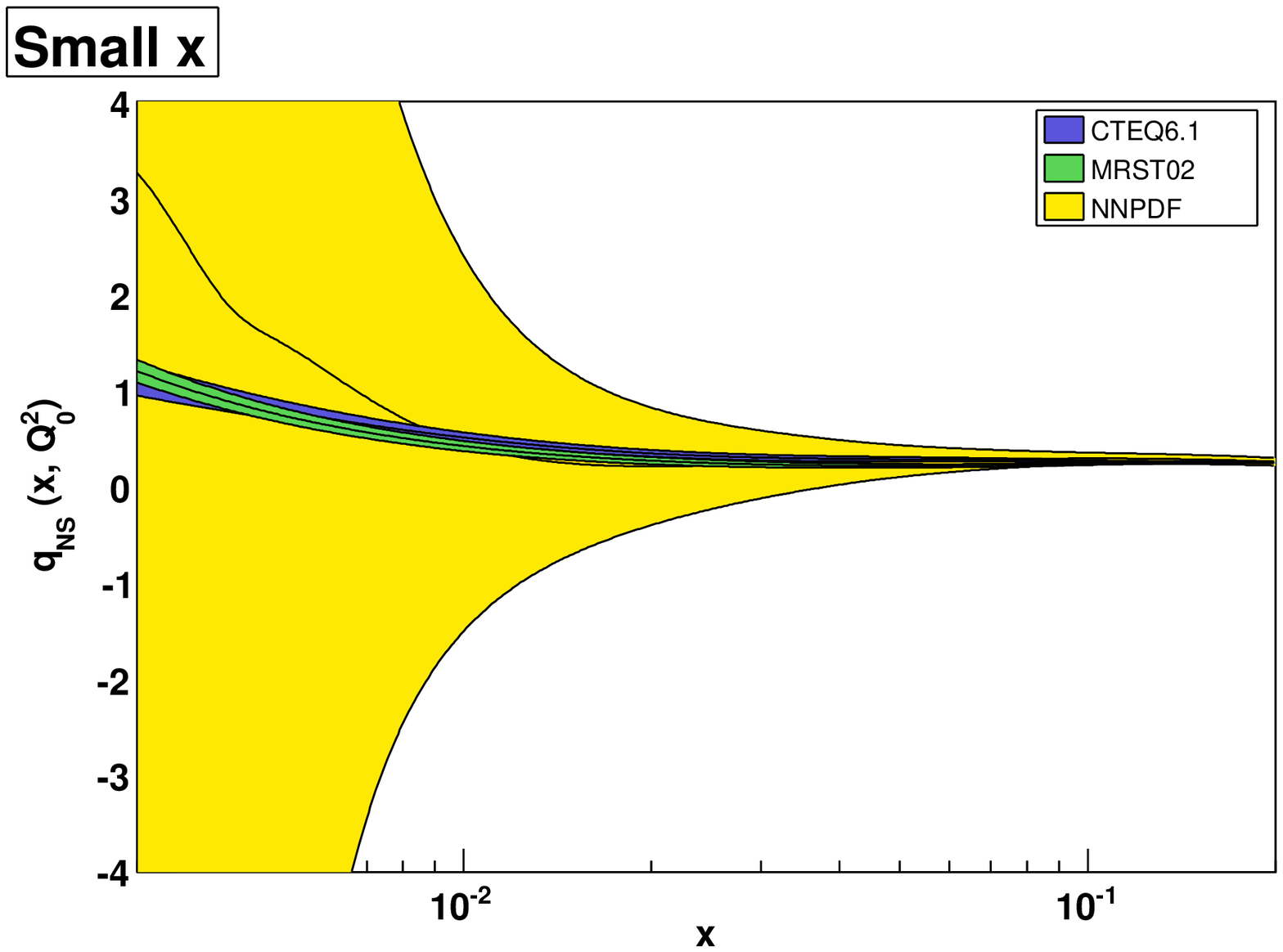}
\caption{}{\small The neural network parametrization of the
nonsinglet parton distribution, compared to the CTEQ and MRST
results, both at large$-x$ (left) and at small-$x$ (right). Note that
uncertainties grow very fast in the small-$x$ region, since
there is no experimental data for $x \lsim 10^{-2}$.}
\label{pdflin}
\end{center}
\end{figure}

\begin{figure}[ht]
\begin{center}
\epsfig{width=0.48\textwidth,figure=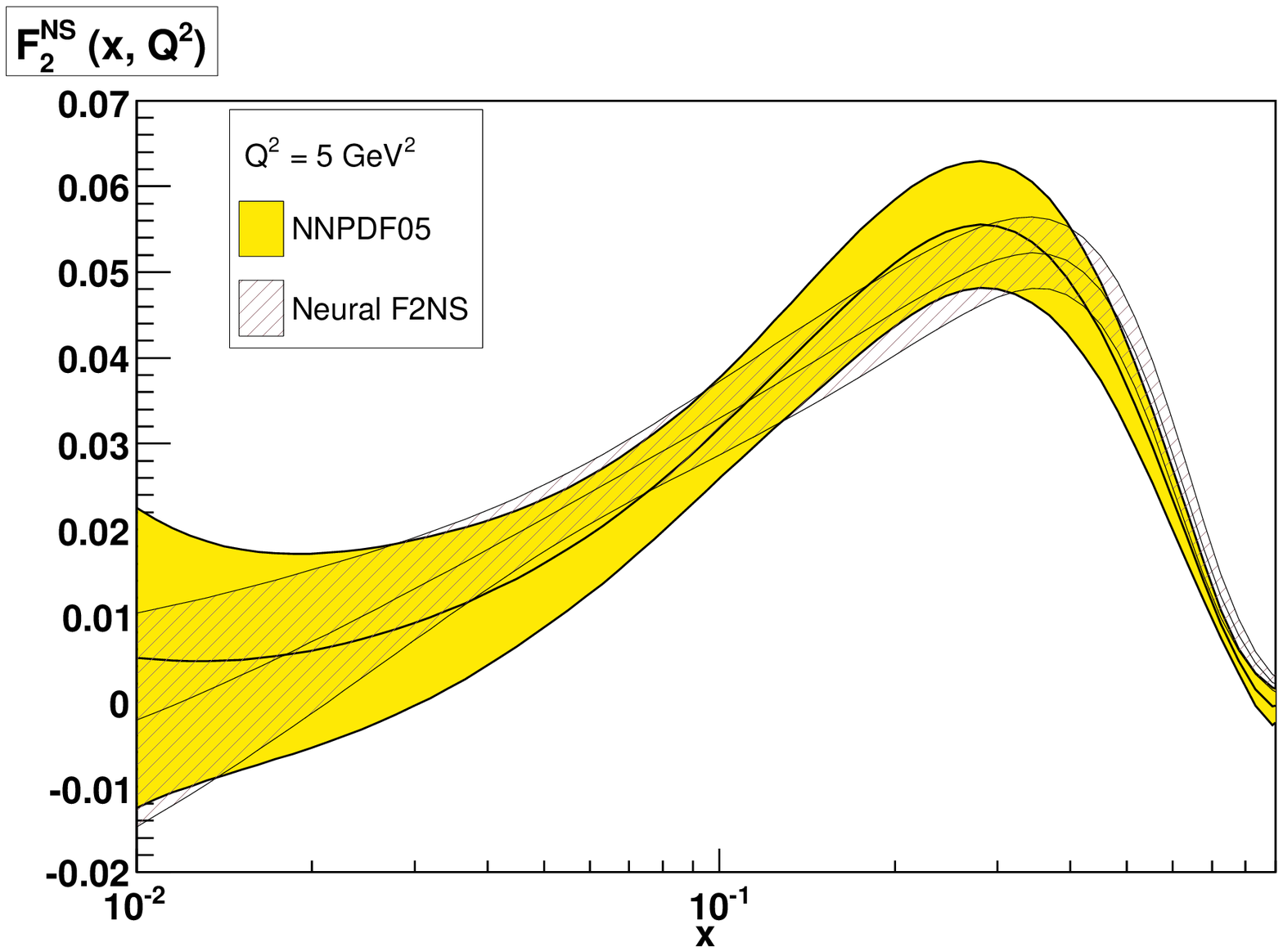}
\epsfig{width=0.48\textwidth,figure=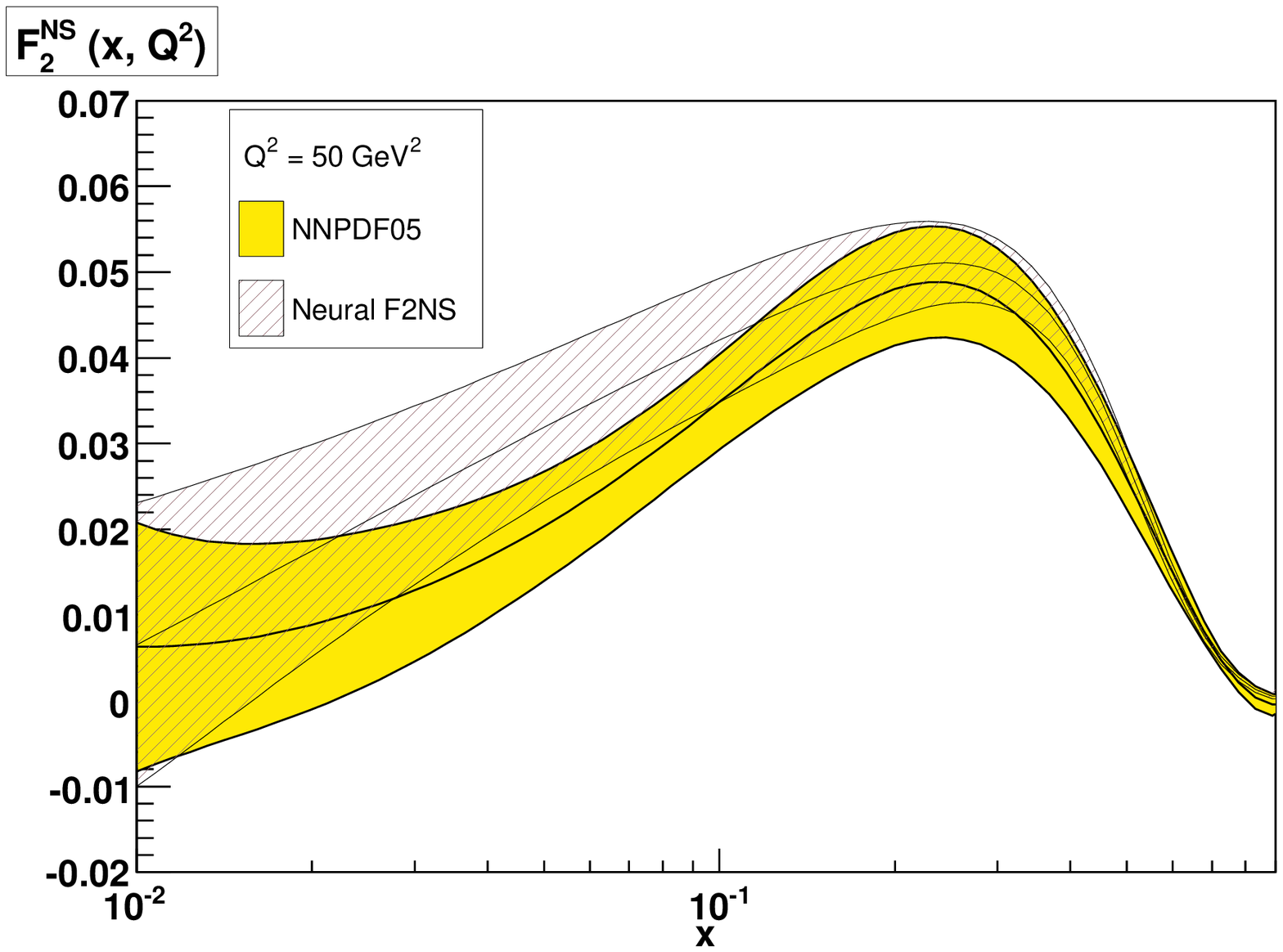}
\caption{}{\small Comparison of the nonsinglet
structure function $F_2^{NS}(x,Q^2)$ as computed
with the parametrization of
$q_{NS}(x,Q_0^2)$ discussed here with the structure
function parametrization of Ref. \cite{f2ns}, 
 for $Q^2=5$ GeV$^2$ (left) and 
$Q^2=50$ GeV$^2$ (right).}
\label{compf2nn1}
\end{center}
\end{figure}

Once the set of $N_{\rep}$ neural networks have been trained with the
strategy described above, the third step of our approach
is to validate these results by means of statistical estimators,
defined in Section \ref{validation}.
Table \ref{resdata} summarizes the most important estimators
for both the total number of data points and the individual
experiments. Note that the same patterns as in Ref. \cite{f2ns}
is observed: errors are reduced, which is sign that the neural network
has found the underlying law from the experimental data, while
correlations are increased, in a way that the
covariances as computed for the probability measure for
$q_{NS}(x,Q_0^2)$ approximately reproduce the corresponding 
experimental quantities. Note also that the
total $\chi^2$ for the two experiments included in the
fit, NMC and BCDMS, is very similar. This result is only achieved
after a detailed study of the optimal weighted training strategy,
as discussed in Section \ref{minimstratt}.

\begin{table}[ht]
\begin{center}
\begin{tabular}{|c|ccc|} 
\multicolumn{4}{c}{
$F_2^{NS}(x,Q^2)$}\\   
\hline
  & Total & NMC & BCDMS \\
\hline
$\chi^2$ & 0.77 & 0.78 & 0.75 \\
$\la E\ra$ & 2.07  & 1.94 & 2.18 \\
$r\lc F^{(\art)} \rc$ & 0.80 &  0.78& 0.95 \\
\hline
$\la \sigma^{(\exp)}\ra_{\dat}$ & 0.011 & 0.017 & 0.006 \\
$\la \sigma^{(\net)}\ra_{\dat}$ & 0.004 & 0.005 & 0.003 \\
$r\lc \sigma^{(\art)}\rc$ & 0.51 & 0.-0.04  & 0.91 \\
\hline
$\la \rho^{(\exp)}\ra_{\dat}$ & 0.18 & 0.04 & 0.16\\
$\la \rho^{(\net)}\ra_{\dat}$ & 0.47 & 0.43 & 0.50\\
$r\lc \rho^{(\art)}\rc$ &  0.31&  0.19 & 0.39  \\
\hline
$\la \mathrm{cov}^{(\exp)}\ra_{\dat}$ & 
$8.6~10^{-6}$ & $1.0~10^{-5}$ & $7.2~10^{-6}$ \\
$\la \mathrm{cov}^{(\art)}\ra_{\dat}$ & 
$9.5~10^{-6}$ & $1.4~10^{-5}$ & $5.5~10^{-6}$ \\
$r\lc \mathrm{cov}^{(\art)}\rc$ & 0.25 &  0.22 & 0.76\\
\hline
\end{tabular}
\caption{\small Statistical estimators for the validation
of the results. }
\label{resdata}
\end{center}
\end{table}

Now we discuss  our final results for the parametrization
of the nonsinglet parton distribution $q_{NS}(x,Q_0^2)$,
and particular attention is paid to the comparison with
the same parton distribution as obtained
with the standard approach described in Section \ref{globalfits}.
In Fig. \ref{pdflin},  the results of our
parametrization of $q_{NS}(x,Q_0^2)$ are presented, both
at large $x$ and at small $x$, where we also compare with
the results of the CTEQ (the CTEQ61  analysis
of Ref. \cite{cteq61}) and MRST (the MRST2001E of Ref. 
\cite{mrst2001e}) collaborations.
The most striking feature of our results is that the size
of the error band at small x is much larger than
in the fits with the standard approach. Note that even
if the global fits of Refs. \cite{cteq61,mrst2001e} 
include much more data than the
present analysis, the low $x$ behavior of the nonsinglet
parton distribution is only constrained by the data 
on the nonsinglet structure function
$F_2^{NS}$ that
is used in this analysis.

We can also analyze the effects that including the information
of QCD parton evolution has on the parametrization
of experimental data. These effects can be seen in
 Figs. \ref{compf2nn1}, where 
the results of the present analysis are compared with the
parametrization of the nonsinglet structure function
$F_2^{NS}(x,Q^2)$ from Ref. \cite{f2ns}.
One observes that the two results are consistent within the respective
uncertainties. Note that due to the effects of QCD evolution,
experimental measurements of the nonsinglet
structure function $F_2^{NS}(x,Q^2)$ for different values
of $Q^2$ correspond to the same measurement of
$q_{NS}(x,Q_0^2)$ repeated many times.

In summary, in this part of the thesis we have
described the first application of the technique
described in Chapter \ref{general} to the
parametrization of parton distribution functions.
The main result appears to be that uncertainties
in parton distributions obtained with the standard
approach are underestimated, specially in the
extrapolation region. The results of this part of the
thesis will be continued in Ref. \cite{nnsinglet},
were the full singlet evolution will be considered.

\newpage

\chapter{Conclusions and outlook}

In this thesis we have described in detail
a general technique to parametrize experimental data in 
an unbiased way with faithful estimation of the
associated experimental uncertainties. This technique
was first introduced  in Ref. \cite{f2ns} and during the course
of this thesis it has been improved and extended to the
analysis of other processes of interest.
In particular we have shown the first application to the
problem that motivated its development, the
parametrization of parton distribution
functions.

The general strategy introduced in Ref. \cite{f2ns} has been
extended in several ways. First of all,
the technique has been applied to 
different situations than the original application,
the parametrization of deep-inelastic structure functions
from a moderately large data set. In these new applications,
for example, we have faced problems with a very large amount
of experimental data from different experiments
and  problems in which the parametrized
quantity is related to the experimental data
in through complicated relations, like
convolutions. Second, new
minimization algorithms for neural network training
have been introduced, in particular genetic algorithms
which are suited for the minimization
of highly nonlinear error functions.
We have also introduced more refined criteria to
assess the optimal point to stop the training of the
neural networks. Additional statistical estimators
to assess several characteristics of the probability
measures have also been introduced. Finally, the
application of the strategy discussed in this thesis
to several different processes, deep-inelastic
structure functions, hadronic tau decays,
semileptonic B meson decays and parton distribution
functions, increases our
confidence on its general validity.

The number of possible applications of the general strategy
to parametrize experimental data described in this
thesis is rather large. The most important one
is to generalize the results of Section 
\ref{nnqns} to the singlet sector \cite{nnsinglet} 
and to produce a full
set of neural network parton distributions with a 
faithful estimation of their uncertainties.
Another promising application is the parametrization of the
nonperturbative shape function
from semileptonic and radiative B meson decays.
Astroparticle physics is another field in which several
applications of the neural network approach have
been envisaged. In particular, there is an ongoing
project \cite{neuralneut} in which 
the general strategy discussed in this thesis is
used to determine  the
atmosferic neutrino flux from experimental data on neutrino
event rates.

\newpage
~
\newpage
\include{conclusiones}

\newpage
~
\newpage

\appendix

\chapter{Elements of statistical data analysis}

\section{Review of probability theory}
\label{probrev}

In this Appendix we review some basic elements of 
probability theory \cite{cowan,pdg}. Let us consider with 
full generality a set of $N_{\dat}$ observables
$F_i$, $i=1,\ldots,N_{\dat}$. One has,
for each observable, $N_{\rep}$ independent
measurements, which will be denoted
by $F_i^{(k)},k=1,\ldots,N_{\rep}$.
As it is clear from the notation, we have in mind
the general strategy to construct a probability measure
in the space of the observable $F$ described
in Chapter \ref{general}, where $N_{\rep}$
stands for the number of generated replicas
of the experimental data.
From this set of measurements, one can
construct the following
statistical estimators for each observable $F_i$:
\begin{itemize}
\item Mean:
\be
\la F_i\ra=\frac{1}{N_{\rep}}\sum_{k=1}^{N_{\rep}}F_i^{(k)} \ ,
\ee
\item Variance of the data:
\be
\sigma_i^2=\la \lp F_i-\la F_i\ra \rp^2\ra=\la F_i^2\ra -\la F_i\ra^2
\ ,
\ee
\item Correlation between data points:
\be
\rho_{ij}=
\frac{\la \lp F_i-\la F_i\ra\rp
\lp F_j-\la F_j\ra\rp
\ra}{\sigma_i\sigma_j}=\frac{\la F_iF_j\ra-
\la F_i\ra \la F_j\ra}{\sigma_i\sigma_j} \ .
\ee
\end{itemize}
Unless otherwise indicated, all the averages are performed
with respect to the $N_{\rep}$ measurements, and we assume
that we are in the limit where $N_{\rep}$ is very large.
The above estimators describe features of the
underlying probability density of the experimental data,
and they approach the true values as the
number of { measurements} $N_{\rep}$
becomes very large.

This result is quantitatively described by the variances
of the different estimators.
These estimator measure the difference of the
values of the mean, variance of the data and correlations
as determined with averages over measurements with respect
to their true values. These variances, written in terms
of moments of the $F_i$, are given by
\begin{enumerate}
\item Variance of the mean:
\be
V\lc F_i\rc= \frac{1}{N_{\rep}}
\la \lp F_i- \la F_i\ra\rp^2\ra=\frac{\sigma_i^2}{N_{\rep}} \ ,
\ee
\item Variance of the error:
\be
V\lc \sigma_i^2\rc=\frac{1}{N_{\rep}}
\lc \la \lp F_i-\la F_i\ra\rp^4\ra-\sigma_i^4\rc
=\frac{1}{N_{\rep}}\lp \la F_i^4\ra-4\la F_i\ra\la F_i^3\ra+
6 \la  F_i^2\ra \la F_i\ra^2-3\la F_i\ra^4-\sigma_i^4 \rp \ ,
\ee
\item Variance of the correlation: 
\bea
\label{varcor}
V\lc \rho_{ij}\rc=
\frac{1}{N_{\rep}}\lc 
\la\lp
\frac{\la \lp F_i-\la F_i\ra\rp  
\lp F_j-\la F_j\ra\rp
\ra}{\sigma_i\sigma_j}
\rp^2 \ra-\rho_{ij}^2
\rc= \qquad  \qquad 
\nonumber
\\
\frac{1}{N_{\rep}}\Bigg(
\frac{1}{\sigma_i^2\sigma_j^2}\Bigg[ \la F_i^2F_j^2\ra
-2\la F_i\ra \la F_iF_j^2\ra
-2 \la F_j \ra \la F_i^2F_j\ra
\qquad \qquad 
\qquad 
\nonumber
\\
+4 \la F_iF_j\ra\la F_i\ra\la F_j \ra
+\la F_i^2\ra \la F_j\ra^2
+\la F_i\ra^2 \la F_j^2\ra -3 \la F_i\ra^2\la F_j\ra^2\Bigg]
-\rho_{ij}^2 \Bigg) \ ,
\eea
which can also be written as
\be
V\lc \rho_{ij}\rc=\frac{1}{N_{\rep}}\lp 1-\la \rho_{ij}^2\ra\rp^2 \ .
\ee
\end{enumerate}
It is clear from the above expressions that the variances of the mean,
the error and the correlations decrease when the number
of measurements is increased. This implies that as statistics
are increased, the measured values of the mean, error and
correlations are closer to their true values.
Therefore, to compare different probability measures, the variances
of the mean, the error and the correlation as
defined above should be used.

\section{The Monte Carlo approach to error estimation}
\label{mcerrequiv}

In this Section we show with a simple example
that the Monte Carlo approach to
error estimation described in 
Section \ref{general} is equivalent to the standard approach,
based on the condition $\Delta\chi^2=1$ for the
determination of confidence levels, with the
assumption of gaussian errors,
up to linearized approximations. Then we present an example
of the application of the Monte Carlo approach to error
estimation in the case of standard functional form fits.

Let us consider two pairs of independent measurements of the
same quantity, $x_1\pm \sigma_1$ and $x_2\pm \sigma_2$ with
gaussian uncertainties.
The distribution of true values of the variable $x$ is a gaussian
distribution centered at
\be
\overline{x}=\frac{x_1\sigma_2^2+x_1\sigma_2^2}{\sigma_1^2+\sigma_2^2} \ ,
\ee
and with variance determined by the $\Delta\chi^2=1$
tolerance criterion,
\be
\label{varmc}
\sigma^2=\frac{\sigma_1^2\sigma_2^2}{\sigma_1^2+\sigma_2^2}\ .
\ee
To obtain the proof of the above results, note that if errors are gaussianly
distributed, the maximum likelihood condition imply
that the mean $\overline{x}$ minimizes the $\chi^2$
function
\be
\chi^2=\frac{\lp x_1-\overline{x}\rp}{\sigma_1^2}+
\frac{\lp x_2-\overline{x}\rp}{\sigma_2^2}  \ ,
\ee
and the variance $\sigma$ is determined by the
condition
\be
\Delta\chi^2=\chi^2\lp \overline{x}+\sigma\rp-
\chi^2\lp \overline{x}\rp \ ,
\ee
which for $\Delta\chi^2=1$ leads to Eq. \ref{varmc}.
Note that these properties only hold for gaussian measurements.

An alternative way to compute the mean and the variance 
of the combined measurements  $x_1$ and $x_2$ is the
Monte Carlo method: generate $N_{\rep}$ replicas of the pair
of values $x_1,x_2$ gaussianly distributed with the appropriate
error,
\be
\label{rep11}
x_1^{(k)}=x_1+r_1^{(k)}\sigma_1, \qquad k=1,\ldots,N_{\rep} \ ,
\ee
\be
\label{rep22}
x_2^{(k)}=x_2+r_2^{(k)}\sigma_2, \qquad k=1,\ldots,N_{\rep} \ ,
\ee
where $r^{(k)}$ are univariate gaussian random numbers.
One can then show that for each pair, the weighted average
\be
\overline{x}^{(k)}=\frac{x_1^{(k)}\sigma_2^2+
x_1^{(k)}\sigma_2^2}{\sigma_1^2+\sigma_2^2} \ ,
\ee
is gaussianly distributed with central value and width equal to the one
determined in the previous case. That is, it can be show that
for a large enough value of $N_{\rep}$,
\be
\la \overline{x}^{(k)}\ra_{\rep}=\frac{1}{N_{\rep}}
\sum_{k=1}^{N_{\rep}}\overline{x}^{(k)}=\overline{x} \ ,
\ee
and for the variance
\be
 \sigma^2=\la  \lp\overline{x}^{(k)}\rp^2\ra_{\rep}-
\la  \overline{x}^{(k)}\ra_{\rep}^2=\frac{\sigma_1^2\sigma_2^2}{
\sigma_1^2+\sigma_2^2} \ ,
\ee
which is the same result, Eq. \ref{varmc}, as obtained
from the $\Delta\chi^2=1$ criterion.
This shows that the two procedures are equivalent in this simple
case. 

The generalization to $N_{\dat}$ gaussian 
correlated measurements is
straightforward. Let us consider for instance that the
two measurements $x_1$ and $x_2$ are not
independent, but that they have correlation $\rho_{12} \le 1$.
To take correlations into account, one uses
the same Eqns. \ref{rep11} and \ref{rep22} to generate
the sample of replicas of the measurements, but this time
the random numbers $r_1^{(k)}$ and $r_2^{(k)}$
are univariate gaussian correlated random numbers,
that is, they satisfy 
\be
\la r_1r_2\ra_{\rep}=\frac{1}{N_{\rep}}\sum_{k=1}^{N_{\rep}}
 r_1^{(k)}r_2^{(k)}=\rho_{12} \ .
\ee
With this modification, the sample of Monte Carlo
replicas of $x_1$ and $x_2$ also reproduces the
experimental correlations. This can be seen
with the standard definition of the correlation,
\be
\rho\equiv \la \frac{ \lp x_1^{(k)}-x_1\rp 
\lp x_2^{(k)}-x_2\rp}{\sigma_1\sigma_2}\ra_{\rep}=\la r_1r_2\ra_{\rep}=
\rho_{12} \ .
\ee
Therefore, the Monte Carlo approach also correctly takes
into account the effects of correlations between measurements.

In realistic cases, 
the two procedures are equivalent only up to linearizations
of the underlying law which describes  the experimental data.
We take the Monte Carlo procedure to be more faithful in that it does
not involve linearizing the underlying law in terms of the parameters.
Note that as emphasized before, the error estimation
technique that is described in this thesis does
not depend on whether
one uses neural networks or polynomials as interpolants.
Conversely, one could derive 1-$\sigma$ errors on the
parameters of the neural network as an alternative
to estimate the uncertainties in  the parametrized function.

As an example of the application of the Monte Carlo
error estimation to standard fits of parton
distributions with polynomial functional forms, we
repeat the nonsinglet fit of Refs. \cite{blum1,blum2}.
We use exactly the same techniques as discussed in Section
\ref{nnqns_appl} but with a functional form to
parametrize $q_{NS}(x,Q_0^2)$ instead of a neural network, which
is taken to have the functional dependence \cite{blum1}
\begin{eqnarray}
\label{qnsblum}
q_{NS} (x,Q^2_0)=\frac{1}{6}\left(u_v - d_v 
- 2 (\bar{d}-\bar{u})_{\mathrm{MRST}}\right)(x,Q^2_0) \ ,
\end{eqnarray}
were we have defined
\begin{eqnarray}
u_v (x,Q^2_0)&=& A_{u_v} x^{a_u}(1-x)^{b_u} 
\left(1-1.108 x^{\frac{1}{2}} + 26.283 x\right) \ , \\
d_v (x,Q^2_0)&=& A_{d_v} x^{a_d}(1-x)^{b_d} 
\left(1+0.895 x^{\frac{1}{2}} + 18.179 x\right)  \ ,\\
(\bar{d}-\bar{u})_{\mathrm{MRST}} (x,Q^2_0)&=& 1.195 x^{0.24}(1-x)^{9.10} 
\left(1+14.05 x - 45.52 x\right) \ ,
\end{eqnarray}
and the $(\bar{d}-\bar{u})$ combination is taken from
the MRST global analysis \cite{mrst_02}.
The normalization constants are fixed by the conservation
of the number of valence quarks
\be \int_0^1 dx\,u_v (x)=2 \ ,\qquad
\int_0^1 dx\,d_v (x)=1 \ .\ee The values of the parameters
obtained from a fit to the experimental data are summarized in
Table \ref{blumtb}, were we compare with the results of the
original fit \cite{blum1}. In particular one observes that the
exponent which governs the small-$x$ behavior of the
nonsinglet parton distribution, $a_u$, is correctly reproduced as
expected,
since at small-$x$ the experimental data that
determines the behavior of $q_{NS}(x,Q_0^2)$ 
is the same in the
two cases.

\begin{table}[ht]
\begin{center}
\begin{tabular}{|c|c|c|c|c|}
\hline
 & $a_u$ & $b_u$ & $a_d$ & $b_d$ \\
\hline
Refs. \cite{blum1,blum2} & -0.686 & 4.199 & -0.587 & 6.190 \\
\hline
NNPDF & -0.705 & 0.844 & 0.384 & 1.035 \\
\hline
\end{tabular}
\end{center}
\caption{}{\small \label{blumtb} The results of a fit to the
nonsinglet structure function $F_2^{NS}(x,Q^2)$ for a parton
distribution with functional dependence given by Eq. \ref{qnsblum},
compared with the results of the fit of \cite{blum1,blum2}.}
\end{table}

\begin{figure}[ht]
\begin{center}
\epsfig{width=0.48\textwidth,figure=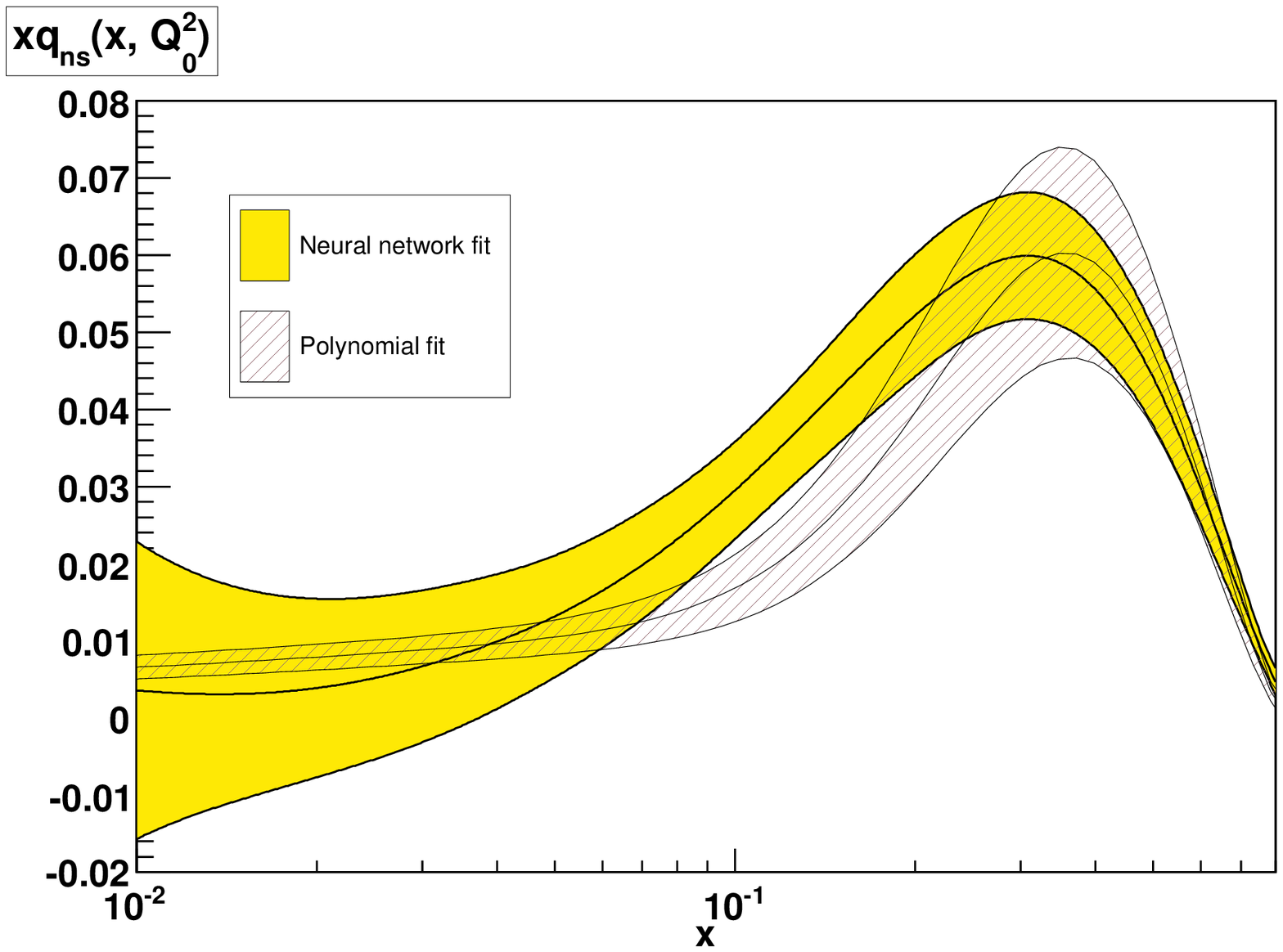}
\epsfig{width=0.48\textwidth,figure=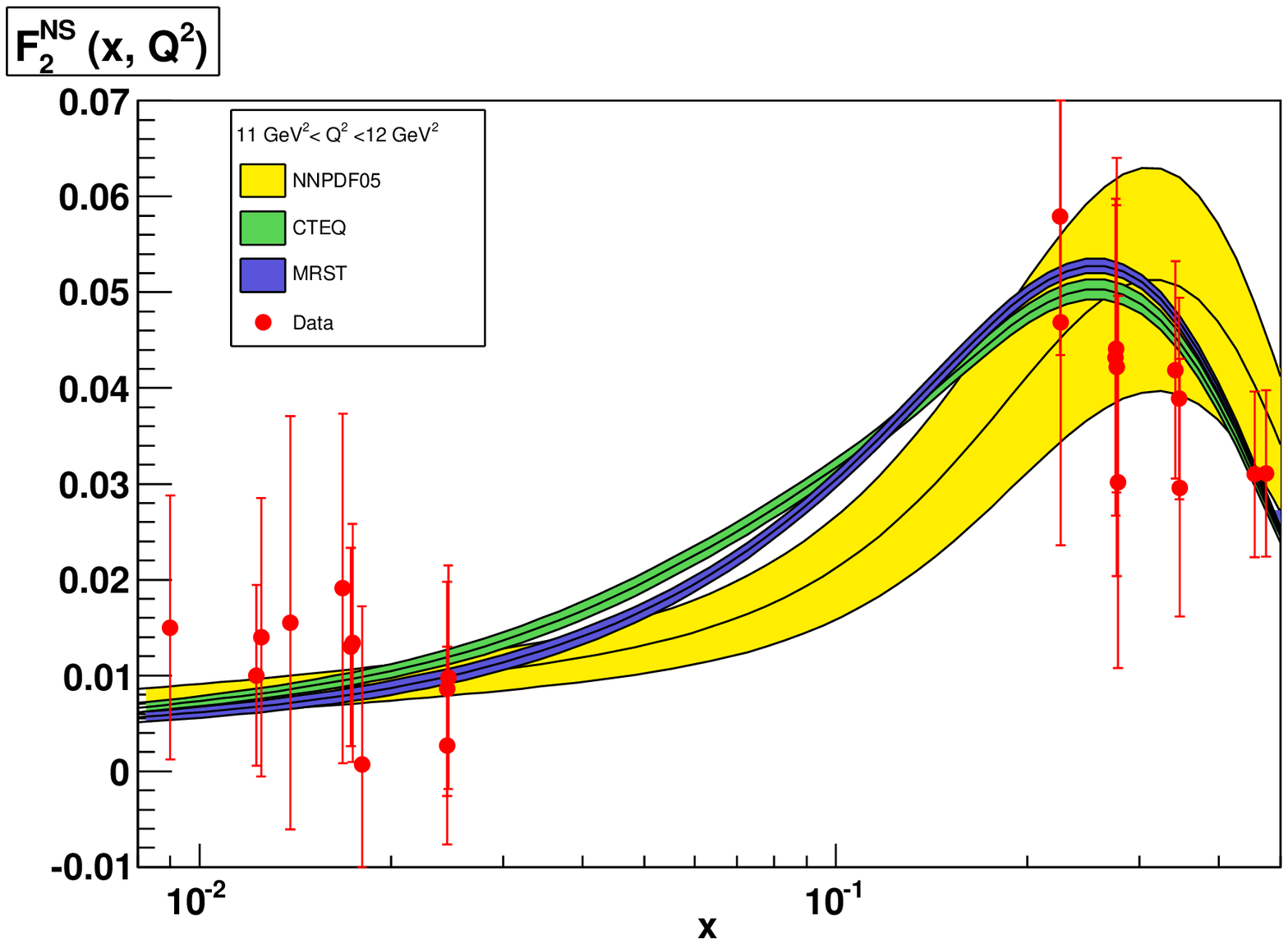}
\caption{}{\small The results of a polynomial fit with
the Monte Carlo method for error estimation with
the parametrization of Ref. \cite{blum1}, compared with
the corresponding fit with neural networks (left) and
with the standard fits of the CTEQ and MRST collaborations
(right).}
\label{blum}
\end{center}
\end{figure}

Note that Refs. \cite{blum1,blum2} 
 have a different parametrization above and below $x=0.3$,
while we take only the one they use for $x<0.3$ and that 
are the large $x$ behavior they use also HERA data.
One observes in Fig. \ref{blum} that the small-$x$
behavior of our polynomial fit coincides 
precisely with the small-$x$ behavior of the
nonsinglet parton distributions from the MRST and
CTEQ global analysis. Note also that at medium and small$-x$
the uncertainties as determined with the 
standard methods introduced in Section \ref{globalfits}
appear to be underestimated.

\section{Correct treatment of normalization uncertainties}

\label{toymodel}

In Section \ref{trainstrat} we have defined different estimators that are
minimized during the neural network training.
To understand the relations between the different estimators we have used,
 we can use a simple  model. 
This  model is simple enough to be solvable
in terms of compact expressions, 
and is therefore suitable to obtain intuition on
the expected behavior of the different error functions.
This exercise will be also useful to understand the effects
of an incorrect treatment of normalization uncertainties.
In the spirit of D'Agostini's analysis \cite{dagostini}, 
we consider the simple case in which two results of the
same physical quantity, $x_1$ and $x_2$ 
are available, and we know the statistical $\sigma_i$, 
the
systematic $\sigma_c$ and normalization $\sigma_f$ uncertainties.
The discussion can be generalized to
the case of additional measurements with different
systematic uncertainties.

The best theoretical prediction in this case will therefore 
correspond to fitting a constant which we will denote by $k$.
The diagonal error, Eq. \ref{er2}, will be given by
\be
\label{sce}
E_2=\frac{(x_1-k)^2}{\sigma_1^2} 
+\frac{(x_2-k)^2}{\sigma_2^2} \ .
\ee
Note that to derive such expression one has to assume that the
experimental measurements are gaussianly distributed.
If normalization uncertainties are neglected, then the
expression for the covariance matrix
error, Eq. \ref{er3}, is given by
 \be
\label{cme}
E_3=E_2\frac{1+\lp x_1-k\rp^2 
\frac{\sigma_c^2}{\sigma_1^2\sigma_2^2}
+\lp x_2-k\rp^2 
\frac{\sigma_c^2}{\sigma_1^2
\sigma_2^2}
-2\lp x_1-k\rp \lp x_2-k\rp
\frac{\sigma_c^2}{\sigma_1^2\sigma_2^2}
}{1+\sigma_c^2\lp \frac{1}{
\sigma_1^2}+ \frac{1}{
\sigma_2^2}\rp 
} \ ,
\ee
and  finally
the full $\chi^2$, including normalization uncertainties, Eq.
\ref{chi2tot}, is given by
\be
\label{cmen}
\chi^2=E_2\frac{1+\lp x_1-k\rp^2 
\frac{\sigma_c^2+x_2^2\sigma_f^2}{\sigma_1^2\sigma_2^2}
+\lp x_2-k\rp^2 
\frac{\sigma_c^2+x_1^2\sigma_f^2}{\sigma_1^2
\sigma_2^2}
-2\lp x_1-k\rp \lp x_2-k\rp
\frac{\sigma_c^2+x_1 x_2\sigma_f^2}{\sigma_1^2\sigma_2^2}
}{1+\sigma_c^2\lp \frac{1}{
\sigma_1^2}+ \frac{1}{
\sigma_2^2}\rp+\sigma_f^2 \lp \frac{x_1^2}{
\sigma_1^2}+ \frac{x_2^2}{
\sigma_2^2}\rp+\lp x_1-x_2\rp^2
\frac{\sigma_c^2\sigma_f^2}{\sigma_1^2\sigma_2^2}} \ .
\ee
Note that for example as $\sigma_c\to 0$ the correlated error
function $E_3$ reduces to the uncorrelated one
$E_2$, as expected.

Once we have defined the different error functions, we
 can compute the values of $k$ for which each of the
different error functions has a minimum. This is
achieved imposing the conditions
\be
\frac{d}{dk}E_2(k)\Bigg|_{k=k_2}=0, \qquad
\frac{d}{dk}E_3(k)\Bigg|_{k=k_3}=0, \qquad \frac{d}{dk}
\chi^2(k)\Bigg|_{k=k_{\chi^2}}=0 \ .
\ee 
For the diagonal error, Eq. \ref{sce} one has
the standard weighted average
\be
\label{naive}
k_{2}= \frac{ x_1\sigma_2^2 +x_2\sigma_1^2}{\sigma_1^2+\sigma_2^2} \ ,
\ee
then for the covariance matrix error, Eq. \ref{cme} 
one has the same
minimum as before
\be
k_{3}=k_{2}=\frac{ x_1\sigma_2^2 +x_2\sigma_1^2}{\sigma_1^2+\sigma_2^2} \ .
\ee
This points to the fact that in a realistic case the result of
a minimization of the diagonal error function, Eq. \ref{er2}
should be rather similar so the corresponding result when 
the minimized error function is Eq. \ref{er3}.
Finally for the full $\chi^2$ with normalization uncertainties 
one has 
\be
k_{\chi^2}=\frac{ x_1\sigma_2^2 +x_2\sigma_1^2}{\sigma_1^2+\sigma_2^2
+(x_1-x_2)^2\sigma_f^2} \ ,
\ee
which as can be rather different
from the naive estimation Eq. \ref{naive} if data are
incompatible and normalization error is sizeable.
This is true even if normalization effects are small
if the measured values are inconsistent, since the effect
of normalization uncertainties is proportional to
$\sigma_f(x_1-x_2)$ are thus can be arbitrarily large.
In particular one has a sizeable effect
if
\be
\frac{\sigma_f^2}{\sigma_1^2+\sigma_2^2}\lp x_1-x_2\rp^2 \gg 1\ ,
\ee
so one can have much larger effects than those naively
expected from the value of $\sigma_f$. 
This shows that the error function with normalization
errors as defined in Eq. \ref{cmen} leads
to completely unexpected and anti-intuitive results.

The quantities that are relevant to compute are the values of the
different error functions at the different possible minima $k_i$.
The first one is the value of the diagonal error when minimizing the
same error, then one has
\be
E_2\lp k_{2}\rp=\frac{(x_1-x_2)^2}{\sigma_1^2+\sigma_2^2} \ .
\ee
Another interesting quantity is the ratio between $E_2$ and
$E_3$ when minimizing either $E_2$ or
$E_3$ (since as long as normalization uncertainties are not included the
minimum is the same for both quantities). 
This ratio is given by
\be
\frac{E_2\lp k_{2}\rp}{
E_3\lp k_{3}\rp}=\frac{1+\sigma_c \frac{
\sigma_1^2+\sigma_2^2}{\sigma_1^2\sigma_2^2}}{
1+\sigma_c \frac{
(x_1-x_2)^2}{\sigma_1^2\sigma_2^2}} \ ,
\ee
This ratio is typically of order 1, showing why both errors
are comparable in fits without normalization error.
In particular if $x_1=x_2$ one has
\be
\frac{E_2\lp k_{2}\rp}{
E_3\lp k_2\rp}=1+\sigma_c \frac{
\sigma_1^2+\sigma_2^2}{\sigma_1^2\sigma_2^2} \ge 1 \ .
\ee
Finally let us consider the case in which we minimize the
full $\chi^2$ with normalizations errors included in the
experimental covariance matrix Eq. \ref{chi2tot}, and we
compute the following ratio at the value $k_{\chi^2}$ for which
Eq. \ref{chi2tot} has a minimum
\be
\label{ratio}
\frac{E_2 \lp k_{\chi^2}\rp}{E_2 \lp k_2\rp}
=\frac{
1+\frac{(x_1-x_2)^2}{\sigma_1^2+\sigma^2_2}\sigma_f^4
\lp \frac{x_1^2}{\sigma_1^2}
+\frac{x_2^2}{\sigma_2^2}+
\frac{2}{\sigma_f^2}\rp}{\lc 1-
\sigma_f^2\frac{(x_1-x_2)^2}{\sigma_1^2+\sigma_2^2}\rc^2} \ .
\ee
Note that the above quantity depends not only on $(x_1-x_2)$
but also on the absolute magnitude $x_1,x_2$ of the 
measurement.
The above ratio always verifies the property
\be
 \frac{E_2 \lp k_{\chi^2}\rp}{E_2 \lp k_2\rp}
\ge 1 \ .
\ee
Therefore including normalization errors in the minimized $\chi^2$ results
not only in a lower value of the fitted parameter $k$, but also
on a larger diagonal error $E_2$, and the two effects arise from the
same source: combination of inconsistent data and normalization
errors. This effect can be very large even if
normalization errors themselves are small due to
the presence of inconsistent data.
One can explicitely check that the same conclusions hold
in the case that the inconsistent data comes from 
different experiments.

\begin{figure}[ht]
\begin{center}
\epsfig{figure=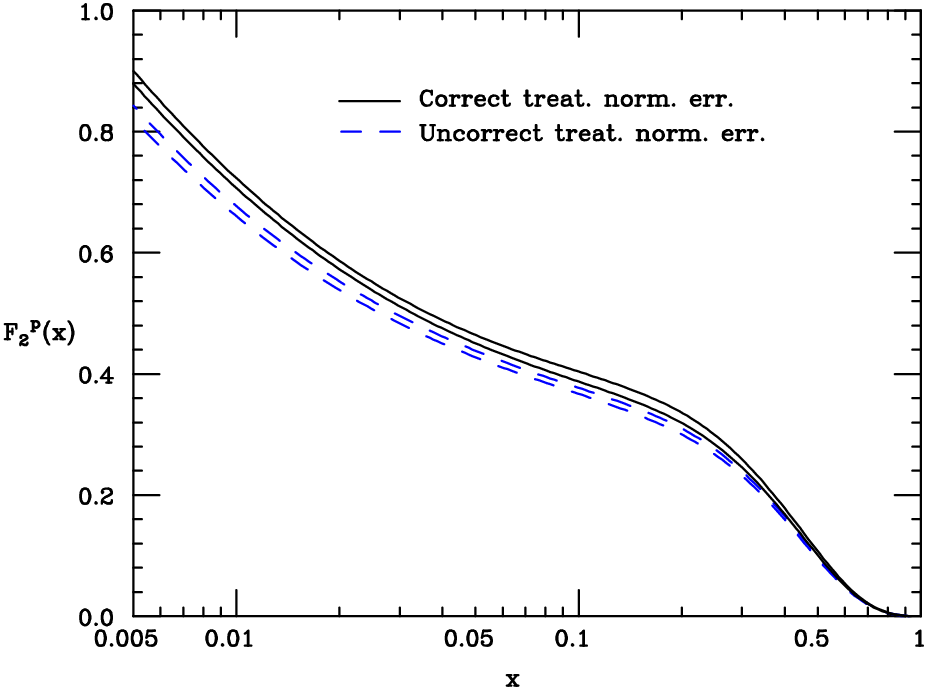,width=0.56\textwidth}
\caption{}{\small Comparison of parametrizations of the
proton structure function $F_2(x,Q^2)$ with the
correct and incorrect treatment of normalization uncertainties.
Note that the effect is much larger than the one
naively expected from the relative size of
normalization errors, $\sigma_N\sim 2-3\%$. }
\label{thesisplot_norm}
\end{center}
\end{figure}

To check the effects of the incorrect treatment of the normalization
uncertainties in a more realistic fit, in Fig.~\ref{thesisplot_norm}
we have repeated the $F_2(x,Q^2)$ parametrization described
in Section \ref{dis_appl} but with the incorrect treatment
of normalization uncertainties, that is, with the
minimization of the error function with the
covariance matrix Eq. \ref{covmat}, rather than with
Eq. \ref{covmatnn}, which does not include the normalization
uncertainties. One observes that the results of the incorrect
treatment of the normalization errors is that the structure
function is systematically lower than the result with
the correct treatment, and this effect is much larger than the
size one would naively expect since normalization errors
are of the order of $2-3\%$.

\newpage
~
\newpage
\chapter{Overview of global parton fits}

\label{pdfstatus}

In this Appendix we summarize the
present status of 
global fits of parton distribution functions.
The standard approach to the determination of parton
distributions from experimental data has been
discussed in detail in Section \ref{globalfits}.
We do not attempt now to review all the huge
available literature in the subject but rather
to provide the reader with a brief description
of the current status of the field.
Much more detailed information can be
obtained from the original references as well
as from proceedings of workshops
like \cite{lh,heralhc}.

During the 1980's and the early 1990's many sets
of parton distributions were developed
to try to describe all the available
hard scattering data \cite{quigg,duke,harriman,diemoz,owens,kwi}.
Today the most commonly used sets
of  parton distributions are those of the CTEQ and 
MRST Collaborations. This is so
because these collaborations
take into account all modern data from a
wide variety of experiments as well as the
progress in perturbative QCD computations, and provide
regular updates of their parton distributions sets. 
Now we review the
current status of the global analysis of these two
groups. Note that even if these two groups release new
versions of their sets rather frequently, in general
these updates are only minor changes of a basic set,
like CTEQ4 or CTEQ5. Note also that all modern
QCD analysis provide  estimations of the uncertainties 
associated to the parton distribution functions.

The MRS(T) Collaboration presented his first global parton fit
in Ref. \cite{mrst_88}. Then this global fit was sequentially improved from
a series of works: \cite{mrst_90,mrst_93,mrst_94,mrst_96,mrst_98,mrst_01,
mrst_02}. One of their latest sets
 of parton distributions is MRST2001 \cite{mrst_01}, which is described in some
detail in the following. 
The experimental data that is used in the fit is given 
by:
\begin{itemize}
\item Neutral current deep-inelastic structure functions
from the H1 and ZEUS experiments at the HERA $e^{\pm}p$
collider.
\item The ZEUS measurement of the charm contribution
to the DIS structure function, $F_2^{c\bar{c}}(x,Q^2)$.
\item The fixed target DIS structure functions measurements from the
CCFR, BCDMS, NMC and E665
 experiments, as well as preliminary data from the
NuTeV experiment, for different types of targets.
\item Inclusive jet cross sections 
from the D0 and CDF  detectors at Fermilab $p\bar{p}$ collider Tevatron.
\item The E866 measurements of the Drell-Yan 
process for both proton and neutron targets, as well
as previous measurements by the E605 experiment. 
\item The measurement of the W-lepton asymmetry from the CDF
detector at Tevatron.
\end{itemize}
Note that the prompt photon data, that was used for some time
in parton fits, is not included any more due to
theoretical problems as well as possible inconsistencies.

\begin{figure}[ht]
\begin{center}
\epsfig{figure=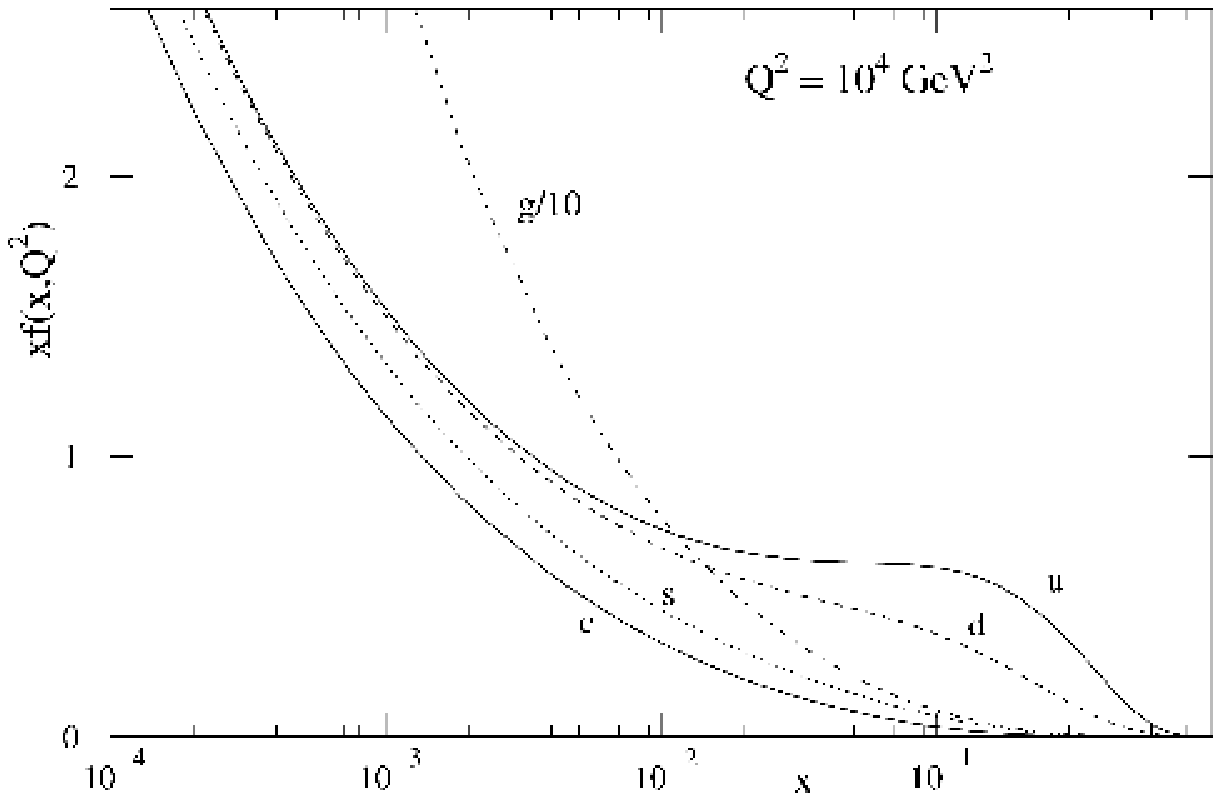,width=0.62\textwidth}
\caption{}{\small The MRST2001 partons at the
scale $Q^2=10^4$ GeV$^2$.}
\label{pdfsetap_mrst}
\end{center}
\end{figure}

MRST2001 is a global NLO QCD analysis with starting evolution scale
$Q_0=1$ GeV, that uses the $\overline{MS}$ renormalization scheme
and the Thorne-Roberts scheme for the treatment of heavy quark mass
effects. The kinematical cuts are given by $Q^2\ge 2$ GeV$^2$
and $W^2\ge 12.5$ GeV$^2$.
The parton distributions at the initial
evolution scale are parametrized by
\be
xu_V(x,Q_0^2)=x\lp u-\overline{u}\rp(x,Q_0^2)=
A_ux^{b_u}(1-x)^{c_u}\lp 1+d_u\sqrt{x}+e_ux\rp \ ,
\ee
\be
xd_V(x,Q_0^2)=x\lp d-\overline{d}\rp(x,Q_0^2)=
A_dx^{b_d}(1-x)^{c_d}\lp 1+d_d\sqrt{x}+e_dx\rp \ ,
\ee
\be
xS(x,Q_0^2)=
A_sx^{b_s}(1-x)^{c_s}\lp 1+d_s\sqrt{x}+e_s x\rp \ ,
\ee
\be
xS(x,Q_0^2)\equiv 2x\lp \overline{u}+\overline{d}+\overline{s}\rp \ ,
\ee
\be
xg(x,Q_0^2)=
A_gx^{b_g}(1-x)^{c_g}\lp 1+d_g\sqrt{x}+e_g x\rp-F_gx^{g_g}(1-x)^{h_g} \ ,
\ee
\be
2\overline{u},2\overline{d},2\overline{s}=0.4S+\Delta,
0.4S-\Delta,0.2S \ ,
\ee
\be
x\Delta=x\lp \overline{d}-\overline{u}\rp=
A_{\Delta}x^{b_{\Delta}}(1-x)^{c_{\Delta}}\lp 1+d_{\Delta}x+
e_{\Delta} x^2\rp \ .
\ee
As well as the parameters that describe the nonperturbative shape
of the parton distributions, for this analysis also the
strong coupling $\aqz$ is fitted, resulting in a value
consistent with the current world average \cite{bethke}.
The associated uncertainties to the above
parton distributions from experimental errors were discussed in detail
in Ref. \cite{mrst2001e} and those
associated to experimental uncertainties, like the perturbative
order, higher twist corrections or $\ln 1/x$ and $\ln (1-x)$
effects,
in Ref. \cite{mrst_th}. In Fig. \ref{pdfsetap_mrst} we show
the parton distributions that result from the global
analysis discussed above at the scale $Q^2=10^4$ GeV$^2$.
Note that at such a large energy scale, the gluon distribution
becomes dominant at medium and small $x$, and the contribution
from heavy quarks becomes sizeable. At large evolution
lengths $Q^2$ the shape of the parton distributions is
essentially determined by perturbative evolution and becomes
less dependent of the initial nonperturbative condition
at $Q_0^2$.

The CTEQ Collaboration 
has performed global analysis
of parton distributions since the early 90s \cite{cteq_90,
cteq3,cteq4,cteq5}
until now. One of their latest work is
named CTEQ6 \cite{cteq6}, that is
summarized in the following. As will be
seen this analysis is very close to that of the MRST2001 analysis
discussed above. CTEQ6 uses as experimental input:
\begin{itemize}
\item Neutral current deep-inelastic structure functions
from the H1 and ZEUS experiments ate the HERA $e^{\pm}p$
collider.
\item The fixed target DIS structure functions measurements from the
CCFR, BCDMS and NMC experiments.
\item Inclusive jet cross sections in several rapidity bins
from the D0 detector at Fermilab $p\bar{p}$ collider Tevatron.
\item The E866 measurements of the Drell-Yan deuteron
to proton ratio, and the E605 measurement
of the Drell-Yan cross section.
\item The measurement of the W-lepton asymmetry from the CDF
detector at Tevatron.
\end{itemize}
For all these experiments, all the information on correlated systematic
uncertainties is available. Note that even if global
QCD analysis succeed in describing a wide variety of
hard-scattering data, the precision DIS structure
function measurements from HERA and fixed target
experiments still provide the backbone
of parton distribution analysis.

\begin{figure}[ht]
\begin{center}
\epsfig{figure=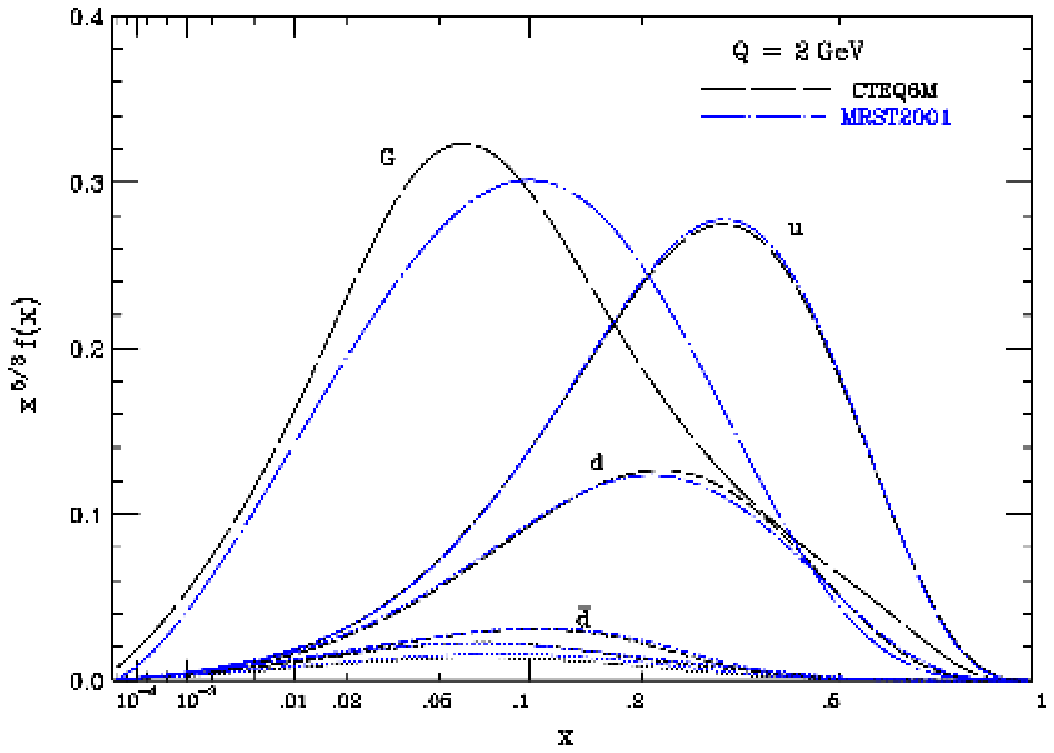,width=0.62\textwidth}
\caption{}{\small A comparison of the CTEQ6M partons
with the MRST2001 partons at the initial evolution scale
$Q=2$ GeV.}
\label{pdfsetap_cteq}
\end{center}
\end{figure}

The nonperturbative input to the parton global analysis, as has been
discussed in Section \ref{globalfits}, are the parametrization
of the parton distributions at a starting evolution scale, which
in the CTEQ6 set is taken to be $Q_0=1.3$ GeV.
Let us recall (see Section \ref{dis_theo}) that the parton
distributions at $Q\ge Q_0$ are determined by the NLO DGLAP
evolution equations. The functional form used in the CTEQ6
analysis is
\be
xf(x,Q_0)=A_0x^{A_1}(1-x)^{A_2}e^{A_3x}\lp 1+e^{A_4}x\rp^{A_5} \ ,
\ee
with independent parameters for the flavor combinations $u-\overline{u}$,
$d-\overline{d}$, g, and $\overline{u}+\overline{d}$.
The strange parton distribution is kept fixed to $s=\overline{s}
=0.2 \lp \overline{u}+\overline{d}\rp/2$.
Also the sea quark ratio 
\be
\frac{\overline{d}}{\overline{u}}=
A_0x^{A_1}(1-x)^{A_2} +\lp1+A_3x\rp\lp 1-x\rp^{A_4} \ ,
\ee
is parametrized. Some parameters are held fixed, for a total
of 20 free parameters to model the nonperturbative
parton distribution shape at the input scale $Q_0$.
From the CTEQ6 global analysis not only the best-fit set
of parton distributions are determined from experimental data,
also the associated uncertainties in the parton distributions
are estimated using some of the methods discussed in Section
\ref{globalfits}, see the original reference \cite{cteq6} for a
more detailed description of the results of this
global analysis.

Note that the two main groups performing global fits
of parton distributions, MRST and CTEQ, use a very similar
set of experimental data, similar assumptions on the
nonperturbative shape of the parton distributions and
quite similar methods to determine the associated errors 
to the parton distributions. This means that the spread of the 
MRST and CTEQ results by itself cannot be taken, even at 
a qualitative level, as a measure of the uncertainty
in the determination of parton distributions. In Fig. \ref{pdfsetap_cteq}
we show the results of the CTEQ6 global QCD analysis
discussed above compared with the MRST2001 partons.
Note the good agreement between the two analysis for the $u(x)$
and $d(x)$ partons, while the difference is larger for the gluon,
as was to be expected since the gluon distribution has rather
large uncertainties.

Finally let us review some other recent determinations
of parton distributions. These global analysis
 are less frequently used in phenomenological studies
than the sets from the CTEQ and MRST collaborations, since
in none of the following cases all the available
hard-scattering data is included.
S. Alekhin has presented several QCD analysis
of deep-inelastic scattering data \cite{alekhin1,alekhin2},
with emphasis in the consistency of the
statistical analysis of the data.
The GRV published a series of parton analysis \cite{grv1,grv2,
grv3,grv4}
whose main feature was that a very low
starting evolution scale $Q_0^2$ for the evolution.
However, since the release of their last set
of parton distributions \cite{grv4}
no updates have appeared. In particular
this group did not produce an estimation of the
associated uncertainties in their parton distributions.
Finally, in the last years different groups have published additional
global fits of parton distributions, each one
with different motivations. For example the
analysis of 
BPZ \cite{bpz} devotes special attention to the
accurate determination of the sea strange parton distribution, while
the  Fermi partons \cite{fermi}, which have some overlap with
our approach, also attempt to construct a probability
measure in the space of parton distributions.

Summarizing, global fits of parton distributions
have been a field of active development in the
recent years. Two sets of parton distributions
(MRST and CTEQ) are nowadays commonly used
in most phenomenological applications, since they
include all available experimental data and provide
regular updates of their parton sets. Note that as
discussed before,
 in both cases the experimental data, the theoretical
assumptions and the technique to assess the uncertainties
are rather similar, and therefore theoretical
predictions for observables using the two sets of
parton distributions are in general in good agreement.

\providecommand{\href}[2]{#2}\begingroup\raggedright\endgroup

~

\newpage

\end{document}

%% file: agradecimientos.tex
\begin{flushright}

~

~

~

~

~

~

~

~

~

~

~

~
{\it A S\`onia}
~

~

~

~

\end{flushright}

\clearpage

~

\newpage

\begin{flushright}
~

~

~

~

Solamente el estupor conoce. \\
{\it San Gregorio de Nisa}

~

~

~

~

Mucho razonamiento y poca observaci\'on llevan al error, \\
Mucha observaci\'on y poco razonamiento llevan a la verdad. \\
{\it Alexis Carrel}

~

~

~

~

Uno no puede evitar asombrarse cuando contempla los misterios de la
 eternidad, \\
de la vida, de la maravillosa estructura de la realidad. \\
 Es suficiente si uno
trata simplemente de comprender un poco de este misterio cada d\'ia. \\
No hay que perder jam\'as la sagrada curiosidad. \\
{\it Albert Einstein}

\end{flushright}

\clearpage

~

\newpage

~

~

~

~
\begin{center}
\bf \Large Agradecimentos
\end{center}

~

~

~

~

La realizaci\'on de una tesis doctoral, m\'as que una obra
individual, es por encima de todo una tarea comunitaria.
Por esto los agradecimientos son el momento en el
cual se tienen en cuenta la contribuci\'on de todas aquellas
personas que con sus aportaciones de todo tipo han
permitido que la presente tesis doctoral haya llegado a buen puerto.

En primer lugar quisiera agradecer a Jos\'e Ignacio Latorre
por acompa\~narme estos cuatro a\~nos como director de esta
tesis doctoral. Desde sus primeras clases de F\'isica de Altas 
Energ\'ias, me cautiv\'o la pasi\'on con la que explicaba, que
es la misma con la que vive la aventura de la investigaci\'on cient\'ifica.
Jos\'e Ignacio ha sido para m\'i no solo un maestro en el
campo de la ciencia sino tambi\'en un modelo por
su honestidad en todos los \'ambitos.

Tambi\'en aprovecho para agradecer  al codirector de
esta tesis doctoral, Stefano Forte. 
Con Stefano he aprendido el rigor necesario
para ser un buen cient\'ifico, as\'i como la
pasi\'on necesaria para afrontar incluso los
problemas m\'as dif\'iciles sin desfallecer. Su ayuda
continua han sido esenciales para llevar a cabo el trabajo
de investigaci\'on desarrollado en esta tesis
doctoral.

Muchas han sido las personas con las cuales durante estos
cuatro a\~nos he compartido el placer de la
investigaci\'on cient\'ifica, aprendiendo continuamente
de todos ellos. De manera especial a los otros miembros
de la NNPDF Collaboration, Andrea Piccione y Luigi Del Debbio,
pero tambi\'en a muchas otras personas: Antonio Pineda, Jorge
Russo, Giovanni Ridolfi, Concha Gonzalez-Garc\'ia, Michele
Maltoni... 
Asimismo, querri\'a agradecer a todas aquellas personas
con las que he compartido siempre instructivas discusiones
sobre f\'isica, y con las cuales he aprendido poco a poco
el arte de ser investigador. La lista es muy larga: Germ\'an
Rodrigo, Joan Soto,  Ignazio Scimemi, Thomas Becher, Einan Gardi ...

Tambi\'en querr\'ia agradecer a todos aquellos compa\~neros del
departamento de Estructura i Contituents de la Mat\`eria  con los que
he pasado  estos a\~nos, compartiendo la aventura de realizar una tesis
doctoral: Xavi, Enrique, Rom\'an, Diego, Dani, M\'iriam, Luca, Jaume, Alex ...
Pero especialmente  a Manel y Arnau, pues desde
que nos conocimos en el primer curso de la licenciatura, nos
hemos acompa\~nado a lo largo de toda la carrera
y el doctorado. Parece que aquel dia tan lejano que
apenas dislumbrabamos cuando hac\'iamos Fonaments de F\'isica
 ahora est\'a cerca para los tres.
Tambi\'en estoy agradecido a Joan Soto y Nuria Barber\'an por
comunicarme tanto su pasi\'on como su rigor en la
docencia de las asignaturas  que yo he colaborado a impartir en los
dos \'ultimos a\~nos.
Finalmente, agradecer a Oriol y a Rosa su continua ayuda en 
tantos  detalles pr\'acticos que han surgido en estos
a\~nos.

Son tantos los amigos que a lo largo de estos cuatro a\~nos me
han acompa\~nado en la apasionante aventura de la vida, haciendo
posible renovar continuamente la pasi\'on por todo, incluyendo
la investigaci\'on cient\'ifica, que pido perd\'on por adelantado
por todos aquellos que no tengo presentes.
A Juan Ram\'on y a Roger especialmente por su
inasequible apoyo y su
continua ayuda en el cuidar la no siempre f\'acil vocaci\'on
cient\'ifica. A Llu\'is por comunicarme su pasi\'on
por el conocimiento y por la vida, y por proponerme una
amistad  que dura hasta hoy. A todos aquellos amigos 
que hemos vivido juntos  estos a\~nos: N\'estor, Raquel, Jorge, Xavi,
Alfonso, Anna, Miquel C., Cristina, Miquel, Josep C.,
Josep M.,  Marc, Albert ...,
gracias una vez m\'as por vuestra infatigable compa\~nia
en este camino.

Tambi\'en estoy profundamente agradecido a los amigos
de f\'isica de Mil\'an: Giuliano, Tommy, Betta, Paola,
Maria, y a toda la asociaci\'on cultural Euresis, por
ayudarme a vivir mi vocaci\'on cient\'ifica con un horizonte
abierto a toda la realidad. Tambi\'en aprovecho para
agradecer a los amigos de Madrid de la Asociaci\'on Universitas,
por su empe\~no continuo en vivir la vida acad\'emica
siempre consciente de las razones de la propia vocaci\'on, y por
tanto posibilitando renovar siempre el inter\'es tanto
por la docencia como por la investigaci\'on.

Sin embargo, por encima de todo quer\'ia agradecer 
a mi familia todo el apoyo y la ayuda incansable que me 
han proporcionado a lo largo
todos estos a\~nos. 
Mi padre Eduardo ha sido desde siempre mi referencia, tanto
a nivel acad\'emico como a nivel personal. La pasi\'on y la
alegr\'ia con la que mi padre vive su trabajo en la Universidad
han sido mi mayor motivaci\'on para empezar f\'isica primero,
y decidirme por la carrera acad\'emica despu\'es. 
Mi madre Carmen tambi\'en me ha ayudado siempre a valorar
el estudio y el trabajo, educandome 
en el inter\'es por toda la realidad, algo
que nunca podr\'e agradecer
suficientemente. Mi hermano Ignacio ha sido
siempre modelo para m\'i, por la pasi\'on y seriedad
con la que ha vivido siempre el estudio, y ahora el trabajo.
Tambi\'en estoy muy agradecido a mi
familia pol\'itica, Ra\"ul, Margarita, Olga y Montserrat, por su
continuo apoyo a todos los niveles en la realizaci\'on 
de esta tesis doctoral.

Finalmente, todo mi agradecimiento 
va dirigido a mi mujer, S\`onia.
Gracias a ella, a su ayuda y su est\'imulo,
 he podido realizar la presente tesis doctoral.
Su apoyo infatigable en todas las circunstancias ha sido
lo que me ha permitido empezar cada d\'ia con una ilusi\'on
plenamente renovada, tanto en la investigaci\'on como en 
la docencia.
Por todo ello, no puedo m\'as que agradecerle otra vez
todos estos a\~nos en que nos hemos acompa\~nado en
la apasionante aventura de la vida y el matrimonio.

%% file: introduccion.tex
\chapter{Introducci\'on}

\begin{figure}[ht]
\begin{center}
\includegraphics[scale=0.28]{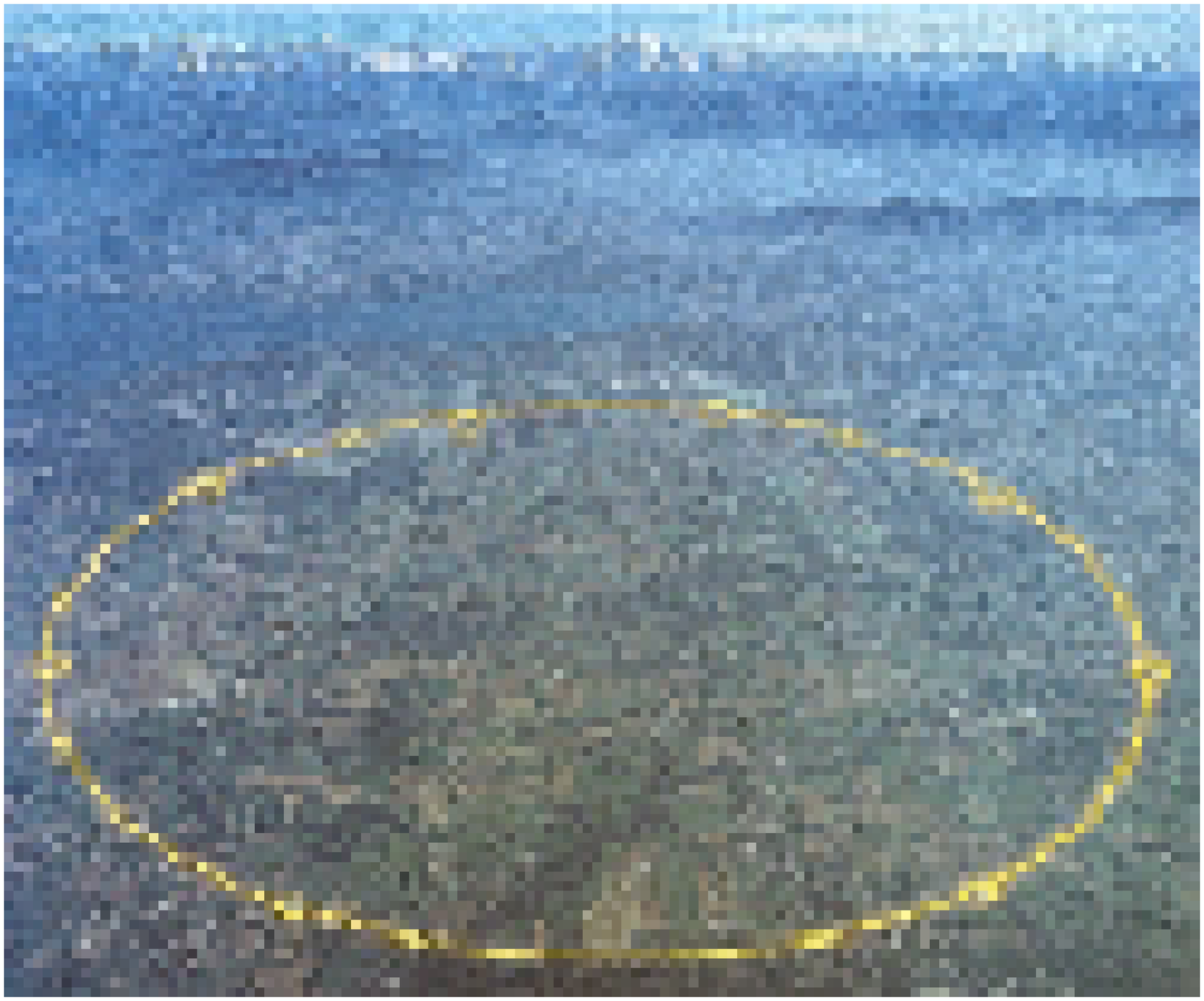}
\epsfig{figure=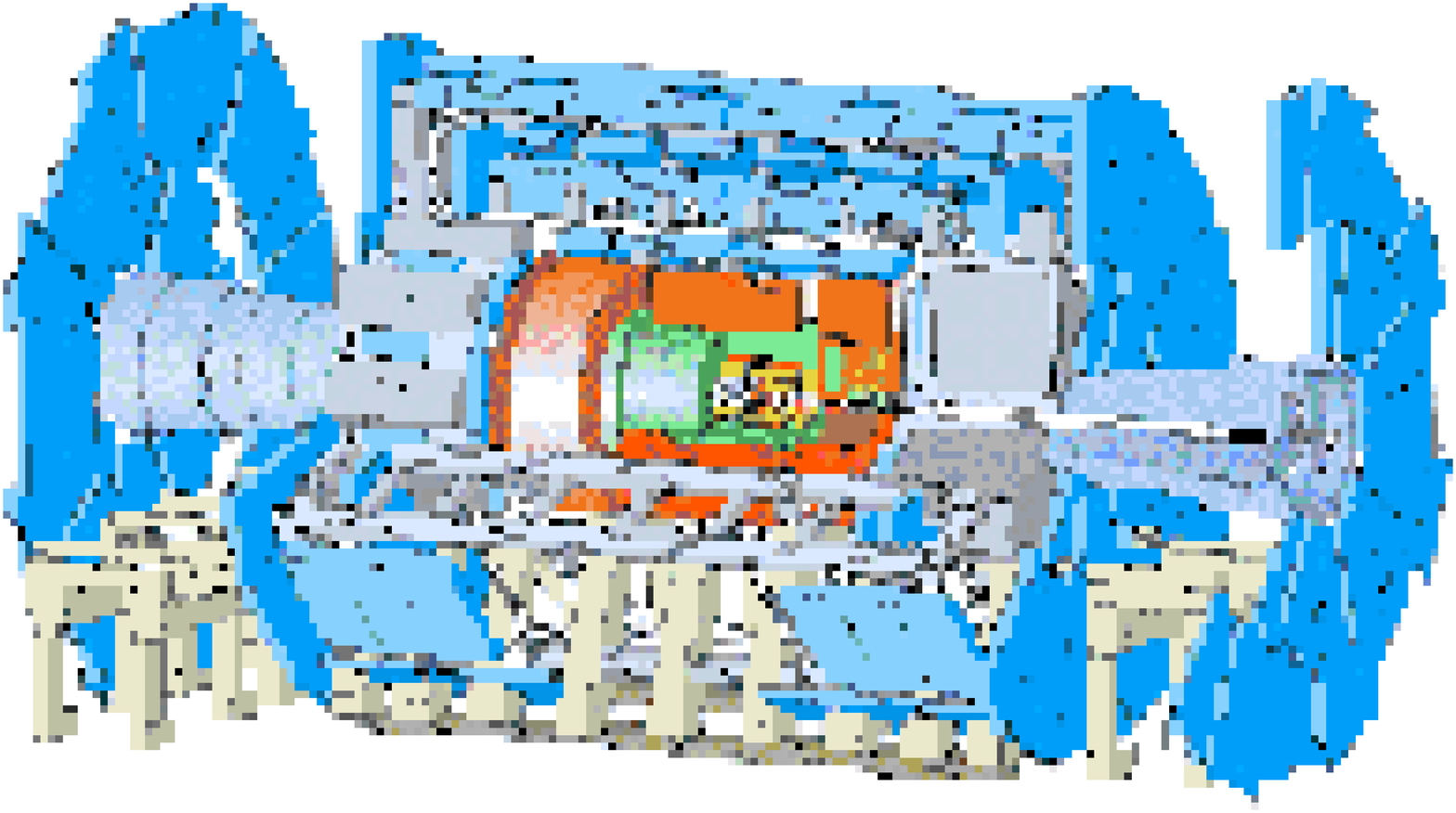,width=0.45\textwidth}
\caption{}{\small La localizaci\'on del t\'unel
de 27 kil\'ometros donde est\'a situado el LHC, cerca de
Ginebra (los Alpes pueden ser vislumbrados detr\'as)
(izquierda)
y uno de sus detectores, ATLAS, que es tan grande
como un edificio de siete pisos (derecha).}
\label{lhc2}
\end{center}
\end{figure}

El Large Hadron Collider (LHC, gran colisionador de hadrones)
es un colisionador de protones a las energ\'ias m\'as elevadas 
conseguidas artificialmente por el hombre, situado
en el Centro Europeo para la Investigaci\'on Nuclear,
el CERN en Ginebra (Suiza). Este acelerador de 
part\'iculas est\'a actualmente en construcci\'on, y se espera
que las primeras colisiones de protones se produzcan antes 
del final del a\~no 2007. Las enormes energ\'ias que se
alcanzar\'an en las colisiones entre protones en el LHC
nos permitir\'an examinar las leyes fundamentales de la naturaleza
a las menores distancias jam\'as investigadas. En particular, se
estudiar\'a en detalle el mecanismo de la ruptura de simetr\'ia
electrod\'ebil, mediada por la part\'icula de Higgs, que es 
la responsable de dar las masas a todas las part\'iculas 
elementales conocidas. Adem\'as, se espera que LHC nos
proporcione  informaci\'on detallada sobre nueva f\'isica
mas all\'a del Modelo Estandard de F\'isica de Part\'iculas, que
ha sido construido y comprobado con enorme \'exito
durante los ultimos 25 a\~nos. En la fig. \ref{lhc2}
podemos ver la localizaci\'on del experimento LHC cerca
de Ginebra, as\'i como uno de sus detectores, ATLAS, donde
se examinan los resultados de la colisiones entre protones.

\begin{figure}[ht]
\begin{center}
\epsfig{figure=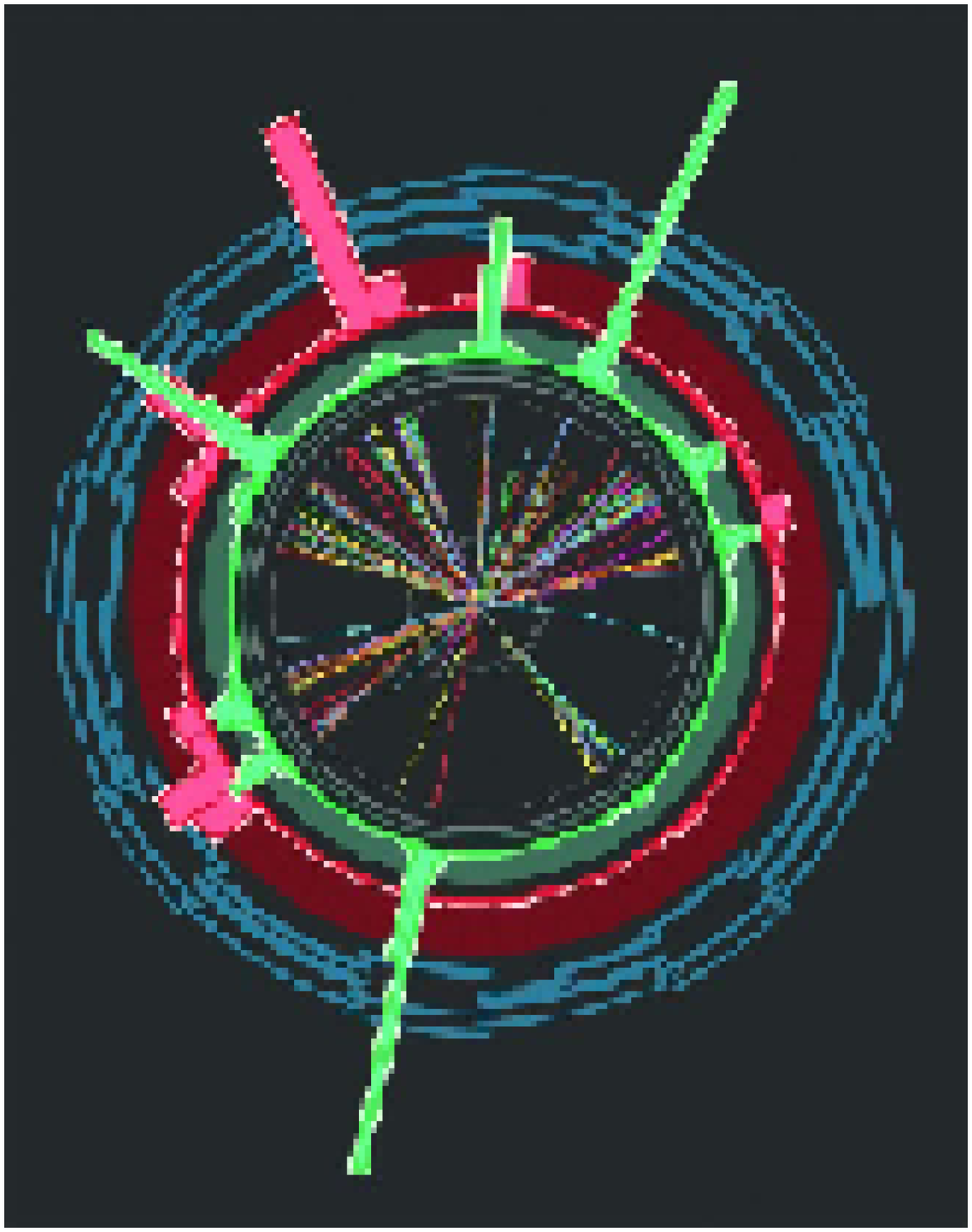,width=0.40\textwidth}
\caption{}{\small Resultado de una colisi\'on t\'ipica entre
protones en el LHC: podemos observar las trazas que dejan los
cientos de part\'iculas producidas en cada colisi\'on, as\'i como
la energ\'ia total que cada una de estas part\'iculas lleva
asociada.}
\label{colli}
\end{center}
\end{figure}

Sin embargo, la extracci\'on de la nueva f\'isica de las
colisiones prot\'on-prot\'on a altas energ\'ias que se producir\'an
en el LHC es una tarea extremadamente complicada, por
una serie de motivos que detallaremos a continuaci\'on.
El principal de estos motivos es que esta nueva f\'isica estar\'a
escondida entre una multitud de procesos de f\'isica conocida,
debido a la interacci\'on fuerte entre quarks y gluones
(que son las part\'iculas elementales que componen los
protones) determinadas por el Modelo Estandard, en
particular por la teor\'ia conocida como
Cromodin\'amica Cu\'antica (Quantum Chromodynamics, QCD).
Estos procesos ser\'an mucho m\'as frecuentes que las colisiones
en donde se produzcan los efectos buscados de nueva f\'isica.
Por lo tanto, el potencial de descubrimiento del LHC, as\'i
como su habilidad para realizar medidas de precisi\'on de
las propiedades de esta nueva f\'isica, dependen de una manera
crucial de nuestra comprensi\'on quantitativa de los procesos
de la interacci\'on fuerte y las incertidumbres que
estos llevan asociados. En la fig. \ref{colli}
podemos ver el resultado de un proceso caracter\'istico
de colisi\'on entre protones en el LHC.
Es necesario recalcar que de los 
miles de millones de colisiones como esta que se producir\'an
cada a\~no en el LHC,
ser\'a necesario extraer aquellas pocas que contienen
informaci\'on aut\'enticamente relevante.

Entre las diversas fuentes de incertidumbre asociadas
a los procesos que involucran part\'iculas
con interacci\'on fuerte, una de las de mayor
importancia es debida a las
distribuciones de partones 
(Parton Distribution Functions, PDFs) en el
interior del prot\'on.
Estas distribuciones son una medida de la cantidad
de la energ\'ia total del prot\'on que lleva cada uno
de sus diversos componentes, los quarks de diferente
sabor y los gluones. Estas distribuciones de partones
no pueden ser calculadas en teor\'ia de perturbaciones, sino
que necesitan ser extraidas de los datos experimentales
que provienen de una gran cantidad de procesos de f\'isica de altas
energ\'ias, como por ejemplo
las colisiones profundamente inel\'asticas
(Deep inelastic scattering, DIS) entre leptones
(part\'iculas elementales sin interacci\'on fuerte)
y protones.

\begin{figure}[ht]
\begin{center}
\epsfig{figure=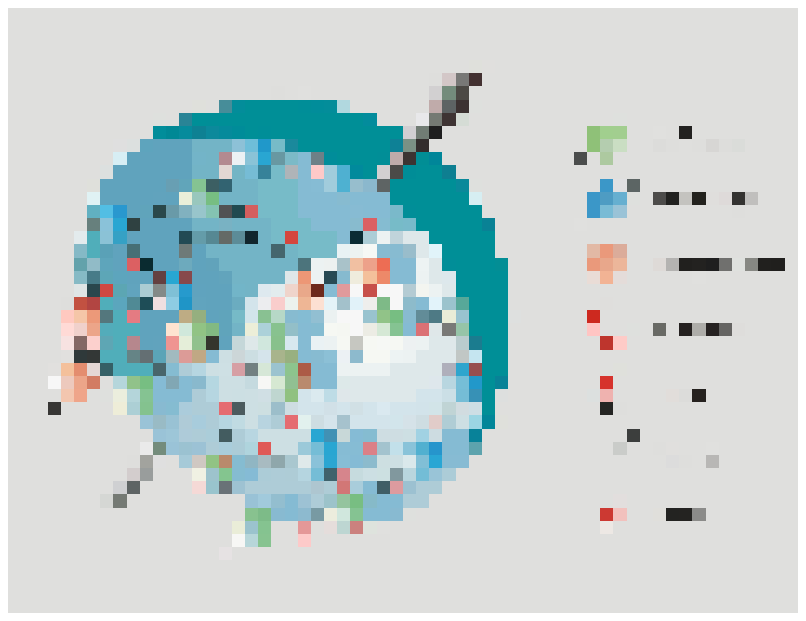,width=0.60\textwidth}
\caption{}{\small El interior de un prot\'on 
se compone de quarks y antiquarks de diferentes sabores,
junto con gluones que mediante la interacci\'on fuerte mantienen
a estos quarks confinados en el interior del prot\'on.}
\label{protonint}
\end{center}
\end{figure}

Las distribuciones de partones las denotaremos por
\be
q_i(x,Q^2) \ , \quad i=1,\ldots,2N_f+1 \ ,
\ee
donde $Q^2$ es la energia t\'ipica del proceso
de colisi\'on, la variable $x$ denota la fracci\'on de la
energ\'ia total del prot\'on que lleva el
parton $i$, y tenemos una distribuci\'on de partones
independiente para cada quark y cada antiquark
de diferente sabor, m\'as una para los gluones.
 Hay que tener en cuenta que en QCD las
energ\'ias son grandes o peque\~nas en comparaci\'on con la masa
del proton $M_p$, que constituye la escala caracter\'istica de la teor\'ia.
Estas distribuciones de partones pueden ser determinadas
gracias a la existencia del teorema conocido
como el Teorema de la Factorizaci\'on en Cromodin\'amica
Cu\'antica. Segun este teorema, cualquier secci\'on eficaz
de colisi\'on
(que mide la probabilidad de que dos part\'iculas
colisionen y interaccionen) en procesos que
involucren la interaci\'on fuerte puede ser separada
en el producto de dos t\'erminos: un coeficiente que
depende del proceso en cuesti\'on, que podemos calcular
en teor\'ia de perturbaciones, y un conjunto
de distribuciones de partones que son universales,
es decir, que son independientes de los detalles
particulares del proceso. En la fig.~\ref{protonint}
observamos un esquema del interior de un prot\'on,
con los diferentes quarks interaccionando entre si
mediante los gluones, y donde la flecha indica el
{\it spin}, el momento angular interno de cada part\'icula.
El movimiento de estos quarks y gluones en el prot\'on viene
dictado por estas distribuciones de partones.

Por ejemplo, las
funciones de estructura en colisiones
profundamente inel\'asticas pueden escribirse como
\be
\label{structesp}
F(x,Q^2)=C_i\lp x,\aq\rp \otimes q_i(x,Q^2) \ ,
\ee
es decir, como la convoluci\'on entre un coeficiente $C_i(x)$
que depende del proceso
y las distribuciones de partones $q_i(x)$, que son id\'enticas
para todo proceso de colisi\'on a altas energ\'ias que involucre
part\'iculas con la interacci\'on fuerte en el estado inicial.
Solamente es necesario determinar las distribuciones de partones
a una escala inicial  $Q^2_0$ relativamente 
peque\~na, $Q^2_0\sim M_p^2$, pues su dependencia con la energ\'ia
(denominada {\it evoluci\'on})
viene dictada por la teor\'ia de perturbaciones de la 
Cromodin\'amica Cu\'antica. Es necesario tener en cuenta
que en general diferentes combinaciones de distribuciones
de partones contribuyen a cada observable de
manera diferente.
Aunque la mayor fuente de informaci\'on experimental
sobre las distribuciones de partones proviene de las
medidas de alta precisi\'on en los procesos de colisi\'on profundamente
inel\'asticos descritos anteriormente,
 otros procesos son esenciales como producci\'on
de {\it jets} (conjuntos de hadrones, es decir, de
part\'iculas que interaccionan fuertemente)
en colisiones de protones o el proceso de Drell-Yan,
esto es, la producci\'on de parejas de leptones tambi\'en en colisiones
entre hadrones.

\begin{figure}[ht]
\begin{center}
\epsfig{figure=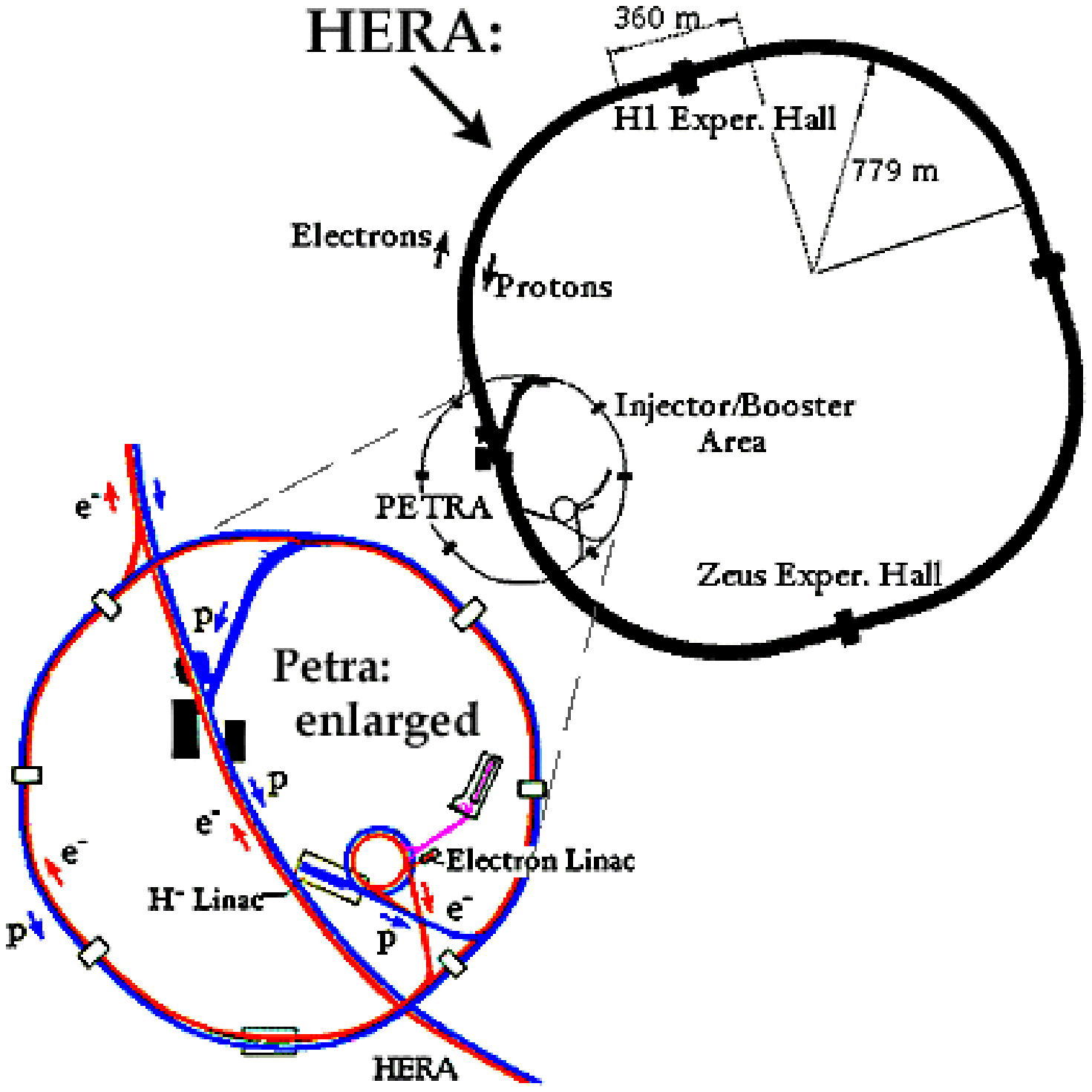,width=0.60\textwidth}
\caption{}{\small Esquema del experimento
de colisiones profundamente inel\'asticas
HERA en Hamburgo.}
\label{disexp}
\end{center}
\end{figure}

El requisito de poder realizar f\'isica de precisi\'on en colisionadores
de protones, como el futuro LHC, implica que es necesario
determinar con la mayor exactitud posible no solamente las
diferentes distribuciones de partones en el prot\'on,
$q_i(x,Q_0^2)$, sino tambi\'en las incertidumbres que estas
tienen asociadas. Estas incertidumbres provienen del hecho
de que puesto que las distribuciones de partones
se extraen de  datos experimentales, que tienen una precisi\'on
finita, tambi\'en estas tendr\'an a su vez
una precisi\'on finita, es decir,
un error experimental asociado. Para ver la importancia
capital que las distribuciones de partones tendr\'an en el LHC, es
necesario notar que una secci\'on eficaz de colisi\'on t\'ipica
tiene la forma 
\be
\label{sigmaexp}
\sigma(x,Q^2)= C_{ij}\lp x,\aq\rp \otimes q_i(x,Q^2) \otimes
q_j(x,Q^2) \ ,
\ee
es decir, que depende de dos distribuciones de partones, una para
cada de los partones que involucran las colisiones prot\'on-prot\'on.
Este problema es particularmente grave pues la region cinem\'atica
que cubrir\'a el LHC ha sido solamente parcialmente cubierta
por aquellos experimentos que han sido usados
para determinar las distribuciones de partones con
anterioridad, como HERA y SLAC
(ver fig. \ref{disexp}), y por lo tanto
ser\'a necesario extrapolar las distribuciones de partones
a regiones donde nunca han sido medidas. Es importante,
por lo tanto, controlar de manera muy precisa las
incertidumbres asociadas a este proceso de extrapolaci\'on.

\begin{figure}[ht]
\begin{center}
\epsfig{figure=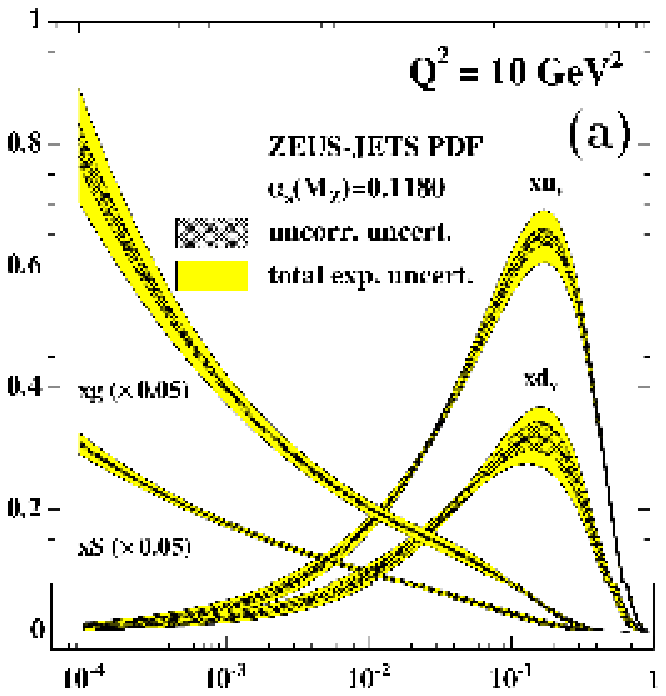,width=0.50\textwidth}
\caption{}{\small Un ejemplo de una determinaci\'on
reciente de distribuciones de partones. Notemos que
cada distribuci\'on tiene asociada una banda de incertidumbre,
que es una medida de los errores de cada una
de las distribuciones de partones.}
\label{zeuspdfs}
\end{center}
\end{figure}

El problema principal, en vista de todo lo descrito anteriormente,
es determinar los errores de una funci\'on continua como son
las distribuciones de partones, es decir, una densidad de probabilidad
en el espacio de estas funciones, a partir de un conjunto finito
de datos experimentales.
Existe toda una serie de t\'ecnicas estandard para determinar las
distribuciones de partones a partir de las medidas
experimentales, y diversas estrategias han sido usadas
para estimar de maneras diversas adem\'as los errores asociados
a estas distribuciones. Todas estas estrategias
han sido  \'utiles para estimar de manera
aproximada el tama\~no de estas incertidumbres, pero sin embargo
sufren de un cierto n\'umero de problemas. Con las
t\'ecnicas estandard, las distribuciones de partones
se parametrizan con funciones relativamente simples,
t\'ipicamente polinomios de la forma
\be
\label{pdfesp}
q_i(x,Q^2_0)=A_ix^{b_i}\lp 1-x\rp^{c_i}\lp 1+d_ix+e_ix^2\rp \ ,
\qquad i=1,\ldots, N_{f}+1 \ ,
\ee
donde los parametros $A_i,b_i\ldots$ se determinan a partir
de un {\it fit} de los datos experimentales. El primer problema
aparece de immediato: nos estamos restringiendo
al espacio de distribuciones de partones parametrizadas
por la ec.~\ref{pdfesp}, lo cual est\'a claramente injustificado
pues no hay  en la Cromodin\'amica Cu\'antica, la teor\'ia que
en principio determina la forma de estas distribuciones, 
nada que implique
 que las distribuciones de partones han de tener la forma
funcional tan espec\'ifica que podemos observar en 
la ec.~\ref{pdfesp}. Segundo, si se quieren propagar
los errores que la ec.~\ref{pdfesp} lleva asociados a otros
observables, como secciones eficaces de la forma
de la ec~\ref{sigmaexp}, son necesarias aproximaciones
de linealizaci\'on, que como es bien conocido no son v\'alidas
en un amplio rango de situaciones.
Finalmente, el tercer inconveniente que presenta la t\'ecnica estandard
es que en presencia de datos experimentales que provienen
de experimentos diferentes, la presencia de incompatibilidades
entre los datos (por ejemplo, que la funci\'on de estructura,
ec.~\ref{structesp}, medida en dos experimentos
distintos sea muy diferente) implica que la condici\'on para determinar
los errores a partir de la funci\'on de error $\chi^2$ no
es $\Delta\chi^2=1$, como indica la estad\'istica b\'asica,
sino $\Delta\chi^2=100$. En la pr\'actica esta elecci\'on implica
que los errores de las distribuciones de partones
no tienen ning\'un significado estad\'istico 
riguroso pues
dependen de un par\'ametro arbitrario $\Delta\chi^2$.
En la fig.~\ref{zeuspdfs} tenemos un ejemplo
de una determinaci\'on reciente de  distribuciones
de partones junto con las incertidumbres asociadas.
Notemos que hay dos tipos de errores asociados: los errores
estad\'isticos habituales (debido a disponer de un n\'umero
finito de medidas) y los errores sistem\'aticos,
que dependen en general del proceso de medida y que
est\'an correlacionados entre diferentes datos
experimentales.

Por todas las razones descritas hasta ahora, y en vista
de la importancia crucial que una estimaci\'on fidedigna
de las incertidumbres asociadas a las distribuciones
de partones tiene para la f\'isica de precisi\'on en el LHC, es claramente
deseable investigar estrategias alternativas 
para la determinaci\'on
de estas distribuciones que permitan superar
y mejorar los problemas de la t\'ecnica estandard descrita anteriormente.
En \cite{f2ns} una novedosa t\'ecnica fue presentada consistente
en la determinaci\'on de la densidad de probabilidad en el espacio
de las funciones, que fue aplicada a la parametrizaci\'on de
funciones de estructura, como la ec.~\ref{structesp},
en procesos de colisi\'on profundamente inel\'asticos entre
leptones y protones. Esta novedosa t\'ecnica usaba una 
combinaci\'on de m\'etodos Monte Carlo para la construcci\'on
de {\it samplings} de los datos experimentales junto con
redes neuronales artificiales como interpolantes universales.
El uso de redes neuronales artificiales en lugar de funciones
polin\'omicas como la eq.~\ref{pdfesp} en la parametrizaci\'on
de las distribuciones de partones permite eliminar
la dependencia de los resultados en la forma funcional
escogida arbitrariamente. Por su parte, la t\'ecnica de
los {\it samplings} Monte Carlo permite una estimaci\'on estad\'isticamente
rigurosa de las incertidumbres asociadas a la funci\'on que
estamos parametrizando, y adem\'as la propagaci\'on de estos
errores a otros observables se puede realizar con toda
generalidad, sin necesidad de aproximaciones de linearizaci\'on.

En la presente tesis doctoral hemos extendido los resultados
presentados en \cite{f2ns}, aplicados a las
funciones de estructura en colisiones profundamente
inel\'asticas, en diversas direcciones.
En primer lugar hemos aplicado esta t\'ecnica general
para parametrizar datos experimentales a 
otros procesos de inter\'es en f\'isica de altas energ\'ias.
Los procesos que han sido estudiados son las desintegraciones
hadr\'onicas del lepton {\it tau}, las desintegraciones
semilept\'onicas del mes\'on B, y las colisiones
profundamente inel\'asticas desde dos puntos de vista:
a nivel de funciones de estructura, incluyendo
todos los datos experimentales a nuestra disposici\'on, y
a nivel de distribuciones de partones {\it non-singlet}, es 
decir, cuando el gluon se ha desacoplado del resto
de distribuciones. En cada una de estas aplicaciones, la relaci\'on
entre los datos experimentales y la funci\'on a parametrizar
era diferente, demostrando que la t\'ecnica descrita en esta tesis
es completamente general, v\'alida para un gran n\'umero
de situaciones. 

Adem\'as, en la presente tesis hemos extendido
la estrategia original descrita en \cite{f2ns} mediante
la introducci\'on de nuevos algoritmos
para entrenar las redes neuronales artificiales, en
particular los conocidos como Algoritmos
Gen\'eticos. Estos algoritmos son necesarios
para entrenar redes neuronales mediante funciones
de error altamente no lineales, como nos sucede
en las diferentes aplicaciones que describiremos
en esta tesis. Finalmente, se han mejorado las
t\'ecnicas estad\'isticas utilizadas para la validaci\'on
de los resultados obtenidos, esto es, 
de la densidad de probabilidad contruida en los diferentes casos.
En el siguiente Cap\'itulo describimos con cierto
detalle los contenidos de la presente tesis
doctoral y los resultados que han sido obtenidos.

%% file: resumen.tex
\chapter{Resumen}

La presente tesis doctoral est\'a organizada de la manera que
se describe con cierto detalle a continuaci\'on.
En el Cap\'itulo 4 hacemos un resumen de los elementos
b\'asicos de la Cromodin\'amica Cu\'antica, que como hemos
explicado con anterioridad es el sector del Modelo
Estandard de f\'isica de part\'iculas que gobierna la llamada
interacci\'on fuerte entre part\'iculas elementales.
Asimismo, describimos tambi\'en aquellos procesos de f\'isica de
altas energ\'ias en los cuales utilizaremos la estrategia
general para parametrizar datos experimental que constituye el principal
objeto de esta tesis. Junto con este resumen, presentamos tambi\'en
una descripci\'on detallada de la t\'ecnica estandard, discutida
en el Cap\'itulo anterior,
que se usa com\'unmente para extraer las distribuciones de partones
con sus errores asociados a partir de un conjunto
de datos experimentales.

A continuaci\'on, en el Cap\'itulo \ref{general} describimos con todo lujo
de detalles la t\'ecnica general para
construir la densidad de probabilidad de una funci\'on
a partir de medidas experimentales, esto es,
una t\'ecnica para parametrizar datos experimentales, sin
necesidad de hacer hip\'otesis alguna sobre la forma
funcional de la funci\'on a  parametrizar y con una
estimaci\'on fidedigna de las incertidumbres asociadas, que
permite una propagaci\'on de los errores 
a observables arbitrarios sin necesidad de aproximaciones
lineales. Este Cap\'itulo est\'a divido en tres partes, 
m\'etodos Monte Carlo, redes neuronales artificiales y
estimadores estad\'isticos,
que
procedemos a describir a continuaci\'on.

 En la
primera parte de este Cap\'itulo 
se describen los m\'etodos Monte Carlo que usamos para
construir un {\it sampling} de los datos experimentales que
contenga toda la informaci\'on que nos proporcionan los experimentos,
incluyendo los errores y las correlaciones. Asimismo,
introducimos un conjunto de estimadores estad\'isticos que nos
permiten estimar quantitativamente como este {\it sampling}
Monte Carlo reproduce las caracter\'isticas de las
medidas experimentales. Esta t\'ecnica nos permite
estimar de una manera fidedigna los errores asociados
a la funci\'on a parametrizar, y demostraremos que 
es equivalente a los errores definidos a partir
de una funci\'on de error $\chi^2$ cuando las
aproximaci\'on lineales en la propagaci\'on de los errores
son suficientes.

\begin{figure}[ht]
\begin{center}
\epsfig{figure=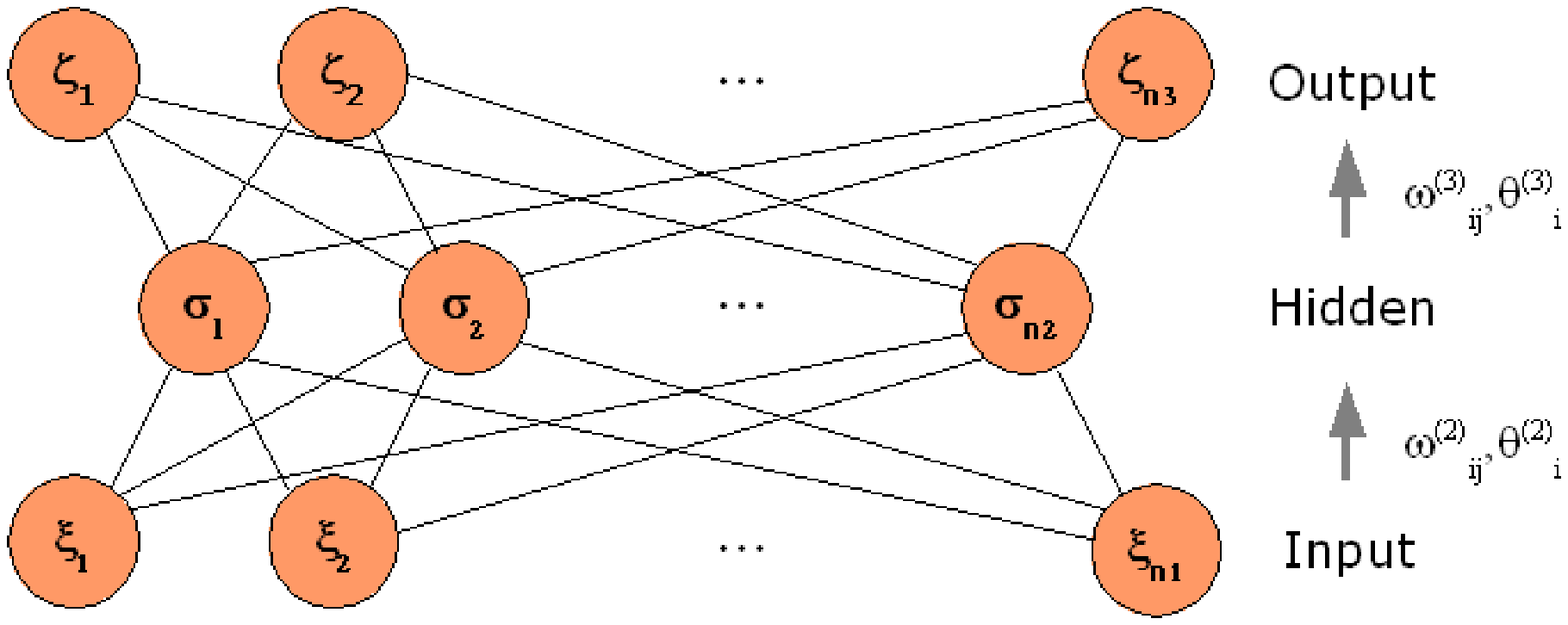,width=0.75\textwidth}
\caption{}{\small 
Diagrama esquem\'atico de una red neuronal
artificial multicapa del tipo {\it feed-forward}.
 \label{nndiagesp}}
\end{center}
\end{figure}

En la segunda parte del Cap\'itulo \ref{general} introducimos
las redes neuronales artificiales, que utilizaremos 
como interpoladores universales, as\'i como
los diferentes m\'etodos que usaremos para entrenar
estas redes neuronales, esto es, para que
aprendan los {\it patterns} presentes en los datos
experimentales. Las redes neuronales
artificiales son una herramienta habitual
en diversos campos cient\'ificos, desde
la biolog\'ia a la computaci\'on, y en
particular se usan con asiduidad en f\'isica
experimental de altas energ\'ias, en
aplicaciones como clasificaci\'on de 
eventos en funci\'on de sus propiedades. 
En la fig. \ref{nndiagesp} mostramos una red
neuronal artificial de la clase que utilizaremos
en esta tesis doctoral, conocida como
{\it multi-layer feed-forward perceptron}.
Asimismo describiremos
como nuestra t\'ecnica permite la incorporaci\'on
de informaci\'on te\'orica de maneras muy diversas, como
reglas de suma o condiciones implicadas por la cinem\'atica
del proceso. 

Las redes neuronales artificiales tienen la interesante propiedad
de que es es posible demostrar que cualquier funci\'on continua, 
independientemente de lo complicada que sea y del n\'umero
de par\'ametros que tenga, puede ser 
representada en t\'erminos de una {\it multi-layer feed-forward}
red neuronal artificial. Una segunda propiedad importante
de las redes neuronales artificiales es que estas son muy
eficientes en combinar de una manera \'optima la
informaci\'on experimental que proviene de diferentes medidas
de una misma cantidad. Esto es, cuando la separaci\'on entre
datos experimentales en el espacio de variables de entrada
es m\'as peque\~na que una determinada longitud
de correlaci\'on, entonces la red neuronal 
combina de manera eficiente esta informaci\'on experimental, de 
manera que el correspondiente {\it pattern} de salida
es mas preciso que las medidas experimentales individuales.

La utilidad de las redes neuronales artificiales es debida a la
existencia de diversos algoritmos de entrenamiento.
Este proceso se llama aprendizaje, pues no requiere
un conocimiento {\it a priori} de la forma funcional
que describe los datos experimentales.
En particular, en la presente tesis doctoral hemos
introducido los llamados algoritmos gen\'eticos
para el entrenamiento de redes neuronales artificiales.
Estos algoritmos gen\'eticos tienen un gran n\'umero 
de ventajas respecto a los m\'etodos de minimizaci\'on deterministas
que los hacen preferibles para problemas, como los
que nos ocupan en la presente tesis doctoral, altamente no
lineales y con un enorme espacio de par\'ametros.

Las ventajas de los algoritmos gen\'eticos, que como
su propio nombre indica est\'an inspirados en los
mecanismos que se observan en la naturaleza sobre
la evoluci\'on y la selecci\'on natural, se pueden
resumir en tres. Primero de todo, estos algoritmos trabajan
simult\'aneamente en poblaciones de soluciones, lo que
les permite explorar regiones diferentes del espacio 
de par\'ametros al mismo tiempo. Segundo, no necesitan
ninguna informaci\'on extra de la funci\'on a minimizar,
como el gradiente de esta. Finalmente, estos
algoritmos tienen una mezcla de elementos
estoc\'asticos aplicados bajo reglas
deterministas, que mejora su eficiencia en problemas
con muchos extremos locales, pero sin la
p\'erdida de efectividad que una b\'usqueda
meramente aleatoria implicar\'ia.

Finalmente, en la tercera parte del
Cap\'itulo, analizaremos aquellas herramientas
estad\'isticas que nos permiten validar
el resultado obtenido, esto es, determinar
de una manera quantitativa como la 
densidad de probabilidad que hemos construido
reproduce las carater\'isticas de los
datos experimentales, as\'i como su dependencia con respecto
diversos par\'ametros, como por ejemplo el n\'umero de
redes neuronales empleadas en la
parametrizaci\'on. Estas t\'ecnicas estad\'isticas
permiten tambi\'en detereminar a partir de criterios
s\'olidos caracter\'isticas de la densidad de probabilidad
como la longitud del entrenamiento de las redes neuronales
o el valor \'optimo de la funci\'on de error $\chi^2$.

El conjunto de redes neuronales artificiales entrenadas
en el conjunto de replicas Monte Carlo de los datos
experimentales para una funci\'on $F$ definen
la densidad de probabilidad en el espacio de funciones $F$
que est\'abamos buscando. Con esta
densidad de probabilidad, podemos obtener los
valores esperados para funcionales arbitrarios
de la funci\'on $F$, $\mathcal{F}\lc F \rc$, a
partir del conjunto de redes neuronales de la 
manera siguiente,
\be
\la \mathcal{F}[F]\ra\equiv \int \mathcal{D}F~
 \mathcal{P}\lc F\rc \mathcal{F}[F]
=\frac{1}{N_{\rep}}\sum_{k=1}^{N_{\rep}}\mathcal{F}[F^{(\net)(k)}]
 \ ,
\ee
como con las distribuciones de probabilidad habituales.
De esta manera podemos determinar la media de $F$, su variancia
y su correlaci\'on, utilizando las definiciones habituales
de estos estimadores estad\'isticos.

\begin{figure}[ht]
\begin{center}
\epsfig{width=0.60\textwidth,figure=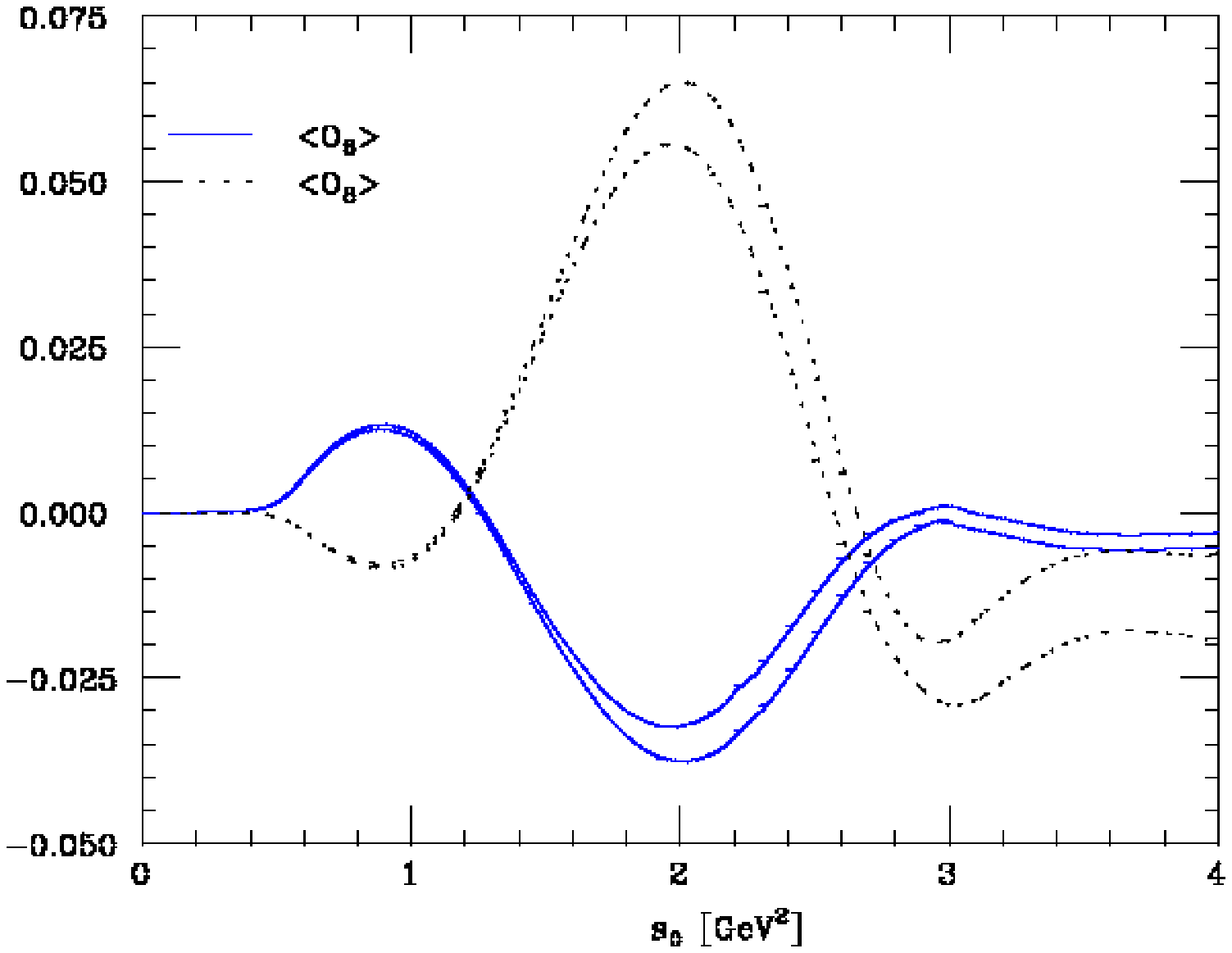}
\caption{}{\small Los resultados para
los condensados de vacio de la
Cromodin\'amica Cu\'antica obtenidos
en la presente tesis doctoral, como
funci\'on del par\'ametro de integraci\'on 
$s_0$. Las bandas de error corresponden a 
incertidumbres de 1-$\sigma$ que provienen de la
parametrizaci\'on de $\rho_{V-A}(s)$.}
\label{thesis_spec}
\end{center}
\end{figure}

El Cap\'itulo \ref{appl} describe con detalle cuatro
aplicaciones diferentes de la t\'ecnica introducida
en el Cap\'itulo \ref{general}. En primer lugar analizamos
la parametrizaci\'on de la funci\'on espectral $\rho_{V-A}(s)$,
esto es, la diferencia entre las funciones espectrales
correspondientes a los canales vectoriales y axiales,
en las desintegraciones hadr\'onicas
del lepton $\tau$,
que han sido medidas con gran precision en el experimento
LEP (Large Electron Positron collider, gran colisionador de
electrones y positrones), el antecesor de LHC en el CERN.
Como producto de este
an\'alisis determinamos los condensados
de vacio de QCD, $\la \mo_k\ra$, que son par\'ametros
no perturbativos que deben extraerse
de los datos experimentales. La determinaci\'on de estos
condensados a partir de los datos experimentales
nos proporciona informaci\'on sobre aspectos
fundamentales de la Cromodin\'amica Qu\'antica, como
el mecanismo de la ruptura de la simetria quiral,
y ha sido objeto de intenso estudio en los
\'ultimos a\~nos. En la fig.~\ref{thesis_spec}
representamos los valores obtenidos
para  los condensados
de vacio de QCD, $\la \mo_k\ra$, como se describe en la
secci\'on correspondiente de la tesis doctoral.

\begin{figure}[ht]
\begin{center}
\epsfig{width=0.60\textwidth,figure=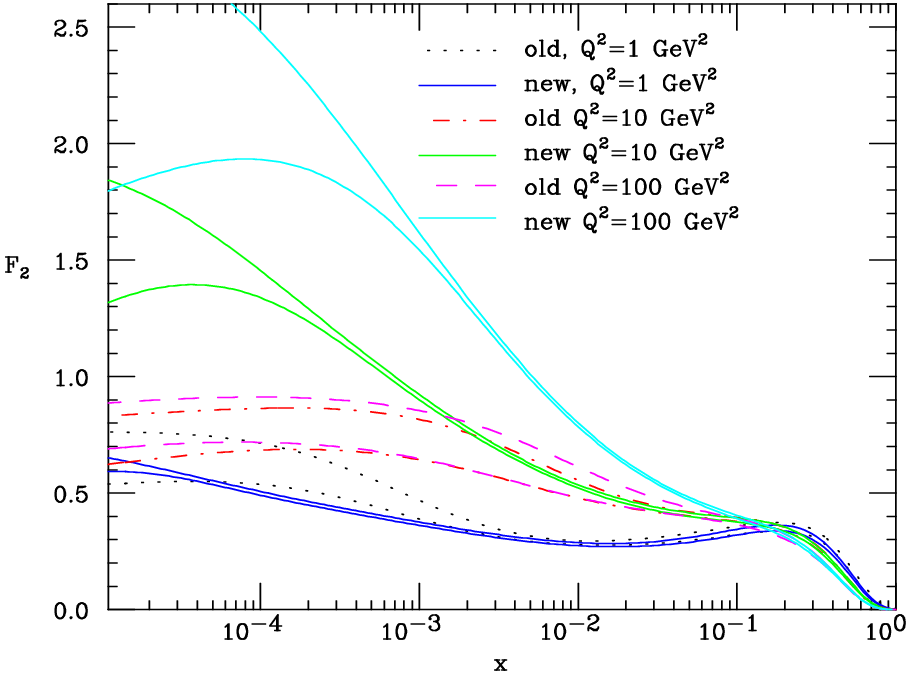}
\caption{}{\small Los resultados para la funci\'on de
estructura del prot\'on $F_2^p(x,Q^2)$ obtenidos
en la presente tesis doctoral comparados con los
resultados originales \cite{f2ns}.}
\label{f2compesp}
\end{center}
\end{figure}

En la segunda de las aplicaciones descritas en el Cap\'itulo
\ref{appl},
generalizamos los resultados
de \cite{f2ns}, referidos a la parametrizaci\'on 
de funciones de estructura $F(x,Q^2)$, construyendo
una parametrizaci\'on de estas funciones
en colisiones profundamente inel\'asticas del prot\'on
 que incluye todos los datos experimentales
de que se dispone, en particular las
medidas de alta precisi\'on del experimento HERA,
especialmente en la regi\'on de peque\~no $x$.
En la fig.~\ref{f2compesp} podemos
ver el resultado de este an\'alisis comparado
con los resultados originales obtenidos en \cite{f2ns}.
Notemos como en la regi\'on cinem\'atica donde los
datos experimentales incluidos en los dos
an\'alisis se solapan, que es la que
corresponde a valores grandes de la variable $x$,
los dos resultados son perfectamente consistentes,
como era de esperar.

\begin{figure}[ht]
\begin{center}
\epsfig{width=0.60\textwidth,figure=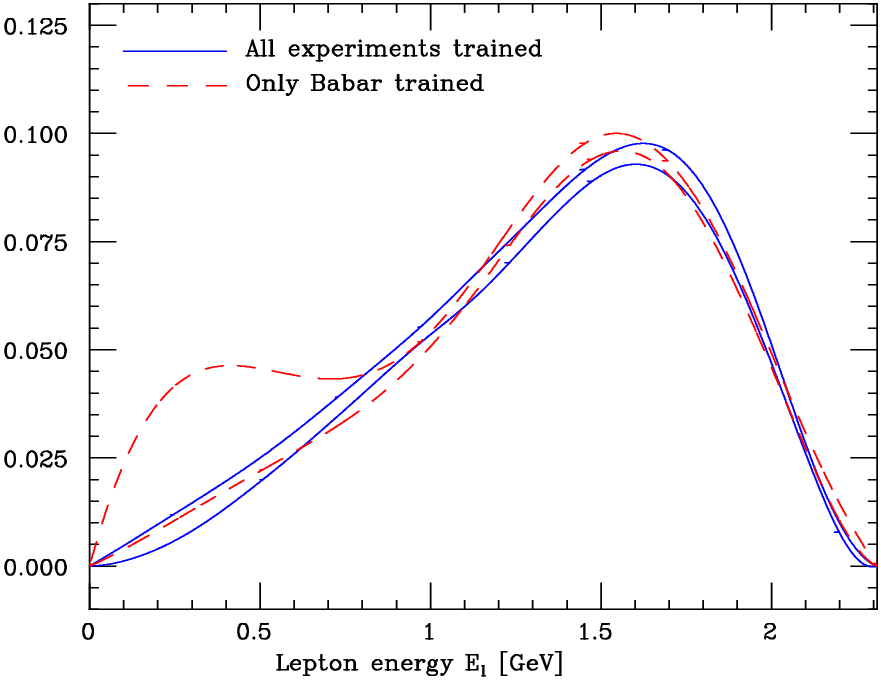}
\caption{}{\small Comparaci\'on de los resultados
para la parametrizaci\'on del espectron energ\'etico
lept\'onico con todos los experimentos incorporados
en el {\it fit}, con el caso en que solamente los datos
del experimento Babar han sido considerados.}
\label{bmeson_esp}
\end{center}
\end{figure}

En tercer lugar,
estudiamos las desintegraciones semilept\'onicas
del mes\'on B, comparamos nuestros resultados con
una serie de resultados te\'oricos y extraemos de la
parametrizacion del espectro energ\'etico del
lept\'on la masa del quark b, $m_b$. 
La determinaci\'on de par\'ametros no perturbativos,
como las masas de los quarks pesados o los
elementos de la matriz de CKM, son una de las principales motivaciones
para los an\'alisis te\'oricos y experimentales de las
desintegraciones de los mesones B.
En la Fig.~\ref{bmeson_esp} mostramos los
resultados obtenidos para la parametrizaci\'on del
especto lept\'onico, para diferentes combinaciones
de experimentos incluidos en el {\it fit}, como
se describe en detalle en la secci\'on \ref{bdecay_appl}.

Esta aplicaci\'on demuestra como nuestra t\'ecnica general
para parametrizar datos experimentales se puede
aplicar a la reconstrucci\'on de una funci\'on si
la \'unica informaci\'on experimental accesible procede
de integrales truncadas de esta funci\'on.
Esto permite una comparaci\'on m\'as general
de las predicciones te\'oricas con los datos
experimentales, sin necesidad de hip\'otesis
adicionales y con una estimaci\'on fidedigna de los
errores experimentales.
Asimismo, el desarrollo de estas t\'ecnicas permitir\'a
en un futuro aplicarlas a otras situaciones
de inter\'es en f\'isica de mesones B, como
por ejemplo la parametrizaci\'on de la
{\it shape function} $S(\omega)$, que
contiene los efectos no perturbativos dominantes
en una serie de procesos como las desintagraciones
radiativas de los mesones B.

\begin{figure}[ht]
\begin{center}
\epsfig{width=0.63\textwidth,figure=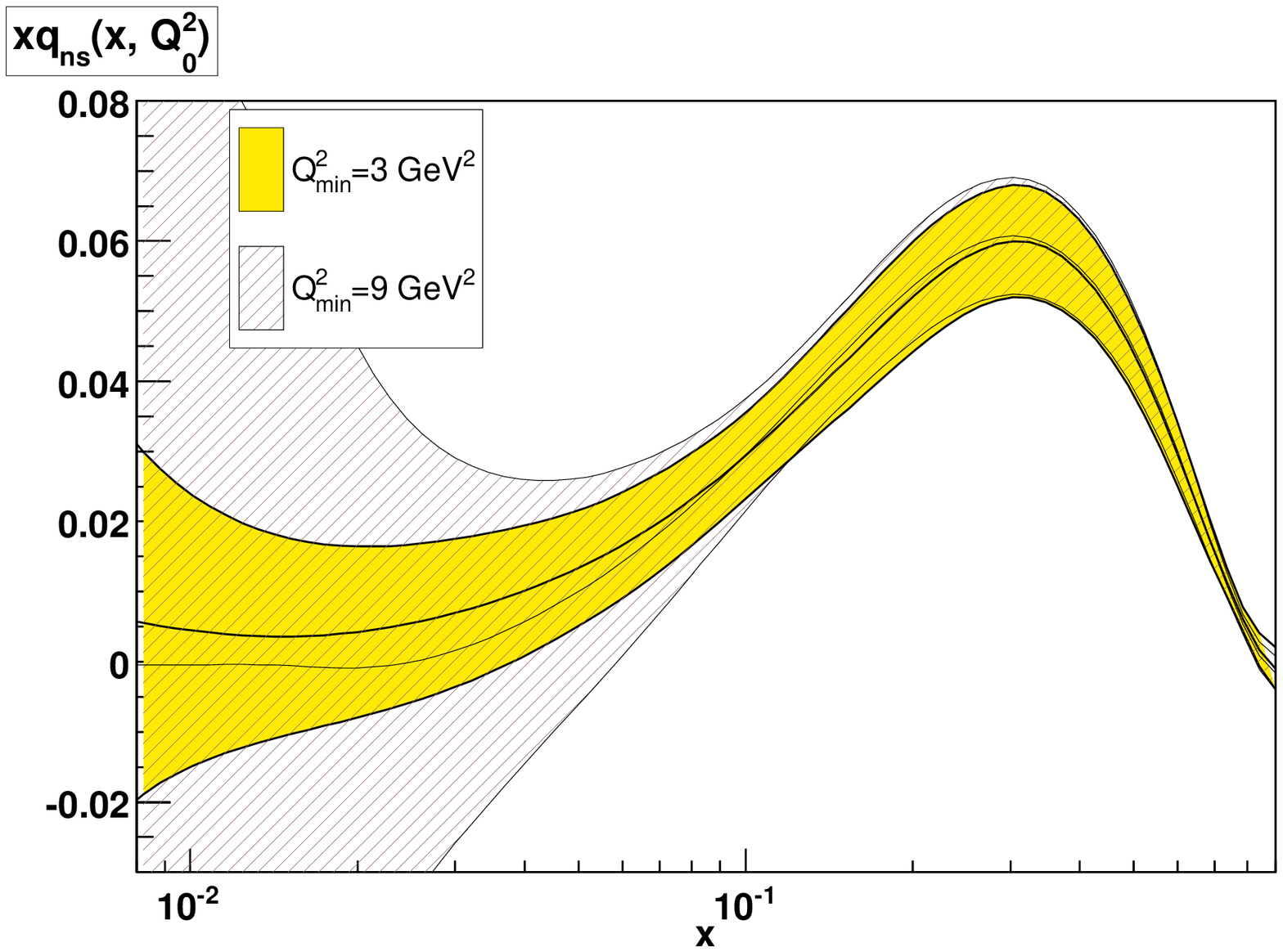}
 \caption{}{\small Comparaci\'on de los
resultados para la distribucion de partones $xq_{NS}(x,Q_0^2)$ 
para dos valores diferentes del
corte cinem\'atico: el de referencia $Q^2\ge 3$ GeV$^2$, 
con uno m\'as conservativo $Q^2\ge 9$ GeV$^2$. }
  \label{kincutsesp}
\end{center}
\end{figure}

La aplicaci\'on m\'as imporante de todas es descrita
en la \'ultima secci\'on del Cap\'itulo \ref{appl}: la parametrizaci\'on
de distribuciones de partones. Primero describimos
una nueva t\'ecnica para implementar la dependencia
con la energ\'ia $Q^2$ de estas distribuciones, y seguidamente
describimos la construcci\'on de la densidad de probabilidad
en el espacio de las distribuciones de partones {\it
non-singlet}, a partir de datos experimentales de funciones
de estructura. 
En la Fig.~\ref{kincutsesp} podemos observar los
resultados para la distribuci\'on de partones
{\it nonsinglet} obtenida en la presente tesis doctoral
con dos valores diferentes $Q_0^2$ del corte cin\'ematico.
Este corte cinem\'atico es
debido a que solo los datos experimentales
de la funci\'on de estructura $F_2^{NS}(x,Q^2)$ con
$Q^2$ suficientemente grande pueden tratarse
mediante la teor\'ia de perturbaciones de la Cromodin\'amica
Cu\'antica. 

\begin{figure}[ht]
\begin{center}
\epsfig{width=0.84\textwidth,figure=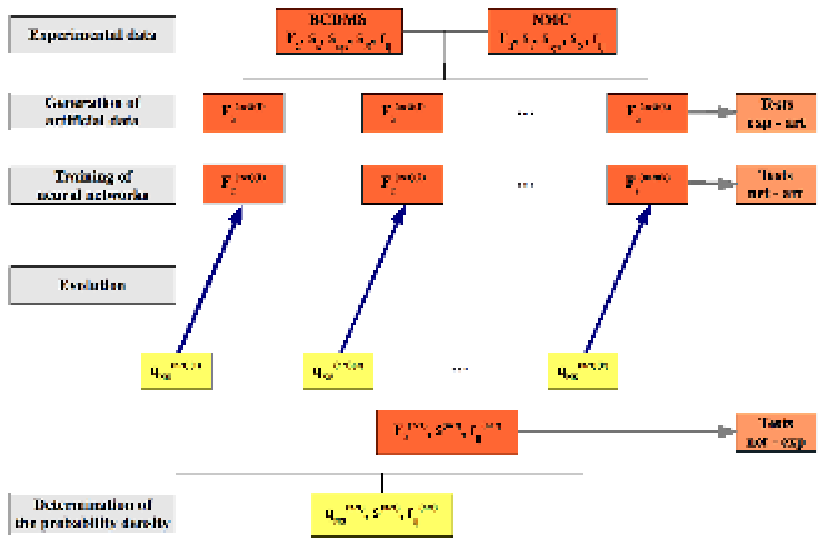}
\caption{}{\small 
Estrategia general seguida para la 
parametrizaci\'on de la  distribuci\'on de partones {\it non-singlet}
 $q_{NS}(x,Q^2_0)$ a partir de datos
experimentales de la funci\'on de estructura $F_2^{NS}(x,Q^2)$.}
\label{gen_stratt_exp}
\end{center}
\end{figure}

En la Fig.~\ref{gen_stratt_exp} tenemos
el esquema del proceso utilizado en la parametrizaci\'on
de la distribuci\'on de partones {\it non-singlet}, a partir
de datos experimentales de las funciones de estructura
en colisiones profundamente inel\'asticas.
Recordemos los tres pasos principales de nuestra t\'ecnica
descrita anteriormente, como pueden verse en la 
 Fig.~\ref{gen_stratt_exp}: generaci\'on Monte Carlo de 
replicas de las medidas experimentales, entrenamiento
de redes neuronales artificiales que parametrizan
la distribuci\'on de partones {\it non-singlet} y
finalmente la validaci\'on estad\'istica de los resultados.

Las t\'ecnicas descritas en la Secci\'on \ref{nnqns_appl}
proporcionan la base para la realizaci\'on de una
parametrizaci\'on de todas la distribuciones de partones,
incluyendo el gluon, que son necesarias para aplicaciones
fenomenol\'ogicas generales. En particular, es directo
extender el formalismo de evoluci\'on que
describiremos en el caso de las distribuciones {\it nonsinglet}
al caso general con combinaciones arbitrarias de 
distribuciones de partones. De la misma manera, los
datos experimentales que se usan en este caso
son los del an\'alisis de la Secci\'on \ref{dis_appl}, esto
es, la parametrizaci\'on de la funci\'on de estructura
del prot\'on $F_2^p(x,Q^2)$. Por lo tanto, con los resultados
obtenidos en la presente tesis doctoral se tienen todos
los ingredientes necesarios para obtener un {\it set}
de todas las distribuciones de partones con la t\'ecnica
descrita en el Cap\'itulo \ref{general}.

Finalmente, tras las conclusiones de la presente tesis
doctoral, dos ap\'endices incluyen material de refer\'encia
sobre an\'alisis estad\'istico de los datos
experimentales y sobre el estado
actual de las determinaciones de las distribuciones de
partones. En el primer ap\'endice, dedicado al tratamiento
estad\'istico de los datos experimentales, describimos
con un modelo sencillo que se puede resolver exactamente las
propiedades de los diferentes estimadores usados
para el entrenamiento de las redes neuronales, as\'i
como los efectos de un tratamiento incorrecto de los
errores de normalizaci\'on. En el segundo
ap\'endice resumimos el estado actual de
las determinaciones globales de 
distribuciones de partones, analizando con
detalle las caracter\'isticas de los
an\'alisis de las dos colaboraciones
m\'as importantes, CTEQ y MRST, junto
con los datos experimentales y las
parametrizaciones de las distribuciones de
partones utilizadas en cada caso.

%% file: conclusiones.tex
\chapter{Conclusiones}

En la presente tesis doctoral hemos descrito
en detalle una novedosa t\'ecnica general para
construir la densidad de probabilidad de una funci\'on
a partir de medidas experimentales, esto es,
una t\'ecnica para parametrizar datos experimentales, sin
necesidad de hacer hip\'otesis alguna sobre la forma
funcional de la funcio\'n a  parametrizar y con una
estimaci\'on fidedigna de las incertidumbres asociadas, que
permite una propagaci\'on de los errores 
a observables arbitrarios sin necesidad de 
asumir aproximaciones
lineales.
Esta t\'ecnica fue introducida en \cite{f2ns}
y durante el transcurso de esta tesis doctoral
ha sido mejorada en diferentes aspectos y extendida
mediante su aplicaci\'on a otros procesos de inter\'es. 
En particular hemos mostrado su aplicaci\'on al proceso
que motiv\'o originariamente el desarrollo de
esta  t\'ecnica, la parametrizaci\'on de distribuciones
de partones en el prot\'on.

La t\'ecnica general ha sido extendida en diversas direcciones.
Primero de todo, hemos demostrado la generalidad de esta
t\'ecnica mediante su aplicaci\'on a cuatro
procesos de inter\'es, cada uno de ellos 
diferente a la aplicaci\'on original descrita
en \cite{f2ns}. Por ejemplo, hemos tratado
problemas con un gran n\'umero de datos
provenientes de diferentes experimentos, as\'i
como problemas en los que la relaci\'on entre los
datos experimentales y la funci\'on que estamos
parametrizando viene dada por una serie de convoluciones.
En segundo lugar, nuevos algoritmos para el entrenamiento
de redes neuronales han sido introducidos, en particular
los conocidos como algoritmos gen\'eticos, que son
necesarios para la minimizaci\'on de funciones de error
altamente no lineales. Finalmente, hemos
ampliado el conjunto de t\'ecnicas estad\'isticas usadas en la
validaci\'on de los resultados, esto es, en la comprovaci\'on
quantitativa de como la densidad de probabilidad construida
reproduce las caracter\'isticas de los datos experimentales.

Las posibles aplicaciones de la estrategia general
para parametrizar datos experimentales descrita en la presente
tesis doctoral son ciertamente numerosas.
La m\'as imporante de estas es la generalizaci\'on de los
resultados decritos en la Secci\'on \ref{nnqns}
al sector {\it singlet} de las distribuciones de partones,
para obtener de esta manera un conjunto completo
de distribuciones de partones parametrizadas con redes neuronales con
estimaci\'on fidedigna de las incertidumbres asociadas.
Otra aplicaci\'on prometedora es la parametrizaci\'on
de la {\it shape function}, una funci\'on que
contiene los efectos no perturbativos dominantes en un cierto
tipo de desitegraciones del mes\'on B. Finalmente, otra
aplicaci\'on interesante consistir\'ia en la parametrizaci\'on del
flujo de neutrinos atmosf\'ericos a partir de datos experimentales
de detecciones de neutrinos en experimentos como
Super Kamiokande.
